%% file: 0204018.tex
\documentclass[a4paper,12pt,final,oneside]{./brunelthesis}

\usepackage[dvips]{graphicx}

\usepackage{amsmath}
\usepackage{amsfonts}
\usepackage{amssymb}
\usepackage{amsthm}
\usepackage{amscd}
\usepackage[mathscr]{eucal}

\usepackage{myHyperref}

\newcommand{\myPart}[1]{
  \part{#1}
  \pagestyle{myheadings}
  \markright{Eaves, Ph.D\hfill2000\hfill\ Page }
  }

\newcommand{\myRef}[1]{\S\ref{#1}}

\newcommand{\superscript}[1]{\raisebox{.6ex}{\scriptsize #1}}
\newcommand{\myDash}{\ ---\ }

\newcommand{\etc}{\textit{etc.}}
\newcommand{\eg}{\textit{e.g.}\,}
\newcommand{\ie}{\textit{i.e.}\,}
\newcommand{\vs}{\textit{vs.}\,}
\newcommand{\etal}{\textit{et al.}\,}

\newcommand{\viz}{\textit{viz.}\,}
\newcommand{\visavis}{\textit{vis \`a vis}\,}
\newcommand{\nb}{\textit{NB.}\,}
\newcommand{\versus}{\textit{vs.}\,}
\newcommand{\viceversa}{\textit{vice versa}\,}

\newcommand{\Aidan}{\emph{Aidan}\,}

\newcommand{\Java}{\emph{Java}\,}
\newcommand{\Jigsaw}{\emph{Jigsaw}\,}
\newcommand{\PostgreSQL}{\emph{PostgreSQL}\,}

\newcommand{\mySrc}[1]{\textsf{#1}}

\newlength{\facewd} \newlength{\faceht}

\newcommand{\markOver}[1]{
  \settowidth{\facewd}{#1}
  \settoheight{\faceht}{#1}
  \raisebox{\faceht}[0pt]{
    \makebox[0pt][l]{\hspace{.15\facewd}$\blacktriangledown$}}#1}

\newcommand{\markUnder}[1]{
  \settowidth{\facewd}{#1}
  \settoheight{\faceht}{#1}
  \raisebox{-\faceht}[0pt]{
    \makebox[0pt][l]{\hspace{.15\facewd}$\blacktriangledown$}}#1}

\newcommand{\myImport}[1]{\markOver{\mySrc{#1}}}

\newcommand{\myExport}[1]{\markUnder{\mySrc{#1}}}

\newcommand{\myAttr}[1]{\mySrc{#1}}

\newcommand{\myEnv}[1]{\texttt{#1}}

\theoremstyle{plain}
\newtheorem{theorem}{Theorem}[section]
\newtheorem{lemma}{Lemma}[section]
\newtheorem{corollary}{Corollary}[section]

\theoremstyle{definition}

\newtheorem{definition}{Definition}[section]
\newtheorem{notation}{Notation}[section]
\newtheorem{remark}{Remark}[section]
\newtheorem{mrule}{Rule}[section]

\newcommand{\droptext}[1]{\ensuremath{
    \text{\ #1\ }}
  }

\newcommand{\grantRight}[2]{\ensuremath{
    \operatornamewithlimits{grant}_{#1 \droptext{to} #2} 
    }
  }

\newcommand{\typeOf}{\ensuremath{
    \operatorname{type}
    }
  }

\newcommand{\cost}{\ensuremath{
    \operatorname{cost}
    }
  }

\newcommand{\initiator}{\ensuremath{\operatorname{initiator}}}

\newcommand{\acceptor}{\ensuremath{\operatorname{acceptor}}}

\newcommand{\eqdef}{\ensuremath{\triangleq}}

\newcommand{\myR}{ \ensuremath{ \mathrel{R} } }

\newcommand{\support}{ \ensuremath{ \operatorname{Supp} } }
\newcommand{\range}{ \ensuremath{ \operatorname{Ran} } }
\newcommand{\domain}{ \ensuremath{ \operatorname{Dom} } }

\newcommand{\graphpath}{ \ensuremath{ \operatorname{Path} } }

\newcommand{\myClassName}[1] {\ensuremath{
    \boldsymbol{#1}
    }
  }

\newcommand{\myModelName}[1] {\ensuremath{
    \mathscr{#1}
    }
  }

\newcommand{\myModel}{\ensuremath{
    \myModelName{U} =
    \langle
    {\mathcal W}, \dots , {\mathcal P }
    \rangle
    }
  }

\newcommand{\myModelForByAt}[2]{
  \ensuremath{ \mathrel {
    \overset{#1}{\underset{#2}{\models}}
    }
  }
}

\newcommand{\myModelForAt}[1]{ \myModelForByAt{\myModelName{U}}{#1} }

\newcommand{\myModelFor}{ \myModelForByAt{\myModelName{U}}{x} }
\newcommand{\myModelValid}{ \myModelForByAt{\myModelName{U}}{} }

\newcommand{\myClassValid}{ \myModelForByAt{}{\myClassName{C}} }

\newcommand{\necessity}{\ensuremath{\mathop{\square}}}
\newcommand{\possibility}{\ensuremath{\mathop{\lozenge}}}

\newcommand{\isModal}{\ensuremath{
    \mathop{\left[ \possibility \mid \necessity \right]}
    }
  }

\begin{document}

  \title{Trust--brokerage systems for the Internet}

  \author{Walter D Eaves}

  \degreeyear{2000}
  \copyrightyear{2001}
  \copyrightname{Walter Dieter Eaves}

  \degree{Doctor of Philosophy}

  \prevdegrees{B.Eng(Hons), M.Sc}

  \researchfield{}
  \chair{}
  \othermembers{}

  \campus{, Uxbridge}


  \date{\today}

  \maketitle

\begin{abstract}
  This thesis addresses the problem of providing trusted individuals with
  confidential information about other individuals, in particular, granting
  access to databases of personal records using the World--Wide Web. It
  proposes an access rights management system for distributed databases
  which aims to create and implement organisation structures based on the
  wishes of the owners and of demands of the users of the databases. The
  dissertation describes how current software components could be used to
  implement this system; it re--examines the theory of collective choice to
  develop mechanisms for generating hierarchies of authorities; it analyses
  organisational processes for stability and develops a means of measuring
  the similarity of their hierarchies.
\end{abstract}

\begin{frontmatter}

\copyrightpage
\tableofcontents
\listoftables \listoffigures

\begin{preface}
  This thesis attempts to use the principles of the design of political
  systems for information processing systems. Political systems are the
  means by which governments order the affairs of states and a political
  system in a liberal democracy uses elections as a feedback system which
  allows the subjects of the political system to direct it in the
  long--term.

  The political systems that govern commercial businesses are not as well
  evolved{\myDash}the only feedback they have formalised is from their
  shareholders. There should be inputs from customers, suppliers and
  government agencies. The information processing systems used by
  businesses reflect this lack of controlled feedback. Most information
  systems are designed to solve the problems of a particular management
  strategy for a business organisation.

  The goal of this thesis is to liberate the management of information from
  enterprises and return it to the people who are its subjects and
  \textit{should} be its owners. If ordinary people can control who has
  access to their personal information then, the hope is, it will not be as
  easily abused. Further than that, the behaviour of honest people should
  be the norm, but it is more often the case in modern society that honest
  and trustworthy people must prove themselves to be so, because most
  organisations have no means of distinguishing between the honest and the
  dishonest. Such information should be made available, so that genuinely
  honest people would find it easier to function in society than those who
  are not.

  My hope is that this thesis will stimulate research into and development
  of information systems that rank people and institutions according to
  different metrics: how solvent they are, what areas of expertise they
  have, and, generally, how trustworthy they are. And from this, be able to
  give suitably--qualified people more influence over different aspects of
  policy. It is, of course, unlikely that this would be the immediate
  result and it would seem more sensible to prove this technology with
  direct research towards managing resources where the ethical issues would
  not cause so great an obstruction. Computer and telephone network
  resources would be one such example of a good proving ground.
  
  I must express my thanks to some of the people who have helped me over
  the years: in particular, from the Electrical Engineering department: Dr
  Clarke, my research supervisor, Mrs Margaret Saunders and Mrs Valerie
  Hayes, our secretaries, who have helped me through the administrative
  maze. Dr Robert Zimmer of Information Systems and Computer Science must
  be thanked because he first introduced me to the more formal areas of
  computer science. Mr Callum Downie of the Faculty of Technology and Mr
  Theodorous Georgiou of Electrical Engineering for keeping the outstanding
  computing services of Brunel University ticking over.
  
  This research was supported by the Engineering and Physical Sciences
  Research Council, but only thanks to the efforts of Bob Thurlby and Bill
  O'Riordan of \textsc{ICL} and Russell--Wynn Jones of the Chorleywood
  medical practice.

  I would like to dedicate this thesis to the memory of Dr.\ Alan MacDonald
  and Frau Marie Kohl \textit{n\'ee} Schneider.
\end{preface}

\end{frontmatter}

\myPart{Trust Broking}

\chapter{Overview} 
\label{cha:ovw}

The original brief for this project was to develop a software system that
would allow a community of individuals to access each other's information
with the owner's consent and yet to do so in such a way that:
\begin{itemize}
\item That consent could be delegated but never forced
\item and there would be no prejudice against an individual for choosing
  not to release information
\end{itemize}

The immediate goal was to realize a means whereby medical practitioners
based in surgeries and hospitals could access the medical records they held
on each other's patients. Such facilities are already available in Germany
\cite{blobel01:clinic} and Iceland \cite{anderson98:decode}, but neither
system provides assurances that consent will be obtained, nor that records
will remain confidential.

It became apparent that many of the problems of controlling the release of
this personal information would need to be resolved by a joint decision of
the owner of the database holding the information, the person who wants to
use the information and the person, or persons, who are referred to by it.
The decision would resolve whether the person wanting to use the
information was entitled to access it and, if so, how much of it.

This is a fundamental business process and most people would recognise it
as one they take part in all the time. Even more fundamental than the
process is the commodity that is traded when the decision to grant access
is made. The sole criterion that both the database owner and the subjects
of the records held within it must feel is satisfied is simply: ``Can this
new user be trusted not to misuse the information contained in our
records?'' and \emph{trust} is the commodity that is exchanged by all three
parties. Trust is a belief that someone else will keep their promises:
\begin{itemize}
\item The subjects of the records trust the owner of the database and the
  user of their records.
\item Because:
  \begin{enumerate}
  \item The owner of the database promises the subjects to release
    information on them to mutually agreed users.
  \item The owner of the database promises the subjects to release only the
    information they have agreed to release.
  \item The user of the records promises not to misuse the information
    contained in the records.
  \end{enumerate}
\end{itemize}

The goal then is to develop systems which will provide individuals with the
degree of trust they require from each other: it is because of that, that
the systems and mechanisms proposed by this research are described as,
jointly, providing a trust--brokerage system. 

Broadly, this research is in the area of computer--supported co--operative
working and it is an active field, but has concerned itself with
environments where there is enough implicit trust between all the parties
involved that confidentiality safeguards can be largely omitted in system
designs.  Hubermann has developed a system known as \textit{Beehive}
\cite{Hubermann:beehive} which has been employed as the basis for
computer--aided engineering systems. There are also proposals for the joint
management of investment portfolios \cite{cscw:portfolios} and there have
been conferences discussing a number of digital library projects
\cite{it:diglib:medoc} which provide on--line texts from a number of
sources and access is only slightly restricted. All of these systems are to
some extent predicated on the existence of virtual organisation\myDash %
an organisation that has been created and designed to fulfil a function
within one real organisation or, more usefully, across a number of them. An
interesting paper that describes how a virtual organisation can be created
to fulfil a need is given in \cite{Miller:evolving}.

Within the medical profession such systems are not as mechanised, but are
developing a trust model within national organisations
\cite{anderson96:secclinic}, \cite{denley99:privacy2nd} and standards
for security mechanisms specific to healthcare are being developed within
Europe and elsewhere\footnote{\cite{Anderson:security}, ``Standards''}.

\paragraph{This Dissertation and Its Structure}

This dissertation concentrates on providing co--operative working
environments built upon databases of information. The databases must
preserve confidentiality, so it divides naturally into two parts:
\begin{itemize}
\item Secure Distributed Processing
\item Virtual Organisation for Resource Management
\end{itemize}

\subparagraph{Secure Distributed Processing}

This part re--examines the basic theory of data and database security and
applies it to a distributed processing environment. A practical
architecture for the secure management of access to any number of databases
is developed and a prototype implementation is discussed.

In the context of medical information systems, this part of the documents
describes mechanisms that must be implemented---such as those proposed by
the European working group \cite{cen:tc51}.

\subparagraph{Virtual Organisation for Resource Management}

Originally this section concerned itself with generating a virtual
organisation for access control to databases, but it is now more general:
it addresses the issue of how to manage access to any resource and the
problem of how a virtual organisation for the management of authorisation
hierarchies can be evolved from the needs of owners and users. This
introduces three research issues:

\begin{enumerate}
\item Forming, analysing and quantifying hierarchies
\item Resolving conflicts within hierarchies
\item Ensuring the stability of self--organising hierarchies
\end{enumerate}

For medical information systems, this follows the argument of Anderson
\cite{Anderson:security} that the medical profession operates under a
collegiate structure: policies for allocating access rights are applied
within autonomous organisations, but must be enforced within co--operating
peer organisations.

\paragraph{This Chapter}

The remainder of this chapter provides a further introduction to the
research by providing some examples of how it might be used.


\section{Application: Universally Accessible Personal Information}
\label{sec:personal}

In this section, the way in which the affairs of people are managed will be
described.

The difficulty most people face is that mechanised information processing
systems rely upon them to provide them with input{\myDash}basically people
have to fill in forms.  To add further annoyance, people have to collect
the information that the information processing systems generate about them
to be able to fill in more forms{\myDash}apply for a bank account, be given
a bank account number and a sort code, then arrange a money transfer by
quoting your bank account and bank sort code.

It would be much simpler, and less error--prone, if people had a repository
of their personal information that was already in machine--readable format,
but which could be projected into a human--readable form that people could
reorganise. A simple drag and drop environment would be very attractive.
For example, log on to a system, this generates an identity object which
appears as an icon, go to a folder which represents your bank, make a new
cheque, find the identity icon of the person you want to give the cheque
to, drop his identity and one's own identity icon into the cheque icon to
sign and address the cheque and drop the cheque icon into an e-mail and
send it.  The e--mail application will read the cheque's identity icons,
lookup the e--mail addresses associated with them and address the e--mail
correctly.

Such a system would greatly enhance the operation of information systems.
There are already prototype banking systems that operate in a manner
similar to that described above and it is hoped that more of them will be
developed. The following discussion looks at the difficulties of
mechanising the access and use of personal information. The discussion
begins with how people manage their confidential information in the
paper--based world we occupy now and then moves on to how they might
organise their information in an electronic environment.

\subsection{People}

\subsubsection{Paper Lives}

Most people will have a collection of papers in their possession that, 
more or less, defines the person they have been and how they stand now:
\begin{description}
\item[Certificates] a birth certificate, possibly a marriage
  certificate and, ultimately, a death certificate will complete the
  collection.
\item[Qualifications] there will be school--leaving certificates,
  examination results, degree certificates, driving licences.
\item[Earnings] Employers will have provided the \textit{Inland
    Revenue's} \textit{P60}.
\item[Status] Notifications of tax codes, a passport, valid visas,
  employment permits.
\item[Finances] Bank account and credit card statements. Direct debits,
  standing orders.
\item[Bills] Receipts showing bills have been paid, statements from
  suppliers, which, for most people, will be the utility bills they
  have settled.
\item[Memberships] Libraries, clubs, professional organisations.
\item[Properties and Contracts] One may own a property, rent one or
  hold a mortgage, similarly for cars. There may be contracts with
  managing agents, rental companies and service company
  warranties.
\item[Histories] There will be one's medical history; perhaps details
  of legal cases in which one may have been involved.
\item[Addresses and Contacts] Just about everyone has a list of addresses
  and phone numbers listing all the companies and organisations who supply
  all of the information held on them. There will also be contact addresses
  for oneself and for friends and business associates.
\end{description}

If one were to bring all this information together it would amount to a
complex inter--related bundle and would suffer from all the typical
problems of data collections:

\begin{description}
\item[Replication] If one were asked to prove one's identity: one
  could use one's passport, driving licence, cheque card. To prove
  one's address: bank statements, driving licence or utility bills.
\item[Specialist Knowledge] Tax codes can only be deciphered by
  someone familiar with tax regulations; leasehold agreements need
  contract lawyers. X--rays, radiographers.
\item[Inconsistent] Driving licences can hold details that are
  out--of--date. Membership cards can have mis-spellings.
\item[Unsubstantiable] One piece of information can be useless without 
  another to substantiate it. For example, if one possesses a national 
  insurance card, it is quite possible to pose as the person whose
  card it is; there is no substantiation of the holder's identity.
\item[Location] A piece of paper is easily lost, or one does not have
  it when one needs it.
\end{description}

\subsubsection{Electronic Lives in a Honest World}

If we were to take all of the information used to lead one's life and make
it available electronically, then software applications could be developed,
like the chequing account mentioned above, which would require no
paper--based input. The information would be easier to manage and to
access.

In an entirely honest world all of our personal information could be placed
on a web--site and we, and others, could freely access it.

\paragraph{A Person's Web--site}

\begin{figure}[htbp]
  \begin{center}
  \includegraphics{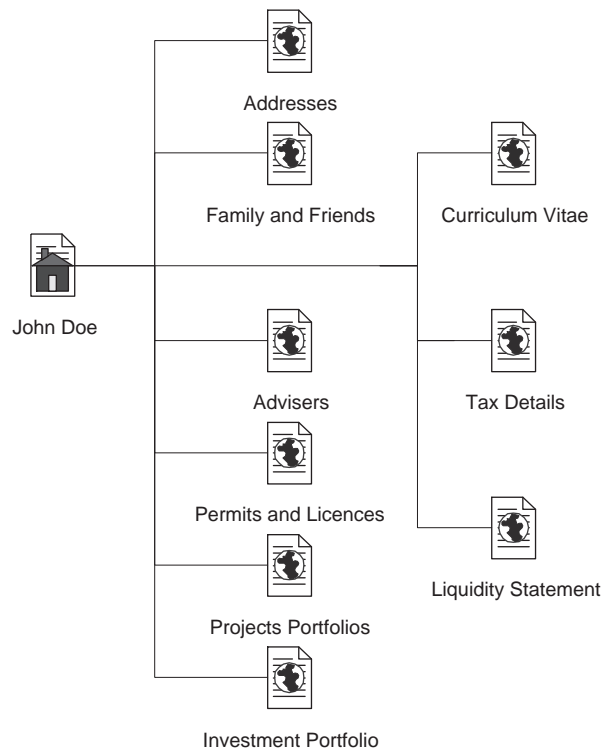}
  \caption{Web--site used to provide personal information}
  \label{fig:web-site}
  \end{center}
\end{figure}

The World--Wide Web could be used to host a web--site which would act
as an organised repository of all these documents in electronic
form. They could then be organised to fulfil their different
purposes. A diagram of how such a web--site might be organised is
given in figure \ref{fig:web-site}. It is owned and administered by an 
individual who has had a full professional working life: working in a
number of countries. The web--site could contain the following pages.

\begin{enumerate}

\item Addresses 

  The current contact addresses of the web--site owner would be
  held. These would reference his residential property address and his 
  employer's address.
  
\item Family and Friends

  The owner might provide references to his wife's and children's
  web--pages; friends and former business colleagues. He might include his
  employment and academic referees here.

\item Advisors

  The owner might provide references to his doctor, solicitor,
  accountant and others.

\item Permits and Licences
  
  The owner might store references to his driving licences, employment and
  residency permits and proof of certain tax exemptions.

\item Projects Portfolios

  The owner may have decided to record his work to help in finding
  suitable employment, so he would keep web--pages detailing his work, 
  his employers and so forth.

\item Investment Portfolio

  The web--site owner may have made some investments and might want to 
  have a financial adviser manage them for him, so he might construct
  a web--page that contains references to the current value of his
  investments, where they are held and the account details.
  
\item Curriculum Vitae

  This composite document would reference to other pages, or part of
  their contents:

  \begin{itemize}
  \item Addresses
  \item Projects Portfolios
  \item Permit and Licences
  \item Family and Friends
  \end{itemize}
  
  and would also reference educational qualifications, memberships of
  professional organisations.

\item Liquidity Statement
  
  This shows the owner to be solvent and would be used to obtain
  credit or accounts. It would reference bank statements, accounts
  held elsewhere (credit cards and utility accounts) and property
  ownerships. It would of course refer to:

  \begin{itemize}
  \item Investment Portfolio
  \end{itemize}

\item Tax Details

  This would be used by the web--site owner's accountant to pay his
  taxes. It would reference statements of his earnings,
  tax--deductible outgoings and would contain a reference to the
  address of his current tax office and possibly his past offices as
  well.

\end{enumerate}

\paragraph{Providing Up--to--Date Information}

A desirable enhancement would be for the web--site to provide current
information: bank account statements could be updated with each
transaction, as could the investment portfolio and any other records
that change frequently.

\begin{itemize}
\item Either: provide a link to the provider of the information with
  the owner's reference number to provide an index look--up in a
  directory service at the information providers web--site.
\item Or: have a dynamic web--page that would create itself on demand
  and carry out the look--ups and the formatting of records itself.
\item Or: have the information provider issue a new page by e--mail
  and have that page replace or be appended to the existing one.
\end{itemize}

The first of these two require the same operation: being able to make a
remote query on a database. As an example, consider looking up the owner's
driving licence in the ``Permits and Licences'' web--page. The page could
simply contain a reference to the issuing authority of the driving licence,
with the index look--up\footnote{The URL used here is fictitious, but most
  insurance brokers are able to acquire this information and some make it
  available in web--based quotation systems.}:

\begin{verse}
  \myURL{http://www.open.gov.uk/dvla/drivers.htm/driver_number?EAVES60762WD9AK} \\
\end{verse}

\begin{figure}[htbp]
  \begin{center}
  \includegraphics{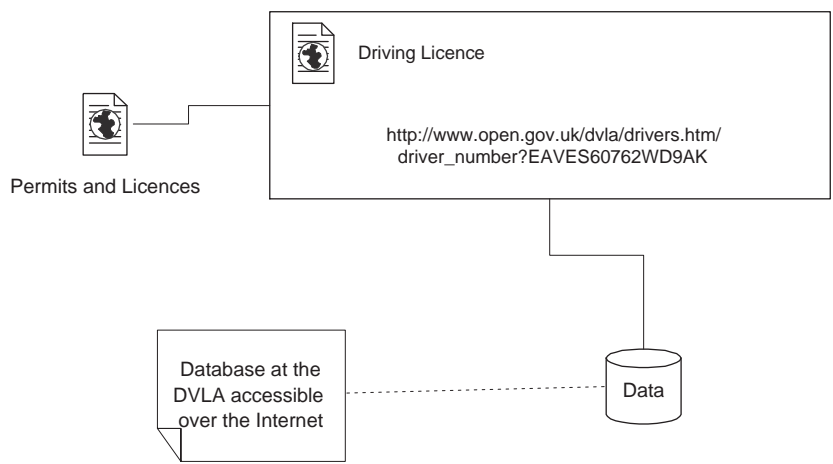}
  \caption{Using the UK's Driver Vehicle Licensing Agency}
  \label{fig:web-site-access-1}
  \end{center}
\end{figure}

A diagram for the interaction is given in figure
\ref{fig:web-site-access-1}.

\paragraph{Using the Information}

With this scheme an individual's personal information could be kept
up--to--date and available over a universally accessible medium. If someone
wanted to join a library they need not fill in any forms, but could just
present the address of their web--page and have that recorded. The library
could then see if the web--page owner provides enough information for its
needs and could take whatever information it needed from the individual
whenever it needed it.

The library would then notify the web--page owner of his new account
at the library and send the address of the remote database that could
be queried for his account details with them.

\paragraph{Security Problems}

Of course, no--one should have a web--site like this because there is no
protection of one's privacy. For example, someone visiting this site could
learn the owner's credit card details and commit fraud with them.

\subsubsection{Electronic Lives in the Real World}
\label{sec:e-lives}

\paragraph{Access Control Mechanisms}

What is is needed to make the web--site secure are a number of access
control mechanisms

\begin{itemize}
\item Placed at the entrance to each web--page
\item Placed at every remote database that can be queried
\end{itemize}

The former restricts access to the page, the latter is a restriction
put in place by the owners of the remote database as to who may access 
the data held on the owner. The owner would probably be entitled to
see the record held on himself, but that need not be the case. If the
owner is able to access his record, he may decide to allow other
people to see it as well.

\begin{figure}[htbp]
  \begin{center}
  \includegraphics{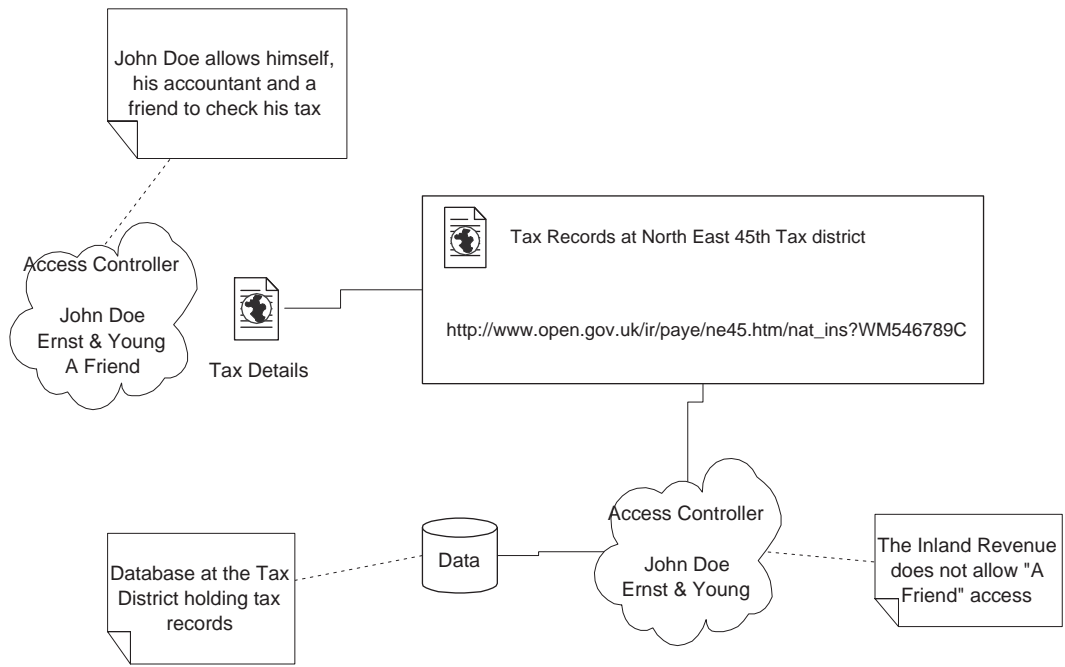}
  \caption{Access control mechanisms in place for Tax Details}
  \label{fig:web-site-access-2}
  \end{center}
\end{figure}

For example, only the owner of the web--page for ``Tax Details'' and
his accountant would be allowed to access it, but he may also decide
to allow his accountant access to the tax details held at his
different tax offices, see figure \ref{fig:web-site-access-2}.

\paragraph{Granting Access Rights}

There would also need to be a mechanism in place to grant access
rights to the web--page owner's record held in remote database. These
issues have to be resolved:

\begin{itemize}
\item Does the web--page owner have the right to grant access to his
  record to someone else?
\item Does the owner of the database where the record is held want to
  allow access to the person the web--page owner proposes?
\end{itemize}

The latter issue might appear contentious, but the data held at the remote
database may have an intellectual copyright attached to it{\myDash}there
may be design documents in the ``Project Portfolios''{\myDash}or it may
give rise to a conflict of interest{\myDash}it may contain a company's
information that should not be released to a stockbroker who might be
involved in a rival bid for the company.

The web--page owner and the owner of the database must come to some kind of
agreement and in so doing they would want to be as well--informed as
possible about the individual to whom they are proposing to grant the
access rights to. If that individual were known to the web--page owner then
he should appear in his ``Advisors'' page or perhaps in his ``Family and
Friends'' pages and if he is there, then his web--pages could be accessed
and it should be possible to gain enough information about him to make a
decision. There would be a system of implicit and explicit permissions.
Company directors would implicitly grant rights to each other to view
information that is common to them in the course of their business, but
some rights may need to be granted explicitly: the right to sign cheques
would be explicitly granted to the finance director, for example.

The mechanism for granting access rights could be anything: e--mail or
a secure web--form. All three parties should be notified and record
the rights granted and this information should also be made available
through a database.

\subsection{Organisations}

\paragraph{Responsibilities}

So far only the information needs of individuals have been discussed.
The organisations that own the databases that hold individuals'
records would also want to use the data. Going back to the example of
the library the web--page owner had joined, it will have access rights
to some of the ``Addresses'' information so that it can send out
statements to the web--page owner, but, if it has access to the address
information for all of its borrowers, it can construct a mailing list
and sell it to a direct--mail company and more junk--mail is almost
certainly something that the library's borrowers would not want.

On the other hand, the library may decide to do something useful with
the address data, it may use it for planning where to build a new
library. It is a matter of intent and a requirement that can be made
is that whenever data is required from an individual, a statement of
intent should be made with the request.

A statement of intent is no protection against abuse of the data held in
the individual's web--site unless one can show that the data has been
properly used and the only way to do this is for the library to show that
it has fulfilled its intent. This would involve proving to owners of the
information that the data has been used correctly.

This is simple enough: if the library accesses an individual's address
record, it should leave a token with him saying that it will be used to
send a statement, when the statement arrives, it will contain a reference
to the token. The token and its reference can then be reconciled. In
effect, every piece of information can be tracked to its source: one can
think of this process as like recording the progress of a note of currency
in the economy by tracking its serial number. There are already mechanisms
for this form of transaction processing in most modern information
processing systems: in particular, web servers have a system of
transferring ``cookies'', unique session identities that can be 
attached to the transactions an individual undertakes.

\paragraph{Possibilities}

The possibilities for co--operative information sharing between responsible
organisations and individuals are enormous. The health-care industry in
particular could benefit by having near instantaneous access to medical
records. Insurers would be able to evaluate risks better; law enforcers
could isolate groups of people who may demonstrate a propensity to become
offenders and take preventive measures; they could also determine which
groups of people are most likely to have crimes committed against them. Of
course such systems would need to protect civil liberties: if someone were
determined as being a potential offender, it must be possible for him to
appeal against that classification.


\myPart{Database Security on the World--Wide Web}

\chapter{Requirements and Analysis}
\label{cha:reqs}

The original brief for this project was to develop a software system that
would allow medical practitioner's databases to be accessed securely and
safely. In this first part of the dissertation, a functional design is
developed based on existing technology which could be used for a system
that could provide personal information in the manner described in
\myRef{sec:personal}.

There are already systems in use that have similarities to that proposed:
\textit{OncoNet} \cite{blobel01:clinic} provides oncological information to
a high degree of data security because it is operated and used by one
well--managed organisation.

When access is more open and management more collegiate, then such systems
are more akin to digital libraries
\cite{it:diglib:medoc,proj:medoc,proj:diglib}. These are well--funded and
have reached a high degree of sophistication, but, it will be seen, they
have a simple internal organisation, which effectively allows only one
level of access.

There are other research projects for the health-care industry; these, too,
are better developed{\myDash}the \textit{LIOM} project
\cite{healthcare:liom}, for example. Although that project aims to reflect
the more complicated internal organisation of the health-care industry, it
is not intended to be self--administering, which this system aims to be.
The \textit{LIOM} project, like others in this field, uses a meta--data
model which has to be maintained \cite{healthcare:metadata:roantree}. This
is feasible on a small--scale (100 users or so), but the coordination
effort needed would probably be excessive for larger systems (1000 users).

In the security model proposed by Anderson \cite{anderson96:secclinic}
there are recommendations for technology, ITSEC standards for operating
systems and databases. There are also recommendations for access control
schemes, variants of Role--based access control, \cite{Sandhu98}, appear to
be most adaptable to the collegiate federation of organisations proposed by
Anderson. There is an implicit argument for a Public Key Infrastructure,
PKI, which could support cross--certification, such as that analysed in
\cite{Maurer96c}.

\paragraph{Secure Distributed Computing}

The system design is also complicated by its attention to secure
distributed computing
\cite{sec:distributed:schiller,sec:distributed:birman}, which is a
relatively mature field. Secure computing requires that processes have
proven implementations and execute on a safely constructed computer system
with the least privilege needed to complete successfully: the principal
requirement is that it should not be possible for other non--privileged
processes to access any of the information produced or consumed by the
secured processes.

In distributed secure computing, this problem is doubly difficult. One
safely constructed computer system, system $A$, may hold confidential data,
$d$, and another safely constructed computer system, system $B$, may hold a
proven implementation of the process, $p$, to be used with the data. $A$
may be able to pass $d$ to $B$ securely, but $B$ cannot be trusted not to
compromise it. $B$ can however send its process implementation, $p$ to $A$
where it could execute and process the data, but $A$ must be sure that $p$
has no means available to it which would allow it to communicate $d$ by a
covert channel as well as be sure that $p$ does not compromise the
integrity of $A$. This interaction is fairly simple: more problems ensue if
the data from different sources has to be merged.

\paragraph{Open Distributed Computing}

The analysis and design process is within the framework of an Open
Distributed Processing, ODP, system, \cite{ODP:intro}, which is to address
the information processing problem from these five perspectives:

\begin{itemize}
\item Enterprise: what has to be achieved
\item Information: what information is needed to do it
\item Computational: how can that information be obtained or deduced
\item Engineering: what quality of service can be achieved in providing it 
\item Technological: what technology exists that could be used to achieve
  it
\end{itemize}

ODP merely recommends that the system design be addressed from all these
perspectives and that each one \textit{could} be modelled, if need be
{\myDash} see for example this discussion of working within the ODP
framework \cite{odp:requirements:caneschi}.

The technological model is usually a given because it is the dominant
technology at the time of design. The remaining four models can be traded
off against one another. This chapter will be just a first pass over design
issues and does nothing more than describe how such a system might work.
Usually one begins with a sketch of the technology and enterprise models,
then one sketches the information and computational models from the other
two models to allow one to produce an engineering model, which is a set of
interacting agents and more or less defines the operation of the system.
The engineering model is then the basis for another iteration of design,
where the enterprise, information and computational issues for each agent
are addressed.

\section{Technology Model: The World--Wide Web}

This section describes the technology available at the time of writing. It
will be seen, as the system is analysed, that the technology is fully
capable of achieving what is required of it. The real design problem is to
establish policies for authorising access and showing that they have been
followed.  Distinct aspects of information security should be clarified
because they are addressed by different technologies.

\begin{itemize}
\item Secure access: the information is protected against indiscriminate
  access.
\item Safe access: the information is protected against indiscriminate use
  by those who are allowed to access it.
\end{itemize}

Incidentally, the reason the World--Wide Web \cite{www:www} has been chosen
as the communications medium ought to be stated.

\begin{itemize}
\item Wide access: the information can be accessed from as widely available
  a medium as is available.
\end{itemize}

\paragraph{Wide Access{\myDash}Web Technology}

As far as end--users are concerned the World--Wide Web has only:

\begin{itemize}
\item Web--browsers
\item Web--servers
\end{itemize}

It is worth mentioning that the Internet protocols underlying the World--Wide
Web can also offer secure electronic mail delivery \cite{smime:rsa}.

Probably the most useful piece of web technology for system designers is
\Java \cite{web:java}. This is an object code interpreted language that
runs on a virtual machine; it can be constrained to only use specific
operating system resources (files, sockets and so forth). This programming
language allows system designers to load software from anywhere on the
World--Wide Web and run it on a designated host in a safe environment. This
is absolutely ideal for agent--based software.

\paragraph{Encryption and Authentication Products{\myDash}Secure Access}

The capabilities of this web technology are widely--known and some
specifications can be found in \cite{netscape:servers}. Most web--browsers
can establish secure connections with suitably enabled web--servers. (The
latest version of the secured socket protocol is called \textit{Transport
  Level Security} and is discussed in \cite{tls:eaves}.)

This hinges upon public--key cryptography and secure repositories for
public--key certificates. The standard governing this is \textit{X.509},
\cite{CCITTConsult88b}, and the certificates are consequently known as
\textit{X.509} certificates. They are available widely, at a charge, from
certification authorities such as \textit{Thawte}, \cite{ca:thawte}. There
is some discussion of their limitations in \cite{Rosche95}.

There are some well--evolved software security products producing
public--key infrastructures\cite{web:rsa}.

\paragraph{Operating Systems and Database Products{\myDash}Safe Access}

Operating systems are now relatively safe. ITSEC \cite{sec:itsec} grades
them and, currently, there are a number of products that have reached the
acceptable security levels recommended by Anderson
\cite{Anderson:security}: E3 and above, \cite{sec:itsec:grades}.

Database systems are also graded by ITSEC and there are a number of
suitable products for the system proposed
\cite{Illustria} and it is possible to integrate these
with web--servers \cite{netscape:servers}.  Most databases support
\textit{SQL}, the Structured Query Language \cite{North:1996:UOS}, which
provides a set of access control mechanisms which, it will be seen later,
are adequate for the system proposed.

\section{Enterprise Model: Contracts}
\label{sec:sys:proposals}

The aim is to propose a suite of protocols that will allow access to
databases to be strictly controlled and thereby allow more and
qualitatively better information to be distributed \emph{and} to simplify,
standardise and partly mechanise the procedure whereby individuals are
granted access. This section will describe the relationships between the
parties as a set of contracts in the style of an enterprise modelling
language as described in \cite{ODP:intro} and, at slightly greater length, in 
\cite{scf:eaves}.

A key argument in Anderson's model for the security of clinical information
systems is that individual information systems are assumed to be
well--managed and align to the structure of the organisation they serve.
The access control mechanisms of these information systems may use either a
Bell--Lapadula \cite{BL:military} or Clarke--Wilson \cite{CW:commercial}
mechanism for determining rights.  They form part of a collegiate system
which has some federal infrastructure which manages access control lists
for the component systems. Anderson's argues that the access control system
for the access control lists can only be Clarke--Wilson in form.

This section attempts to develop r\^oles that could be used within the
federal system, so that they might be used in a r\^ole--based access
control system with constraints such as RBAC2 described in \cite{Sandhu98}.

In this section, these r\^oles will be specified using the principles of
deontics \cite{deontic:meyer}, which aims to reduce difficult contractual
relationships to sets of rules concerning rights and duties.

\paragraph{Adjudication and Loss}

Contracts describe expected behaviour{\myDash}usually, the formalisation of
an existing behaviour. Each party to a contract believes that the expected
behaviour will be forthcoming because it would be too costly to do
otherwise. Either because it is practically too expensive not to behave as
required, (it might change existing procedures), or, that penalties will be
incurred by the party who breaches the contract. The latter requires that
an adjudication service be available to determine if one party has not
complied to the terms of the contract and that that party be punished and
the other recompensed for the loss suffered. The operation of an
adjudication service is quite sophisticated, but is discussed, in outline,
in chapter \ref{cha:norms}; recompense for loss suffered is achieved by
surety or insurance. The insurance industry already has some policies for
disclosure of information, but it would be desirable if they were able to
give cover at the time a contract is formed and it should be a precondition
that cover be arranged before granting access. The insurance and surety
process is capable of being mechanised \cite{econ:inet:licensing}; this
paper also discusses how licences could also be issued for people offering
services through web--servers.

\subsection{Parties}
\label{sec:parties}

There are four types of party to the contracts. These are specified with
respect to the rights and duties they must possess and should fulfil. These
are the entities that must follow the principles given in Anderson's model
\cite{Anderson:security} and would appear as r\^oles within the federal
superstructure of the collegiate organisation.

\pagebreak[2]

\begin{itemize}

\item Subjects
  
  \nopagebreak[3]

  Subjects are the people (or organisations) on whom information is kept by
  the owners of databases.

\item Custodians
  
  These individuals are appointed by the subjects; the appointment is
  usually \textit{de facto}, a person's doctor is obliged to act as their
  medical representative and is therefore the custodian of their medical
  information. It should be possible to make the appointment of a custodian
  explicit and a subject should know who there custodian is. It may also be
  possible for a subject to be his own custodian.  The function of a
  custodian is to make access control decisions for the subjects.  It will
  almost certainly be necessary that custodians do this, because:
  \begin{enumerate}
  \item Subjects will not usually have the specialist knowledge needed to
    assess access requests.
  \item Custodians can act on behalf of groups of subjects which have
    similar interests.
  \end{enumerate}
  
  Subjects will usually delegate decision--making to one (or more)
  custodians. If they choose to delegate to a group of custodians, then the
  subject can choose from a number of decision processes{\myDash}\eg veto,
  unanimity, majority vote{\myDash}how the decision will be taken.
  
  Custodians are responsible for the safety of information. They are not
  responsible for the security of data storage and transfer of the
  information. That is the job of the owner of the database, or databases,
  upon which the information is stored and the facilities by which it is
  communicated. In short, custodians specify the policies for information
  use, storage and transfer; owners execute these policies. Very often the
  custodian and the owner will be the same person acting in two r{\^o}les.
  Within a medical practice, a doctor will make decisions regarding
  information safety when he decides what to include in a letter of
  referral and will make decisions regarding data security when he chooses
  to send the letter of referral by electronic mail.
  
  Custodians should have some legal responsibility to the subjects. Most
  custodians will be subject to legislation, such as, in the United
  Kingdom, the Data Protection Act \cite{law:dpa}. It may be useful to
  think of legislation, and other policies a custodian should adhere to, as
  having a custodian, which could be made an active part of the system.
    
\item Owners
  
  These people (or organisations) own the databases and the communications
  infrastructures that hold and distribute the information held on the
  subjects. They follow policies from custodians

  The owners will add their own information to that the subjects have
  provided. The owners will want to protect this information in the same
  way that subjects will want to protect theirs. In this respect they can
  thought of as a subject who is its own custodian for all the records in
  the database.

  \nopagebreak[2]

\item Accessors
  
  These are the people (or organisations) who access the databases held by
  the owners. Accessors will be assigned a security clearance class and
  each of these will have a membership panel of trusted peers who should be
  known to appropriately qualified custodians.
  
  It may prove expedient for accessors to copy those parts of databases
  that interest them, add their own interpretations to the data and
  republish the data and, in so doing, they become custodians.
  
\end{itemize}

\paragraph{Closed Relationships}

A simple set of relationship rules might help clarify the entities' r\^oles
with respect to one another. The aim here is to ensure that the
relationships are closed, so that the system can be self--governing.

There are four sets of rules. 

\begin{enumerate}
\item Database management and composition

\item Subject--Record--Custodian

\item Multiple r{\^o}les

\item Accessor permissions
\end{enumerate}

They are expressed as class relationships of the \textit{HAS--A} and
\textit{IS--A} kind. \textit{HAS--A} relationships can be by aggregation or
by reference. Aggregation means that one entity is a composition of the
others. The reference relationship means that one entity knows about the
existence of the others and there is some association between them, usually
ownership or delegation or use.

\textit{IS--A} can be of two kinds: by generalisation and by template. The
latter is best called the \textit{IS--KIND--OF} relationship and means that
the two classes have the same meta--class, but have distinct identities and
behave autonomously.  The \textit{IS--A} is usually implemented by
inheritance and allows one class to be used in the same way as the other
sharing a more abstract identity and cannot always act autonomously.

\subparagraph{Database management and composition}

This first pair of rules state that an owner manages a database, which is
composed of a set of records. The relationships are one to many. An owner
may manage more than one database.

(Database should be a more general concept because the rules describe control
relationships. The more general concept is one of a \textit{Resource}. This
would include networking resources. An example of which might be a port
number on a host computer. Only databases have been described here because
they refer back to subjects directly.)

\begin{displaymath}
  \begin{aligned}[t]
    \begin{CD}
      Owner @>{manages}>> Database \\
      Database @>{is-composed-of}>> Set of records
    \end{CD}
  \end{aligned}
  \droptext{and}
  \begin{aligned}[t]
    \begin{CD}
      Record @>{describes}>> Subject \\
      Subject @>{has}>> Custodian
    \end{CD}
  \end{aligned}
\end{displaymath}

\subparagraph{Subject--Record--Custodian}

The second pair state that each record maps to a subject and that each
subject has a custodian. This is the most important rule: it associates data
with information.

All the relationships can be one to many. In particular, a subject may have
more than one custodian. There might be a custodian responsible for policy
regarding data encryption of medical records and another responsible for
the policy regarding the disclosure of confidential information.

One to many also implies that if you have obtain a record, it will have a
subject and {\viceversa}.

\subparagraph{Multiple r{\^o}les}

The following \textit{IS--A} relationships closes the system, so that all
the entities introduced must have at least one custodian. There are three
cases to cover:

\begin{itemize}
  
\item The database is a record which must describe a subject{\myDash}this
  would be itself and the record would be the database's meta--data. Each
  subject must have a custodian.

\item An owner is a subject and must therefore have a custodian.

\item A custodian is also a subject and so must have a custodian.

\end{itemize}

These rules encompass the relationships as they so often arise in practice.
That an owner also acts as a custodian, but in different r{\^o}les{\myDash}the
example of a doctor sending a letter of referral by e--mail. It also allows
a custodian to be his own custodian. This is useful for expressing supreme
legal relationship: governments are only answerable to themselves.

\begin{displaymath}
  \begin{aligned}[t]
    \begin{CD}
      Database @>{is}>> Record \\
      Owner @>{is}>> Subject \\
      Custodian @>{is}>> Subject \\
    \end{CD}
  \end{aligned}
  \droptext{and}
  \begin{aligned}[t]
    \begin{CD}
      Accessor @>{has}>> Permission \\
      Permission @>{has}>> Record \\
      Set of custodians @>{creates}>> Set of permissions \\
    \end{CD}
  \end{aligned}
\end{displaymath}

\subparagraph{accessor permissions}

The last three \textit{HAS--A} relationship state how accessors are
involved. They stand outside the system, because there are no practical
means of enforcing any behaviour upon them. They access records via
permissions. Permissions are created by custodians. 

Each permission has a record so the custodians act as a linking entity
between records (and their corresponding subject) and the set of
permissions.

\paragraph{Summary}

The important points are:

\begin{itemize}

\item That a database of records has custodians for each of the records
  contained in it \emph{and} for the database as a whole.

\item Subjects, owners and custodians themselves all have one or more
  custodians. 

\end{itemize}

The latter point makes the system closed and self--governing.

This set of relationships is very similar to the architecture of
per-formative agents proposed by \cite{ODP:intro} and also described in
\cite{scf:eaves}.

\subsection{Grades of Anonymity}

A well--known problem with personal information is that anonymity is no
real protection if it is possible to obtain an identity and a profile from
one database and use the profile to isolate some other confidential
information from another database \cite{infr:schlorer}. Anonymity can prove
to be an obstacle to legitimate use of data. A grading of degrees of
anonymity might prove useful in specifying access contracts. This grading,
see table \ref{tab:identity}, is illustrated with reference to the medical
profession, but can be used elsewhere.

\begin{table}[htbp]
  \begin{center}
    \begin{tabular}[left]{|l|l|}
      \hline
      Class & Name to Identity Relationship
      \\ \cline{2-2} 
            & Examples \\
      \hline\hline
      Synonymous & Person is named
      \\ \cline{2-2} 
      & Personal Physician \\
      \hline
      Pseudonymous & Person goes under an assumed name 
      \\ \cline{2-2} 
      & Secondary Physicians \\
      \hline
      Anonymous & Person is unnamed.                     
      \\ \cline{2-2}
      & Researchers and Administrators \\
      \hline
      Eponymous & Group to which person belongs is named.
      \\ \cline{2-2} 
      & Researchers and Administrators \\
      \hline
    \end{tabular}
    \caption{Name to Identity Relationships for Medical Information}
    \label{tab:identity}
  \end{center}
\end{table}

Pseudonymous identities are already widely--used in medical research; it
allows a particular patient to be referred to consistently and a thread of
discussion can be developed around that identity.

Eponymous identities are subtly different from anonymous ones, because an
individual is tagged as belonging to a particular group. Usually, genuinely
anonymous subject data is partitioned over and over until all the subjects
have eponymous identities. \emph{If} it is possible to write back to the
original record that a particular subject has been designated as belonging
to a particular group, \emph{then} the subject has an eponymous identity.
Under some methods of inference control, in particular \textit{random
  sample queries} \cite{Denning:1980:SSD}, it is not possible to attach any
deductions, and therefore eponymous identities, to particular sets of
records.

Some accessors may make local copies of datasets from databases and would
add their own classifications, if these are re--published then the
identities would be eponymous{\myDash} because it is possible to use the
original database to establish a profile and, by inference using that
profile, obtain the classification made in the copied and augmented
database.

\subsection{Views of Records}

When a record of a subject is released, it should not be the entire record,
but rather a restricted view of the record that contains enough information
to allow the accessor to do their own processing. This is a principle more
or less enshrined in most information system security texts: ``The
principle of least information''\footnote{See for example,
  \cite{Denning:1976:LMS}.}.

Having a restricted view of a record does allow identities to be restricted
to eponymous, but anonymity{\myDash} or at least the anonymity granted by
using random sample queries{\myDash} requires an additional mechanism.

\subsection{Duties of a Custodian to a Subject}

Custodians are usually practicing professionals in a particular field.  The
relationship between a custodian and their subjects is usually governed by
an accepted code of practice from a professional body, which is the
collective identity of the custodians. These professional bodies usually
grant licences to their members to practice their profession and there is
usually an adjudication procedure to determine if a member has acted
improperly. This is the only contract between custodian and subject that is
needed. The duties of the custodian are to:

\begin{enumerate}
\item Grant minimal views to accessors.
\item Inform subjects of:
  \begin{enumerate}
  \item Classes of accessors granted access to their records.
  \item View of record given to those accessors
  \item Evaluation of risk and insurance of surety gained on their behalf
  \end{enumerate}
\end{enumerate}

\subsection{Duties of the Owner to the Custodian} 
\label{sec:custodian}

Owners are instructed by custodians as to whether or not the records for
the custodian's subjects can be released. The duties of the owner are to:

\begin{enumerate}
\item Release no record without prior approval from a Custodian

  This is actually required by the Data Protection Act \cite{law:dpa}.

\item Provide Stated Degree of Anonymity

  If a custodian allows a subject's record to be published then the degree
  of anonymity must be upheld.
  
\item Minimal View of the Record

  The custodian will state the view of the record that can be granted to a
  class of accessor and will expect no more than that to be released.

\item Minimal Set of Accessors
  
  A custodian will grant access to a particular security class of
  accessors.
  
  It is usually the case that a \textit{lattice model}
  \cite{Denning:1976:LMS} of secure information flow is in
  force\footnote{Lattice models are described in section
    \myRef{sec:lattices}. For now they can be thought of as hierarchical
    organisation structures.}. This would allow accessors who have a
  ``higher'' security clearance to be given \emph{de facto} access as well,
  without having to negotiate with the custodians. The lattice model is a
  generalisation of the Bell--LaPadula access control hierarchy
  \cite{BL:military}, which could be described as ``read--down, write--up''
  or an accessor at a particular security clearance can read everything
  graded below his own grade, but what he writes can only be read by those
  above his grade.
  
  Whether \emph{de facto} access is granted should be open to negotiation
  between custodians and accessors as well.

\end{enumerate}
  
\subsection{Duties of the Accessors to the Custodian}

The main problem is that accessors may need the right to re--publish the
data they have acquired from a database owner. An accessor could be an
organisation and might need to re--publish the data internally or the
accessor may decide to re--publish to a wider audience.

\begin{enumerate}
\item External Re--publication

  The accessors must ensure that other accessors be vetted and approved by
  the custodians in the same way that they were vetted and approved for
  access.
  
\item Internal Re--publication

  It may be possible for the accessor to show that his organisation's own
  data security procedures are good enough to provide the custodian with
  enough assurance to forgo vetting procedures for every internal accessor.

\end{enumerate}

\subsection{Duties of the Custodian to the Accessors}

If an accessor is \emph{denied} access then they have a right to know how
they can put themselves in a position whereby they may be granted access.

\begin{itemize}
\item Provide justification for denial of access

  If an accessor fails to meet particular security clearance requirements
  they should be told which so that they can change their clearance.
  
\end{itemize}

\section{Enterprise Model: Adaptability Issues}
\label{sec:sys:adaption}

There is enough tested and approved technology available to implement a
system that could implement these relationships and provide the means for
each entity to fulfil the duties imposed upon it, the challenge is to
design it in such that a way that authorisation policies can formulated in
a semi--automated way and that the system could be almost wholly
self--governing. The use--case scenario for accessors would be something
like this:

\begin{enumerate}
\item Reference to a database appears at a secure web--server
\item An accessor requests a particular set of records using secure e--mail
  from the relevant custodians
\item Custodians make their decisions and return them by secure e--mail
\item Based on the replies: a set of views for the records is generated and
  the owner of the database is instructed to publish it for the accessor's
  eyes only
\item Accessor is notified of the views to use
\item Accessor uses secured connections to submit queries to the database
  on these generated views
\end{enumerate}

There are a number of problems with this. Firstly, the custodians will make
their decisions based on the current state of the records: a particular
epoch of their existence. The accessor should either be restricted to that
epoch or negotiate access for all subsequent epochs. Only a few databases
directly support epochs{\myDash}\PostgreSQL \cite{sft:prod:postgres}
does{\myDash}without that different tables would need to be created for
each epoch and separately maintained.

Secondly, the collation of the replies would need to employ some least
lower denominator for the records, because some custodians may not grant
access to particular fields within the record.

Thirdly, the process would generate a lot of request traffic, which
custodians would be hard--pressed to keep track of and, therefore, to be
consistent when applying their own criteria. It would be useful to employ
precedents, \eg a custodian has granted similar access rights to an
accessor who has been classed at a lower level than the current requesting
accessor, is it permissible to allow this accessor the same rights? This
requires a lattice of information flow, which, it was specified above,
would be the subject of negotiation between custodians and accessors. Using
precedents requires that accessors and records be classified. The process
of generating precedents can be accelerated if custodians are also
classified, so that clearance gained from a custodian who is ranked higher
than another means that there is no need to secure acceptance from the
lower--ranked.

Fundamentally, custodians would either have to specify rules for access
which could be applied by a mechanical agent on their behalf, or a
mechanical agent would deduce rules from their actions and, after checking
them with custodians, add them to a rule base to be used later. Similar
systems to this have been deployed \cite{db:pol:bertino,db:pol:castano}.

\section{R\^ole--based Access Control Systems: Information Model}
\label{sec:informational}

The information model proposed is that used for r\^ole--based access
control systems. These are described in Sandhu \cite{Sandhu98}, in which he
argues that r\^ole--based access control systems have such a sophisticated
information model that they can be constructed to support all the other
important forms of access control system.

Sandhu gives a simple information model, but this has been modernised,
using Booch's notation \cite{des:booch}, and clarified. (A more suitable
notation would \textit{object role modelling}, \cite{halpin95:schema_db},
which is more easily formalised for implementation, but Booch's notation
has been used for consistency.)

\begin{figure}[htbp]
  \begin{center}
  \includegraphics[keepaspectratio=1,width=6in]{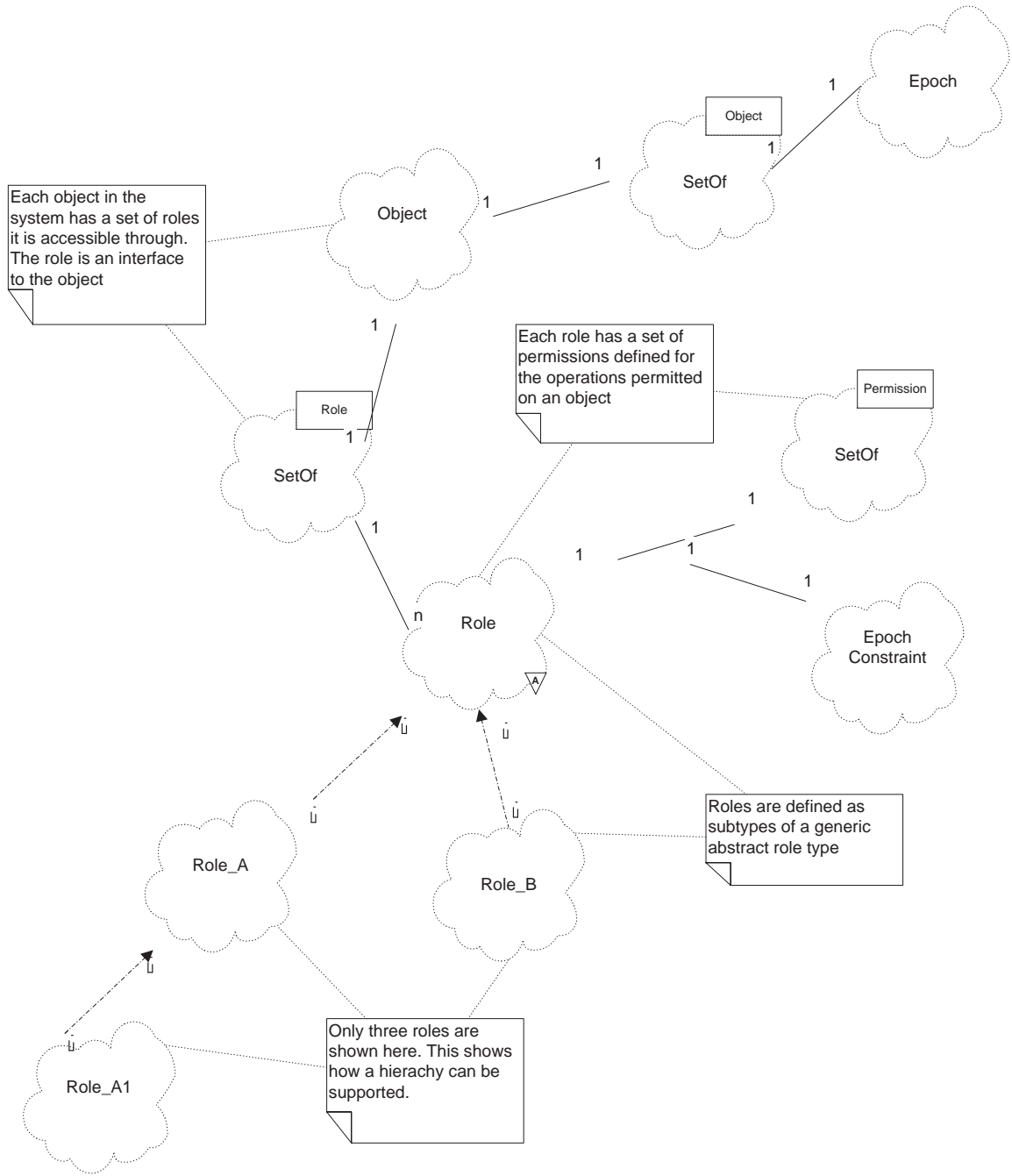}
  \caption{Role--based access control: object information model}
  \label{fig:object-im}
  \end{center}
\end{figure}

\paragraph{Objects} The principal innovation of Sandhu's r\^ole--based
design can be seen in figure \ref{fig:object-im}. Each object in the system
has its own set of r\^oles associated with it. Each r\^ole acts as an
interface to the object.  Each r\^ole has a set of permissions associated
with it. Both relationships{\myDash}object--r\^ole and
r\^ole--permissions{\myDash}have constraints objects associated with them.

\subparagraph{Epochs and Constraints} These permissions can be qualified by
applying a set of constraints, the \textit{Epoch} constraints. 

Although Sandhu specifies that these constraints exist, he does not make it
especially clear what they constitute. This denotation follows the practice
in modern database design that access control rules and data descriptions
can be revised so that the previous generation is still available as a
different \textit{epoch}\cite{Illustria}. It will be seen that a
self--organising access control system will need to remember its previous
state.

\subparagraph{Hierachies} The lattice structure of many access control
systems is effected by allowing r\^oles to have a hierachy. This is
illustrated using the class tree in figure \ref{fig:object-im}. There is an
abstract \textit{role} and this is sub--classed twice for \textit{role A}
and \textit{role B} and \textit{role A} is sub--classed once for
\textit{role A1}. Modern database systems such as \PostgreSQL
\cite{Illustria} have support for sub--typed data classes.

\begin{figure}[htbp]
  \begin{center}
  \includegraphics{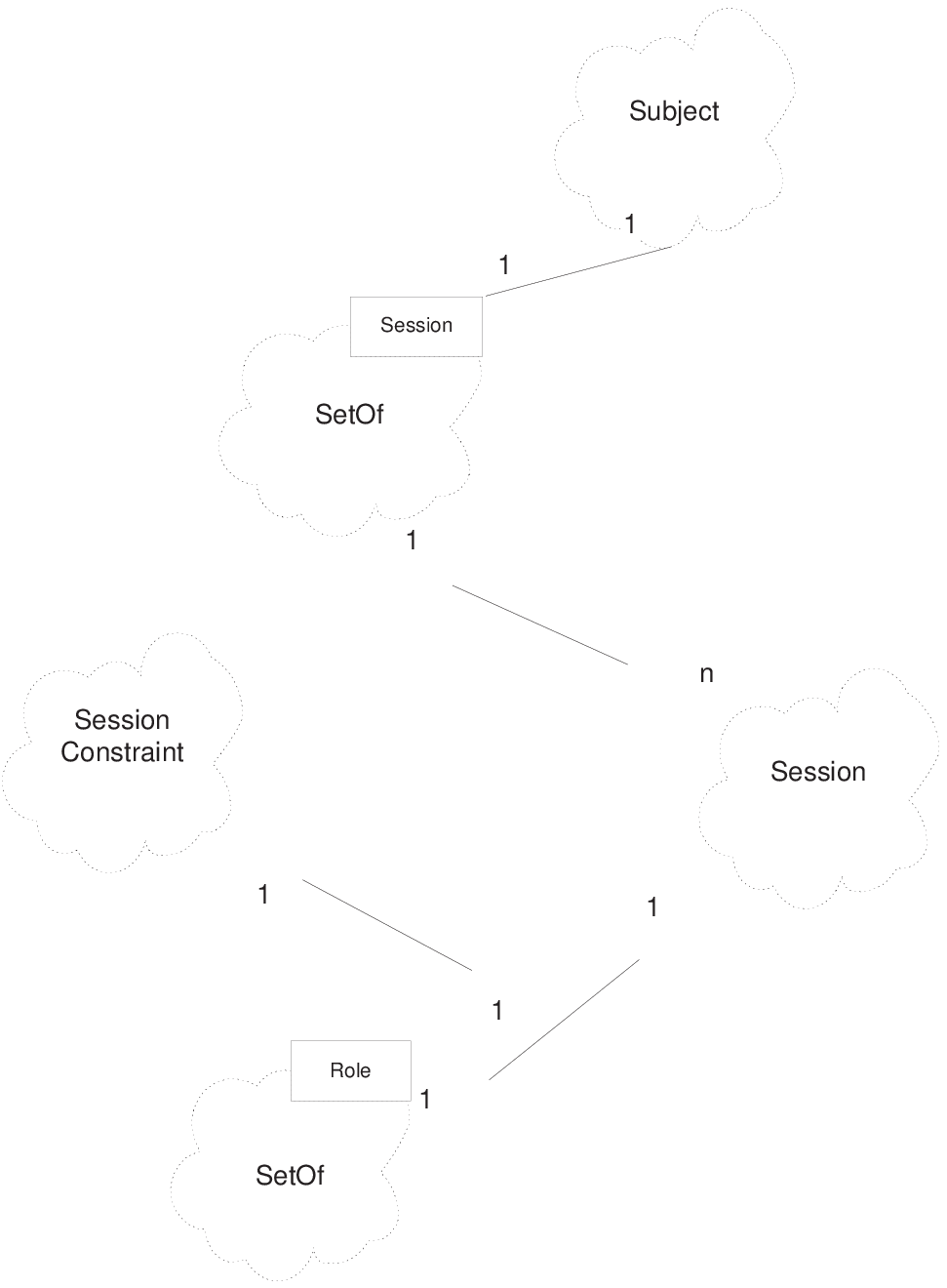}
  \caption{Role--based access control: subject information model}
  \label{fig:subject-im}
  \end{center}
\end{figure}

\paragraph{Subjects} Figure \ref{fig:subject-im} shows how a subject
obtains the set of r\^oles by which he may access the objects. A subject
first obtains a session. A session is an engineering entity that qualifies
what r\^oles may be used by means of a session constraint entity.

\subparagraph{Constraints} The session constraint is dependent upon the
manner in which the session is established and is designed to reflect the
different ways in which the same subject may access the system. Access from
a physically secure local area network will be less constrained than from
an insecure dial--up line. Other constraints may be imposed because of
accepted usage practice: some records may only be available for specified
dates and times.

Each session may have a different r\^ole set. This allows the same subject
to act in the system in a different way.

\paragraph{Constraints} A simple information model for constraints entities
is given. These control the system who may access what within the system.

\begin{figure}[htbp]
  \begin{center}
  \includegraphics{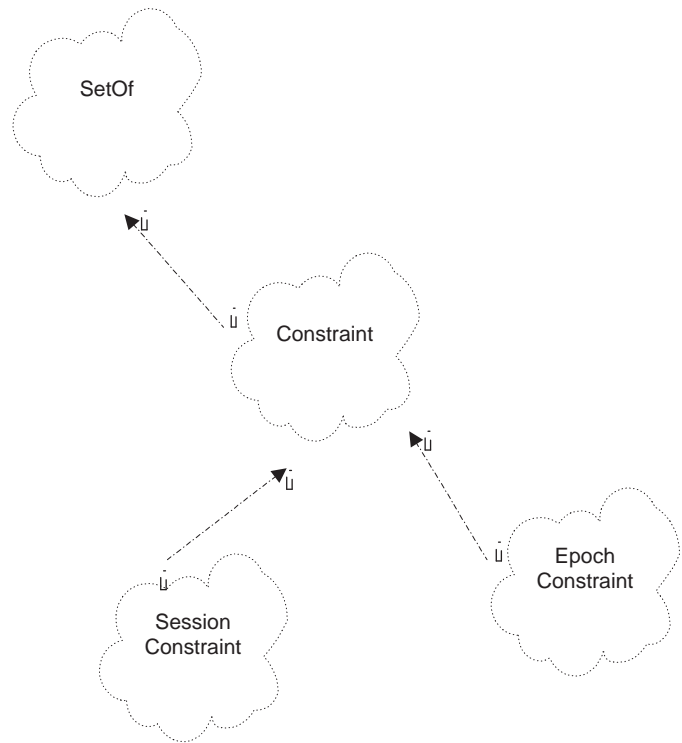}
  \caption{Role--based access control: constraint information model}
  \label{fig:constraint-im}
  \end{center}
\end{figure}

\paragraph{Summary} Sandhu's model for r\^ole--based access control systems
is very useful to this design discussion. It will be seen that most of the
design and analysis for this system will focus on generating the r\^ole
hierachy and a structure for constraints objects and the information that
must be placed in the constraints objects.

\section{Computational Model}
\label{sec:computational}

Most of the computation performed by the system would be to provide its
adaptability:

\begin{enumerate}
\item Generating views of records
\item Classifying accessors
\item Generating access rules
\end{enumerate}

\subsection{Generating Views of Records}
\label{sec:comp:views}

Custodians would deny access to records or restrict access to certain
fields. This suggests two strategies:

\begin{itemize}
\item Either provide a full record with \emph{NULL} put into the field
  values where a custodian has denied access.
\item Or generate a least common denominator view.
\end{itemize}

\paragraph{Nulling fields}

\begin{itemize}
\item Either a new database table has to be created and the modified
  records inserted,
\item Or a set of \emph{triggers}\footnote{\textit{SQL} allows a function to be
    invoked when a record is operated upon, see \cite{sft:sql:iso}.} to be
  generated to insert the \emph{NULL} values where specified.
\end{itemize}

Neither of these is particularly desirable: the former requires a new table
which would need to be separately maintained; the latter requires triggers
to be written which would need to check a profile (specified by the
custodian) for each record for every access of the view, which would
greatly affect performance.

\paragraph{Least Common Denominator View}

The collation procedure to produce the least common denominator view could
employ one of two strategies:

\begin{itemize}
\item Maximum field coverage by minimising records included
\item Maximum record coverage by minimising fields included
\end{itemize}

Both of which could be qualified by the accessor stating percentages of
coverages and ranking fields to be included.

\subsection{Classifying Accessors}
\label{sec:classifying}

Accessors would need to be grouped and then those groups ranked relative to 
one another to produce an authorisation lattice. Clearly, there are many
policies for this: most involve some arbitration outside of the
information system itself between the representatives of the different
entities. 

One procedure would to make use of professional standing within a respected
professional institution. There are many groups extant that could be used
as the basis for accessor control groups. The British Medical Association
is the accrediting professional organisation for practicing doctors in the
United Kingdom. The Law Society for solicitors. Belonging to a professional
group implies that one performs a certain r\^ole. It may be necessary to
enforce members of groups \emph{not} to use their group identity if
operating in a r\^ole not sanctioned by the group.

\paragraph{Grouping Accessors}

This is a proposal for system of grouping accessors together which makes
use of modern certification technology. The aim is that accessors and their
groups would be self--regulating. 

Each accessor would have an \textit{X.509} certificate proving their
identity. They would then need to obtain a proposer and seconder from the
group they wish to join. The proposer and seconder would corroborate the
identity of the applicant and make some recommendation to the membership
committee.

This protocol can be secured using a certificate chain and blind--voting
protocols described in \cite{crypto:schneier}. (Certificate chaining is
just one message encrypted using the private key of one certificate and
then by another. Blind voting allows a vote to be taken which allows each
party to prove to themselves that their vote has been counted, without
knowing who else has voted.)

Once an applicant has been granted membership of a group, they would then
need to be issued with a new certificate which would be their group
membership. This is an \textit{X.509} certificate with the group acting as
the certification authority. In the event that membership is revoked, the
certificate could be made void without inconveniencing any other members of
the group. It also reduces the amount of encryption needed to just one
pass.

(Incidentally, the method proposed above produces a ``Web of Trust''. There
are a number of different mechanisms for achieving this, again see
\cite{crypto:schneier} and also \cite{yahalom.klein.ea:trust-based}).

\paragraph{Ranking Groups}

There are two other functions that need to be developed for
self--organising groups. They must be able divide themselves up and to
merge.  This, combined with an authorisation hierarchy, will allow them to
better define who may access what information.  Groups, in this context,
are abeyant to set theory and what is needed is a defining membership
function: much as one might say, $X$ is the set of all odd dice throws.
This requires a distance measures which would allow someone to say that
under, a particular distance measure, member $x$ is very similar to $y$.
Statisticians and actuaries do this all the time, it just remains to
develop it for professional groupings.

\subparagraph{Deference}

For groups and their members to be ranked: the r\^ole of the group (or the
function of its members) has to be quantified. The principle of deference
is a useful basis since professional groups apply it. Referring again to
the British Medical Association, it has sub--groups: student members,
juniors, general practitioners, consultants and specialists. At the same
time, members of the BMA will have different affiliations to other
organisations, the Royal College of Surgeons, British P{\ae}diatric
Association and so forth.

A family doctor with no special p{\ae}diatric expertise involved in a
p{\ae}diatric case would be expected to defer to another doctor who is a
member of a p{\ae}ediatric association.

\subparagraph{R\^oles and Ontologies}

This can be quantified by a distance measure. Doctors would collect
accreditations and whoever has the most of them over the range of issues
involved would be ranked above the others. The issues effectively determine
the r\^oles. This, again, is a collective choice procedure which will be
analysed in more detail later. Suffice to say, that the members of the
groups would rank themselves within their own groups and rank their groups
with respect to others \emph{with respect to their current r\^ole}.  There
is no objective ranking between groups, or, come to that amongst group
members, because it depends on the issue at hand, which demands that
individuals take certain r\^oles. This concept is explained in more detail
in \cite{RBAC97*153}.

``Issues'' is too imprecise a term, so \emph{ontology}\footnote{This is the
  term used in KIF see chapter \myRef{cha:kqml}.} will be used in place of
it.

Some relationship diagrams might help clarify this. An individual possesses
certain r\^oles. The ontology within which the individuals are operating
will require that certain r\^oles be fulfilled. With regard to accessors
and custodians, these are both types of individual. This relationship
analysis is applicable to both accessors and custodians, because mappings
between the two sets using a common ontology will help in allocating access
views.

\begin{displaymath}
  \begin{aligned}
    \begin{CD}
      Accessor @>{is-kind-of}>> Individual \\
      Custodian @>{is-kind-of}>> Individual \\
    \end{CD}
  \end{aligned}
  \quad \droptext{and} \quad
  \begin{aligned}
    \begin{CD}
      Individual @>{has}>> Roles \\
      Ontology @>{requires}>> Roles \\
    \end{CD}
  \end{aligned}
\end{displaymath}

Given the r{\^o}les and the ontology, an ordering of individuals for an
ontology can be formed.

\begin{displaymath}
  \begin{CD}
    (Roles, Ontology) @>{orders}>> Individual \\
  \end{CD}
\end{displaymath}

\subsection{Classifying Records and Fields}

Initially, views of records, and the fields within them, will be classified
by the custodian for each accessor given the ontology. The accessors will
themselves be classified and it should be the case that certain classes of
accessors will require certain types of record view. Consequently,
classifications for record views will evolve for different ontologies. This
is another important requirement of the system so that it can be
self--organising: if views are ordered relative to group memberships then
it will be possible to recommend that groups be sub--divided to match
information protection requirements. Conversely, it can be used to simplify
access rules by merging similar groups.

These relationship diagrams might help to make clear how records can be
ordered. A record will have a number of views. Each ontology would require
certain views.

\begin{displaymath}
  \begin{aligned}
    \begin{CD}
      Record @>{has}>> Set of views \\
    \end{CD}
  \end{aligned}
  \quad \droptext{and} \quad
  \begin{aligned}
    \begin{CD}
      Ontology @>{requires}>> Set of views \\
    \end{CD}
  \end{aligned}
\end{displaymath}

Individuals have been ranked relative to one another for a particular
ontology, so if an individual is given access to a view, then granting that 
permission effectively orders the views of the records in that ontology.

\begin{displaymath}
  \begin{CD}
    (Individual, Ontology) @>{orders}>> Views \\
  \end{CD}
\end{displaymath}

\paragraph{Discussion}

There are three classification processes at work; the last is a corollary
of the first. This assumes that all the individuals are working within the
same ontology.

\begin{enumerate}

\item Individuals classify one another

\item Individuals classify views

\item Custodians classify accessors

\end{enumerate}

If the system is bootstrapped by a number of carefully deliberated
classification decisions, then more specific access rules can be generated.
When it is not possible to apply a rule, it will be resolved by another
classification decision by a custodian and a new rule can be added.

\subsection{Generating Access Rules}
\label{sec:comp:access-rules}

This process relies upon the generation of an information flow model
\cite{Denning:1976:LMS,sec:denning}. Without going into detail, an
information flow model requires:
\begin{itemize}
\item A security classification scheme that classifies:
  \begin{enumerate}
  \item All views, \emph{and} 
  \item All the accessors
  \end{enumerate}
\end{itemize}

Then, for a given set of views, an accessor must have a security
classification that is greater than or equal to the least upper bound of
all the views demanded. So, one can conclude, that computationally it is
relatively simple to determine access rights, if accessors and views are
graded.

\begin{figure}[htbp]
  \begin{center}
  \includegraphics[angle=-90]{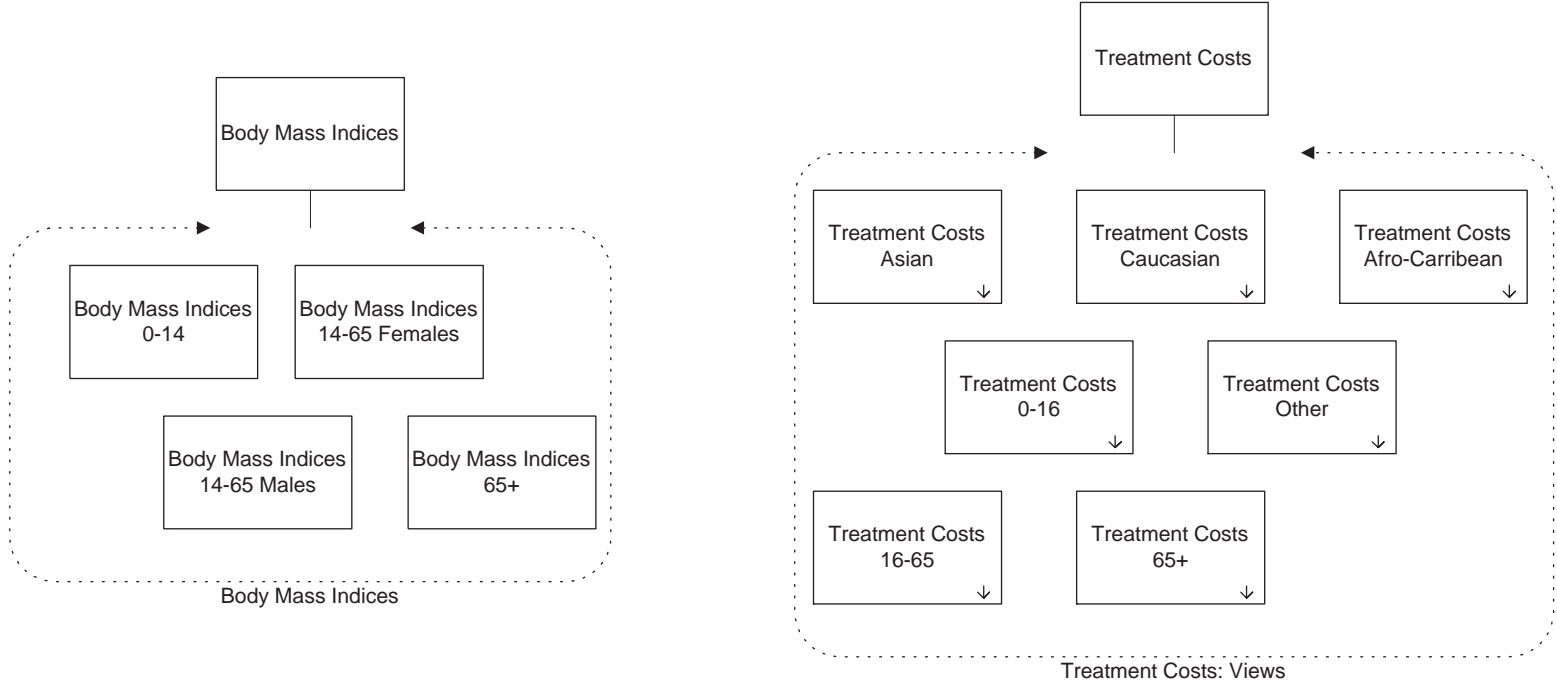}
  \caption{Some example of record views}
  \label{fig:views}
  \end{center}
\end{figure}

In figure \ref{fig:views}, two sets of views are presented: Body Mass
Indices, BMI, and treatment costs. There is a choice of sub--views for
each. \myRef{sec:comp:views}, paragraph ``Least Common Denominator View'',
stated that there are two parameters that a custodian varies in generating
a view for an accessor: the fields in the view and the range of
records. The fields here are the BMI entries in an historical medical
record and the treatment costs. The ranges varied are the age and ethnic
groups.

\begin{figure}[htbp]
  \begin{center}
  \includegraphics{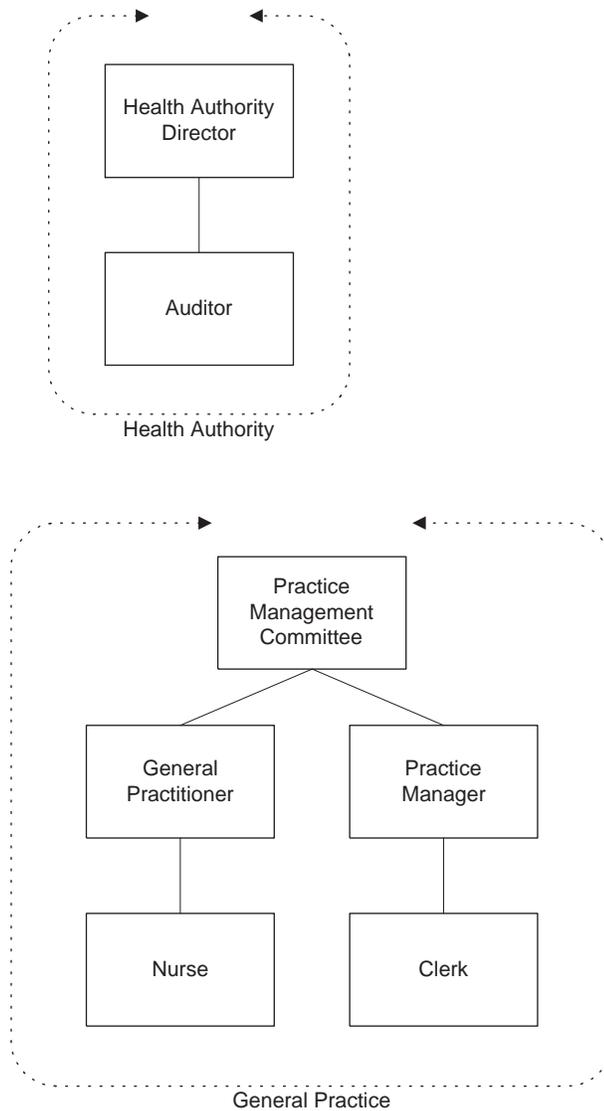}
  \caption{Medical practice and health authority}
  \label{fig:practice}
  \end{center}
\end{figure}

A simple medical practice is shown in figure \ref{fig:practice}. The
practice has two organisational functions: medical and administrative.
Referring to the views available, the medical staff, doctors and nurses,
would be given write access to Body Mass Index data, but not to treatment
costs; the administrative staff would be given write access to treatment
costs, but not the BMI data. The owner of the treatment cost data is the
administrative arm of the practice, the owner of the BMI data is the
medical arm.

The health authority which reimburses the medical practice for treating
people in its catchment area would need access to treatment costs records
for all medical practices in its area.

In administering the practice{\myDash}aligning it to the needs of the
health authority{\myDash}it would be necessary to value the cost of
taking a BMI reading and this would be discussed at a meeting of the
Practice Management Committee. In the classification of the information
held by the practice, the least--upper--bound of the medical and
administrative arms of the practice is the practice management committee.

\begin{figure}[htbp]
  \begin{center}
  \includegraphics{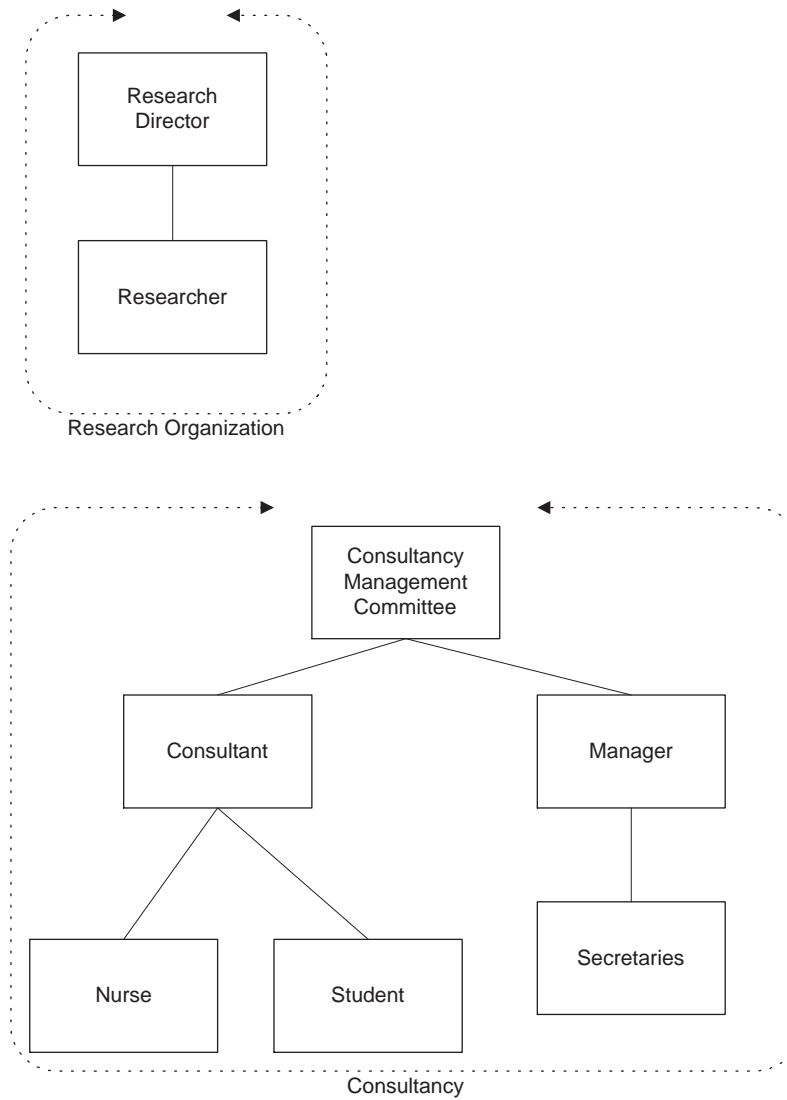}
  \caption{Consultancy and research body}
  \label{fig:research}
  \end{center}
\end{figure}

A medical consultancy is shown in figure \ref{fig:research}. It has a
similar structure to a medical practice, see figure \ref{fig:practice}, but 
would also undertake the training of students, who would be answerable to
the consultant. A medical consultancy would undertake research and would be 
answerable to a research organisation for any funding it receives.

The difficulty is to join the two organisations' structures. This would
need to be performed in the appropriate ontology.

\begin{enumerate}
\item Consultancy

  The most typical scenario is, of course, that patients would be referred
  to consultants. The consultant would obtain his own information on the
  patient. The patient's practitioner would expect to be informed of the
  consultant's findings. The consultant might want to use the information
  he has obtained from the patient for his research; the consultant should
  obtain the patient's consent from his custodian the patient's medical
  practitioner.

\item Research

  If the medical practice decided to make available the BMI information it
  holds, the medical practice and the health authority would place
  stipulations on its release.

\end{enumerate}

\section{Information Model{\myDash}Views and Constraints}
\label{sec:info}

As stated, the technology to realize this system is already available, so
it is not necessary to detail all of the information the databases and
web--servers would need to operate. The information that is of concern is
that needed to provide the newer features: 

\begin{enumerate}
\item Granting access to views
\item Provide assurances that each party is keeping its contract with the
  other.
\item Facilitate classification of accessors and views.
\end{enumerate}

\subsection{Granting Access to Views}
\label{sec:req:granting}

There are two cases to consider:

\begin{itemize}
\item Either an access rule exists and can be applied
\item Or there is no applicable access rule
\end{itemize}

(Bear in mind, that an access rule may actually deny access.)

It is only necessary to consider the latter case, the procedure is simply a
custodian receives a request from an individual who wants to access some
specified records.

Clearly, to do this the accessors will need to know who the custodians are
and how they can be contacted.

The custodians will need to know:

\begin{enumerate}
\item Accessor's identity and proof of group memberships
\item Ontology under which the accessor is operating
\item View of the record they require
\item Precedents set by other custodians
\item Precedents set by access rules
\end{enumerate}

The accessor will then be informed of the custodian's decision. The
decision could then be formalised as a precedent upon which an access rule
could be based.

\subsection{Proofs of Contract Compliance}
\label{sec:compliance}

Most of the information that needs to be retained by the system will simply
show that the duties of each party are being fulfilled. All of this
information would be available from the log files of the databases and
web--servers used.

\begin{enumerate}
\item Custodian Actions
  
  Views of which records granted by a custodian to which accessors or groups
  of accessors. Whether the view is re--publishable by the accessor and
  whether any access rules are in force which would allow unvetted access.
  
  If a custodian rejects an accessor's request for access, then it must
  retain a justification for that denial.

\item Owner Actions

  Log all transactions by each accessor stating views and records accessed.

\item Accessor Actions

  Retain notifications of access rights granted.

\end{enumerate}

This is the information that would need to be reconciled to show a subject
that either his custodian or a database owner has acted improperly.

\subsection{Adaptability Support}

Enough information has to be retained to classify accessors, and to group
them, and to classify the record views they are granted.

\paragraph{Classifying Accessors}

Each accessor has to retain the membership certificates they have been
granted by the groups they are members of. The group membership committees
should also retain their justifications of why each member was given
membership.

\paragraph{Classifying Record Views}

This information can be derived: as custodians grant views to accessors,
their group memberships and the ontology under which they are working will
be known. Consequently, the views can be classified.

\section{Engineering Model}
\label{sec:engineering}

The engineering of an information system usually concerns itself with how
services can perform best. This is usually a choice between resilience and
speed. This information system has very different primary requirements:
security and safety, and, as noted above, this is a secure distributed
processing problem. There are two functions that must be engineered safely
and securely: the protocol and formatting of messages and the processing of
them.

\begin{enumerate}
\item Security
  \begin{itemize}
  \item Integrity
  \item Confidentiality
  \item Authentication
  \end{itemize}
\item Safety
  \begin{itemize}
  \item Authorisation
  \item Information Flow
  \item Inference Control
  \end{itemize}
\end{enumerate}

There is no problem with message security, well--proven protocols, like
Transport Level Security \cite{tls:eaves} and its predecessors, the Secure
Socket Layers \cite{draft:ssl} are supported by web--servers and
web--browsers. It is the safety of the processing that needs to be
considered. The most important point is that a process is effectively an
accessor and any process that operates on secured data should have the
appropriate security clearance. A security clearance would be required both
for the implementation of the process and the host machine it runs on. This
has long been recognised in secure processing, see \cite{sec:sesame} for
more explanation on the difficulties this gives rise to.

Security clearance for implementations is needed because they may possess
covert channels of communication, see, for example,
\cite{TsaiGligorChandersekaran90} and security clearance for the host
machine is required because the process owner, or a corrupt systems
administrator, could trace the process as it runs and capture any
information it holds.

Secure process implementations and secure execution environments are
collectively known as a Secure Computing Base; the need for which has been
well--known for some time \cite{sec:denning}; the difficulties of
developing a secure computing base for mobile agent--based systems are
discussed in \cite{Spreitzer93}.

\subsection{Functional Specifications}

The database access system proposed has the following functions to fulfil:

\begin{itemize}
\item Memberships accreditation
\item Database broking
\item Access negotiation
\item Query formulation
\item Query delivery
\item Results delivery
\item Results presentation
\end{itemize}

\paragraph{Memberships Accreditation}

The accessors would organise group memberships amongst themselves at a
web--server which supports secure transactions. This would require that the
professional organisations that accessors would belong to have a
Certification Authority, CA, available. (\textit{Thawte} \cite{ca:thawte},
for example, offers a cross--certification service.)

\paragraph{Database broking}

At a web--server, access to which may also be secured, accessors would see
which databases are available to them to negotiate access to. This is
essentially a trading service, \cite{ODP:trader}. This would seem the
sensible point to put them in contact with the database owners and begin
the process of access negotiation.

\paragraph{Access Negotiation}

There are three processes that could be followed:

\begin{itemize}
\item Access by rule
\item Access by custodian consent
\item Both of the above
\end{itemize}

If there is a rule that can be followed, then it is simply a matter of
checking the accessor's credentials (group memberships) with the
requirements of the rule for the given database and the stated ontology of
the accessor.

The other two require that either all the custodians be contacted or those
custodians who have not delegated to a rule.

Some agent has to be put in place that can:
\begin{itemize}
\item Apply access rules
\item Obtain access decisions from custodians
\end{itemize}

\paragraph{Query formulation}

Most accessors would not want to formulate queries using standard
\textit{SQL}. They would probably use some kind of forms interface, as is
common with most commercial web--based databases \cite{Illustria}, but this 
might be specific to their ontology and may be recommended by their
professional organisation.

\paragraph{Query delivery}

The query, once formulated as \textit{SQL}, would be encrypted and submitted
to the database back-end. The query should be archived as evidence in the
event of misuse.

\paragraph{Results delivery}

The results of the query may need post--processing to minimise the
opportunities for inference\footnote{See \cite{sec:denning} for inference
  control mechanisms.}. The results would also need to be archived as
evidence. 

\paragraph{Results presentation}

Again, it is unlikely that accessors would be able to use the results in
the format returned by the database and some post--processing may be
required to present the results in a usable format for them.

\subsection{Agents}
\label{sec:engineering:agents}

The agents for the system can now be specified following the functional
specifications:

\begin{enumerate}
\item Accessor Memberships Agent
\item Database Trader
\item Access Negotiator
\item Query Formulator
\item Query Delivery Agent
\item Results Delivery Agent
\item Results Presentation Agent
\end{enumerate}

\section{Enterprise Models: System Agents}

As stated there are three parties to each access contract:
\begin{itemize}
\item Owner
\item Accessor
\item Custodians (Owner is a custodian of the database as a whole)
\end{itemize}

The functions performed by each agent have been listed and briefly
described. It is now necessary to specify who has responsibility for
providing each agent's secure computing base and who uses it and to whom
the secure computing base must provide assurances. Table \ref{tab:secclear}
clarifies this. \textit{Supplier} indicates which of the parties should
provide the secure computing base, \textit{Users} is a list of the parties
who would use the agent, \textit{Assurances to} is the list of those it
must provide assurances to. (Bear in mind, again, that the owner of the
database is also the custodian of it as a whole.) The concept of a
supra--organisation has to be introduced{\myDash}Supra--accessors
\etc{\myDash}they vet their own members or act, collectively, on their
behalf.

\begin{table}[htbp]
  \begin{center}
    \begin{tabular}[left]{|l|l|l|l|l|}
      \hline
      System & \multicolumn{2}{l|}{Secure Computing Base} & \\
      Agent & Suppliers & Users & Assurances to \\
      \hline
      Accessor Memberships Agent &
      Supra--accessors &
      Accessors &
      Owners, Custodians \\
      
      Database Trader &
      Supra--owners &
      Accessors &
      Owners, Custodians \\

      Access Negotiator &
      Supra--custodians &
      Accessors &
      Custodians \\

      Query Formulator &
      Supra--accessors &
      Accessors &
      Accessors \\

      Query Delivery Agent &
      Owner &
      Accessors &
      Custodians \\

      Results Delivery Agent &
      Owner &
      Accessors &
      Custodians \\
      
      Results Presentation Agent &
      Supra--custodians &
      Accessors &
      Custodians \\
      \hline
    \end{tabular}
    \caption{Ownership, trust and use relationships for system agents}
    \label{tab:secclear}
  \end{center}
\end{table}

What is unusual about these ``ownership, use and trust'' relationships is
that they are tri--partite. Most relationships between entities in systems
are bi--partite. All bi--partite system interactions can be reduced to be
(produce, consume), but tri--partite relationships have to introduce a
second interaction which is to observe the produce--consume interaction:
(produce, consume) and ((produce, consume), observe). Observation is
achieved by having the producer and consume both sign the information they
produce and consume and using that as a product the observer consumes. The
observer would reconcile the information produced and consumed.

As Spreitzer \cite{Spreitzer93} has pointed out, one secure computing base
will do for all parties, \emph{if} they are all satisfied that the
computing base is secure enough for all of them.

\section{Engineering Model}
\label{sec:engineering:model}

It is now possible to specify the system. Booch \cite{des:booch} object
interaction diagrams are used here. (The class relationship diagrams are
not given.) The interaction diagrams are easy to understand. 

\begin{enumerate}

\item Objects are described by $Name:  $ {\myDash} the name may be omitted.
  The attributes of the object are listed under the name and class. Very
  often one of the attributes an object possesses is a back--reference to
  the object that contains it.

\item The short arrows are method invocations by one object onto another.
  The arrows terminated by a small circle are the return values of the
  method.

\item The lines connecting objects denote that the object sending a message
  has the object reference to the recipient \textit{a priori}: it is part
  of the object's state when the interaction starts. Object references can
  be qualified by an \textsf{F} if the object is a field of the one
  referring to it.

\item New object references are obtained as the interaction proceeds by
  returning an object reference as the result of a method invocation.

\item Objects can create other objects. The creation call is the name of
  the class with the construction parameters. An arrow to the object
  indicates which has been produced.

\end{enumerate}

In the text, classes are denoted by this style \textsf{Class}, objects by
this $A$.

\subsection{Accessor Memberships Agent}
  
Every accessor is initially an object of class \textsf{Member}. Each one of
which has been issued an \textit{X.509} certificate, encapsulated in an
object of class \textsf{Credential}.

\begin{figure}[htbp]
  \begin{center}
  \includegraphics{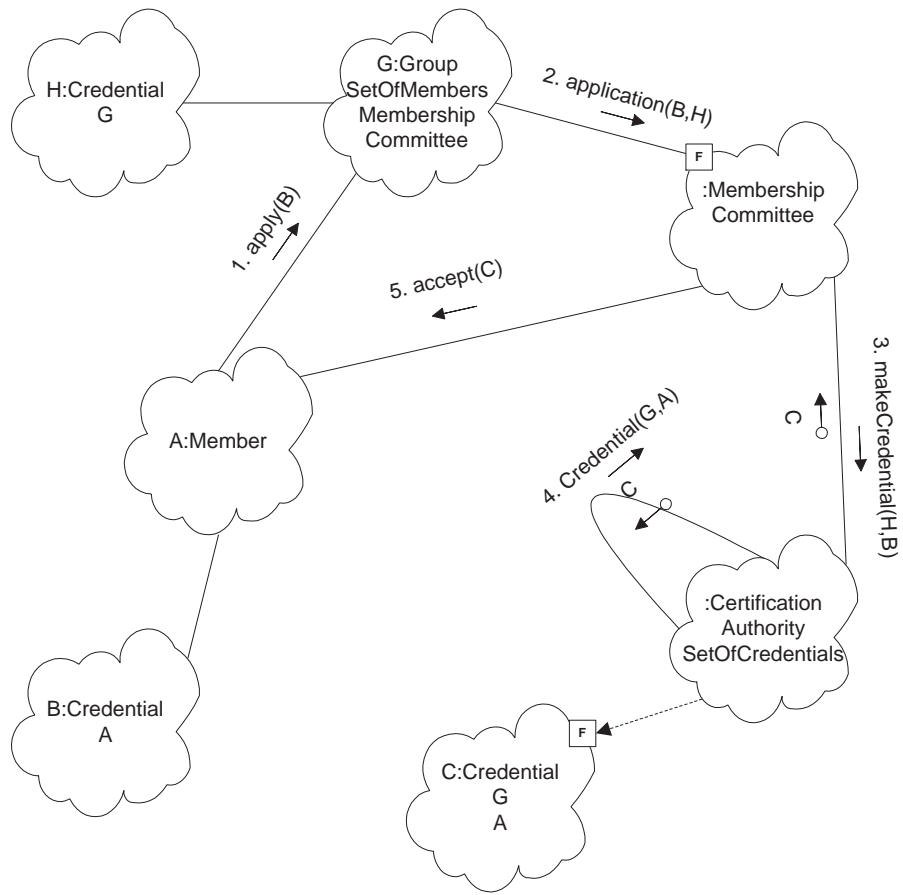}
  \caption{Accessor Memberships Agent}
  \label{fig:membership}
  \end{center}
\end{figure}

When an \textsf{Member} object applies to join a group, \textsf{Group}, it
follows the object interactions shown in figure \ref{fig:membership}.
Object $A$ has a credential $B$ and applies to join group $G$ having
credential $H$. The group has a membership committee which vets the
application and, if successful, asks a certification authority to create a
new credential specifying $C$ that states $A$ is a member of $G$. This
credential is then passed back to $A$, who accretes it for its own use.

(Note in the interaction diagram, the \textsf{Membership Committee} object
has a reference to the prospective \textsf{Member}. This is just shorthand.
Ordinarily, the new credential would be sent back to $A$ by $G$.)

\subsection{Database Trader}

This system agent is best provided by some collective agency for the owners
of the databases{\myDash}Supra--owners.
  
Custodians will want to prevent database owners from publishing their
databases indiscriminately, because it will mean they will have to vet
too many access requests and possibly incur greater risks of
disclosure. The database trader has to provide assurances to them.

Database owners will want to specify which type of accessors be allowed to
\emph{know} what databases they possess.

(This is not specified in the interaction diagrams: the database trader
should also provide a justification for any denial of access to an
accessor.)

It should be apparent that publishing the database at a trader is a similar
access control problem to that of determining whether access to the records
to a particular accessor is allowed.

\begin{figure}[htbp]
  \begin{center}
  \includegraphics{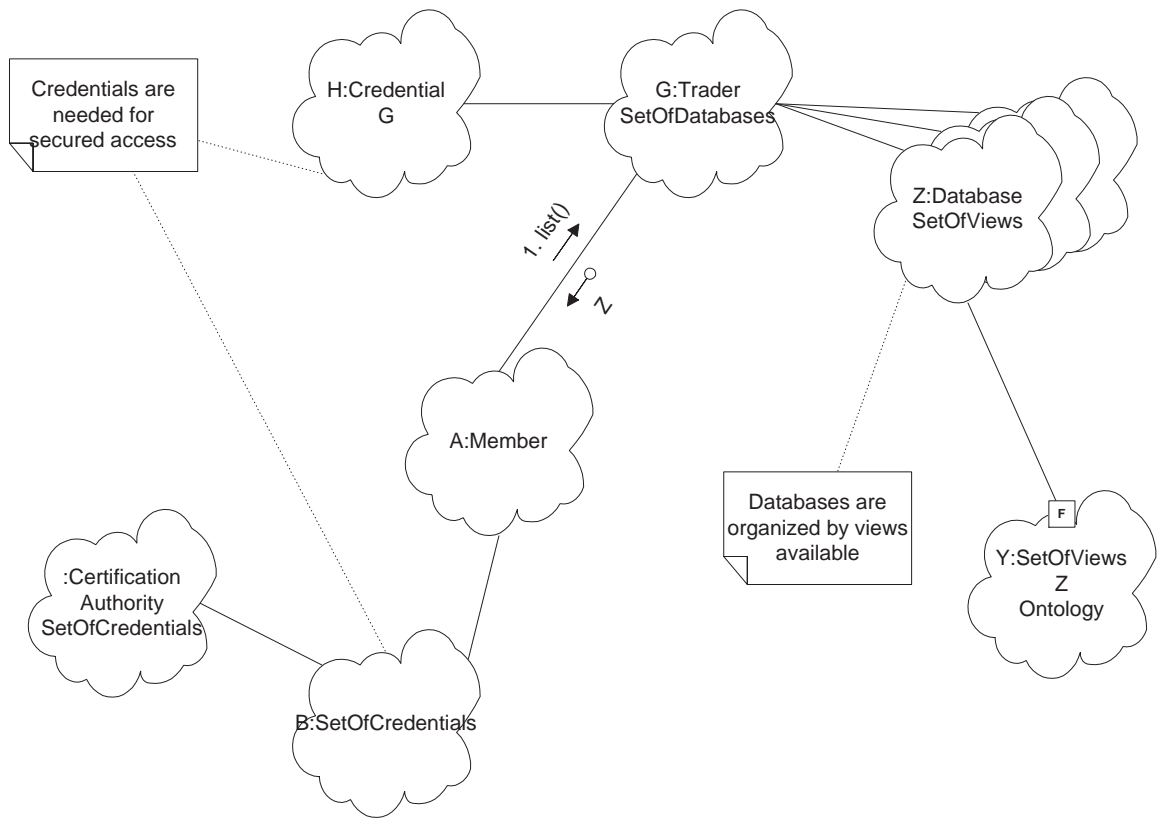}
  \caption{Trader Agent}
  \label{fig:trader}
  \end{center}
\end{figure}

The interaction shown in figure \ref{fig:trader} shows how a member would
obtain a list of databases. Each database has a set of views and each set
of views contains a statement of its relevant ontology.

\begin{figure}[htbp]
  \begin{center}
  \includegraphics{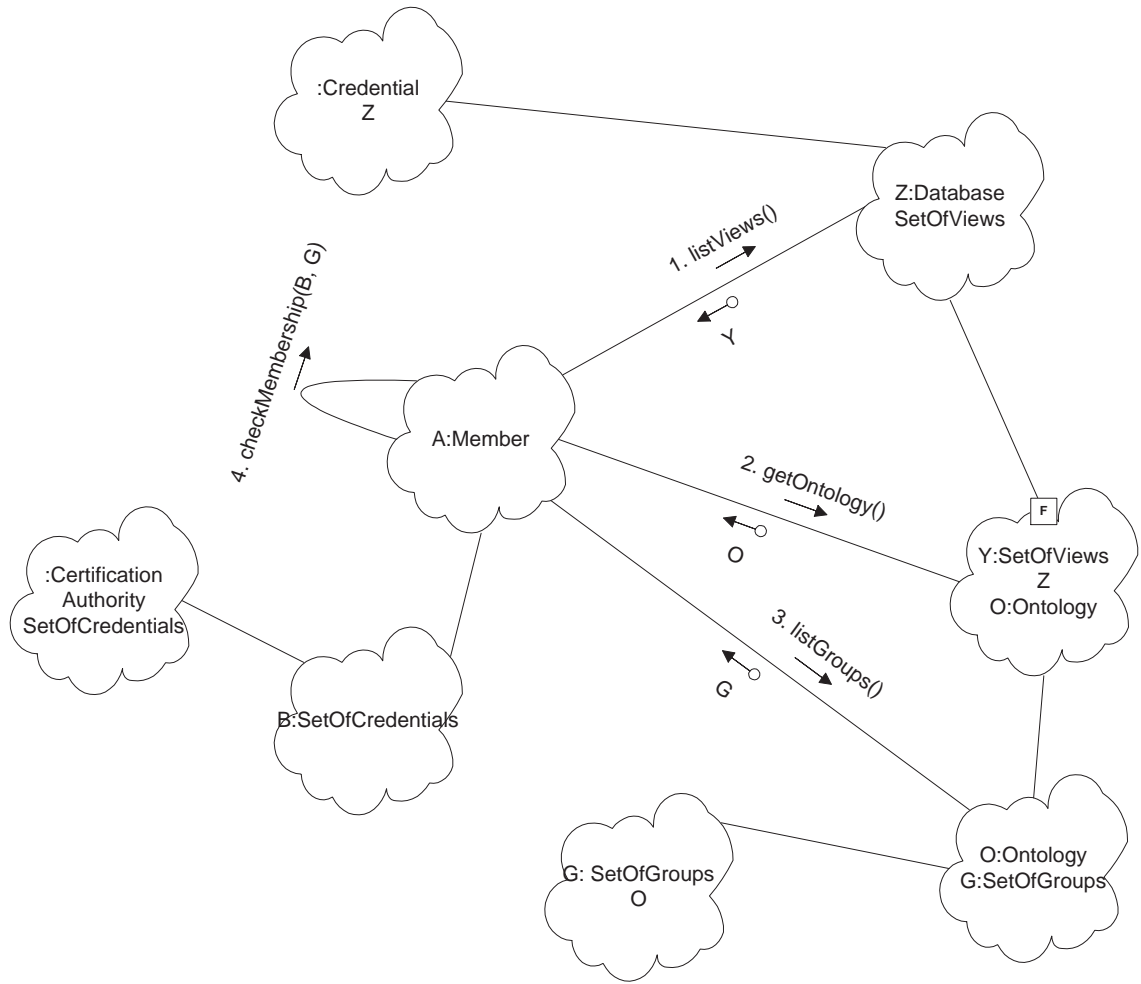}
  \caption{Database, Views, Ontology and Groups}
  \label{fig:access-neg1}
  \end{center}
\end{figure}

Once the member has obtained a reference to a database, it can interrogate
it to see what ontologies it can be used for and the current set of groups
who have designated as existing in that ontology. This is illustrated in
\ref{fig:access-neg1}.

\subsection{Access Negotiator}

The \textsf{Access Negotiator} agent is best provided by a collective
agency acting on behalf of the custodians {\myDash} Supra--custodians. It
services the requests of the accessors and provides assurances to the
custodians.

\begin{figure}[htbp]
  \begin{center}
  \includegraphics[angle=-90]{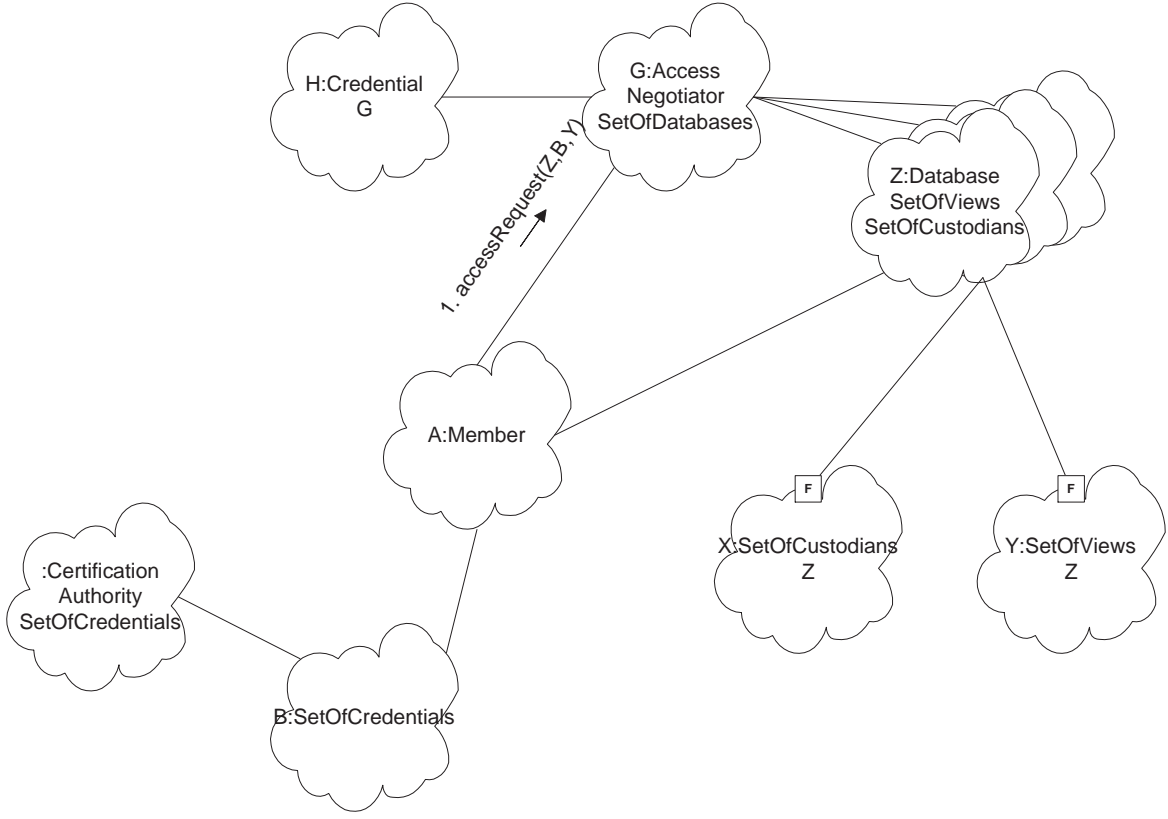}
  \caption{Access Negotiator Agent}
  \label{fig:access-neg2}
  \end{center}
\end{figure}

Figure \ref{fig:access-neg2} shows the arrangement of the objects, before
access negotiation begins in earnest. $A$ submits a request to the
\textsf{Access Negotiator} $G$ stating his credentials, the database $Z$
and set of views $Y$ $A$ wishes to access. Because $A$ has been able in
figure \ref{fig:access-neg1} to interrogate the database directly, it can
supply a subset of its credentials $B$ which it knows will satisfy the
criteria.

The \textsf{Access Negotiator} would be empowered to determine if $A$ has
supplied a set of credentials $B$ which do entitle $A$ to access the
requested views $Y$ in $Z$ by referring to a \textsf{SetOfPrecedents} for
the \textsf{SetOfViews} $Y$. This is not detailed in an interaction
diagram. In this activity, the access negotiator simply acts as an access
control list enforcer.

\begin{figure}[htbp]
  \begin{center}
  \includegraphics[angle=-90]{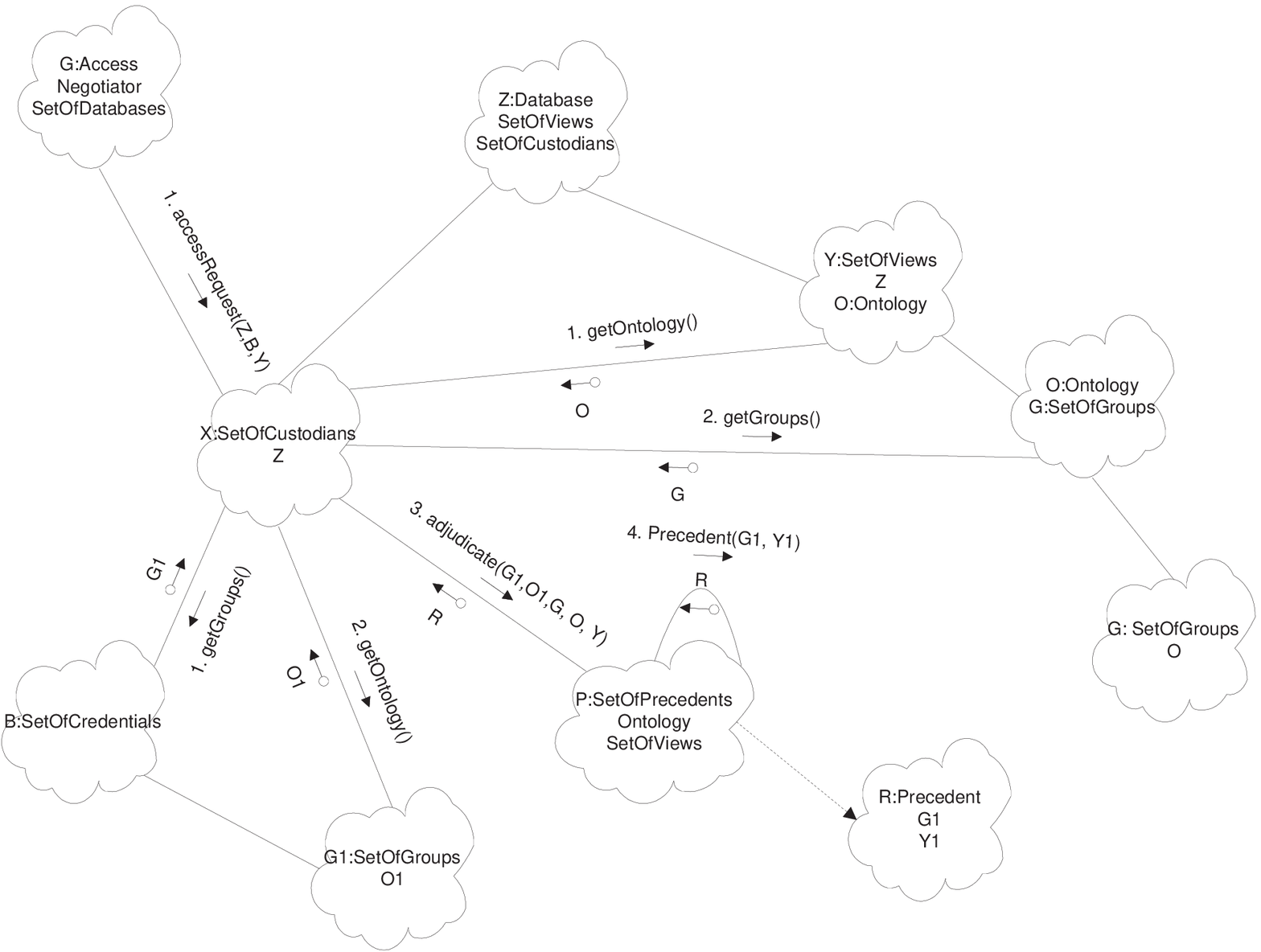}
  \caption{Custodians establishing new precedents}
  \label{fig:access-neg3}
  \end{center}
\end{figure}

The real work of access negotiation is shown in figure
\ref{fig:access-neg3}. The \textsf{Access Negotiator} passes the access
request to the \textsf{SetOfCustodians}. Each of the custodians would
obtain the ontology and the groups working within that ontology for the
views requested by the \textsf{Member} object making the request \emph{and}
the ontologies and groups of the \textsf{Member} by referring to the
\textsf{SetOfCredentials} supplied by the \textsf{Member} object.

If the \textsf{SetOfCustodians} collectively agree that the \textsf{Member}
object should be allowed access they would create a new \textsf{Precedent}
object allowing members of group $G$ to access views $Y$.

Incidentally, it may be necessary to create a new group with corresponding
credentials to allow a particular accessor a set of views.

\subsection{Query Processing}

A set of four agents form a call and reply chain. The incomplete object
interaction diagram appears in figure \ref{fig:delivery}. The
\textsf{Member} object instructs a \textsf{Factory} object to create the
objects using the \textsf{SetOfCredentials} $B$ to access the database $Z$,
the views may need to be provided if the accessor has a choice of views
available to him. The remainder of the object interaction is not
diagrammed, but it would consist of formulating a query, which would then
send it on to $Q2$, which would encrypt it correctly for the database, and
would probably share an encryption key with $R2$. $R2$ would send the
results on to $R1$ which would then return them to the accessor.

Queries submitted and results delivered would need to signed and returned
to the respective originators as part of the observation procedure. Again,
this is not detailed in the interaction diagram.

\subsubsection{Query Formulator}

The view granted to the accessor will provide meta--data describing its
contents. Although this information could be construed as being sensitive,
there seems little point in protecting this, so the query formulator can be
wholly owned by the accessor.

\subsubsection{Query Delivery Agent}

This agent is responsible for delivering the query securely to the database
that can answer it. This agent is responsible for collecting the query from
the query formulator. Constructing a message containing the query. Having
that query signed as originating from the accessor and sending it. On
receipt, the server at the database would check the signature and so be
able to check that the accessor submitting the query has the right
privileges to do so.

The database owner knows best how to do this, but the query has to be
logged, should evidence of a breach of trust between owners and custodians
be needed.

\subsubsection{Results Delivery Agent}

The results will contain classified information and they need to be logged
to prove breach of trust if needed.  As pointed out above, there may be a
need to perform post--processing, so the results delivery agent should be
owned by the custodians.

\subsection{Results Presentation Agent}

The results are classified, so the presentation agent has to be owned by
the custodians. The accessor would collect the results from the agent using
a secure channel.

\begin{figure}[htbp]
  \begin{center}
  \includegraphics[angle=-90]{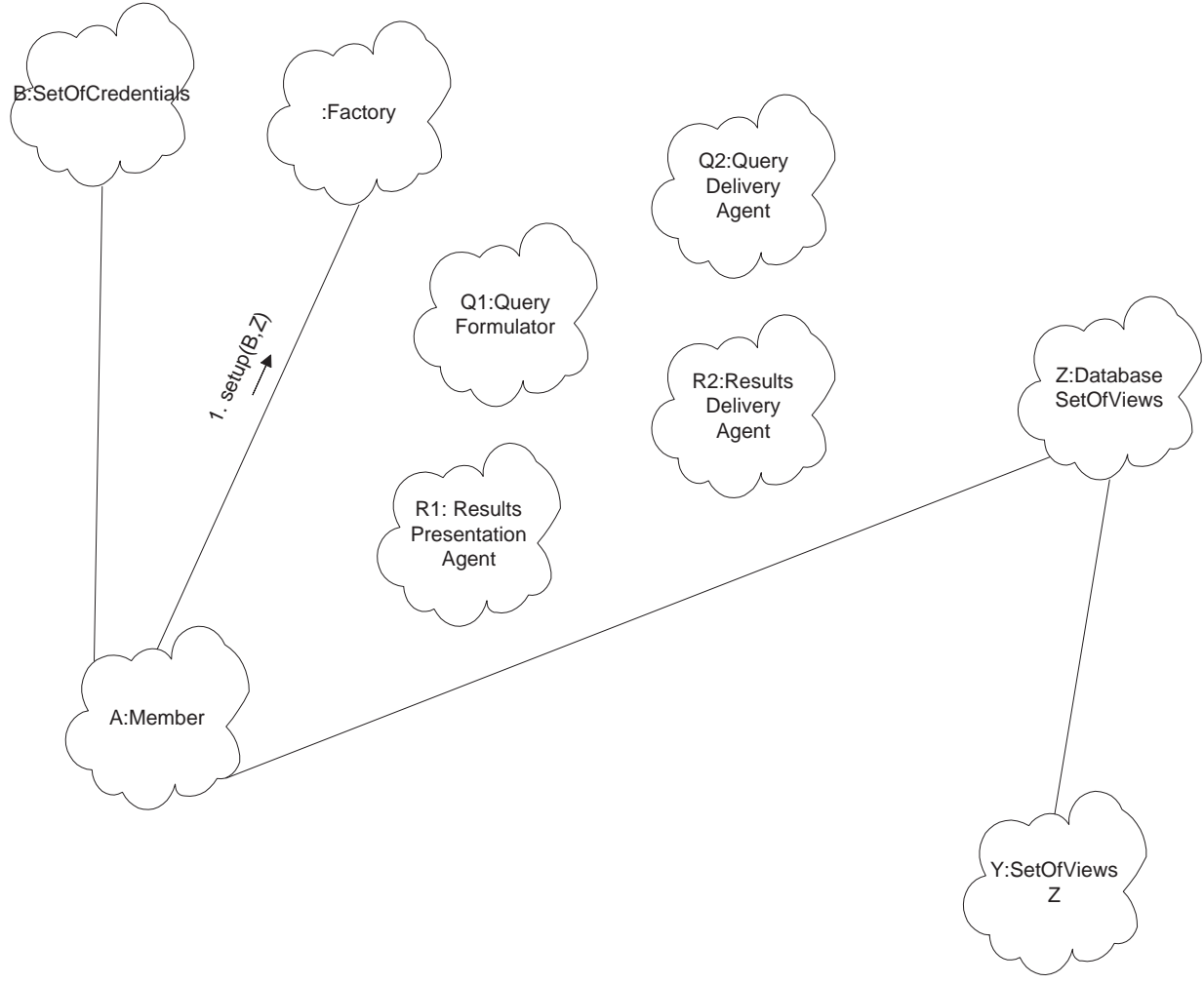}
  \caption{Query Formulation and delivery; results delivery and
    presentation}
  \label{fig:delivery}
  \end{center}
\end{figure}

\section{Summary}

The design issues that have arisen from this analysis are that lattices of
information flow are needed.

\begin{itemize}
\item Each accessor would require a set of security clearances assigned to
  them
\item Each view of the records available in the databases must be assigned
  a security clearance
\item Each of the system agents must have a security clearance
\item Each secure computing base must have a security clearance.
\end{itemize}

It should be possible to generate these lattices by interpreting the
following information:
\begin{itemize}
\item Hierarchy of group memberships produced by accessors'
  supra--organisations
\item Adjudications by custodians of views granted to accessors
\item Hierarchy of secure computing bases.
\end{itemize}

Methods of generating lattices will be addressed later.


\chapter{Access to a Database} 
\label{cha:dbaccess}

Chapter \ref{cha:reqs} took the functionality{\myDash}and the security
features{\myDash}of databases very much for granted. This chapter sets that
right, with a positive result: \textit{SQL} databases do have the
functionality to support an adaptable information service. The previous
chapter proposed a system architecture, this chapter is a technical
analysis of the capabilities of the most important component of that
system.

\section{Issues}

All that is required is that a database can be shared by two sets of users:
``native'' local users and World--Wide Web-based remote users.  This
produces its own set of issues:
\begin{itemize}
\item Capabilities
\item Ownership and Autonomy
\item Portability and Adaptability
\item Security and Safety
\end{itemize}

The first of these addresses how different databases grant or deny
access{\myDash}what information and computations can be performed by them.
The second addresses: how much can a database owner allow a foreign
administrator to operate upon the database. Thirdly, how flexible is
database access technology, can it adapt to a client's environment or must
the client adapt to it. Finally, can databases be secured and how safe are
they.

\subsection{Capabilities of Databases}

What follows is a brief summary of the differences in design between a
number of databases\footnote{Oracle, Informix, Postgres and
  Illustria\cite{Illustria}.} that support \textit{SQL}. The differences
considered are how may the database be secured. Currently, \textit{SQL} is
at version 2.0 or that standardised in 1992. Unfortunately, there are three
types of \textit{SQL} conformance: full, intermediate and entry.

\begin{description}
  
\item[Meta--data] \textit{SQL}--conformant databases have different
  internal architectures, since it depends on the type of \textit{SQL}
  conformance as to whether catalogues must be provided. Catalogues and
  schema are implemented as tables, so it is possible to emulate them. Only
  in full--\textit{SQL} is it required that the full set of procedures to
  manipulate catalogues be available. Catalogues contain schema.  If
  catalogues are supported then at least the Information Schema must be
  contained in each catalogue. The information schema contains meta-data
  about tables, views and procedures. It will contain names and
  descriptions of fields and the behaviour of procedures. All persistent
  objects named by a schema are associated with the authorisation
  identifier of the schema.  When an \textit{SQL}--session is started a
  cluster of catalogues is assigned to the session.
  
\item[Relations] Relations are tables or views and all \textit{SQL}
  databases support both.
  
\item[Accessing] Two aspects to this: security and loading. Most
  \textit{SQL} databases support the access control primitives.  These are
  ``Revoke'' and ``Grant'' for a named user or group and are only
  applicable to relations. Rights are not associated with procedures. As
  for loading, most databases can limit the number of clients that can
  simultaneously access it, but not all of them allow clients to be
  differentiated between internal and external users.
  
\item[Rules] Some databases support an additional ``rules'' or ``triggers''
  feature which is relatively recent and should be required in the next
  issue of the \textit{SQL} standard. A rule allows one to specify actions
  to carry out in addition to \emph{or instead of} the invoker's action to
  select, update, delete or insert a record. One of those actions is to do
  nothing. The rule concept was originally introduced to allow indices to
  be updated or for exported keys to be updated. A ``where'' clause is
  permitted which allows one to perform any tests on or with a user's
  identity.  So this mechanism could be used to check whether a record can
  be released to a given user or not.  Unfortunately, the behaviour of
  rules is difficult to specify and, consequently, implementations vary.
  The main point of debate is whether a rule is to be applied to a table or
  a record{\myDash}with records inheriting rules from tables.

\end{description}

To secure \textit{SQL} databases potentially every accessor would need to
be given an information schema of the set of views and procedures to use.
And there would need to be a corresponding set of grant and revoke commands
issued on those views.

The only rights that can be granted or revoked are select, insert, update
or delete. There is no means of preventing the execution of a procedure,
but there is no need to, since one can only operate upon relations and
access to them is constrained.

In addition, full \textit{SQL} con-formant systems support the propagation
of grant rights by allowing a user to grant the ``WITH GRANT OPTION'' to
another. This particular feature is very useful for delegation and
re--publication and is discussed in more detail later \myRef{sec:safety}.

\subsection{Ownership and Autonomy}

Most organisations regret that their own database administrators have
complete access rights to their information and so are unlikely to extend
those rights to an external administrator.  Further to that most databases
are so complex, it is unlikely that any administrator would allow an
external administrator to create views for each group of external
accessors, but both of these requirements are a necessity for the system
proposed.

\subsection{Portability and Adaptability} 
\label{sec:portability}

Portability addresses how easily a system can be used in a different
technological environment. Adaptability: how easily that system can be
modified for a different end-user.

\paragraph{Portability}

A residual problem when remotely accessing a database has been the lack of
any standard for the database driver. This problem has been eliminated by
the adoption of the Open Database Connectivity, ODBC, standard, see
\cite{North:1996:UOS} for references. It provides an addressing scheme that
can locate databases on remote hosts and specify how access should be
gained (user-name and password). It does not propose a standard protocol.
ODBC drivers are still specific to the databases they drive{\myDash}a
client side needs to be installed for each type of database.  There is
directory service, one simply has to know the correct form of the address.

It is now possible to use a platform independent database querying engine,
namely the \Java Database Connectivity package. It relies upon each database
having a Uniform Resource Locator, URL, and the JDBC manager attempts to
load a driver class having a specified relationship to the URL. The driver
class can be loaded over the network, this means that client machines can
be configured for accessing a particular database with no down-time. Also,
because URLs are used to locate databases, the directory service is a
world-wide browser and one can use catalogues of URLs to locate the
resources needed.

\paragraph{Adaptability}

Because a URL is used to specify the driver class and the driver class can
be loaded over the network, it is possible to load a different driver class
for the same database: one that might have different access rights
available or access to different catalogues.

\subsection{Security and Safety}

\subsubsection{Security}

This topic covers practical aspects of database access. Can the queries and
results sent to and received from a database be confidential to the client?
The simple answer is yes, but most databases do not support secured
channels directly, it is necessary to have an encrypting and decrypting
agent placed in the communications channel between client and server
database. Using the JDBC package, it is possible to write a custom driver
which encrypts queries and sends to them its decrypting counterpart which
would then forward the query to the database and then encrypt the results.
More sophisticated protocols can be implemented by using remote procedure
calls between the encrypting and decrypting agents, such as the scheme
described in \cite{ch-o:eaves}. These could negotiate keys, add sequence
numbers and perform time--stamping transparently to the client and server.

\subsubsection{Safety}
\label{sec:safety}

\paragraph{Views}

One of the assumptions made in chapter \ref{cha:reqs} was that accessors
would use negotiated views of records and that these could be made safe by
granting rights to a set of accessors and revoking rights from all others.

There are two authorities which justify this: Minsky
\cite{Minsky:intentional} and Denning \cite{sec:denning:2}. Also one needs
to consider how access rights might be propagated.

\subparagraph{Branding, Tickets and Capabilities}

Minsky \cite{Minsky:intentional} describes a concept called
\textit{branding}. Essentially, every data type is branded and each user
has a set of brands that can be accessed. This is easily realized for a
database in the following way:

\begin{displaymath}
  \begin{aligned}[t]
    \begin{CD}
      Data Type @= View \\
    \end{CD}
  \end{aligned}
  \quad \text{and} \quad
  \begin{aligned}[t]
    \begin{CD}
      Brand @= Username \\
    \end{CD}
  \end{aligned}
\end{displaymath}

Although this might seem facile, \textit{SQL} is one of the few data access
languages that supports branding. It is not possible to brand objects in
most other programming languages because there is no access control
mechanism which could require a brand. It is possible to implement branding
in object--oriented programming languages that support a call--back
mechanism, but this is cumbersome. Only \Java has institutionalised it with
the \textsf{SecurityManager}, \textsf{GuardedObject} and the
\textsf{Permission} classes of their security architecture
\cite{sec:java:security,sec:gong}. An object--oriented system presents an
additional problem for branding, because it might be desirable to brand a
base class and allow access to it, but not to branded derived classes.
\Java avoids this problem by only allowing implementation classes to be
extensions of \textsf{GuardedObject}{\myDash}not interfaces or abstract
base classes.

Finally, it should be said that object--oriented database access languages,
such as those proposed by the Object Database Management Group
\cite{sft:odmg}, have not really addressed security issues in their
language proposals. It is well--known that safe and secure programming
needs to be implemented in the programming language\footnote{Again, see
  Denning's discussion of flow control in \cite{sec:denning}.}. Denning in
\cite{sec:denning:2} advocated a database system known as
\textit{System--R}, developed by \emph{IBM}, which supported branding and
had other attractive features. The query language developed for
\textit{System--R} was the prototype for \textit{SQL}.

It should be added that \textit{branding} is now considered to be a variant
of ticket--(or capability--)based authorisation. An accessor must be in
possession of a valid ``ticket'' to access an object \cite{db:pol:bertino}.

\subparagraph{Subjects, Objects and Rights}

On the occasions that authorisation systems are discussed it will be
necessary to use some special terminology. This is a pr\'ecis of a
description found in \cite{auth:jones}. Users of an information system are
usually designated as \textit{subjects}\footnote{This kind of subject is
  different from the subject that was described in the requirements chapter
  \ref{cha:reqs}. Subject in an authorisation context would be an accessor
  in the context of the system requirements.} and the data entities they
access as \textit{objects}. Which subjects may access which objects is
specified by a lattice model: this acts as an organisation hierarchy,
lattice models have useful properties which are discussed in
\myRef{sec:lattices}.

Every subject and object has a specified access classification. It is this
that determines whether access is granted. There are just two access
rights{\myDash}read and write{\myDash}and one special right: invoke and two
meta--rights, take and grant. These latter two are discussed in detail in
the next section, but it should be pointed out that the grant right is
something of a misnomer, it does include the ability to deny a right as
well. There is one relationship which is ownership.

The create and destroy rights can be thought of as the right to invoke the
create operation on a factory object or the destroy operation on an object
itself.

The creator of an object is its initial owner. Ownership can be transferred
and shared. Subjects are not owned by any other subject or object. All
objects are owned.

Subjects can create and destroy objects and objects can create and destroy
objects if they are the owner of the object or the owner grants permission
to invoke the destroy operation. Subjects can only be created and destroyed
by some special means.

A subject can also take on the r\^ole of an object, but an object cannot
take on the r\^ole of a subject. Subjects may try to access other objects,
objects other objects and, because subjects are also objects, subjects may
try to access other subjects and objects subjects.

Because every subject or object may access every other subject or object,
then, as far as data access is concerned, they can all be thought of as
objects. This makes the rule for granting access easier to express:

\begin{quotation}
  Access is only granted if the accessing object has a high enough security
  clearance for the object it wishes to access for the specified right.
\end{quotation}

There are two other rights which are more subtle in their operation. These
are discussed next.

\subparagraph{Taking and Granting Rights}

Denning in \cite{sec:denning} gives a fairly complete discussion of the
difficulties of taking and granting rights to data objects. In particular,
the ``Take--Grant'' model of Jones, Lipton and Snyder \cite{auth:jones}.
This model is constructed as follows: between a subject $S$ possessing
certain access rights and a data object $O$ which requires particular
access rights, there must exist a path of actions to take or grant rights
from and to other agents before $S$ can access $O$.

\begin{description}
\item[Take] A take action is performed by $S$, or an entity acting on
  the behest of $S'$, and takes access rights from others.
\item[Grant] A grant action is performed by another agent $R$ who grants
  access rights to $S$ or an agent, $S'$, acting for it.
\end{description}

\begin{figure}[htbp]
  \begin{center}
  \includegraphics{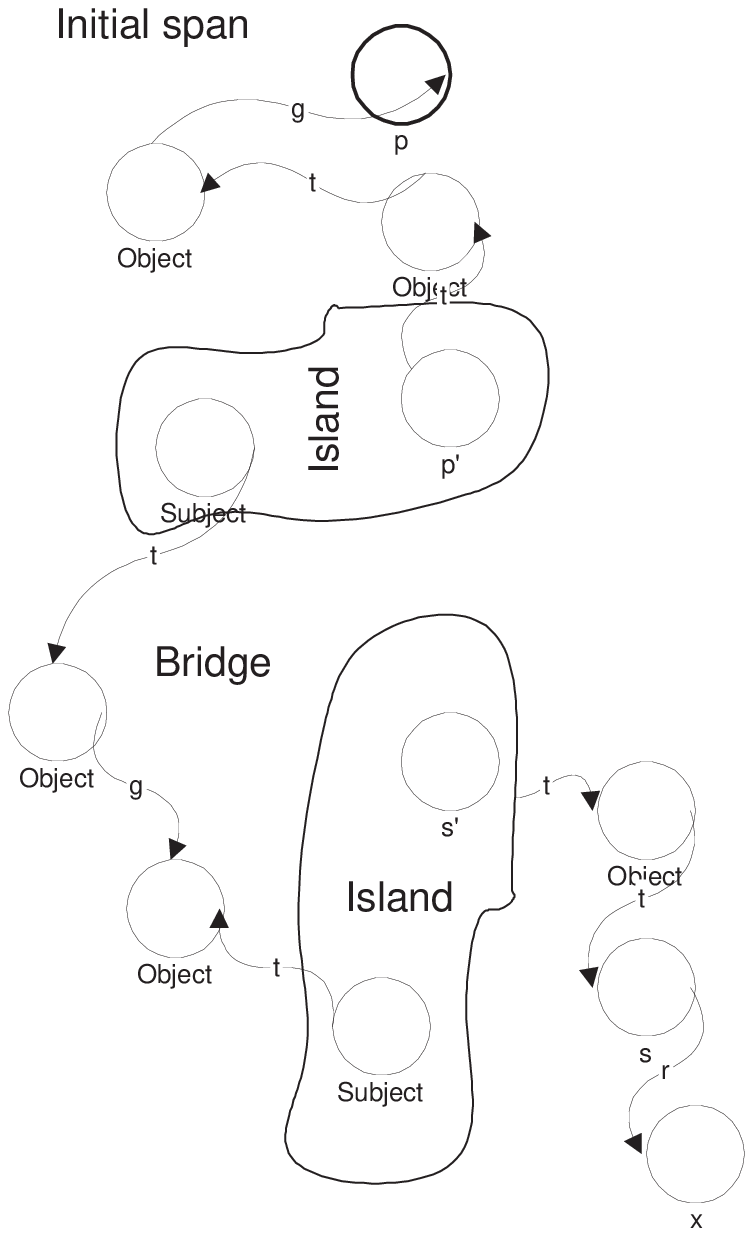}
  \caption{A take--grant path across islands and bridges}
  \label{fig:take-grant}
  \end{center}
\end{figure}

Clearly, if $S$ is to access $O$, then $S$ must find a $tg$--connected ($t$
for ``take'', $g$ for ``grant'') path to $O$. Jones, Lipton and Snyder show
that this path can be found in linear time. They introduce the concepts of
\textit{bridges} and \textit{islands}.  Islands are maximal $tg$--connected
subgraphs of subjects only, where everyone may take whatever rights the
others (on the island) have. A bridge is a $tg$--connected path which gains
access to an island{\myDash}a chain of take actions is an example of a
bridge. An initial span is a bridge to an island from a user. Figure
\ref{fig:take-grant} shows a principal $p$, using an initial span to reach
an island where $p'$ has a $tg$--path across a bridge to another island
where a terminal span from $s'$ to $s$ gains access to an object $x$.

This is exactly what people do every day when they make use of computer
systems. They log-on{\myDash}the initial span{\myDash}they are placed in the
island of their group and may access particular objects because of that.

As for databases, \textit{SQL} only supports a ``GRANT'' action{\myDash}
there is a corresponding ``REVOKE'' action{\myDash}and it is quite
difficult to conceive of a system that actually employs a ``TAKE'' action.
Bishop \cite{Bishop81} argues that take actions are only performed by
privileged users{\myDash}the superuser in a \textit{Unix} system,
\textit{Administrator} under \textit{Windows} and the \textit{DBA},
database administrator, under most \textit{SQL} systems{\myDash}because they
are able to grant rights to whomsoever they wish. It could be argued that
if a user in one role, such as system administrator, grants to himself in
another role, an ordinary user, a right that he would not normally have,
then this constitutes a take action. It would seem then, that a take action
is a grant action that does not require inducing the owner of the right to
grant it.

In which case, there are many examples of take actions. If one inspects the
\textit{CERT Coordination Centre's} archives, see \cite{sec:dns} for
example, it is all too clear that there are many ingenious ways to take
rights.  A common method is to force an overrun on a statically allocated
command buffer, which, if correctly formatted, will overrun into another
more privileged command buffer which can then make an illegal grant
command. These are implementation errors which one can only hope would not
appear in well--designed software{\myDash}\textit{CERT} do issue guidelines
for application programmers.

\textit{SQL} users can only grant access rights, this requires the owner's
permission which must be negotiated. If that negotiation is subject to peer
review, such as the group membership committees, then one must hope that
such membership committees will not grant membership rights lightly.

This, still, may not be a sufficient safeguard, because it may be the case
that members of a group have access rights they do not use, but, were they
to make use of them, they would have access to data they should not
have.

It is therefore necessary to require that as security clearance lattices
are constructed, it must be proven that there is no access path from lower
classes to information accessible to only higher classes and as Jones \etal 
make clear, this is possible in linear time.

\subparagraph{Republishing Databases}

There are two aspects to republication that need to be considered. Firstly,
if a user has legitimate access to the database through a safe system, what
prevents him from republishing it through an unsafe system. Secondly, it
would be desirable if a group can decide to grant a limited right of
republication to a sub--group, under the {\ae}gis of one of its members,
and possibly to allow access to other accessors who may not have been
vetted by the group, but who are vouched for by the member to whom the
republication rights have been granted.

The former can be dismissed quickly. It is practically impossible to
prevent republication of material gathered electronically, but it can be
costly to do. Some web--browsers now have the ability to deny the user the
ability to print or save a web--page \cite{web:hotjava} (and not store the
page in a local cache), but it is possible to capture the page by other
means. 

It is the cost of republishing without permission that makes republication
with permission so attractive that system designers should it make easy to
do. \textit{SQL} fortunately makes provision for this facility, but does
not require that it be implemented: it is the ``GRANT'' ``WITH GRANT
OPTION'': it grants to a user the right to grant the named right. This is
exactly what is needed to give to one trusted accessor the ability to name
his own group of accessors without having to duplicate the database.
Unfortunately, it will probably be the case that the people he wants to
grant access to will not be members of any groups known to the database
owner. However, if it is possible for the trusted accessor to give
certificates to the members of his group of accessors, then it should be
possible to give them a security clearance.

\chapter{A Prototype System}
\label{cha:simple}

This chapter describes a prototype database access system which is a proof
of concept development for many of the ideas and analysis discussed so far.

In the chapter discussing requirements, chapter \ref{cha:reqs}, emphasis
was placed upon making the relationships between the different parties
closed, see \myRef{sec:parties}, so that the management of access rights
could be self--governing.

The prototype developed implements some key processes described in
\myRef{sec:engineering:model}.

\begin{itemize}
\item Access negotiator agent in figure \ref{fig:access-neg2}.
\item Custodians establishing new precedents in figure
  \ref{fig:access-neg3}.
\end{itemize}

The latter of these is the self--governing process and the former triggers
its operation.

Some other processes were implemented for convenience: the database trader
agent figure \ref{fig:trader} and there is a simple means of submitting
queries and receiving formatted replies figure \ref{fig:delivery}.

The implementation is far from complete. The procedures implemented might
allow one to claim the system is safe, but it is by no means secure. There
is no encryption and no certificate technology has been deployed; so, data
transfer is all in cleartext and authentication is rudimentary.

\section{Technological Components}

The prototype system contains just two components:

\begin{itemize}
\item A Web--server \Jigsaw \cite{sft:prod:jigsaw}
  
  The web--server is unusual because it is wholly implemented in \Java.
  This makes the use of software agents much simpler{\myDash}they can
  loaded into the web--server and work within its secure environment. The
  web--server acts as the secure computing environment for all users
  {\myDash}accessors, owners and custodians.

  Referring to \myRef{sec:engineering:agents} the agents that were
  implemented in the prototype were implemented as \Java classes which were
  loaded and run by the web--server.
  
\item A Database \PostgreSQL \cite{sft:prod:postgres}
  
  This database system does provide some advanced features not usually
  found in similar products. It did not prove necessary to use them. The
  real attraction of using this product is that the source code for the
  database and for the \Java Database Connectivity driver
  \cite{sft:jdbc:postgres} is freely available.

\end{itemize}

The prototype system's security features are very limited. The requirements
for the system and whether they were implemented are listed in table
\ref{tab:prototype} and \ref{tab:prototype:1}.

\begin{table}[htbp]
  \begin{center}
    \begin{tabular}[left]{|p{1 in}|p{0.5 in}|p{2 in}|}
      \hline
      Requirement & Status & Notes \\
      \hline
      \begin{flushleft}
        Secure web--server access
      \end{flushleft}
      &
      \begin{flushleft}
        No
      \end{flushleft}
      & 
      \begin{flushleft}
        There is currently no web--server written in \Java that can
        implement secured channels. Even if there were, it would be
        difficult to obtain a licence for the cryptographic technology.
      \end{flushleft}
      \\

      \begin{flushleft}
        Secure database access        
      \end{flushleft}
      &
      \begin{flushleft}
        No
      \end{flushleft}
      & 
      \begin{flushleft}
        Again the web--server cannot support secured channels and, because
        the JDBC driver is also written in \Java, neither can the driver.
      \end{flushleft}
      \\

      \begin{flushleft}
        Safe database access
      \end{flushleft}
      &
      \begin{flushleft}
        Yes
      \end{flushleft}
      & 
      \begin{flushleft}
        Only makes use of user-name and password pair and the basic
        authorisation mechanism available within the hypertext transfer
        protocol.
      \end{flushleft}
      \\
      
      \hline
    \end{tabular}
    \caption{Prototype system: implementation \vs requirements (I)}
    \label{tab:prototype}
  \end{center}
\end{table}
      
\begin{table}[htbp]
  \begin{center}
    \begin{tabular}[left]{|p{1 in}|p{0.5 in}|p{2 in}|}
      \hline
      Requirement & Status & Notes \\
      \hline
      \begin{flushleft}
        Certificate Technology
      \end{flushleft}
      &
      \begin{flushleft}
        No
      \end{flushleft}
      & 
      \begin{flushleft}
        Certification Authority servers are costly and are effectively just
        a modern replacement for user-names and passwords.
      \end{flushleft}
      \\

      \begin{flushleft}
        Re--publication
      \end{flushleft}
      &
      \begin{flushleft}
        No
      \end{flushleft}
      & 
      \begin{flushleft}
        \PostgreSQL does not support the ``GRANT'' ``WITH GRANT OPTION''.
      \end{flushleft}
      \\
      
      \begin{flushleft}
        Adaptive Capabilities
      \end{flushleft}
      &
      \begin{flushleft}
        Yes
      \end{flushleft}
      & 
      \begin{flushleft}
        This is the whole point of the prototype and a simple rule--adding
        system was implemented.
      \end{flushleft}
      \\
      
      \hline
    \end{tabular}
    \caption{Prototype system: implementation \vs requirements (II)}
    \label{tab:prototype:1}
  \end{center}
\end{table}

\section{System Agents}

Regarding the implementation status of the agents
\myRef{sec:engineering:agents}:

\begin{enumerate}

\item Accessor Memberships Agent
  
  Not implemented. There are suitable systems available, for example, a
  moderated e--mail list would suffice for some applications. The prototype 
  system used the configuration feature of the \Jigsaw web--server to add
  new members.

\item Database Trader
  
  Implemented. It is possible for database owners to post the URLs of their
  databases to the \Jigsaw web--server.
  
\item Access Negotiator
  
  This is the key feature and both of its parts, figures
  \ref{fig:access-neg2} and \ref{fig:access-neg3}, have been implemented,
  and there no adaptive access control.

\item Query Formulator
  
  Not implemented. There are suitable systems available. There was a
  JDBC--based database query agent freely available, but this has been
  superseded by a commercial product. In the prototype, one can only submit
  queries on the name fields and one receives the whole record, from the
  assigned view, in return.
  
\item Query Delivery Agent

  Implemented. A query delivery agent is part of the database driver that
  is supplied by the database owner to the accessor.
  
\item Results Delivery Agent

  Implemented in the same way as the results delivery agent.

\item Results Presentation Agent

  Implemented. The accessor can choose which results format is used. The
  classes to present the data are posted to the web--server.
  
\end{enumerate}

Most of the information needed to prove that the contracts between the
parties have been adhered to \myRef{sec:compliance} is available from the
web--server's logs. These would need to be reconciled with the database
logs of the queries submitted and results returned.

\section{Adaptive Components}

The systems needed to provide adaptability support,
\myRef{sec:computational}, are not part of the prototype. They are more
experimental and need a surer mathematical basis before they can be
implemented.

\section{Software}

This section describes which parts of the system were implemented by which
component of the software. The software is described in more detail in
appendix \ref{cha:aidan}. The software used by the web--server to implement
the functions of the system has a different engineering configuration. 

\subsection{Implementation Engineering}

\begin{enumerate}
\item Database driver

  This provides the following system agents:
  \begin{itemize}
  \item Query Delivery Agent
  \item Results Delivery Agent
  \end{itemize}

\item Queriers

  This provides:
  \begin{itemize}
  \item Query Formulator
  \end{itemize}

\item Formatters

  This provides:
  \begin{itemize}
  \item Results Presentation Agent
  \end{itemize}

\end{enumerate}

The following agents have not been implemented in a way in which they can
be recognised as agents.

\begin{itemize}
\item Access Negotiator{\myDash}the access request component is implemented
  as a web--form; if access is granted, it is carried out using the
  configuration tools of the web--server.
\item Database Trader{\myDash}appear as web--pages within the web--server
  describing the databases available.
\end{itemize}

\subsection{Design Engineering}

\begin{enumerate}

\item Accessor Memberships Agent
  
  This was implemented by using the configuration features of \Jigsaw and
  therefore custodians interacted with a set of web--pages. As new users
  (or accessors) were added to the system, they were made part of the
  \mySrc{UserRepository}. Each user is allocated to a ``Realm'' which
  approximates to their ontology. (Unfortunately, a user can have only one
  realm in this implementation of \Jigsaw.) The \mySrc{UserRepository}
  actually makes use of a database as well. The idea underlying this is
  that each custodian, or supra--group for custodians, would provide a
  database of their members. Access to this database would be subject to a
  contract between the web--server provider and the supra--group
  administrators.

\item Database Trader
  
  This was implemented using an HTML form within the web--server known as
  the \mySrc{ResourceAdder}. A database owner fills in the form on the
  web--server describing the database he plans to make available. He must
  provide URLs for a suitable JDBC database driver and for compatible query
  formulator agent{\myDash}\mySrc{QueryByNames}.

  The JDBC database driver provides the two delivery agents (query and
  results). The query formulator provided is a simple one that only allows
  a single query by a patient name to be submitted.
  
\item Access Negotiator
  
  This has been implemented as a web-page form which generates an e--mail
  which is processed by an e--mail filtering program operated by the
  custodian. If the custodian grants access, an e--mail is sent to the
  owner who then creates the view granted, a user identity and posts a new
  database resource to the web--server. The \mySrc{UserRepository} is then
  modified by the owner using the configuration editor of the
  web--server. Periodically, e--mail messages were sent to users to tell
  them which database resources were available to them.
  
  There is also some simple rule application. When the e--mail from the
  web--server is sent to the custodian, the custodian is informed in that
  message who else from which realms has been granted what views. The
  custodian can reply to the e--mail specifying which of these is
  applicable can be applied to the incoming access request.
  
\item Query Formulator

  A simple version of a query formulator was implemented and appears as the 
  resource \mySrc{QueryByNames} in the web--server.
  
\item Query Delivery Agent

  This is provided by the database owner when it posts the URL of the
  database into the database trader.
  
\item Results Delivery Agent

  Same as the Query Delivery Agent.

\item Results Presentation Agent

  This is posted with the database driver at the database trader. It need
  not be provided by the database owner, it is a set of text--processors
  which are compatible with the output of the database.
  
\end{enumerate}

\section{Testing and Discussion}

\subsection{Tests}

A fairly large database (1400 records) was made available. Two sets of
tests: functional and performance.

\paragraph{Functional}

One set of tests showed that it was possible for accessors to send e--mail
messages to custodians who could then instruct database owners to add views
and that the users were notified of the views now available to them.

Another set of tests showed that the queries presented at the web--server
were satisfied in exactly the same way as they would have been had there
been no web-server interceding.

\paragraph{Performance}

The test conditions are summarised below:

\begin{enumerate}
\item Machines
\begin{table}[htbp]
  \begin{center}
    \begin{tabular}[left]{|l|l|r|}
      \hline
      Sub-system & Machine & Load \\
      \hline
      Client Web-Browser & Sun SparcStation 50MHz & less than 10\% \\ 
      Web-server & Sun Ultra 5.10 & 30\% \\ 
      Database & Sun SparcStation 50MHz & less than 10\% \\ 
      \hline
    \end{tabular}
    \caption{Test: Machines}
    \label{tab:test:machines}
  \end{center}
\end{table}

\item Query Parameters
  \begin{enumerate}
  \item Select on indexed key
  \item Display 38 field record
  \end{enumerate}

\item Database Parameters
  \begin{enumerate}
  \item 1356 records in table
  \item Query satisfied in 4.5 seconds with text formatting by the database
    acting alone.
  \end{enumerate}

\end{enumerate}

The results were:

\begin{table}[htbp]
  \begin{center}
    \begin{tabular}[left]{|l|l|}
      \hline
      Quality & Time (seconds) \\
      \hline
      Startup: Worst & 45 \\
      Startup: Best & 25 \\
      Worst & 30 \\
      Best & 10 \\
      \hline
    \end{tabular}
    \caption{Database Query Access Times}
    \label{tab:results}
  \end{center}
\end{table}

\subparagraph{Some analysis}

The web-server system is a lot slower than the database alone: up to ten
times slower on start-up and a constant 3 times slower when running.

\begin{enumerate}
\item Implementation: The performance has the characteristic of \Java
  systems, which is that first time operation is particularly slow, since
  the \Java virtual machine must load all the classes needed. Thereafter,
  access can be faster, but Java memory management is non-deterministic
  because classes are unloaded if space becomes cramped.
  
\item Communication Overheads: the browser is connected to the web-server
  which is connected to the database twice, once for the meta-data and once
  for the query. \PostgreSQL operates by receiving a query on one
  process{\myDash}the postmaster{\myDash}and forking another process for
  the connection and processing the query. When using the database's own
  front--end, the forking has been performed already and the same process
  is used for both meta--data and data enquiries.
  
\item Extra Processing: the web-server implementation formats the query
  results into an HTML table which the \PostgreSQL query frontend does
  not. The implementation of this formatting code is quite a general
  parser, so it is fairly inefficient.
\end{enumerate}

\subparagraph{Some discussion}

There is unlikely to be much improvement in performance from running the
processes on the same machine, since most operating systems treat internal
pipes in the same way as remote sockets.

Despite there being three processes involved there is not much parallelism
to exploit: the only opportunity might be with sparse queries in a very
large database, one could could be retrieved and formatted while the next
is sought at the database.

There is one very real reason why performance suffers:

\begin{itemize}
\item Looking up catalogues on each query submission
\end{itemize}

HTTP is a stateless protocol{\myDash}the server holds no state{\myDash}the
client must present all credentials on every access. The client only
presents his identifier when accessing the server, so a call has to be made
to the database to collect the catalogue and the view and then another call
to submit the query and the results.  When one considers this, it becomes
clear why the best performance of the system is at least twice as slow that
of the \PostgreSQL frontend.

The obvious improvement is to cache the catalogues against each accessor.
This is quite feasible in \Java, it would be possible to hold catalogues
against a \emph{cookie}\footnote{Cookies are just randomly generated
  signatures which both a client and a server hold, effectively a session
  identifier.} associated with the client at the web--server.
Unfortunately, caching is troublesome to implement. It might be the case
that a view is revised, in which the catalogues held in the client's cache
at the server would be out--of--date. This might lead to a security
breach.

Lately, \Java has added a better database access facilities under the
\textit{Enterprise Beans} framework. This manages cookies and reuses a pool
of database connections. Future implementations of web--based database
access products should use this technology.

\subsection{Discussion}

The proposed system's use of agent technology has been proven in concept
and it only remains to consider how it might be improved and extended to
provide all of the functionality given in \myRef{sec:engineering}.

\paragraph{Improved Implementation}

The performance can be improved with a cache for catalogues and improved
implementations of the formatting functions. However, as always with
software systems, it is best not to concentrate any coding effort on
improving performance until the design is complete, but the need for
improved performance must be recognised in the design.

\section{Realizing the Proposed System}

The prototype system is quite similar to the proposed system in its use of
agent technology: the database driver contains the query and results
delivery agents, which do the bulk of the work. The query formulator and
results presentation agents are pieces of software that the accessor
runs. The former uses the catalogues generated for the accessor, the latter 
uses the results when delivered. The prototype system does not implement
re--publication, but it would be relatively easy for it to do so.

To secure the system: encryption and certificate technologies have to be
used and each agent implementation has to be signed with a manifest so that
accessors, owners and custodians can be sure that the agent implementations
are authorised. Encryption is available, certificates can be bought and
there are application development libraries that allow the two to be used
together.  Application signatures under certificates are already available.
The only difficulty is to implement and integrate it all.

Integrating the application signatures for the agent implementations of the
query formulator and results presentation agents would take the form of
presenting them on the the web--pages where they are selected. The server
at the database can check the signature of the query delivery agent and the
web--server can check the signature of the results delivery agent.

All of the agents{\myDash}query and results delivery, query formulator and
results presentation{\myDash}would need to be assured that they have been
invoked on a mutually acceptable secure computing base, (SCB). This SCB
would need to sign the messages going between client, web--server and
database as well.

There is currently no well--evolved technology to do this. The only method
that is appropriate is secured remote procedure calls. There is a proposal
to extend the remote procedure call system of \Java to support this
\cite{ch-o:eaves}. When there is a solid technology on which to base these 
interactions then the system will be fully realizable.

The prototype does not perform any adaptive work{\myDash}it does not
attempt to classify views granted, secure computing bases or accessors.


\myPart{Political Control Mechanisms}


\chapter{Political Control of Access Privileges}
\label{cha:pol:1}

The first part of this dissertation has described sophisticated mechanisms
for securing data and making information safe, all of which would require
that policies be specified stating what information may be accessed by
whom, \myRef{sec:comp:access-rules}{\myDash}when and where data may be
accessed would also need to be made plain. It is the notion of policies and
the formulation of them that prompts one to consider using political
procedures to control them.

As will be seen, this thinking is not entirely new but it does not appear
to have been openly acknowledged that policy--making for system management
is a political process. Add to that, that politics is not as amenable to
quantitative analysis as economics.  There are many excellent mathematical
analyses of the operation of auctions, see \cite{econ:auction:mccabe}, for
example, but, in comparison, there are relatively few that describe how an
organisation structures itself.

The remainder of this chapter describes how authorisation systems can be
thought of as political systems and introduces the quantifiable concept of
norms. It then briefly describes some more suitable languages for
communicating rules and concludes with a discussion of the enterprise: its
goals and norms.

\section{Authorisation Policies for Databases}
\label{sec:auth:db}

Access control policies will be discussed in this section. The same
terminology used in \myRef{sec:safety} in the paragraphs \textit{Subjects,
  Objects and Rights} and \textit{Taking and Granting Rights} will be used.
The rules described there will be changed here to demonstrate how different
access control policies are implemented.

\subsection{Mandatory and Discretionary Access Control Policies}

The formulation of authorisation policies can be carried out in, broadly,
two ways\footnote{This is a summary of a longer, illustrated argument in
  \cite{scf:eaves}}: mandatory or discretionary access control.

Before describing the differences between them, it is best to describe what
they have in common. Both systems are under control of an
administrator. The administrator is the only entity that can create
subjects. In access control system design, the systems are idealised. It is
not possible to copy an object and use it in place of the original. This is
form of object protection is cryptographically feasible see
\cite{sec:gong}.

\paragraph{Mandatory access control}
  
If the lattice of information flows between subjects and objects cannot be
changed by the subjects while the system is operational, then access
control is mandatory. This is the case with military systems which use
variants of the Bell--LaPadula model \cite{BL:military}.  

The rights management rules for mandatory access control systems are as
follows: 

\begin{enumerate}
  
\item \label{sec:mac:owning} Ownership may not be transferred or shared
  by any subjects or objects. Only the administrator can change ownership.
  When ownership is changed, the object can be thought of as being
  destroyed and created anew. One would then apply \myRef{sec:mac:modify}.
  
\item \label{sec:mac:creation} When a subject creates an object, the
  object's access rights are fixed in that state and can only be changed by
  the administrator.
  
\item \label{sec:mac:modify} When a subject reads an object and modifies
  it, a new object is created and rule \myRef{sec:mac:creation} is applied.
  
\item The take and grant rights are denied to all except the administrator.
  
\end{enumerate}

Usually, the delete right is denied to all except the administrator.

\paragraph{Discretionary access control}

The alternative is to allow enough information to flow within a system so
that subjects and objects can interact with one another. They will then
evolve a lattice of information flows by specifying security clearances for
the objects they create between themselves{\myDash}Discretionary Access
Control. This is the case in most operating systems and \textit{SQL}
databases. A system administrator creates an initial set of subjects and
objects and the initial lattice flows by allocating the subjects and
objects to groups. As each subject or object creates another object, it can
specify which groups may access it. Only the system administrator can
create subjects and give them group memberships.

\begin{enumerate}
  
\item Rules \myRef{sec:mac:owning}, \myRef{sec:mac:modify} from mandatory
  access control are applicable.
  
\item \label{sec:dac:creation} When a subject creates an object, the object's
  access rights may only be changed by the owner.
  
\item \label{sec:dac:grant} All owners possess the grant right, but the
  grant right may only be possessed by owners. It may not itself be
  granted. The take right is denied to all except the administrator.
  
\end{enumerate}

Usually the delete right is available to owners and may be granted to
others.

\paragraph{\textit{Unix} System V release 4}

This system of rights is the same as that found in \textit{Unix} operating
system since System V release 4, (SVR4). System V release 3 and prior
versions allowed owners to transfer ownership which proved to be a major
security flaw. It has made inter-working between subjects more difficult.

\paragraph{Berkeley Standard Distribution $5.2$ \textit{Unix}}

The BSD of \textit{Unix} overcame the security problems of transferring
ownership in a much more flexible way. It does so by applying set semantics
to access control. Each subject has a group of his own, of which he is the
sole member and there are additional groups of which he is also a member.
When a user changes group ownership of a file, it can only be to a group of
which he is a member of; in this way subjects can act in a limited way as
an administrator.

BSD also allows designated areas to belong to a particular group. The
\textit{setgid} bit of a directory can be set and this ensures that every
file in that directory belongs to the group of the directory, which was set
by the subject when he created the directory.

Only the administrator can create subjects and groups, but the subjects now
have a means of administering group membership of objects indirectly. What
subjects designate as belonging to a group defines the membership of the
group. For example, if there are three users $a, b, c$ and four groups $W$,
$X$, $Y$ and $Z$, such that $W=\{a,b\}$, $X=\{a,c\}$, $Y=\{b,c\}$ and %
$Z=\{a,b,c\}$. If $a \grantRight{r}{Z} o$ then he allows everyone to read
it, but if $a \grantRight{r}{W} o$ then only $a$ and $b$ may read $o$.

(The problem with conventional operating systems is that $b$ may now copy
$o$ and make it available to $c$, which may not have been what $a$
intended. This is not allowed in an idealised system.)

\paragraph{SQL}

Finally in this section, the discretionary access control system of SQL
must be analysed: it is essentially the same as \textit{Unix SVR4}, but
allows a grant right. The grant right can be constrained to grant only read
or write from objects. In effect, this gives to table owners in databases a
\textit{setgid} bit on any of the views they create from the table.

\subsection{Authoritarian and Self--governing Access Control}
\label{sec:sgac}

\paragraph{Authoritarian}

Both mandatory and discretionary access control policies are effectively
authoritarian because of the privileged position of the administrator.
Under the more liberal schemes used in BSD and SQL, it is possible for the
subjects to determine the information flow, but the administrator controls
group membership and subject creation.

\paragraph{Self--governing}

For a system to be self--governing the privileged r{\^o}le of the
administrator must be removed by distributing it amongst the subjects. They
will vote collectively to specify group membership and whether new subjects
can be admitted.

Subjects could then choose to migrate to groups which are able to accept
them and would be able to provide them with better information. The act of
migrating to a different group with a different collective administration
is very similar to electing a political leader. It is, in fact, a
generalisation of the election process. This is a simplification of
political control that is used in the analysis of a well--known economic
policy issue: the Tiebout model \cite{econ:tiebout:tiebout}. It can be
thought of as an adaptive optimisation problem \cite{Miller:tiebout} where
voters cluster around norms of their expectations.

\subsection{Adaptive Discretionary Access Control Policies}
\label{sec:access:adaptive}

There are some system proposals which attempt to discern norms of behaviour
and use them to constrain the information flows between the subjects and
objects of a system. Minsky proposes a system of laws under which software
systems would work \cite{minsky89a,db:pol:minsky}.  A paper by Rabitti
\etal \cite{Rabitti:1991:MAN} describes additional authorisation generation
mechanisms to support the lattice model for an object--oriented database.
The innovation of the system is that it generates its authorisation policy
as it operates by generalising current behaviour to form norms.
Authorisation is viewed as having three dimensions:
 
\begin{description}
\item[Expression] Authorisations specified by users, which are known as
  \emph{explicit} and those that are derived by the system, known as
  \emph{implicit}.
\item[Direction] An authorisation can be \emph{positive}, stating what may
  be done, or \emph{negative} stating what may not be done.
\item[Strength] An authorisation may be \emph{strong}, in which case it may
  not be overridden, or \emph{weak}, in which case it can. Strong
  authorisations can be thought of as axiomatic.
\end{description}

This model has been extended \cite{db:pol:bertino} and a recent
contribution by Castano \cite{db:pol:castano} introduces metrics that can
be used to generate norms, including:
\begin{itemize}
\item Operation compatibility
\item Subject similarity co--efficient
\item Authorisation compatibility
\item Semantic correspondence
\item Clustering of subjects
\end{itemize}

All of which seem to be methods of ascertaining if different subjects (or
objects) belong to the same type. If the subjects and objects are typed
already, then some concrete questions can be resolved by using abstract
rules. If a subject $z$ is able to grant rights to an object $o$, if $z
\grantRight{r}{x} o$ and $\typeOf(x) = \typeOf(y)$ then $z
\grantRight{r}{y} o$ is implied.

\subsection{Preferential Logics and Operators}

If two organisations are to share information, then a new organisation is
formed which contains the authorisation hierarchy of both. This requires
that the two information flow lattices be combined and this, in turn,
requires a well--defined logic to do so. There has been some research into
preferential logics \cite{schobbens:preference} and some useful operators
have been defined. This work is based on graph--theoretical analysis of the
flow lattices which is something taken up later in this dissertation.

\section{Suitable Languages}

If preferential logics are to operate on authorisation hierarchies of
organisations using databases, the combined hierarchy will need to be
communicated to all interested parties. There are already some suitable
languages for this purpose. 

\begin{description}
\item[KQML] The Knowledge Query Manipulation Language is an
  agent--communication language and is described in \cite{KQML}. It is part 
  language and part protocol.
\item[KIF] The Knowledge Interchange Format is an ontology definition
  language defined in \cite{KIF}. 
\end{description}

Both of these languages are explained in more detail in appendix
\ref{cha:kqml}. For the time being, \textit{KIF} would be used to define
the boundaries within which agents may operate, (see
\myRef{sec:classifying}, paragraph \textit{Ranking Groups}.) \textit{KQML}
would be used to communicate \textit{KIF} descriptions of group membership
lists and rules to the different databases.

The attraction of \textit{KQML} is that it is a more enterprise--oriented
language and has been proposed for governing the interactions of
agents. \textit{KIF} is a formal language which would define the
information model for a set of collaborating organisations.

\section{Enterprises, Goals and Norms}
\label{sec:enterprises}

In chapter \ref{cha:reqs} the Open Distributed Processing modelling
perspectives were introduced. The least understood of these is the
enterprise model. It is considered to be a statement of the goals that an
enterprise wishes to achieve{\myDash}it would include ``The Mission
Statement''{\myDash}and the management structure that coordinates the
enterprise to achieve its mission. An enterprise is a network of
\emph{performative}, \emph{constative} and \emph{normative} agents.

\begin{description}
\item[Performative] A performative agent can claim that it has carried out
  some action: Executive function.
\item[Constative] A constative agent can judge if the action has been
  carried out: Judicial function.
\item[Normative] A normative agent determines which performative agent
  should do what task and which constative agent should verify it:
  Legislative function.
\end{description}

Most information processing systems have performative and constative
agents which are machine--processes, but the normative agents are usually
human.

\paragraph{Example: A Library}

System designers allow policies to be specified for the number of books
that may be borrowed simultaneously for particular classes of
borrowers. They also allow policies for fines to be specified. The policies 
actually in place at a particular library will vary.

Implicit in the design of the system are norms of behaviour expected from
borrowers. The local policies can be tuned to the behaviour of the
borrowers at a particular library when the library system is installed.

The enterprise model for the library system has been specified in part by
both the designers of the library systems (the policies that may be
effected) and its administrators (how those policies are put into effect).
The library system will have many performative and constative agents: when
a borrower takes a book out a performative agent initiates a process which
will invoke a constative agent to determine if the borrower would exceed
his quota. The only normative agents are the library administrators who
determine the local policies. It may be possible to have adapting normative
agents which are programs{\myDash}they might, for example, set the level of
fines relative to a cost index, for example.

But if there are new norms of behaviour: can the enterprise model's system
of normative agents cater for them? A simple case might be a new class of
borrower, the system designers may have made it possible to add new classes
of user. A normative agent, in the form of one of the library
administrators, will then add a new class of user and allocate people to
it.

But some norms of behaviour may be not be so easy to cater for. For
example, if borrowers feel that fines are too high and borrowing periods
are too short, they might organise some collective action: they choose to
take their full quota of books out and return them on the same day. This
appears to have no effect on the library system other than an increase in
turnover, but it is very annoying for the library staff.

Under these circumstances, would it be possible to cater for this new
behaviour by the borrowers? Would it be possible, for example, to charge
for re--shelving if books are returned too soon? Would it be possible to
isolate the borrowers who are taking part in the collective action and
enforce the re--shelving cost on them alone? Would it be possible to invoke
a new process altogether, for example, preventing users from entering the
library if they appear to be taking part in the collective action. If that
were the case, how could a borrower appeal against the decision to bar
him. 

Most library systems, like most information systems, do not have the degree
of flexibility needed to integrate new policies based on new norms of
behaviour without significant re--engineering. There are even fewer
information systems that are able to generate and instigate new polices
without human intervention.

\section{Summary}

Political processes cater for changes in norms of behaviour: the judiciary
reports increased numbers of adjudications, the executive reports increased
workloads, the legislative modifies the law to accommodate the changes in
behaviour observed by the judiciary and legislature.

By contrast, information systems are usually incapable of changing to
accommodate new norms of behaviour. One of the reasons for this is that it
was, until recently, very difficult to re--engineer an information system.
As can be demonstrated with the prototype system described earlier, chapter
\ref{cha:simple}, processes can be specified by the interactions of agents
and these agents can be replaced, upgraded and relocated without any loss
of service.

The remainder of this dissertation addresses norms of behaviour and how to
isolate them.



\chapter{Preference Aggregation}
\label{cha:top}

The database access system proposed should provide some adaptive
discretionary access control partly supported by automated deduction based
on a rule-base of precedents \myRef{sec:auth:db}. There will be some
relevant information available within the system on which to base these
decisions \myRef{sec:req:granting}, but it is unquantified. This chapter
concentrates on finding general quantification methods{\myDash}preference
aggregation and collective choice procedures.

\section{Generating Security Hierarchies} 
\label{sec:hier}

Generating security hierarchies is an exercise in classification. This
section describes the information that needs to be classified. This is a
generalisation of the descriptions and examples given in
\myRef{sec:comp:access-rules}.

\paragraph{Information}

The information available is derived from the relationships in
\myRef{sec:parties} and \myRef{sec:classifying}:

\begin{itemize}
\item Databases hold records
\item Views are made up of sets of fields
\item Views are made up of sets of record ranges
\item Custodians make views
\item Accessors hold views
\item Accessors have group memberships
\end{itemize}

Group membership is effected by issuing certificates as described in the
engineering model \myRef{sec:engineering:model}, in particular in figure
\ref{fig:membership}.

\paragraph{Rules}

There are information--ordering rules which are derived from the ranking of
individuals and their membership of groups, see \myRef{sec:classifying}
paragraph \textit{R\^oles and Ontologies}.

\begin{enumerate}
  
\item A set of fields may be accessed by an ontology.
  
\item There is some set of group memberships which maps an accessor to an
  ontology.
  
\item Within each group, members rank one another by seniority.
  
\end{enumerate}

and because of this, information can be ordered.

Two consistency rules can be added.

\begin{itemize}
\item The most senior member of an ontology is allowed access to a maximal
  set of records for the view.
  
\item The most senior member of an ontology is allowed access to a maximal
  set of fields in the view.
\end{itemize}

\paragraph{Hierarchies}

Seniority hierarchies are needed for the following:
\begin{itemize}
\item Within groups
\item Within an ontology, \ie across groups.
\end{itemize}

\paragraph{Voting Processes}

Votes will need to be taken within an ontology to determine:
\begin{itemize}
\item Which fields, and 
\item Which records
\end{itemize}
it would be desirable to access. Accessors vote by making requests of
custodians.

Votes will need to be taken from custodians to determine:
\begin{itemize}
\item Which fields, and 
\item Which records, by
\item Which accessors
\end{itemize}
can be accessed. Custodians vote by granting (or denying) requests.
Custodians may also create views to meet accessors' requirements.

\paragraph{\textit{Note Bene}}

Groups are usually of a broad professional concern, \eg British Medical
Association, Royal College of Surgeons; ontologies will usually be
project--related, \eg ``study of cellular immune responses against KSHV in
HIV infected patients during anti-retroviral treatment''.

Only groups rank, ontologies inherit the rankings from groups, but groups
have to be ranked relative to one another or seniority levels within groups
have to be equated with those in other groups. This may give rise to
anomalies, it will be seen later that these anomalies should appear in the
preference hierarchies and, hopefully, will be detected before the combined
preference hierarchy is put in place.

\section{Preferences, Values and Norms}

In section \myRef{sec:sgac}, it was argued that a self--organising access
control system would allow subjects wishing to access objects a choice of
groups to join. These groups would emerge around norms of behaviour. In
section \myRef{sec:classifying}, it was argued that an ordering for
individuals could be used to form new groupings which would better align
with the ordering of views of data in different ontologies.

What is needed is a means of converting the preference hierarchies derived
from classifying individuals and data views into norms of behaviour which
would then be used to reclassify groups. To do this, a metric is needed.

That there is a means to do this is partly justified by Swanson \etal
\cite{Swanson:entropy}. Swanson, a theoretical accountant, analyses the
function of money in economies and notes that it is used as an indicative
measure of abstract quantities such as: ``worth'', ``liquidity'' and
``earnings potential'' in real instances. Anyone who analyses systems
within an object--oriented programming paradigm knows the distinction
between a class and an object: an object is an instance of a class. In the
real world, classes are an artificial construct and there are only objects,
so given an object, how does one know to which class it belongs in a
particular context. So, for example, a particular dog would belong to the
class \textsf{Canine} and possibly the class \textsf{Pet}. A metric is
needed to measure how many of the qualities of \textsf{Pet} exist in this
dog.

A class is an expected norm of behaviour and classification involves
ranking behaviours relative to one another. If one could say that a class
has a set of behaviours with \emph{expected} rankings, then it would be
possible to state, with some statistical certainty, if an object could
belong to a class if its actual rankings for its behaviours are close to
the expected rankings of the class.

\subsection{Norms from Preferences}

Norms are expressed by means of preferences. Table \ref{tab:choice:1} is a
simple voting procedure that illustrates a difficulty in collective choice
theory \cite{scf:eaves}, which is known as \emph{Susceptibility to
  Irrelevant Alternatives}.

\begin{table}[htbp]
  \begin{center}
    \begin{tabular}[left]{|l|r|r|r|r|}
      \hline
      & \multicolumn{4}{c|}{Policy and Ranking} \\
      Voter & w & x & y & z \\
      \hline
      i & 4(3) & 3(-) & 2(2) & 1(1) \\
      j & 4(3) & 3(-) & 2(2) & 1(1) \\
      k & 1(1) & 2(-) & 4(3) & 3(2) \\
      \hline
      & 9(7) & 8(-) & 8(7) & 5(4) \\
      \hline
    \end{tabular}
    \caption{The Borda ``Preferendum''}
    \label{tab:choice:1}
  \end{center}
\end{table}

There are three voters: $i, j, k$ and they are asked to rank:
\begin{itemize}
\item Four policies: $w, x, y, z${\myDash}rankings in plain text
\item Three policies: $w, y, z${\myDash}rankings given in brackets
\end{itemize}

Using the \emph{Borda Preferendum} voting system to aggregate their
preferences, the results are:
\begin{itemize}
\item Four policies: $w > x, x = y, y > z$
\item Three policies: $w = y, y > z$
\end{itemize}

When there were four policies, it was quite clear that $w$ was preferred to 
$y$, but with only three policies $w$ and $y$ are equally favoured. An
anomaly like this undermines confidence in political choice.

\subsection{Values from Preferences}

Voters $i$ and $j$ rank $w>x>y>z$ and voter $k$ $y>z>x>w$. They all agree
that $y>z$, but not on $w>x$. If one were to group $y$ with $z$ to form
$y'$ and $w$ with $x$ to form $w'$, then they disagree on the merits on
$w'>y'$. This seems to indicate that the two groups of voters $\{i,j\}$ and
$\{k\}$ have different values at a higher level of abstraction.

To give this example a little more intuitive credibility, the three voters,
$i$, $j$ and $k$, have been asked to rank four individuals $w,x,y$ and $z$
on their ability to fulfil a task. This poll of their opinions has actually
revealed that one voter has a different idea (or value) of what is the most
important ability needed to fulfil this task: for example, $i$ and $j$ may
have concurred that ``trustworthiness'' is the most important ability.  $k$
feels that ``trustworthiness'' is important, \emph{but} that some other
quality such as ``ability to take the initiative'' is \emph{more}
important.

\subsection{Values are Relative}

Put simply, $\{i, j\}$ have assessed $\{w,x,y,z\}$ with a different set of
values from $\{k\}$. Discerning the underlying values of voters provides
two courses of action, which could be used to eliminate the effect of an
irrelevant issue, the courses are:

\begin{itemize}
\item Either: refine the issues
\item Or: refine the voters
\end{itemize}

\paragraph{Refining Issues}

Refining the issues would require conditional questions to be posed. For
example: ``Rank $w,x,y,z$ in order of their trustworthiness'' and ``Rank
$w,x,y,x$ in order of their ability to take the initiative''.

\paragraph{Refining Voters}

Refining the voters is simply discarding or downgrading the rankings of
voters who do not meet one norm. Although this might seem undemocratic,
most proportional representation voting systems do this.

\subsection{Trading Goals}

In the management of information systems, it is \emph{not} appropriate that
more abstract goals be traded against one another. For example, some users
may require an information system to be ``safe'' and some require it to be
``fast''. Safe and fast are abstract. The users will request quite specific
features: the general goal of safe may appear as a wish-list of a dozen or
more safety features; the general goal of fast as another dozen or so. If
these were presented on a combined wish-list, and a vote taken, then if the
voters were equally split between those who want a fast system and those
who want a safe system, then the resulting system would have some fast
features, some safe, but fulfil the requirements of neither.

The more fruitful approach is to find those who want the system to be
safe and those who wish it to be fast and separate them so that they
use different, relatively autonomous information systems.

\section{Organisational Structure}

\subsection{Managing Database Access}

To relate this work to the management of database access hierarchies,
Castano \etal proposed a number of metrics to determine similarities of
usage between subject and operations. They are described in section
\myRef{sec:access:adaptive}.  These metrics could be used to inform a
custodian of a record if the accessor requesting access was behaving
normally for someone with the accessor's interests.

All that is known of a potential accessor is his group memberships and the
view of the records he wishes to see. There are also precedents set by
others which could be used.

\begin{table}[htbp]
  \begin{center}
    \begin{tabular}[left]{|l|p{1 in}|p{2 in}|p{2 in}|}
      \hline
      Comparor & Comparee(s) & Qualities & Notes \\
      \hline

      Custodian & Accessor &
      Group memberships &
      \begin{flushleft}
        Subject similarity: specifies how similar the custodian is to the
        accessor
      \end{flushleft} \\

      Accessor & Accessor's Peers in Group(s) &
      Group memberships &
      \begin{flushleft}
        Subject similarity: specifies how similar the accessor is to other
        members of his own group(s)
      \end{flushleft} \\
    
      Accessor & Accessor's Peers in Group(s) &
      Views &
      \begin{flushleft}
        Authorisation compatibility: specifies how similar the
        accessor's requested view is to that used by others in the same
        group(s)
      \end{flushleft} \\
    
      Accessor & Other accessors with Access &
      Group memberships &
      \begin{flushleft}
        Authorisation compatibility: specifies how similar the accessor is
        to those who have access
      \end{flushleft} \\
    
    \hline
  \end{tabular}
    \caption{Comparative metrics for Security Clearance Class Membership}
    \label{tab:distances}
  \end{center}
\end{table}

There would also need to be a set of measures that would rank the
trustworthiness of different secure computing environments.

\subsection{Lattices as Organisational Structures}

\label{sec:lattices}

The metrics referred to above and described in section
\myRef{sec:access:adaptive} are just some of those that might prove to be
useful in helping a custodian decide if access should be granted to a set
of records. There will no doubt be other metrics that might prove to be
useful. What has to be addressed is how one should go about quantifying the
choices that custodians have made \visavis those who have been allowed
access to records.

The metrics will be distance measures which effectively measure two
orthogonal qualities:

\begin{itemize}
\item Trustworthiness
\item Relevance
\end{itemize}

These are used to form lattices: the vertical dimension indicates the
degree of trustworthiness, the horizontal dimension indicates the degree of
relevance to an ontology. Referring again to the medical profession:
seniority should be a reliable trustworthiness measure and the field of
medical work would be a suitable relevance measure: so a General
Practitioner conducting private research into cardiology would be ranked
lower than a consultant cardiologist with respect to cardiology, but the
cardiologist would be ranked lower than the GP with respect to general
practice issues.

\begin{figure}[htbp]
  \begin{center}
  \includegraphics{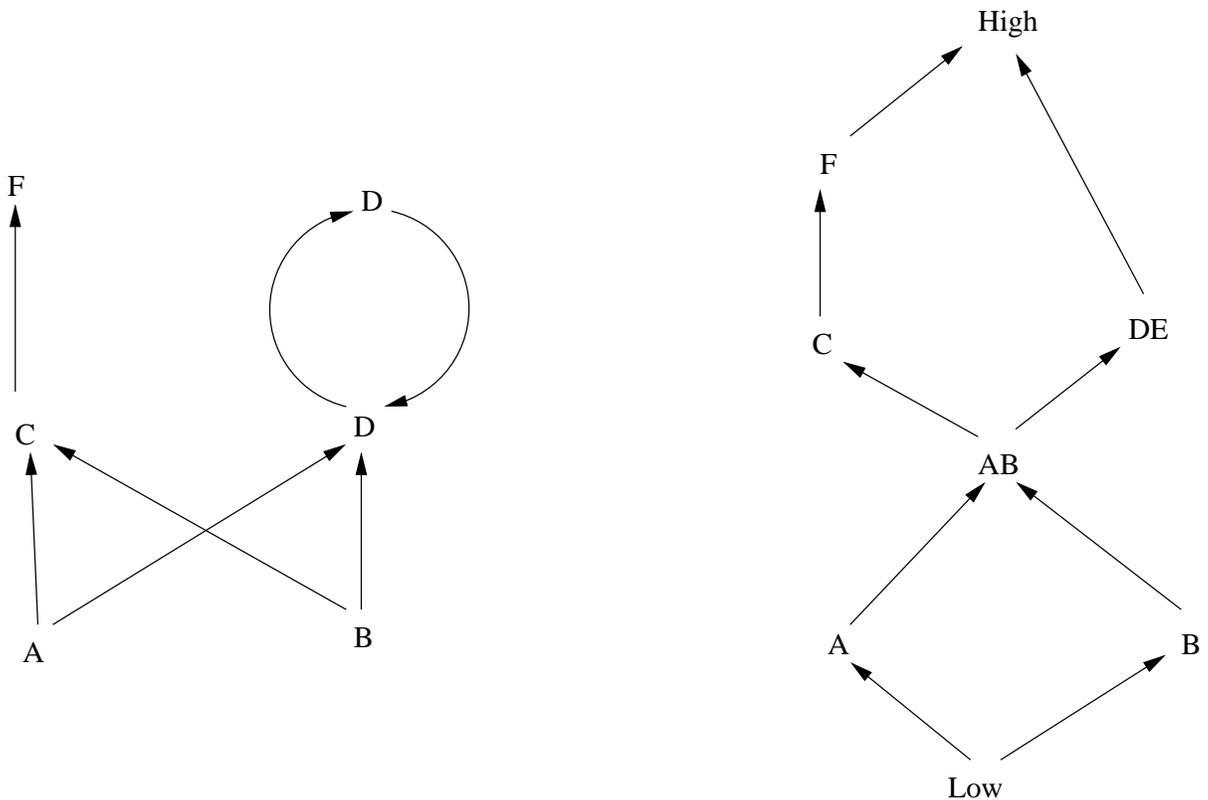}
  \caption{A lattice produced from a set of preferences}
  \label{fig:lattices}
  \end{center}
\end{figure}

These lattices are the basis for the structure of an organisation (or
community): they form an organisation chart of the relationships that exist
between groups. Organisation charts take the form of multi--way trees and
lattices can be forced to take the same structure. This is done by
introducing lowest and highest common points to all of the branches and
creating joining groups where cycles or multiple choices exist. An example,
is given in figure \ref{fig:lattices}

The custodians will be called \emph{voters} when discussing preference
aggregation, since their choices are effectively votes. Each
custodian/voter will have their own lattice of preferences for the different
types of accessors. The problem is to aggregate their personal lattices
with those of their peers to form an aggregate lattice.

\pagebreak[5]

\section{About Preference Aggregation}

\subsection{History and Importance}
\label{sec:pref:history}

A leading contributor to the field of decision theory (to which preference
aggregation belongs) feels that computer--aided decision--making would
take the following form.

\begin{quotation}
  Computers will play an increasingly important role in applications during
  the next century. Along with routine tasks of data compression and high
  speed analysis, computers will have ever more sophisticated programs to
  ferret out interactively the most salient features of decision problems
  and structure problems accordingly. A few well--directed questions about
  values, acceptable risks and probabilities will yield a proposed solution
  or menu of solutions. Programs will discern the most likely directions
  for improvements and determine their promise by means of challenge
  questions. Sensitivity analyses that account for vagueness in preferences
  and probability judgements, and tend to discount marginal improvements,
  will be standard\footnote{\cite{scf:fishburn:hist} This paper is also an
    excellent summary of key results and the directions that research in
    decision theory is taking.}.
\end{quotation}

There are a number of problems with preference aggregation: it is a
collective choice procedure and whenever there are more than two choices
from which voters can choose there is no method of choosing that cannot be
subverted by sophisticated voting strategies\footnote{Based on the theorems
  of Arrow, \cite{inst:arrow}, and summarised in \cite{scf:eaves}.}. This
does not invalidate collective choice procedures that decide between more
than two options, it means that one must be cautious when interpreting the
results.

\subsection{Structure and Notes}

Before moving onto the analysis of preference aggregation, it is best to
clarify some terms. This notation is the same as that used in
\cite{scf:eaves}. The mechanics of the analysis will use graph theory,
unfamiliar terms can be found in the appendix \ref{cha:graph}. There are a
number of examples of the operations, these are also given in the
appendix. 

\subsubsection{Preferences and Indifferences}

\paragraph{Indifferences}

A weak ordering allows voters to express their indifference between two
choices. A strong ordering does not allow indifferences. A weak ordering is 
a tri--state logic. Indifference is represented by $=,\quad x =
y$. Preference is, incidentally, represented by $> \droptext{or} <$.

\paragraph{Transitivity}

It is usually assumed that preference relations are transitive, \ie $x>y$
and $y>z$ then $x>z$.

\subsubsection{Lattices}

\paragraph{Acyclic}

\begin{figure}[htbp]
  \begin{center}
  \includegraphics{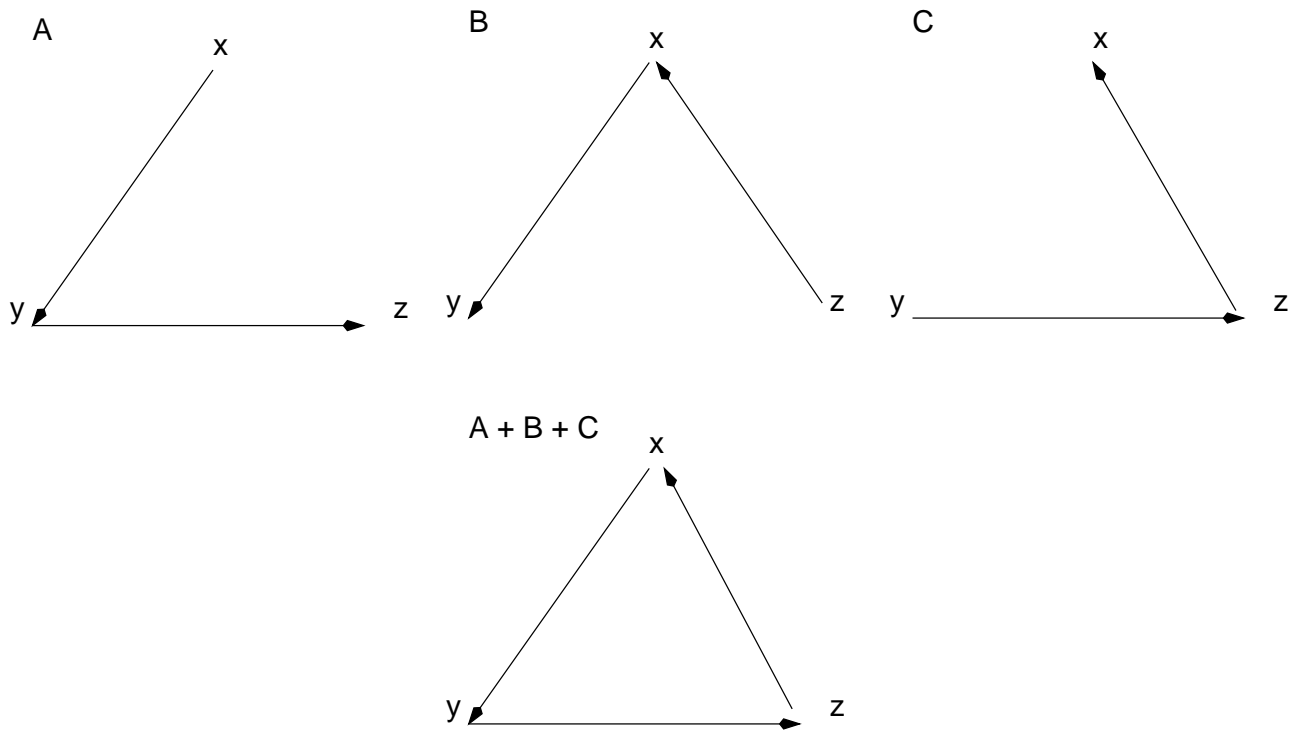}
  \caption{The voting paradox}
  \label{fig:v-paradox}
  \end{center}
\end{figure}

A lattice of preferences must be an acyclic structure. It is usually
assumed that individuals will not have a cycle in their own lattice of
preferences, but when aggregated it is possible a cycle will arise. The
voting paradox, see table \ref{tab:v-paradox} and figure
\ref{fig:v-paradox}, is a simple, and irreducible, example of this.

\begin{table}[htbp]
  \begin{center}
    \begin{tabular}[left]{|l|r|}
      \hline
      Voter & Ranking \\
      \hline
      $A$ & $x>y>z$ \\
      $B$ & $z>x>y$ \\
      $C$ & $y>z>x$ \\
      \hline
    \end{tabular}
    \caption{The voting paradox}
    \label{tab:v-paradox}
  \end{center}
\end{table}

Referring to the figure showing a lattice produced from a set of
preferences, figure \ref{fig:lattices}, it can be seen that cycles can be
removed by grouping classes together. In the case of the voting paradox,
this is not possible, since all classes are equally highly--rated.

\paragraph{Unanimities}

These are very useful. A unanimity is a preference upon which everyone
concurs. A unanimity can be said to express a
\textit{Pareto--optimal}\footnote{See \cite{scf:eaves} for the original
  definition in the context of economic welfare.} choice and there are
degrees of \textit{paretian} choice.

\subsection{Methods and Representations}

The remainder of this chapter looks at methods and representations that
could be used to form, manipulate and quantify the ordering of lattices of
preferences. Essentially, individuals preference ordering superimposed upon
one another give rise to a connected graph, which has to be reduced to a
spanning tree which will be the collective preference hierarchy. This is a
well--analysed task of graph theory \cite{graph:christo}, but there are a
choice of spanning trees for a graph, the preference hierarchy has to be
the most preferred.

To simplify preference hierarchies to spanning trees, a number of graph and
set manipulation techniques have to be used and some distance measures
developed.

\begin{enumerate}
\item Cycles
  
  These are the principal indicator of an anomaly in choice. They have to
  be detected and, in some way, eliminated.

\item Unanimities

  It will be seen that unanimities can be used to partition lattices and
  can therefore be used to simplify them, which will allow anomalies to be
  avoided.

\item Chains and Anti--chains
  
  Another useful structural indicator is the length of each chain in a
  lattice and the number of anti--chains in the lattice. Chains and
  anti--chains are described in appendix \ref{cha:graph}, but they can be
  thought of the branches of a lattice. Each chain has at its head an
  anti--chain. There will always be at least as many anti-chains as chains.
  A small number of chains of similar length will give rise to fewer
  anomalies than a large number of chains of dissimilar lengths.

\item Distance measures

  A useful distance measure will be introduced which will allow lattices to 
  be compared to one another. This, combined with a knowledge of the
  chain/anti--chain composition of the lattice will allow less anomalous
  but sufficiently similar lattices to be chosen.

\item Deciding Sets
  
  There are alliances between voters on issues. A method of determining the
  underlying values of voters will be introduced which will allow issues to
  be grouped together within lattices.

\end{enumerate}


\section{Cycles and Topological Entropy}
\label{sec:cycles-entropy}

The most obvious sign of an anomaly of choice is a cycle in a preference
lattice. An aggregate preference lattice should be free of cycles or the
effect of cycles should be controlled.

\begin{enumerate}
  
\item Complete Cycles
  
  A \emph{complete cycle} can be interpreted as the voters assessing
  policies with values which are entirely different. The voting paradox,
  figure \ref{fig:v-paradox} is an example of a complete cycle.

\item Incomplete Cycles
  
  A \emph{cycle} that is not complete is a statement that a group of people
  differ on the merits of some subset of policies. This may be due to the
  policies being too similar or too different. An example of this is the
  Borda Preferendum anomaly in \ref{tab:choice:1}. In the former case, the
  ``conflict'' represented by the cycle may be manageable by eliminating
  one of the policies. In the latter case, it would be best to partition
  the voters and allow them to resolve their differences using some
  arbitration process.

\end{enumerate}

The following two methods are recommended for detecting and quantifying the
effect of cycles.

\subsection{Using an Adjacency Matrix}

Meyer and Brown \cite{scf:meyer} have developed a measure which they call
the \emph{topological entropy} of voting preferences\footnote{Topological
  entropy is more formally defined in \cite{math:top:ent}.}. It enumerates
cycles and their length.

Using the \emph{adjacency matrix} representation for each individual's
preferences{\myDash}the graphs of which and the matrices themselves are
given in appendix \ref{cha:graph}{\myDash}these can be added together and
normalised by dividing by $\#I = n = 3$ to give:
\begin{equation*}
  F = \frac{1}{n} \sum_{i=1}^{n} F_i = \frac{1}{n} (F_A + F_B + F_C) =
  \begin{pmatrix}
    1 & \frac{1}{3} & \frac{2}{3} \\
    \frac{2}{3} & 1 & \frac{1}{3} \\
    \frac{1}{3} & \frac{2}{3} & 1
  \end{pmatrix}
\end{equation*}

Following Meyer and Brown \cite{scf:meyer}, the topological entropy of a
choice matrix $F$ can be expressed thus:
\begin{equation*}
  \begin{split}
    \Lambda_{F} & = \max ( \text{Eigenvalue of}\  F) \\
    S(F) & = \log \Lambda_{F}
  \end{split}
  \label{eq:choice:entropy}
\end{equation*}

The value of $\Lambda_{F}$ gives the length of the longest cycle were one
to generate $F^n$. If logarithms are taken to the base $n$ then an entropy
of 1 indicates a policy cycle of length $n$ and an entropy of 0 a
policy cycle of length $1$, \ie only each policy with itself. For this
example, $S = 0.690759$. (This is different from Meyer and Brown's
formulation as they had applied simple majority rule to the matrix $F$,
effected by rounding up to 1 or down to 0.) By \emph{not} applying the
social choice function, one can analyse how the choices of the individuals
would be interpreted by a social choice function.

Unfortunately, this is not as useful a measure as one might hope. The
graph representation, and thus the matrix form, does not handle statements of
indifference particularly well. For example, $a>c, b>c$ and $a>c, b>c, a=b$
have different representations:
\begin{equation*}
  \begin{aligned}[t]
    F_1 = 
    \begin{pmatrix}
      1 & 0 & 0 \\
      0 & 1 & 0 \\
      1 & 1 & 1 
    \end{pmatrix}
  \end{aligned}
  \droptext{,}
  \begin{aligned}[t]
    F_2 = 
    \begin{pmatrix}
      1 & 0 & 1 \\
      1 & 1 & 0 \\
      1 & 1 & 1 
    \end{pmatrix}
  \end{aligned}
\end{equation*}
and correspondingly different entropies: $S(F_1) = 0, S(F_2)=0.63093$
because the largest eigenvalue for each is 1 and 2 respectively.

\subsection{Using a Transition Matrix}

Probably better is to follow the approach used in games theory
\cite{DN:games} and use transition matrices\footnote{This approach has been
  used by at least one other author to show that ergodic Markov processes
  are in fact voting processes \cite{prob:voting:matloff}, so that would
  substantiate its use here.}. In the language of probability theory,
either voter $A, B$ or $C$ will get their way{\myDash}assuming they are
statistically independent, which means that they would vote sincerely. One
transition matrix can be formed by addition, if one assumes they are
equally probable to influence the election.  The formation of the
transition matrices is not difficult but is long--winded, so it is
described in the appendix \ref{cha:graph}.  Normalisation here simply
ensures that the sum of the probabilities in each row\footnote{Some authors
  prefer column sums to be 1 and so use the transpose of the matrix.}
continues to be 1. The final result is very easy to interpret: the
probability of the system reaching a state where $a$, $b$ or $c$ is more
dominant than the other is exactly equal $\frac{1}{3}$ so $a=b=c$. More
formally, using the formulation of the matrices described in the appendix
\ref{cha:graph}.

\begin{equation*}
  F = \frac{1}{n} \sum_{i=1}^{n} F_i = \frac{1}{n} (F_A + F_B + F_C) =
  \begin{pmatrix}
    \frac{1}{3} & \frac{1}{3} & \frac{1}{3} \\
    \frac{1}{3} & \frac{1}{3} & \frac{1}{3} \\
    \frac{1}{3} & \frac{1}{3} & \frac{1}{3}
  \end{pmatrix}
\end{equation*}

Although it is now self--evident in this example that no one policy is
preferred over the others, the method is to find the eigenvalue that has
the value 1 and the most probable final state of the system will be
described by the corresponding eigenvector, which specifies the probability
of each policy being in force in the infinitely long--term. An entropy
measure can then be generated from this steady state probability vector in
the usual way.

The transition matrix representation is more intuitive for indifferences,
for $a=b>c$, the matrix would be $\Bigl(
\begin{smallmatrix}
  \frac{1}{2} & \frac{1}{2} & 0 \\
  \frac{1}{2} & \frac{1}{2} & 0 \\
  \frac{1}{2} & \frac{1}{2} & 0
\end{smallmatrix}
\Bigr)$, \ie indifference means equiprobable. Unfortunately, this
representation is still not satisfactory, since mutual indifference has the
same representation as a mutual contra--position, \ie if $a>b>c$ and
$c>b>a$ for two voters and $a=b=c$ for another two voters both would yield
the same steady state eigenvector and entropy, but, with the former, the
two voters are in conflict over the relative merits of $a$ and $c$ and, in
the latter, they are in agreement.

Nonetheless, this formulation of topological entropy is still quite useful
for examining the effect of irrelevant alternatives, see table
\ref{tab:choice:1}, which has a transition matrix representation as follows
\begin{equation*}
  \begin{aligned}[t]
    F_A, F_B =
    \begin{pmatrix}
      1 & 0 & 0 & 0 \\
      1 & 0 & 0 & 0 \\
      0 & 1 & 0 & 0 \\
      0 & 0 & 1 & 0
    \end{pmatrix}
  \end{aligned}
  \droptext{,}
  \begin{aligned}[t]
    F_C =
    \begin{pmatrix}
      0 & 1 & 0 & 0 \\
      0 & 0 & 0 & 1 \\
      0 & 0 & 1 & 0 \\
      0 & 0 & 1 & 0
    \end{pmatrix}
  \end{aligned}
\end{equation*}
in the first, four option case and when the second--ranked option is
removed, the matrices are
\begin{equation*}
  \begin{aligned}[t]
    F_A, F_B =
    \begin{pmatrix}
      1 & 0 & 0 \\
      1 & 0 & 0 \\
      0 & 1 & 0 
    \end{pmatrix}
  \end{aligned}
  \droptext{,}
  \begin{aligned}[t]
    F_C =
    \begin{pmatrix}
      0 & 0 & 1 \\
      0 & 1 & 0 \\
      0 & 1 & 0
    \end{pmatrix}
  \end{aligned}
\end{equation*}

\begin{table}[htbp]
  \begin{center}
    \begin{tabular}[left]{|l|r|r|}
      \hline
      & Four Options & Three Options \\
      \hline
      Borda Order & $w>x=y>z$ & $w=y>z$ \\
      \hline
      Markov Order & $w>x>y>z$ & $w>y>z$ \\
      \hline
      Probabilities & $\frac{12}{23}, \frac{6}{23}, \frac{3}{23},
      \frac{2}{23}$ & $\frac{6}{11}, \frac{3}{11}, \frac{2}{11}$ \\
      \hline
      Entropy & 0.843 & 0.905619 \\
      \hline
    \end{tabular}
    \caption{Topological Entropy for the Borda ``Preferendum'' with and
      without an irrelevant alternative}
    \label{tab:choice:3}
  \end{center}
\end{table}

Referring to table \ref{tab:choice:3}, it is clear that the order produced
by the transition matrices preserves the supremacy of option $w$ over $y$.
Neither method ranks $x$ as being the same as $y$ \myDash which is another
anomaly of the \textit{Borda} Preferendum \myDash but it preserves the
ordering of $y$ over $z$.

(A \emph{Mathematica} package is available that calculates the topological
entropy using transition matrices \cite{sft:aidan:math}.)

Incidentally, this entropy measure as described only yields the entropy of
the largest cycle: it may be the case that there are a number of lesser
cycles, and that would mean that the entropy of the system is greater,
since there is more confusion over its content.

\section{Unanimities} 
\label{sec:lattice:p}

The preceding section looked at measures that detected and quantified the
length of a cycle in an aggregation of a set of lattices, where the
aggregation was achieved by simple addition to produce a likelihood of a
particular policy being chosen amongst all others. A number of difficulties 
arose from that discussion. These are addressed now and, it will be seen,
that the exploitation of any unanimous choices can be used to isolate
cycles so that they can resolved.

\begin{enumerate}
  
\item Policies or Policy Preference 
  
  It will prove to be more useful to determine which policy preference is
  most often, or unilaterally, stated. For example, in the discussion of
  table \ref{tab:choice:3}, it was clear that all parties preferred $y$
  over $z$.  If this were represented in a graph, it would be clear that
  one edge is traversed more than any of the others. This can be used as a
  measure of how well--ordered a preference hierarchy.
  
  If an edge (policy preference) is always traversed in one direction,
  then, it is a pareto--optimal preference: one policy, $x$, is always
  preferred over another, $y$, but other policies may be preferred to $x$.
  
\item Similarities or Contrasts
  
  Under a strong ordering, two statements by voters $A$ and $B$ of $x>y$
  and $y>x$ respectively could indicate:
  \begin{itemize}
  \item Either: a juxtaposition: $A$ and $B$ have entirely different values
    and that $x$ and $y$ are also different from one another and exemplify
    the differences between voters $A$ and $B$.
  \item Or: a similarity: $A$ and $B$ have similar values and that $x$ and
    $y$ both embody that same value, so that $A$ and $B$ are unable to
    sincerely and consistently choose between $x$ and $y$.
  \end{itemize}
  
  This problem does not exist with a weak--ordering, but in an aggregation
  it can be obscured: an equal number preferring $x>y$ and $y>x$ would
  suggest $x=y$ in the aggregate.

\end{enumerate}

Pareto--optimality is a desirable quality for a edge in an aggregate
preference lattice\footnote{Sen \cite{scf:borda:sen} proposed that
  pareto--optimality should be ranked higher than simple majority
  preference and \cite{scf:borda:farkas} has quantified this.} and it is
denoted here as unanimity. In their paper, Batteau \etal
\cite{scf:theory:batteau} defined two forms of Paretian choice: weak and
strong. The former was defined as the case when all voters agree on $x > y$
for one, or more, $y$ and the latter was defined as the case where all
agree on $x > y$ for all possible $y$.  It will be seen that both forms can
be discerned using methods described here (a strong pareto choice is, in
fact, a \emph{source}).

What is needed is an additional fitness measure that highlights if a set of
preference lattices contain Pareto--optimal statements of preference.

If one can find some unanimities, so much the better, but, if there are no
unanimities, one simply has to remove enough voters or enough policies to
produce one (or more). Both of which are quite meaningful ways of
partitioning the two sets, since a unanimity is a value.

\begin{definition}[Preference and Indifference Graphs] \label{def:p-graph}
  This simple innovation makes use of two graphs to represent the lattice.
  One graph will represent the strong orderings between the vertices and
  will be the transitive closure of the directed graph. The other will be
  an undirected graph expressing the indifferences, which will, usually,
  contain mostly isolated vertices.

  The directed graph will be the preference graph \emph{P--graph} and the
  undirected graph will be the indifference graph \emph{I--graph}.
\end{definition}

\subsection{P--graph aggregates}

The preference lattices will be aggregated in some optimal way for a
particular voting rule. Aggregate weights will be assigned to each directed
edge{\myDash}the aggregates of the di-graphs is a 2--di-graph, \ie a
multi--graph with at most two directed edges between each pair of vertices,
going in opposite directions.

\begin{definition}[For and Against--Weights] If the P--graphs are
  aggregated then for any pair of policies $u$, $v$ two weights will be
  assigned to each directed edge between the pair of policies. The greater
  will be known as the \emph{for--weight} and the lesser will be known as
  the \emph{against--weight}. The vertex having the greater for--weight
  with respect to another vertex will be said to \emph{dominate in
    aggregate} the other vertex.
\end{definition}

\subsection{Unanimities, Sources and Sinks}

\begin{definition}[Unanimity] When an edge within an aggregate of the
  P--graphs has a zero against--weight and a non--zero for--weight then it
  will be called a \emph{unanimity}.
\end{definition}

A complete cycle does not prohibit or invalidate a unanimity, since the
latter is a statement of preference between just two (or more) policies.  The
other policies, which cause the cycle, can be dismissed as irrelevant
alternatives, if need be.

\begin{definition}[Simple unanimity] A unanimity $\vec{uv}$ is called
  \emph{simple} if it is the only unanimity which involves either $u$ or
  $v$.

  Note that a unanimity can be a compound simple unanimity, \eg
  $\vec{uvw}$, if $\vec{uv}$ and $\vec{vw}$ are both simple
  unanimities.
\end{definition}

\begin{definition}[Complex unanimity] A complex unanimity $\vec{uv}$ is one 
  where either $u$ or $v$ is not unanimously linked to another unanimity of
  the other. For example: $\vec{uv}$, $\vec{uw}$ are both unanimous, but
  not $\vec{vw}$.
\end{definition}

A unanimity may prove to be a \emph{sink} or a \emph{source}.

It may arise that each vertex of an incomplete cycle is part of a
unanimity.  It might be helpful to interpret this as follows: all the
voters may agree that $w$ is the best policy, but each voter ranks the
other policies $x$, $y$ and $z$ differently. These definitions for cycles
will be used:

\begin{definition}[A Dominated Cycle] If each vertex
  in an incomplete cycle is dominated by the same vertex, then this
  configuration is a \emph{dominated cycle}.
\end{definition}

The voters are agreed on what is best, but cannot agree on what is
worst. For example, all prefer $w$ to a cycle of $x, y$ and $z$.

\begin{definition}[A Dominating Cycle] If each vertex
  in an incomplete cycle dominates the same vertex, then that cycle is
  known as a \emph{dominating cycle}.
\end{definition}

The voters are agreed on what is worse, but cannot agree on what is
best. For example, all prefer $x, y$ or $z$ to $w$.

It should be apparent that a complex unanimity is a form of cycle, since
two (or more) policies have a policy which unanimously dominates the cycle
or is dominated by it. Complex unanimities can be treated as if they were
one of the cycles.

\begin{definition}[Condensed Aggregate I--Graph] If all individuals have
  the same sub--set of vertices connected in their respective I--graphs then
  that sub--set can be reduced to one ``super--vertex'' and this change can
  be carried over to the \emph{individual} P--graphs.
\end{definition}

In effect, the individuals are unanimous in their indifference between
particular policy pairs.

\begin{definition}[Condensed Aggregate P--Graph] A \emph{condensed graph}
  of the aggregate P--graph can be formed by reducing a sub--set of
  vertices to a single super--vertex if the sub--set of vertices forms one
  of the following configurations:
  \begin{enumerate}
  \item A simple unanimity including compound simple unanimities
  \item A dominated cycle
  \item A dominating cycle
  \end{enumerate}

  A cycle may also be a complex unanimity.
\end{definition}

The point of doing this is that it simplifies the aggregate graph. Simple
unanimity can be replaced by the dominating vertex without loss of
information, but the cycles may not. In effect, the resolution of the cycle 
has been deferred. This appears to be the process followed in
 \cite{sec:denning} when producing a lattice from a set of relationships,
see figure \ref{fig:lattices}.

\subsection{Unanimous Properties}

\paragraph{Common Indifferences}

This simple algorithm can be used to determine if any of the vertices in the
I--graphs contain the same indifferences.

\begin{mrule}[Common Indifferences] The algorithm is as follows:
  \begin{enumerate}
  \item Form the adjacency matrix for the I--graph of each individual.
  \item Form the intersection of the adjacency matrices, $\bigcap_{i}
    A(i)$.  If any element of the resulting matrix is non--zero, then the
    edge represents a common indifference.
  \end{enumerate}
\end{mrule}

\paragraph{For and Against--Weights} \label{sec:unan:for}

\begin{mrule}[For and Against--Weights] For a set of preference lattices:
  let $i$ range over the voters, let $u$ and $v$ range over the policy
  vertices.
  
  \begin{enumerate}
  \item Form the P and I--graphs, $P(i)$ and $I(i)$ for each individual.
  \item Find any common indifferences and carry them from the aggregate
    I--graph to the P--graphs.
  \item Form the \emph{reach matrices} for the P--graphs for each
    individual: $\textbf{Q}(i)$.
  \item Form the sum of the reach matrices, $\Sigma_{i} \textbf{Q}(i)$.
  \end{enumerate}
  
  The resulting aggregate matrix $\textbf{Q}$ will have a central diagonal
  of zeroes and the entries can be evaluated for their properties.
\end{mrule}

All the other elements of the matrix $q_{uv}$ will have a complementary
edge $q_{vu}$, the following conditions apply:

\begin{enumerate}

\item $q_{uv} \ge 1$ and $q_{vu} = 0$

  Then there is a unanimity of $u>v$
  
\item $q_{uv} > q_{vu}$
  
  Then preference $u>v$ has a for--weight of $q_{uv}$ and an
  against--weight of $q_{vu}$.

\item $\Sigma_{v} q_{vu} = 0$ \ie a row is zero

  Then $u$ is a source.
  
\item $\Sigma_{u} q_{vu} = 0$ \ie a column is zero

  Then $v$ is a sink.
  
\end{enumerate}

One can also find just the unanimities by performing a logical conjunction
of all the matrices and locating the entries that remain true.

\subsection{Degree Of Unanimity}
\label{sec:unan:1}

Given that it is possible to find unanimities within aggregate graphs, it
would be useful to have an entropy metric based upon it. Entropy is a
probabilistic measure and it is required that the total number of events
needs to be calculated and also the number of events observed.

Unfortunately, there is no simple calculation for the total number of
different preference orders given a set of policies, because using weak
ordering complicates the formation of the permutations.  The algorithm for
the calculation of the total number of preference orders is relatively
simple however, but there are no tables that can be consulted, so:

\begin{mrule}[Preference Orders] The total number of different weak
  preference orders for $n$ policies can be calculated as follows:
  \begin{enumerate}
  \item Generate all the partitions\footnote{ \cite{graph:skiena}, p. 56} of
    $n$.
  \item Calculate the number of permutations for each partition, call this
    $N(\operatorname{partitions})$.
  \item For each partition find the number of ways in which the policies
    could be allocated to the elements of the partition,
    $N(\operatorname{policies})$.
  \item Multiply $N(\operatorname{partitions})$ by
    $N(\operatorname{policies})$ for each partition and sum them together.
    \begin{equation}
      \label{eq:preferences}
      \Sigma_{\operatorname{partitions}}
      N(\operatorname{policies}) \
      N(\operatorname{partitions}) 
    \end{equation}
  \end{enumerate}
\end{mrule}

A \emph{Mathematica} package is available \cite{sft:aidan:math} that
performs the calculation. Table \ref{tab:orders} lists the total number of
different preference orders for up to 6 policies and clearly shows how
large the search space becomes.

\begin{table}[htbp]
  \begin{center}
    \begin{tabular}[left]{|r|r|}
      \hline
      Policies & Orders \\
      \hline
      1 & 1 \\
      2 & 3 \\
      3 & 13 \\
      4 & 75 \\
      5 & 541 \\
      6 & 4683 \\
      \hline
    \end{tabular}
    \caption{Number of Different Preference Orderings for $n$ policies}
    \label{tab:orders}
  \end{center}
\end{table}

As for the number of events observed (or distinct P--graphs produced by the
voters) the data needed for the entropy calculation is the count of voters
for each distinct P--graph. A suitable entropy metric will be presented in
the next chapter. For now, if an entropy metric is available, then there
are two entropy values for a preference hierarchy that can be calculated.
These will give an indication of the degree of cohesion amongst the voters:

\begin{enumerate}
\item P--graphs before condensing 
\item P--graphs after condensing
\end{enumerate}

The entropy of the former will indicate how varied the opinion of the
voters is with respect to the totality of choices available to them. The
latter is best used in generating a \emph{conditional entropy} amongst the
voters. Just to illustrate why one would need both figures: consider two
sets of votes, $A$ and $B$, the same issues but different equal numbers of
voters. Both yield only two distinct P--graphs with the same proportion of
voters supporting each: in vote $A$, the two P--graphs share \emph{no}
unanimities, while in vote $B$ there are a number of unanimities which can
be exploited which allow both P--graphs to be combined to one. In vote $A$
there is still complete disagreement, but in vote $B$, apart from perhaps
some ``agreements to differ'' in the form of incomplete cycles, there is
enough general agreement to form a single preference lattice.

\section{Chains}

The form and number of \emph{chains} and \emph{anti--chains} is a useful
indicator of the structure of aggregate lattice. These concepts are
described more precisely and references are given in appendix
\ref{cha:graph}. Briefly, one can say that a chain is an arm of the lattice
and an anti--chain can be formed from all those elements in each chain that
are not directly connected.

Under some circumstances, it may prove preferable that there be a few long
chains and one short anti--chain. The anti--chain will contain all the
maximal elements of a preference hierarchy and each chain will contain one
element from the anti--chain.

For example, a company with a manager and a clerk in each of four
departments has four chains and four anti--chains. The length of each chain
is two. If each manager is responsible to a director and the directors meet
together on a board, then there are four chains and no anti--chains. The
anti--chains are removed by the board of directors where all conflicts are
resolved. The length of each chain is now three. 

To eliminate all the anti--chains it is necessary to extend all the chains
by one. If a chain is long, it is more likely to produce a cycle under
preference aggregation.

\subsection{Size of Largest Anti--Chain}

An important theorem regarding the \emph{Decomposition of Partial Orders}
can be used as a fitness measure. (A partially ordered set is a set of
orderings and is the most appropriate mathematical structure to use for the
analysis of preference hierarchies.) It is relatively easy to compute the
maximum anti--chain of a partially--ordered set (and therefore a preference
lattice). The largest anti--chain is the \emph{maximum independent set} of
the order. The theorem is given in appendix \ref{cha:graph}.

\section{Distance Functions}
\label{sec:top:distance}

The problem of preference aggregation has been addressed by researchers in
other fields with different goals. In particular, statisticians have
researched \emph{pairwise comparisons}, there are two papers which have a
direct relevance to preference aggregation as described here: Thompson and
Remage \cite{stats:pairwise:thompson} which deals with generating rankings
from sets of pairwise comparisons each of which form a strong ordering;
and, Singh and Thompson \cite{stats:pairwise:singh} which generates
rankings from weak orderings. Singh and Thompson's paper is the basis of
what follows: all theorems, corollaries and lemmas are due to them.
Unfortunately, Singh and Thompson analyse preferences with the goal of
producing alternative rankings, effectively ``league tables'', of all the
policies rather than an aggregate lattice, but their analysis is also valid
for the latter.

\subsection{Bigraphs}

Singh and Thompson define a \emph{bigraph}, $\langle X, C \cup D \rangle$,
which has a set of vertices $X$, $C$ contains all the statements of
indifference and $D$ all the statements of preference{\myDash}the I-- and
P--graphs respectively as described above \myRef{def:p-graph}. They make a
distinction between \emph{circuits}, which are directed cycles, and
\emph{loops}, which are cycles that may contain undirected edges.  They
also define \emph{semi--completeness}{\myDash}a generalisation of
completeness{\myDash}a lattice can be said to be semi--complete if for
every distinct $x_i, x_j$ in $X$, $i \ne j$, there exists a path between
$x_i$ and $x_j$ or vice--versa. With these they are able to state the
following theorem:

\paragraph{Partial Rank Order}

\begin{theorem}[Partial Rank Order] A partial rank order $P$ is an ordering of
  the elements of $X = \{ x_1, \dots, x_m \}$, so for $P=\{ p_1, p_2,
  \dots, p_m \}$, then each $p_i \in P \equiv x_j \in X$. $P$ is a
  effectively a permutation of $X$. A relation $R$ is a partial rank order
  when $(p_j, p_i) \not\in R$ whenever $j>i$. A partial rank order is
  reflexive, anti--symmetric and transitive.

  \begin{itemize}
  \item A relation $R$ on $X$ determines at least one partial rank order
    iff it is loop--free
  \item A relation $R$ on $X$ determines a unique partial rank order iff it 
    is loop--free and semi--complete.
  \end{itemize}
\end{theorem}

This simply states that for any graph, if there are no loops, then a
ranking of the elements can be imposed. It does not specify how the
elements of $X$ are compared to one another.

\paragraph{Pairwise Comparisons}

Three relations are introduced which permit pairwise comparisons; they are:
$E$, equivalence; $T$, strong order; $W$, weak order.

\begin{definition}[Indirect Relations] \label{def:top:distance:indirect}
  Let $T$ be a strong order relation (asymmetric, anti--reflexive and
  transitive), $E$ be an equivalence relation (symmetric, reflexive and
  transitive) and $W$ be a weak order relation (reflexive and transitive).
  For any pair $(x, y)$ of elements of $X$:

  \begin{enumerate}
  \item $(x, y) \in E$ iff $x=y$ or they are in the same loop of $C \cup
    D$.
  \item $(x, y) \in W$ iff $x=y$ or there is a path from $x$ to $y$ in $C
    \cup D$.
  \item $(x, y) \in T$ iff there is a directed path from $x$ to $y$ in $C
    \cup D$.
  \end{enumerate}
\end{definition}

Using these definitions it is possible to develop the following theorems
for $\langle X, C \cup D \rangle$.

\begin{enumerate}

\item There is at least one partial rank order

\begin{theorem} In which case, the following conditions are equivalent:

  \begin{itemize}
  \item $C \cup D$ is circuit--free.
  \item $T$ is a preference relation.
  \item $T$ is loop--free.
  \item $T$ determines at least one partial rank order on $X$.
  \end{itemize}
  
\end{theorem}

So a preference relation $T$ defines a partial rank order if it is
loop--free.

\begin{corollary} $T$ determines at least one partial rank order, $P=\{
  p_1, p_2, \dots, p_m\}$, on $X$ for $(p_i, p_{i+1}) \in W$ for $i=1,2,
  \dots, m-1$ iff $C \cup D$ is semi--complete and circuit free.
\end{corollary}

\begin{lemma} $W$ is a partial rank order iff $C=\emptyset$ and $D$ is
  circuit--free.
\end{lemma}

\item Exactly one partial rank order

\begin{theorem} The following conditions are equivalent:

  \begin{itemize}
  \item $C = \emptyset$, $D$ is circuit--free and semi--complete.
  \item $W$ is a simple order (transitive, anti--symmetric and reflexive).
  \item $T$ is loop--free and complete.
  \item $T$ determines a unique partial rank order on $X$.
  \end{itemize}
  
\end{theorem}

\begin{corollary} $T$ determines a unique partial rank order, $P=\{ p_1,
  p_2, \dots, p_m\}$, on $X$ for $(p_i, p_{i+1}) \in D$ for $i=1,2,
  \dots, m-1$ iff $C = \emptyset$ and $(p_i, p_j) \in D$ for $i>j$.
\end{corollary}

\end{enumerate}

\subsection{Changing the Orientation of Edges}

Bigraphs are not always circuit--free and it will be necessary to change
the orientation of edges to make them so. There are three ways an edge can
be re--oriented:

\begin{enumerate}
\item Reverse the direction of a directed edge.
\item Assign a direction to an undirected edge.
\item Make a directed edge undirected.
\end{enumerate}

If an aggregate graph is generated, it can be forced to be circuit--free by 
applying a combination of re--orientations. How many of these, and which of
the three they are, can form the basis of a distance metric.

Singh and Thompson prove a theorem which states that it is immaterial if
one deletes edges or re--orients them. This might at first seem
contentious, but if one bears in mind that a partial rank order is a
transitive relation, then deleting an edge would remove a circuit and the
transitivity of the relation would retain some preferences. For any graph
there is a class of maximal circuit--free sub--bigraphs each one of which
forms a partial rank order.

\begin{theorem}
  If $C_1 \cup D_1$ is a maximal circuit--free sub--bigraph of a complete
  bigraph $C \cup D$, then $T(C_1 \cup D_1)$ determines a unique partial
  rank order iff $C_1 = \emptyset$.
\end{theorem}

The maximal circuit--free sub--bigraphs can be enumerated by generating all 
of the \emph{Hamiltonian} paths.

\begin{theorem}
  If $C \cup D$ is a complete bi--graph, then there is a one--to--one
  correspondence between the maximal circuit--free sub--bigraphs $D_1$ of
  $C \cup D$ and Hamiltonian paths in $C \cup D$.
\end{theorem}

\subsection{Maximum Likelihood Preference Relations}
\label{sec:top:max-likelihood}

\paragraph{Probability Function}

Fortunately, Singh and Thompson developed their graph--theoretical analysis
into a probabilistic model. Unfortunately, some more notation has to be
introduced.

\begin{notation}
  \begin{description}
  \item[$X$] the set of policies, $X = \{ x_1, x_2, \dots, x_m \}$. Typical
    elements $x_i$ and $x_j$. Each distinct pair of elements is compared
    in, statistically, independent trials to yield:
  \item[$(x_i, x_j)$] An ordered pair of $X$ which can be either $x_i
    \rightarrow x_j$ or $x_j \rightarrow x_i$ or $x_i = x_j$. Each
    comparison may be carried out $n_{ij} > 0$ times.
  \item[$\mathscr{I}$] is the set of all subscript pairs $(i,j)$ that have
    been compared and $1 \leq i < j \leq m$. The total number of
    comparisons is $n$ and $n \leq \bigl(
    \begin{smallmatrix}
      m \\
      2
    \end{smallmatrix}
    \bigr)$.
  \item[$\pi_{ij}$] is a population parameter and is the probability that
    the voters prefer $x_i$ to $x_j$, \viz $P(x_i \rightarrow x_j)$.
  \item[$\gamma_{ij}$] is a population parameter and is the probability
    that the chooser is indifferent between $x_i$ and $x_j$, \viz $P(x_i =
    x_j)$.
  \item[$s_{ij}$ and $t_{ij}$] are the number of times that $\pi_{ij}$ and
    $\gamma_{ij}$ occur respectively.
  \end{description}
\end{notation}

These can then be used as parameters to a multinomial distribution. (The
multinomial distribution is the binomial but has more than two outcomes.)
This distribution is used to determine the probability that the number of
statements of preference of one policy over another $\pi_{ij}$ is exactly
equal to the number of times of times a preference is stated. It measures
the strength of a preference: how often it is stated against how often the
chooser is indifferent between them.  This equates to the graph--theoretic
notion that traversing an arc between nodes in the same direction is a
better measure of order than counting the number of times a node is chosen,
see \myRef{sec:lattice:p}.

\begin{equation}
  \label{eq:top:distance:1}
  \begin{aligned}
    s_{ij} + s_{ji} + t_{ij} & = n_{ij} \\
    \pi_{ij} + \pi_{ji} + \gamma_{ij} & = 1 \\
    \gamma_{ij} & = \gamma_{ji} \\
    P(\pi_{ij} = \frac{s_{ij}}{n_{ij}},
        \pi_{ji} = \frac{s_{ji}}{n_{ij}},
        \gamma_{ij} = \frac{t_{ij}}{n_{ij}} ) & =
    \Pi_{\mathscr{I}}
    \frac{n_{ij}!}{s_{ij}!s_{ji}!t_{ij}!}
    \pi_{ij}^{s_{ij}} \pi_{ji}^{s_{ji}} \gamma_{ij}^{t_{ij}}
  \end{aligned}
\end{equation}

Because $\pi_{ij}, \pi_{ji} \droptext{and} \gamma_{ij}$ sum to one, it is
possible to eliminate $\pi_{ji}$ and use just $\pi_{ij}$ and $\gamma_{ij}$
as the parameters. In which case, because there are $n$ parameters in
$\mathscr{I}$, the set of all pairings, there are $2n$ parameters in all.

What is needed now is a measure of the most likely preference ordering:
this can be represented as a point $\hat{\pi}$ being the ordered pair $(
\hat{\pi}_{ij}, \hat{\gamma}_{ij} )$. A sequence of these will form a
preference relation $T(\pi)$ in a more constrained portion $\omega$ of the
search space $\Omega$.

\begin{notation}
  \begin{description}
  \item[$\Omega$] is the parameter space, a subset of $2n$ dimensional
    space.
  \item[$\pi$] is a typical point in $\Omega$.
  \item[$\hat{\pi}$] is the maximum likelihood estimate of $\pi$. It will
    have coordinates drawn from: $\hat{\pi}_{ij} = \frac{s_{ij}}{n_{ij}}$
    and $\hat{\gamma}_{ij} = \frac{t_{ij}}{n_{ij}}$.
  \item[$\omega$ and $T(\pi)$] $T(\pi)$ is a preference relation over some
    portion $\omega$ of $\Omega$.
  \item[$\bar{\pi}$] is the maximum likelihood estimate of $\hat{\pi}$
    which is restricted to $\omega$, \ie to where $T(\pi)$ is in force.
  \end{description}
\end{notation}

It is now possible to form a maximum likelihood measure.
(Maximum--likelihood is analogous to the traditional entropy measure
$p\log(\frac{1}{p})$, but does not necessarily have the same properties.
Statisticians use it as a variance measure.)  It is used as estimator, it
can be formulated thus:

\begin{equation}
  \label{eq:top:distance:2}
  L(\pi) = \Sigma_{\mathscr{I}}
  n_{ij}(\hat{\pi}_{ij} \log \pi_{ij} + \hat{\gamma}_{ij} \log \gamma_{ij}) 
\end{equation}

which, when maximised in $\omega$, will yield $T(\bar{\pi})$, which is the
maximum likelihood ordering of the elements of $X$.

\subparagraph{Graphs and Voting Rules}

Referring again to the graph $[ X, C \cup D]$, $C$ is the I--graph of
indifferences and $D$ is the P--graph of preferences. $T(\pi)$ can be
defined thus:

\begin{definition}
  $(x_i, x_j) \in T(\pi)$ iff $\graphpath(x_i, x_j)$ in $C(\pi) \cup
  D(\pi)$, where:

  \begin{description}
  \item[$D(\pi)$] $(x_i, x_j) \in D(\pi)$ iff $\pi_{ij} > \max ( \pi_{ji},
    \gamma_{ij} )$.
  \item[$C(\pi)$] $(x_i, x_j) \in C(\pi)$ iff $\gamma_{ij} > \max ( \pi_{ij},
    \pi_{ji} )$.
  \end{description}
\end{definition}

This definition also defines a particular sub--graph of $[X, C \cup D]$
which will be called $[X, C(\pi) \cup D(\pi)]$.

\begin{figure}[htbp]
  \begin{center}
  \includegraphics[keepaspectratio=1, totalheight=4in]{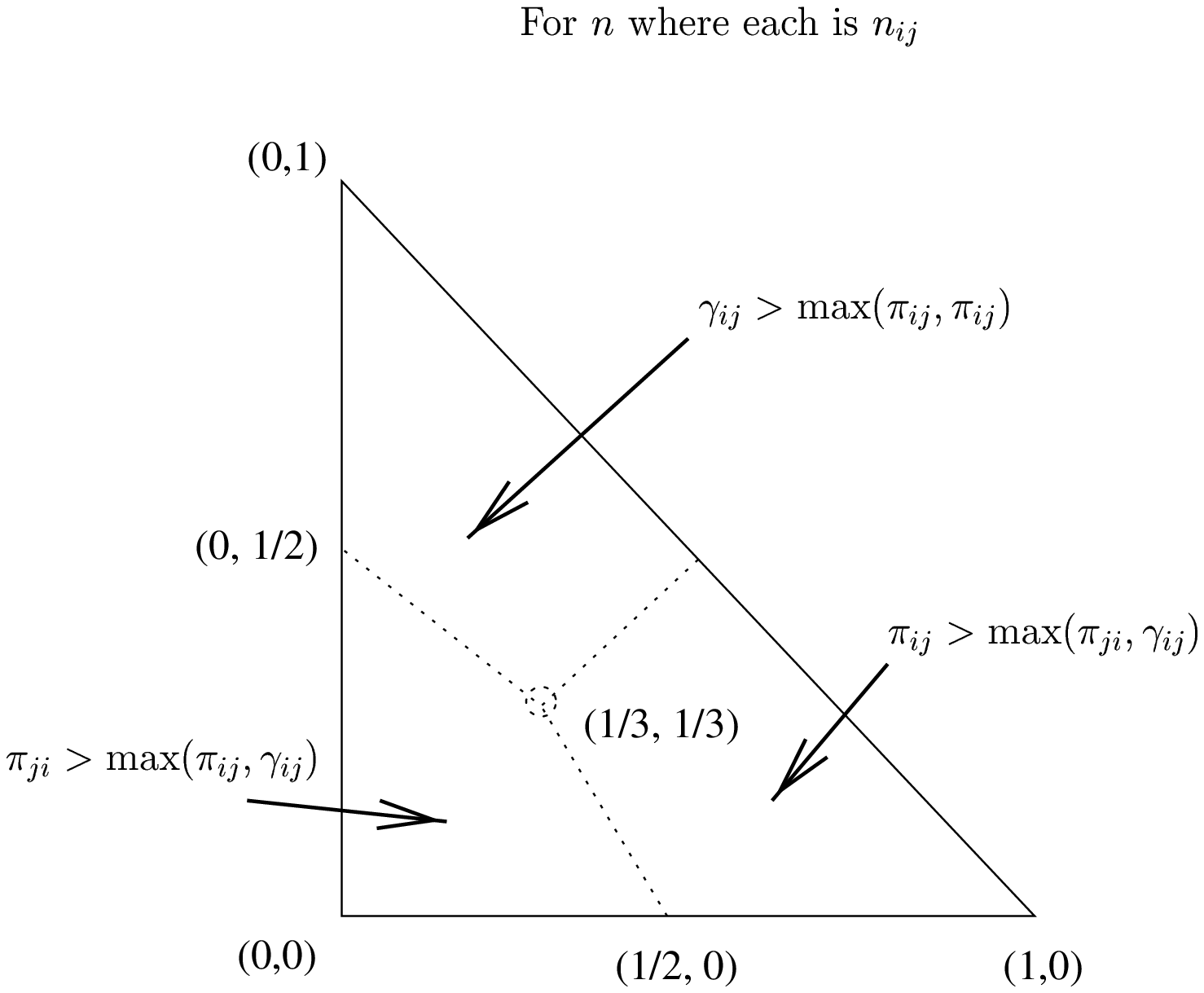} 
  \caption{Simplex for Voting}
  \label{fig:top:distance:1}
  \end{center}
\end{figure}

In effect, this defines a pair of voting rules for this distance function,
which is simple majority, but requires that the ``for''--vote be greater
then the ``don't care''--vote as well. It has similarities to the
normalised simplexes used by Saari \cite{vote:saari}, an illustration is
given in figure \ref{fig:top:distance:1}.

Singh and Thompson's result does hinge upon the definitions of $T(\pi)$ and
$[X, C(\pi) \cup D(\pi)]$, but it \emph{should} be the case that they can
be adapted to different voting rules\footnote{Theorem 10 of
  \cite{stats:pairwise:singh} is the key to this argument since it relies
  on the properties of the voting rules.}.

Under these voting rules, if one requires a \emph{unique} strong ordering
of a \emph{complete} set of comparisons, \ie there exists a $T(\bar\pi)$
that is contained in $D(\pi)$, which is what most theoretical political
scientists want, then it can be safely \emph{calculated} that any such
order:

\begin{itemize}
\item Either: just does not exist, \ie there are no strong orderings at all.
\item Or: a strong ordering exists, but is not more likely than any weak
  orderings. 
\item Or: there is a strong ordering and it is maximally likely, in which
  case it will be unique.
\end{itemize}

\paragraph{Minimising Uncertainty}

An interesting insight by Thompson and Remage \cite{stats:pairwise:thompson}
is as follows: the uncertainty of a single comparison of $x_i$ to $x_j$ is:

\begin{equation}
  U_{ij} = - ( \pi_{ij} \log \pi_{ij} + \pi_{ji} \log \pi_{ji} +
  \gamma_{ij} \log \gamma_{ij} )
  \label{eq:entropy:thompson}
\end{equation}

For all $n_{ij}$ comparisons of $x_i$ to $x_j$ the uncertainty is: $n_{ij}
U_{ij}$ and for all comparisons:

\begin{equation*}
  U(\pi) = \sum_{\mathscr{I}} n_{ij} U_{ij}
\end{equation*}

From which it is clear that $L(\pi) = - U_{ij}$ but in a specified subset
of $\omega$ only, so maximising the likelihood of $L(\pi)$ is equivalent to 
minimising the uncertainty of $U(\pi)$. Thompson and
Remage \cite{stats:pairwise:thompson} make it quite clear by stating that
$U(\pi)$ represents the total number of sample preferences which are
violated by the ranking $T(\pi)$.

\paragraph{An Example}

\begin{figure}[htbp]
  \begin{center}
    \includegraphics{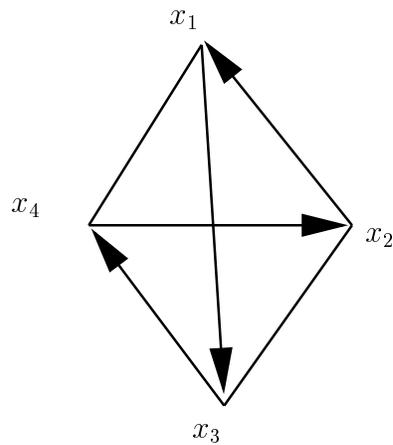}
    \caption{Aggregate bigraph from table \ref{tab:top:distance:2}}
    \label{fig:top:distance:2}
  \end{center}
\end{figure}

Table \ref{tab:top:distance:2} contains some illustrative data for four
options $X = \{ x_1, x_2, x_3, x_4 \}$. From this an aggregate bigraph has
been generated, see figure \ref{fig:top:distance:2}, which is not
circuit--free.

\begin{table}[htbp]
  \begin{center}
    \begin{tabular}[left]{|r|r|r|r|r|}
      \hline
      $(ij) \in \mathscr{I}$ & $s_{ij}$ & $t_{ij}$ & $s_{ji}$ &
      $(\hat\pi_{ij}, \hat\gamma_{ij})$ \\
      \hline
      (12) & 2 & 1 & 3 & $(\frac{1}{3}, \frac{1}{6})$ \\
      (13) & 4 & 1 & 1 & $(\frac{2}{3}, \frac{1}{6})$ \\
      (14) & 0 & 4 & 2 & $(0, \frac{2}{3})$ \\
      (23) & 1 & 3 & 2 & $(\frac{1}{6}, \frac{1}{2})$ \\ 
      (24) & 1 & 2 & 3 & $(\frac{1}{6}, \frac{1}{3})$ \\ 
      (34) & 4 & 0 & 2 & $(\frac{2}{3}, 0)$ \\
      \hline
    \end{tabular}
    \caption{Paired comparisons, $m=4$, $n_{ij}=6$ for all $i$ and $j$}
    \label{tab:top:distance:2}
  \end{center}
\end{table}

\begin{figure}[htbp]
  \begin{center}
    \includegraphics{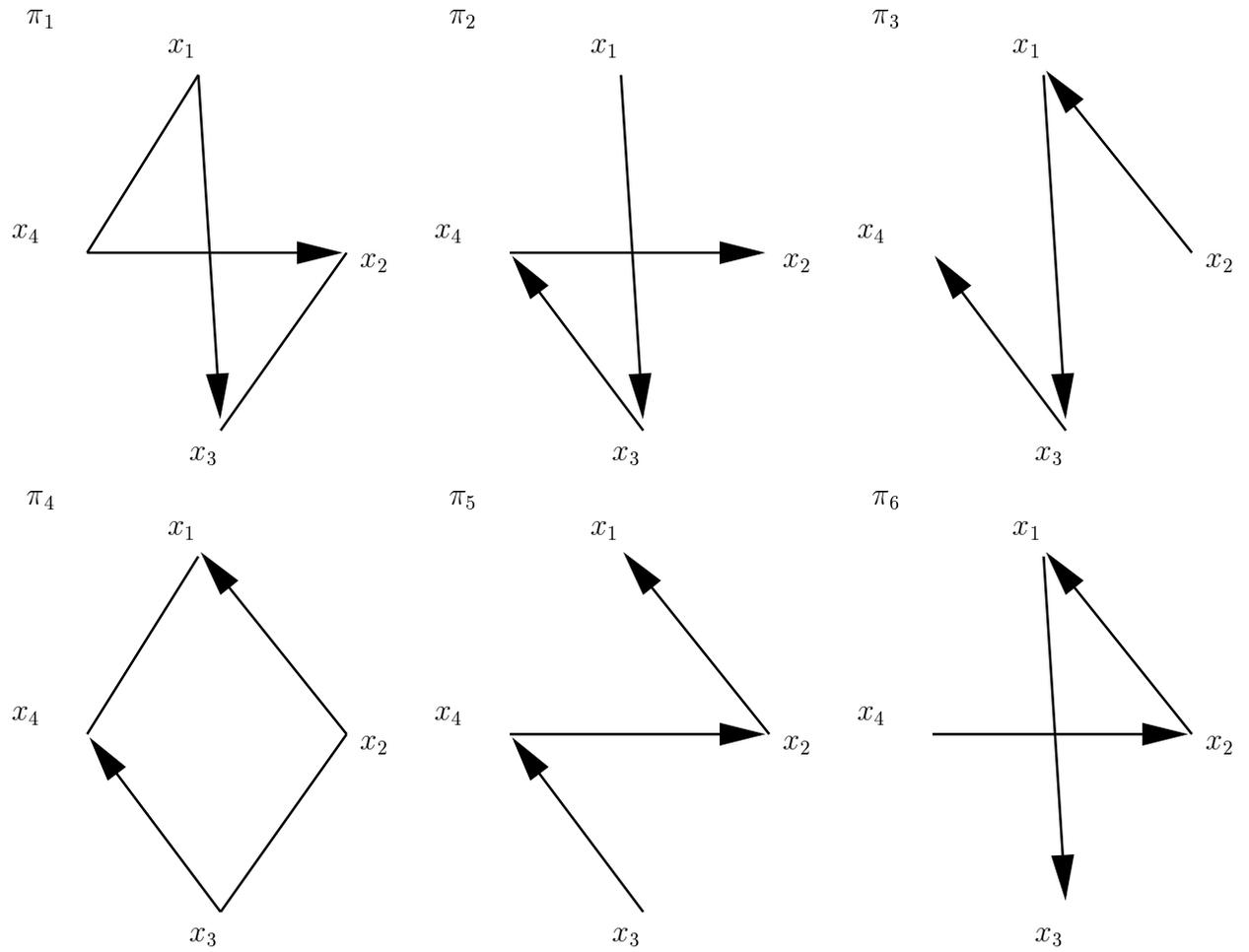}
    \caption{Maximal sub--bigraphs of bigraph in figure
      \ref{fig:top:distance:2}}
    \label{fig:top:distance:3}
  \end{center}
\end{figure}

All of the possible maximal circuit--free sub--bigraphs have been
enumerated in figure \ref{fig:top:distance:3} and each of these has its own
preference order, $\pi_1, \pi_2, \pi_3, \pi_4, \pi_5, \pi_6$.

For each of these the uncertainty has been calculated and is shown in table
\ref{tab:top:distance:3}. From which it is clear that the most certain
order is $\pi_4 \droptext{:} (x_1 = x_4) > (x_2 = x_3)$, but this is a weak
ordering; there are two strong orderings which are equally certain:
$\pi_2 \droptext{:} x_1 > x_3 > x_4 > x_2$
and
$\pi_3 \droptext{:} x_2 > x_1 > x_3 > x_4$,
which differ markedly in their ranking of $x_2$.

\begin{table}[htbp]
  \begin{center}
    \begin{tabular}[left]{|r|r|r|r|r|r|r|r|}
      \hline
      $\pi$ & 
      $(\pi_{12}, \gamma_{12})$ &
      $(\pi_{13}, \gamma_{13})$ &
      $(\pi_{14}, \gamma_{14})$ &
      $(\pi_{13}, \gamma_{23})$ &
      $(\pi_{24}, \gamma_{24})$ &
      $(\pi_{34}, \gamma_{34})$ &
      $U(\pi)$ \\
      \hline

      $\pi_1$ & $( \frac{5}{12}, \frac{1}{6}) $ & $( \frac{2}{3},
      \frac{1}{6}) $ & $( 0, \frac{2}{3}) $ & $( \frac{1}{6}, \frac{1}{2})
      $ & $( \frac{1}{6}, \frac{1}{3}) $ & $( \frac{1}{2}, 0) $ & 2.279224
      \\ 

      $\pi_2$ & $( \frac{5}{12}, \frac{1}{6}) $ & $( \frac{2}{3},
      \frac{1}{6}) $ & $( 0, \frac{1}{2}) $ & $( \frac{1}{6}, \frac{5}{12})
      $ & $( \frac{1}{6}, \frac{1}{3}) $ & $( \frac{2}{3}, 0) $ & 2.286507
      \\ 

      $\pi_3$ & $( \frac{1}{3}, \frac{1}{6}) $ & $( \frac{2}{3},
      \frac{1}{6}) $ & $( 0, \frac{1}{2}) $ & $( \frac{1}{6}, \frac{5}{12})
      $ & $( \frac{1}{6}, \frac{5}{12}) $ & $( \frac{2}{3}, 0) $ & 2.286507
      \\ 

      $\pi_4$ & $( \frac{1}{3}, \frac{1}{6}) $ & $( \frac{5}{12},
      \frac{1}{6}) $ & $( 0, \frac{2}{3}) $ & $( \frac{1}{6}, \frac{1}{2})
      $ & $( \frac{1}{6}, \frac{5}{12}) $ & $( \frac{2}{3}, 0) $ & 2.234382
      \\ 

      $\pi_5$ & $( \frac{1}{3}, \frac{1}{6}) $ & $( \frac{5}{12},
      \frac{1}{6}) $ & $( 0, \frac{1}{2}) $ & $( \frac{1}{6}, \frac{5}{12})
      $ & $( \frac{1}{6}, \frac{1}{3}) $ & $( \frac{2}{3}, 0) $ & 2.348982
      \\ 

      $\pi_6$ & $( \frac{1}{3}, \frac{1}{6}) $ & $( \frac{2}{3},
      \frac{1}{6}) $ & $( 0, \frac{1}{2}) $ & $( \frac{1}{6}, \frac{5}{12})
      $ & $( \frac{1}{6}, \frac{1}{3}) $ & $( \frac{1}{2}, 0) $ & 2.303824
      \\ 
      
      \hline
    \end{tabular}
    \caption{Sub--bigraphs uncertainty rankings}
    \label{tab:top:distance:3}
  \end{center}
\end{table}

Finally, it can be said that Thompson and Remage's work provides a good
analytic method of determining which of a set of orderings is most
preferred and it gives us the ability to choose between popular weak
orderings which are less decisive, in that they cannot differentiate
between as wide a range of choices, and less popular strong orderings which
are more decisive.

A popular weak ordering is less prone to anomalies of choice than a strong
one, but is not as useful in decision--making. Computationally, this
calculation is not particularly difficult. The search space sizes are given
in table \ref{tab:orders}. For this search space of four policies, table
\ref{tab:orders} requires a maximum of 75 running totals to be kept for the
voter population.

\section{Deciding Sets} \label{sec:lattice:q}

Deciding sets are those groups of voters whose support is needed to win a
vote. This concept is clarified (and re--named) in a paper by Batteau
\etal, \cite{scf:theory:batteau}, as the \emph{preventing set}. Most people
would know them as coalitions, but they can be decisive without actually
imposing their preferences for policies. In this sense, a deciding set acts
as a dictator or, more typically, as a vetoer.

Voters usually express their values by grouping sets of issues together and
setting them against other sets of issues. Under some circumstances, a
``Kingmaker'' group \cite{scf:stcred} can emerge, which has an unfair
influence according to Arrow's theorems \cite{scf:eaves}. If one were to
analyse an issue space then it would be necessary to search it for all the
voter alignments that might give rise to ``Kingmaker'' groups emerging. The
space is large, given by \eqref{eq:preferences}. This problem would be a
simpler variant of the bin--packing problem, which is known to be
\emph{NP--complete}. There are some efficient genetic algorithms for the
bin--packing problem \cite{ga:group} and these could be adapted to search
for voter alignments.

\subsection{Spectral Analysis of Ranked Data}

A paper by Diaconis \cite{voting:prob:diaconis} introduced some interesting
analysis of ranked data which is a Borda preferendum, see table
\ref{tab:choice:1}:
$f(\pi) = \bigl(
\begin{smallmatrix}
  1 & 2 & 3 & 4 & 5 \\
  x_1 & x_2 & x_3 & x_4 & x_5 \\
\end{smallmatrix}
\bigr) = n$, \ie $\pi$ is a ranked order $1$ to $5$ of $5$ policies such
that $x_1 > x_2 > x_3 > x_4 > x_5$ and $n$ voters have chosen this
ordering. Diaconis then introduces a suite of first--order functions which
return counts, let $j$ be the index of the policy $x_j$, then:

\begin{equation*}
  \delta_{i\pi(j)} = 
  \begin{cases}
    1 & \droptext{if} \pi(j) = i \\
    0 & \droptext{otherwise}
  \end{cases}
\end{equation*}

Or, simply, the count of all those who place policy $x_j$ in position $i$.

And a suite of \emph{unordered} second--order functions which return counts
thus:

\begin{equation*}
  \delta_{\{i,i^{\prime}\}, \{\pi(j), \pi(j^{\prime})\} } = 
  \begin{cases}
    1 & \droptext{if} \pi(j) = i \droptext{and}
    \pi(j^{\prime}) = i^{\prime} \\
    0 & \droptext{otherwise}
  \end{cases}
\end{equation*}

Or the count of all those who:

\begin{itemize}
\item Either: place candidate $x_j$ in position $i$ and candidate 
  $x_{j^{\prime}}$ in position $i^{\prime}$
\item Or: place candidate $x_j$ in position $i^{\prime}$ and candidate 
  $x_{j^{\prime}}$ in position $i$ 
\end{itemize}

It is also possible to have a set of \emph{ordered} second--order functions 
written thus: $\delta_{(i,i^{\prime}), (\pi(j), \pi(j^{\prime})) }$ which
returns the count of all those who place in the order $x_j$ at $i$ and
$x_{j^{\prime}}$ at $i^{\prime}$.

And it is possible to have higher orders still, unordered and ordered.

In his paper, Diaconis used these functions to analyse the data of an
election that used a proportional representation voting system and
demonstrated that the election contained two main types of voters $A$ and
$B$ who voted on two issue blocks $ x' = (x_1,x_3) $ and %
$ y'= (x_4,x_5) $, in the following way: %
$ A, x' > y'$ and $B, y' > x' $.

This analysis would yield fitness metrics for selection. It would be used
to partition the voters and the issues, so that each partition would
demonstrate fewer cycles and more unanimities. 

The computational complexity of spectral analysis is very high. It requires
repartitioning the search space for each possible combination. If it is
compared to calculation of distance between preference orderings given in
\myRef{sec:top:distance} which required no more than keeping 75 running
totals for a four choice system, spectral analysis would require %
$ 4! \cdot 75 $ totals to be kept. 

\section{In Conclusion}

A comprehensive set of analytic techniques have been presented which should
allow aggregate hierarchies of preferences to be generated from individual
statements of preferences with an option to choose either a strong-- or
weak--ordering and to quantify how acceptable they would be to the
individuals. Other techniques have been described which would allow voters
and issues to be partitioned so that the level of acceptance within those
sub--groupings could be higher.

What this brings to the discussion of collective choice procedures is a
reinterpretation of the limitations of Arrow's \textit{Impossibility
  Theorem} \cite{scf:eaves} using information theory: it is impossible to
design a representative collective choice procedure that can select one
from more than two issues \textit{if the preference orderings have too high
  an entropy}. If the preference orderings are sufficiently well--ordered,
the collective choice cannot be subverted by a perverse and sophisticated
clique.

The principal entropy measure is due to Thompson \etal and is given by
\eqref{eq:entropy:thompson}. This allows a choice between strong and
weak--orderings to be made. It should be possible to develop this to use
Diaconis' spectral analysis of collective choices to generate preference
orderings for different sub--groups of the voters. 

In short, the Thompson metric allows issues to be merged, the Diaconis
method allows voters to be merged. Unfortunately, the computational burden
for this latter measure would be very high, see \myRef{sec:lattice:q}, so
only the Thompson and Remage method will be used in the remainder of the
analysis in this dissertation.


\chapter{Stability of Self--Organisation}
\label{cha:norms}

From the discussion in chapters \ref{cha:reqs} and \ref{cha:dbaccess} it
was made clear that software technology is mature enough to support a
self--organising database access system, but it was also apparent that such
a system would need precisely specified policies in order to operate.
Chapters \ref{cha:pol:1} and \ref{cha:top} showed that these policies would
necessarily be aggregations of preference hierarchies.

Chapter \ref{cha:top} introduced some distance measures which could be used
to compare different preference hierarchies.  This chapter uses distance
measures to produce a self--stabilising system of interacting agents, which
act autonomously but are controlled by their own peers.  Essentially, this
chapter addresses whether a self--organising system based on group
memberships can be active enough to allow policies to emerge, but stable
enough so that the agents do not follow inconsistent policies in a short
time period.

\section{Formation of Cultures}

The system of creating access hierarchies based on group memberships is very 
similar to a series of simulations carried out by Axelrod
\cite{Axelrod:culture}, which were an attempt to elucidate the processes
underlying the adoption of standards in industry \cite{Axelrod:coal}.

Axelrod's analytical method is unusual in that he constructs systems which
have agents that have very simple behaviour. He then allows the agents to
interact in a series of simulations and then analyses the behaviour of the
system as a whole. The hope is that by specifying the behaviour of the
agents, what Axelrod describes as \textit{small--scale behaviour}, it is
possible to control the \textit{large--scale behaviour} of the system.

Axelrod gives a number of common--place examples of the formation of
cultures.  Consider nightclubs: people visiting nightclubs often want to
meet other people going there and they try to emulate one another's style
of dress and manners. A set of features that might be considered important
at nightclubs would be: hair length and cut, style of dress, dancing style
and so on and so forth. So a person who dances in a particular way might
see someone who dances in the same way and would choose to become more
alike to them in the hope they might meet. To do this, that person might
change their haircut. Now another person with the same haircut may choose
to change his dancing style. If this behaviour continues, then either all
the people visiting this nightclub will look the same and dance in the same
way, or cliques will emerge. One group of people will dress and dance in
one particular way and another group will do so in another particular way,
which can be more or less incompatible. If these two cliques are completely
incompatible, they can never become alike to one another, because they have
nothing in common to begin with. It may also be the case, that there are a
number of competing cliques who are slightly incompatible; they interact
with one another only rarely and there is no long--term effect. It may also
be the case that the cliques interact with one another so often that there
is no discernible similarity between the people visiting the nightclub from
night to night.

Surprisingly, many systems where standards have evolved are very similar in
operation. Consider the variation of standards for electricity supply: for
consumers in the United States, 120 volts at 60 hertz is the standard; in
Europe, 240 volts at 50 hertz. In the US, industry uses 380 volts three
phase, in Europe 440 volts. This kind of dissimilarity goes right across
the electricity supply industry. Similar arguments can be made for the
evolution of the metric and imperial measurement systems, or the morphology
of human (and programming) languages. There do appear unique similarities
that cut across all systems: for example, even though people in the United
Kingdom drive on the left and people in the Europe drive on the right, the
system of traffic law is very similar: pedestrian crossings, giving way at
junctions and so on.

To add some formalism, using the evolution of culture in a nightclub as an
example, the way people visiting a nightclub choose to become similar to
one another is the \textit{small--scale behaviour}, the appearance of the
nightclub and its constellation of cliques is the \textit{large--scale
  behaviour}. Each individual's small--scale behaviour is based on a
particular probability distribution which is time--invariant and will be
different from everyone else's. The large--scale behaviour probability
distribution is the superposition of all of the small--scale behaviours and
may aggregate to something quite predictable: in the same way that the sum
of a set of independent random variables can be assumed to behave as a
normal distribution.

Consider the quality control of paint colours. The manufacturing of paint
is a sophisticated process, the amount of dye added can vary because of
changes in the granularity of the dye powder interacting with variation of
tolerances in the injection heads, which can also interact with the
frequency and thoroughness of the head--cleaning procedures. One should
also consider variations in the temperature of the oil resin, the efficacy of
the mixing motors, variations in phase in power supply. All these factors
can be approximated as independent random variables, because the causality
of one set is counteracted by the causality of another.

A statistician if asked to make an estimate of the intensity of a colour of
a tin of paint would not attempt to analyse the whole of the paint
manufacturing process but would simply assume the final probability
distribution is a normal one.

Axelrod's cultural evolution model aims to perform the same simplification:
to determine a simple probability distribution for the large--scale
behaviour of the system, but he allows himself the facility to control one
causal probability distribution for the components of the larger system.

\subsection{Small--Scale Behaviour}

The simulation model employed by Axelrod was deceptively simple and is an
example of the use of cellular automata simulation
\cite{vonNeumann:1966:TSR}. Each agent was given a fixed set of features $A
= \{ a_1, a_2, \dots, a_n \}$, each feature had an enumerated set of
values, known as the traits, for that feature. The traits were initially
randomly generated and assigned to the features.

\begin{definition}[Distance Functions] Axelrod developed a simple
  similarity function $S(\cdot, \cdot)$ which was used to evaluate how
  alike two agents were. If $X$ and $Y$ are their respective sets of
  features, with each feature being $x_i$ or $y_i$ respectively, then if
  the features had the same trait, $x_i = y_i$, a $1$ is scored, otherwise
  $0$.

  This describes the behaviour of the trait comparator.
  \begin{equation}
    \label{eq:adaptive:axelrod:2}
    s(x_i, y_i) = 
    \begin{cases}
      0 & x_i \neq y_i \\
      1 & x_i = y_i \droptext{if features have same trait}
    \end{cases}
  \end{equation}

  This describes the behaviour of the feature set comparator.
  \begin{equation}
    \label{eq:adaptive:axelrod:3}
    S(X, Y) = \sum_{i=1}^{i=n} s(x_i, y_i)
  \end{equation}
  This is Axelrod's similarity function. Clearly, two agents, $X$, $Y$, are
  identical if $S(X, Y) = n$. Therefore a distance function would be
  \begin{equation}
    \label{eq:adaptive:axelrod:4}
    d(X, Y) \eqdef n - S(X, Y)
  \end{equation}

\end{definition}

This can then be used as the basis for an interaction criterion. For
example, if $d(X, Y)$ is equal to at least $1$ the agents can interact, if
$d(X, Y)$ is greater than one they are more likely to interact. If it is
zero, they will be unable to interact.

\begin{definition}[Interaction Criterion] Axelrod used a simple dice--throw
  to simulate human decision--making, the uniform distribution:
  $U(0,1)$. The distance between two agents must then be scaled to fall
  within $[0,1]$, so a factor of $k$ is introduced. An offset of $\epsilon$
  can be set in the range $[0,1]$ to make interactions less likely.
  \begin{equation}
    \label{eq:adaptive:axelrod:1}
    k \cdot d(X, Y) + \epsilon < U(0,1)
  \end{equation}
\end{definition}

The interaction would be that $X$ would choose from $Y$ a feature that was
different from its own and accrete it, \ie set a feature to have the same
trait as $Y$.

It should be clear from the specification of the change procedure that each
agent is biased towards agents that are similar to itself. For agents to
interact, it is required that $k \cdot d(X, Y) + \epsilon > 0$. The choice
of $k$ and $\epsilon$ determine under what small--scale conditions
interactions cease. If $\epsilon$ were zero and $k=\frac{1}{n}$, then no
interaction can take place if the agents are completely dissimilar. But
setting $\epsilon$ to $\frac{1}{n}$ and $k=\frac{1}{n} \cdot ( 1 +
\frac{1}{n})$ would mean there would be no interaction even if the agents
had one identical trait.

\subsection{Local convergence leads to global polarisation}

\label{sec:adaptive:convergence}

The software simulation Axelrod used has been replicated
\cite{sft:aidan:cultural}. The agent to undergo the accretion was randomly
chosen; the parameters of the simulations were initially:
\begin{itemize}
\item 5 features
\item 10 traits per feature
\item 100 agents
\item 4 neighbours
\item Topology was a square field
\end{itemize}

This is quite a testing culture. There are $\binom{10^5}{10}$ different
individuals to choose from. A significant emergent property appeared that
is typical of large systems:

A particular trait would become current within a group, this would make
members of that group more attractive to one another and they would
exchange more traits until their features were identical to one another.
Should such a group encounter another group that had undergone the same
process, it would be relatively improbable that they would be able to
interact. In this way, islands of homogeneity emerged.

\paragraph{Typical Culture}

\begin{figure}[htbp]
  \begin{center}
  \includegraphics{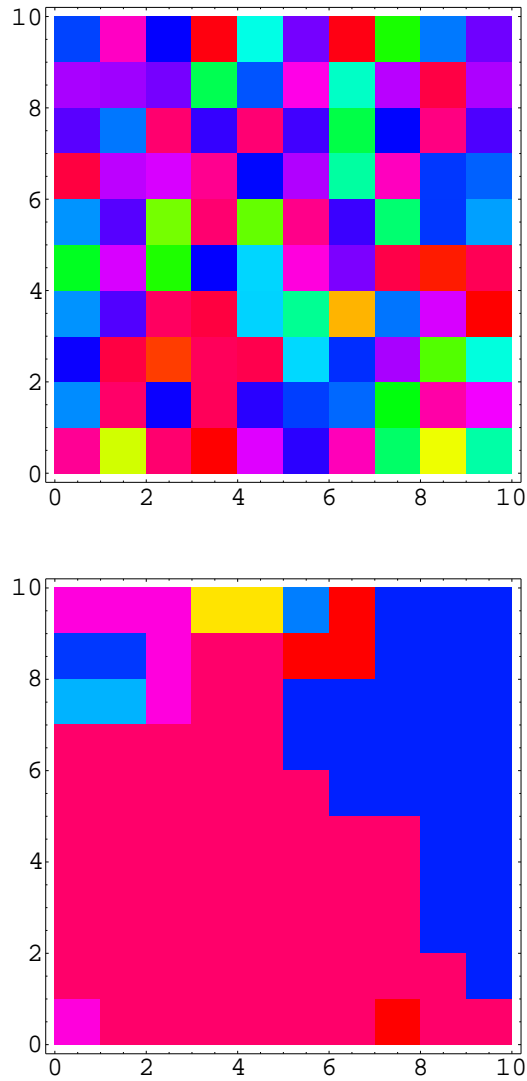}
  \caption{Density plots of the evolving culture of figure \ref{fig:ego-1}}
  \label{fig:ego-1d}
  \end{center}
\end{figure}

\begin{figure}[htbp]
  \begin{center}
  \includegraphics{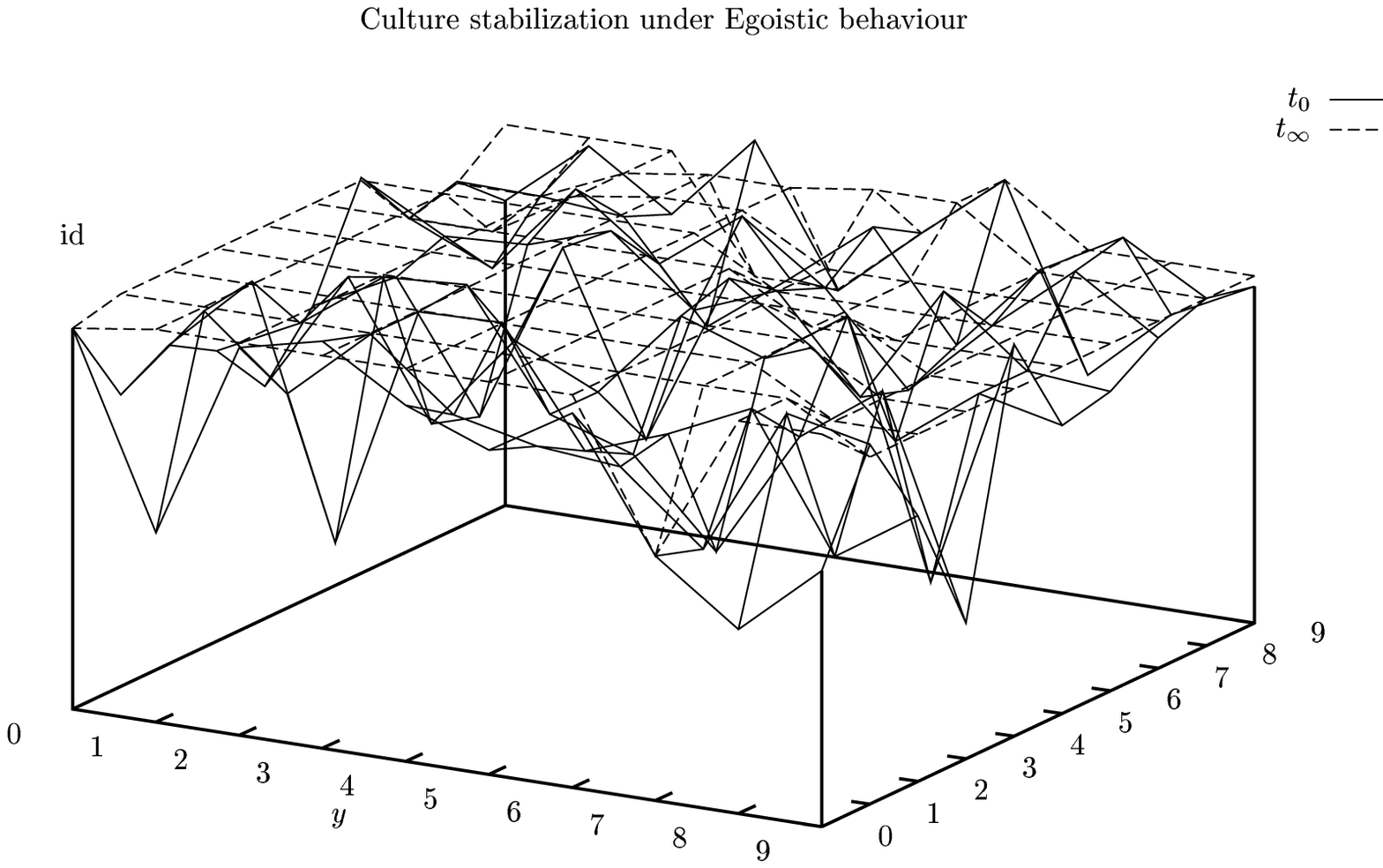}
  \caption{Surface plot of an evolving culture at $t_0$ and $t_\infty$}
  \label{fig:ego-1}
  \end{center}
\end{figure}

This can be seen in the following density and three--dimensional height
plot, see figures \ref{fig:ego-1d} and \ref{fig:ego-1} respectively, both
of which are reasonably typical. The features have been mapped to a
continuous valued metric using the function given by \eqref{eq:adaptive:h1}
and then logged to the base $n$ so that the metric is linear, see
\eqref{eq:adaptive:lnh1}:

\begin{align}
  h(A) & = n^0 \cdot a_0 + n^1 \cdot a_1 + \dots + n^{n-1} \cdot a_n
  \label{eq:adaptive:h1} \\
  \hat{h} & = \frac{\ln h(a)}{\ln n} \label{eq:adaptive:lnh1}
\end{align}

\subparagraph{Density plot}

This is shown as a pair of plots in figure \ref{fig:ego-1d}. It maps the
identity metric to a red--green--blue colour--code. The location of an
agent can be determined from the $x$ and $y$ axes. The colours form into
blocks where the agents are similar. The upper plot is the state of the
system before interaction commences, the lower when it has reached stasis
and no more interaction is possible.

\subparagraph{Height plot}

Colour density plots show where regions of identical agents have formed,
but do not clearly show how incompatible the regions are. It is difficult
to tell if a yellow region is incompatible with a blue. There may be hint
of blue colour in the yellow or \viceversa. This is even more difficult if
the plots are only viewed in grey-scale.

Incompatible regions are easy to see using the height plot in figure
\ref{fig:ego-1}. The \textit{z} axis is marked \textit{id} and the physical
location of the individual is specified by the \textit{x} and \textit{y}
axes as with density plots. The surface marked by the solid lines is the
initial state. The surface marked by the dotted lines is the final state.

If the difference in height between two plateaus is greater than $1$ unit
then the regions are incompatible{\myDash}if $\epsilon=0$ and $k =
\frac{1}{n}$ in \eqref{eq:adaptive:axelrod:1}.

The two types of plot are complimentary. It is difficult to interpret one
without the other. The density plot shows compatibilities clearly, the
height plot incompatibilities. The distribution of the varieties is given
in table \ref{tab:culture:1}.

\subsection{Large--Scale Behaviour}
\label{sec:adaptive:axelrod:large}

The model is a Markov process with absorbing states \cite[p. 396, ``Birth
Processes'']{stoch:papoulis}, so it should settle after some initial
transient behaviour, but may, possibly, possess limit cycles.

In Axelrod's analysis, these following points were considered important:
\begin{enumerate}
\item How many areas of dissimilarity?
\item How many different areas of dissimilarity?
\item How large were they relative to one another?
\item How quickly did the system stabilise?
\end{enumerate}

\paragraph{Metrics}

Axelrod's method of to determine the fluctuation of the areas of
dissimilarity used time--series plotting and peak detection. A set of
metrics were developed that more conveniently measured the qualities of
regions. Some system activity metrics were also introduced.

\begin{description}
\item[$\eta$] is an efficiency measure and is the ratio of the number of
  interactions to the number of selections in each period. This is used
  to measure system activity.

\item[$S(v)$] The entropy of the different varieties of agents on the
  field. This is a single metric for measuring variance, $S(v)$ is a
  normalised entropy, \ie $0 \leq S(v) \leq 1$.

  \begin{equation}
    S(v) = - \frac{1}{\ln N} \Sigma_{v}^{N} p_{v} \ln p_{v} 
  \end{equation}

  $p_v$ is the probability of selecting an agent having variety $v$,
  simply $\frac{n_v}{N}$; $N$ is the maximum number of varieties that can
  exist simultaneously, which, in this case, is the same as the total
  number of agents that can exist simultaneously, in these simulations $N =
  10 \times 10$.  The variety entropy is the measure of the homogeneity of
  the agents in the population as a whole, when it is zero, the population
  has only one variety of agent and no further interaction is possible.

\item[$S(c)$] The compatibility entropy is a measure of how compatible the
  varieties are with one another. The probability on which the measure is
  based is that of an agent of variety $u$ interacting with an agent of
  variety $v$, denoted by $u \wedge v$:

  \begin{align*}
    P(u \wedge v) & = P(u)P(v|u) + P(v)P(u|v) \\
    \droptext{where} & \\
    P(w) & = \frac{n_w}{N} \\
    P(x|w) & = \frac{n_x}{N - n_w}
  \end{align*}

  The event $u \wedge u$ is not acted upon, so it is removed from the
  probability space as are the events $u \wedge v$ when $u$ is not
  compatible with $v${\myDash}the agents can only interact with one another
  if they are similar in one feature \emph{and} have at least one
  dissimilar feature. The events that form part of the entropy measure do
  not cover the entire event space, so they need to be normalised. Once
  that is done, the entropy can be formed in the usual way. The entropy
  metric is itself normalised using the factor $\binom{N}{2}$, \ie the
  maximum number of pairs of different agents it would be possible to have.

  \begin{equation}
    S(c) = - \frac{ 1 }{\ln \binom{N}{2} } \Sigma_{u, v}^{\binom{N}{2}}
    P(u \wedge v) \ln P(u \wedge v)
  \end{equation}

  When the compatibility entropy is zero, every agent is of an incompatible
  variety with every other agent and no further interaction is possible. It
  is possible for a system to have a non--zero compatibility entropy and
  for no further interaction to be possible; this would arise if two (or
  more) ``islands'' of compatible agents are separated by a sea of agents
  with which they are incompatible.
  
\end{description}

\paragraph{Stasis Condition}

Because the system may fall into a limit cycle, a condition needs to be put
into place that will cause the simulation to terminate. Neither of the
entropy metrics is useful for this, so a simple test is to see if the
number of varieties has changed over a certain number of periods. (This
actually needs to be improved upon, because under some short duration limit
cycles the number of varieties does change. This is inavoidable, it will be
argued later that this model shows chaotic behaviour and the length of
limit cycles is a fractal number.)

\paragraph{Typical Metrics and some Characterisation}

\begin{figure}[htbp]
  \begin{center}
  \includegraphics[angle=-90,keepaspectratio=1,totalheight=7in]{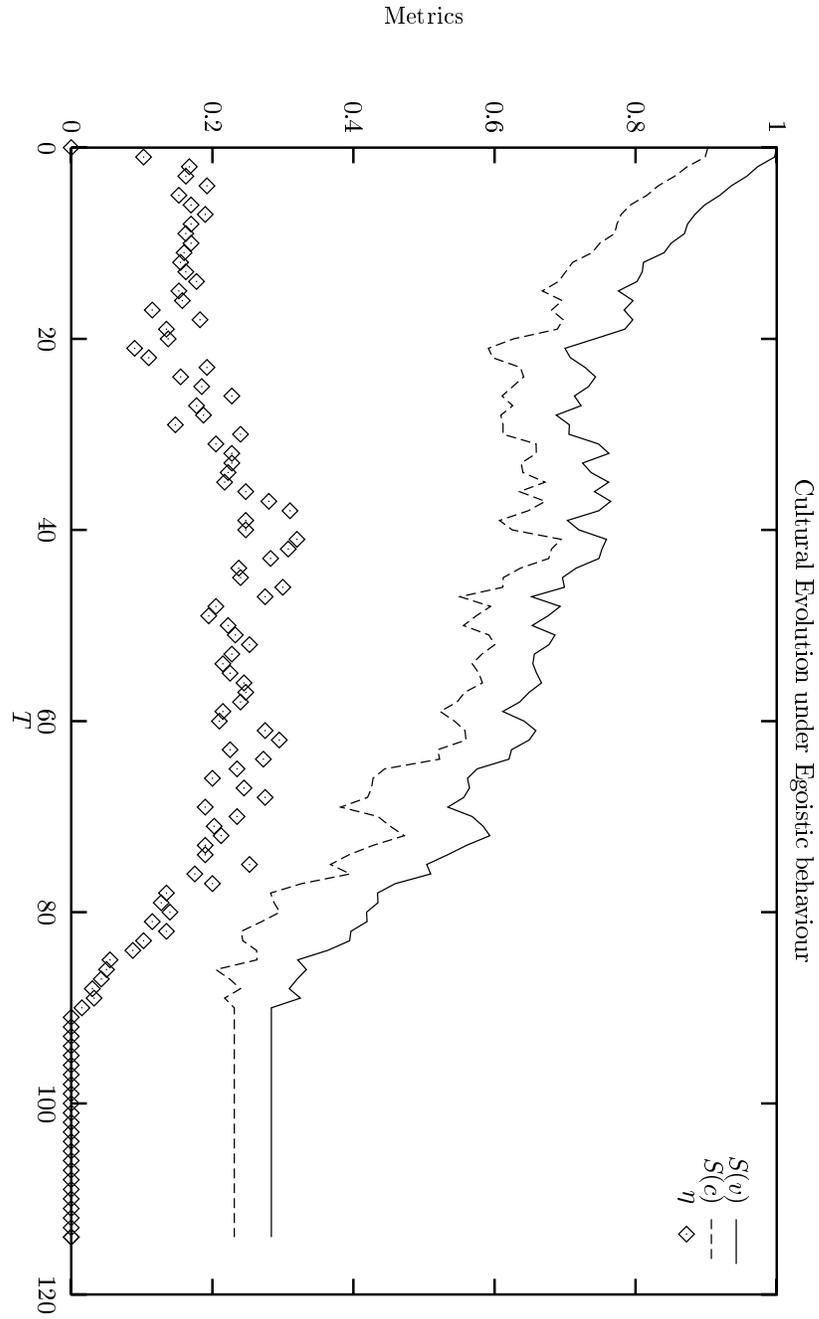}
  \caption{Entropies and activity for the evolving culture in figure \ref{fig:ego-1d}}
  \label{fig:ego-1-a}
  \end{center}
\end{figure}

The plot of these metrics against time for the culture whose initial and
final states are shown in figure \ref{fig:ego-1d} appears in figure
\ref{fig:ego-1-a}. The horizontal axis, marked \emph{T}, represents the
number of cycles.

Neither the variety nor the compatibility entropy has reached zero, but
after 25 periods of no activity and no change in the number of varieties,
the system is static. Entropy drives activity, the higher the entropy of
the system the greater the activity can be. One can divide the activity of
the system into four broad epochs:

\begin{enumerate}
\item Grouping and Simplifying - Anarchy

  Entropy is high and the activity $\eta$ rises quickly and is relatively
  constant for the first epoch $[0, 20]$; the entropies fall rapidly. The
  compatibility entropy is more or less synchronised with the variety
  entropy, meaning that as a new variety is formed it is compatible with
  the majority of other varieties (which one would expect, since there are
  so many varieties around.) The end of the anarchic epoch is characterised
  by an entropy dip. This is due to some traits being annihilated at the
  edges of the square. There is a lull in activity during this dip.

\item Migrating - Collectivism

  During the epoch $[20, 40]$, the activity increases, but the entropies
  remain relatively constant. Critically, the compatibility entropy
  increases and is slightly advanced in phase relative to the variety
  entropy: implying that an interaction between two agents generates
  another agent having a different variety. The traits are migrating across
  the population.

\item Concentration - Oligarchy

  Epoch $[40, 70]$ has a constant level of activity, but both entropies
  begin to fall: no more new varieties are being generated and varieties
  are forming into incompatible groups. The compatibility entropy now falls
  behind the variety entropy in phase and the difference between the two
  increases. This implies that when two agents interact the agent is either
  unchanged or becomes identical to the neighbour with which it interacted.

\item Isolation and Stasis - Authoritarianism
  
  From $[70,90]$ the activity falls as do the entropies. The compatibility
  entropy falling behind in phase and having more pointed peaks than the
  variety entropy. By period $90$, the system is inactive.

\end{enumerate}

At the end of the simulation there are $10$ varieties: these are ordered
and inter--related as shown in table \ref{tab:culture:1}. The first--ranked
is incompatible with all the others and has a numeric majority over all the
others combined. It is now a perennial dictator, it cannot be changed and
has a numeric majority.

\begin{table}[htbp]
  \begin{center}
    \begin{tabular}[left]{|r|l|r|p{1.25 in}|}
      \hline
      Order & Identity & Number & Compatible with \\
      \hline
      1 & 1 8 3 5 8 & 56 & none \\
      2 & 6 7 1 6 5 & 27 & 3, 8, 9, 10 \\
      3 & 8 5 6 6 7 & 5 & 7, 8, 10 \\
      4 & 7 2 2 4 9 & 3 & 5, 6 \\
      5 & 5 3 8 4 2 & 2 & 6, 7 \\
      6 & 3 6 8 4 5 & 2 & 4, 5, 9 \\
      7 & 9 3 5 8 4 & 2 & 3, 5, 8 \\
      8 & 2 7 6 8 1 & 1 & 2, 3, 10 \\
      9 & 6 8 1 1 5 & 1 & 2, 6, 10 \\
      10 & 6 7 7 6 7 & 1 & 2, 3, 8, 9 \\
      \hline
    \end{tabular}
    \caption{Identities in an evolved culture}
    \label{tab:culture:1}
  \end{center}
\end{table}

\paragraph{Some Expected Deductions}

The activity plot in figure \ref{fig:ego-1-a} shows that the system
stabilised in $90$ periods{\myDash}each period allows up to a hundred
interactions. Axelrod was able to substantiate some intuitive deductions:

\begin{enumerate}

\item More features, More interaction
  
  The more features agents possessed the more likely they were to
  interact. The efficiency $\eta$ had a higher average. (There were no
  conclusions as to its expected distribution.)

\pagebreak[2]

\item More traits, Less interaction

  The more traits per feature the less likely agents were to interact. The
  efficiency $\eta$ would be lower in this case.

\item Bigger neighbourhoods, More interaction

  The larger the neighbourhood (that is, the number of adjacent
  neighbours), the more likely agents were to interact.
  
\end{enumerate}

\paragraph{An Unexpected Deduction}

The relative spread of regions of dissimilarity and their distinctness
proved to be less intuitive. Axelrod found that \emph{the larger the
  system, the fewer the number of dissimilar regions}. The explanation for
this is rather subtle: traits migrate across the system on a random walk
\cite[p.  389]{stoch:papoulis}; the more random the system is, the further
they will progress, the system is random for longer if it is larger;
therefore, the larger the system, the wider the spread of a particular
trait, therefore the less likely it is that the trait will be confined to
one isolationist group.

This result can be interpreted as a thermodynamic effect: the system is a
hot liquid that is cooling, substances dissolve in it and are dispersed by
Brownian motion; the greater the volume of liquid the more mixing takes
place.

\begin{figure}[htbp]
  \begin{center}
  \includegraphics{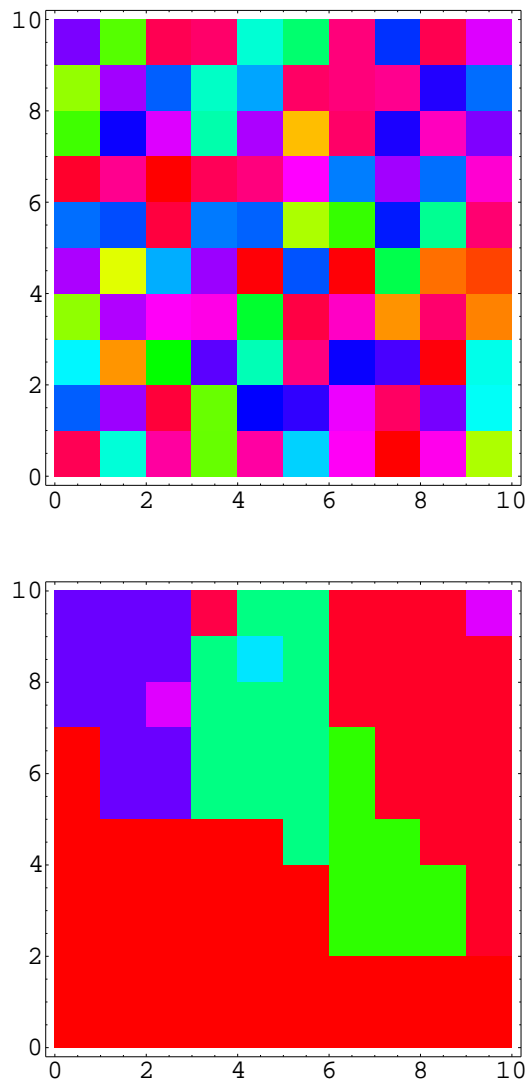}
  \caption{Density plot for competing varieties in a culture at $t_0$ and $t_{800}$}
  \label{fig:ego-lc}
  \end{center}
\end{figure}

\begin{figure}[htbp]
  \begin{center}
  \includegraphics[angle=-90,keepaspectratio=1,totalheight=7in]{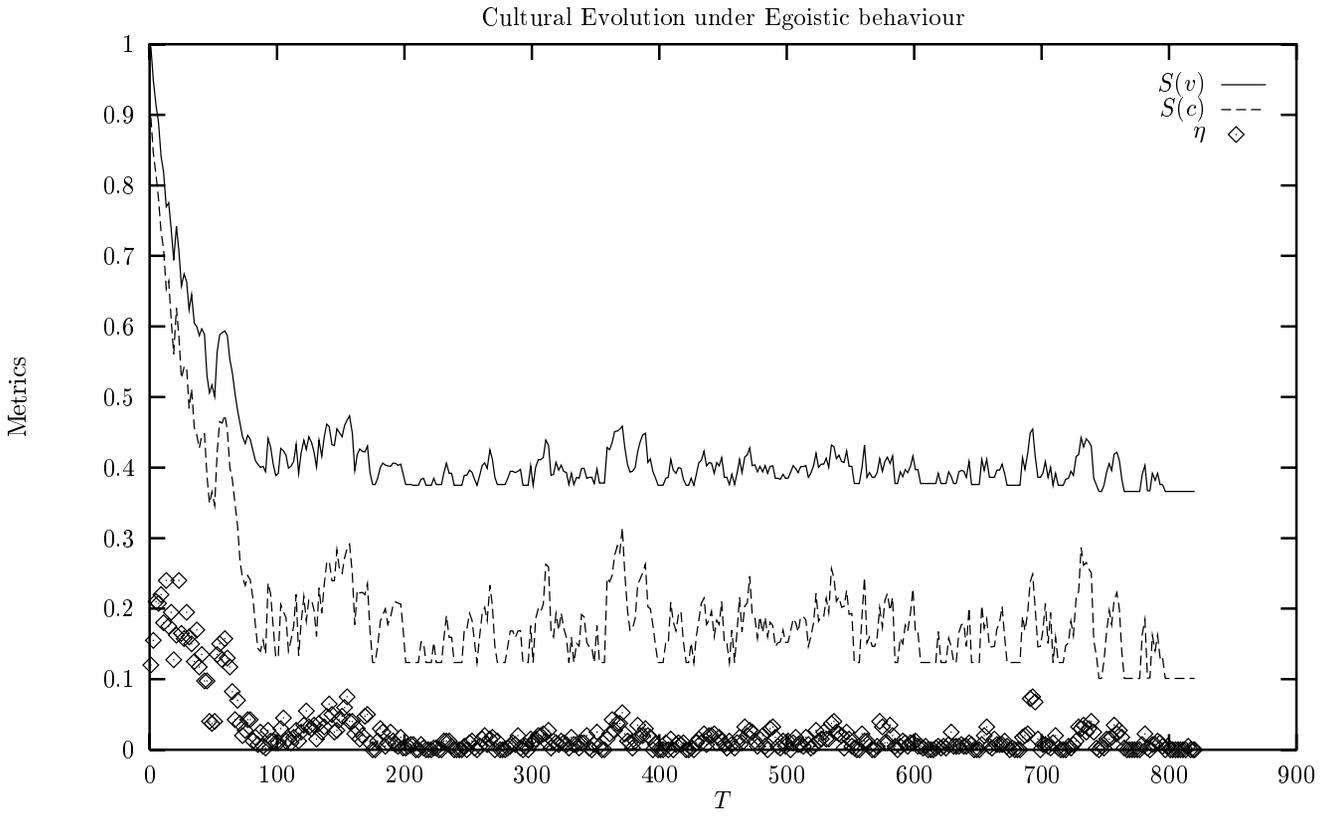}
  \caption{Entropy and activity for the evolving culture in figure \ref{fig:ego-lc}}
  \label{fig:ego-lc-a}
  \end{center}
\end{figure}

This observation also helps in understanding the variation in the number of
distinct regions. On the whole, one variety will tend to dominate all
others: it is probabilistically more likely to reach stasis in this way. If
two blocks of varieties were to form which were, more or less, of equal
size and they were identical in every respect except one, then a limit
cycle would develop. At the border between the two blocks, some would
accrete a trait and join the other block while a similar number would
accrete the other trait and join the other block. Figure \ref{fig:ego-lc}
shows a density plot which has three blocks surrounded by a fourth. The
block of three are compatible with one another and differ very slightly
from the block surrounding. Figure \ref{fig:ego-lc-a} shows the metrics
over time, the system behaves very much as any other would up to period
$200$, thereafter it enters a cycle.

\subsection{The Collective Choice Interpretation}
\label{sec:axelrod:interp}

Axelrod modified the cultural model in a number of ways and drew further
conclusions which will prove to be useful later. The relationship between
features and traits can be interpreted as a collective choice procedure.

\paragraph{Axelrod's Investigations}

Axelrod's simple model does help to explain large--scale system behaviour
given an intuitively appealing small--scale behaviour. Axelrod investigated
the effects of different topologies and different stochastic inputs for
selection{\myDash}in particular, selecting central agents more often,
because traits can be destroyed at the edges{\myDash}and the initial
allocation of traits{\myDash}using a Gaussian distribution, to see how much
the final varieties could vary from a variety of Gaussian averages.

Axelrod chose not to vary the rules agents used to accrete traits. This
does affect how the model can be used to make a collective choice.

\paragraph{Collective Choice Interpretation}

The earlier analysis of preference hierarchies and preference aggregation,
see chapter \ref{cha:top}, gives an insight into the operation of the
cultural model as a collective choice procedure.

Assume there are three issues, $a,b,c$, a system must determine the
relative strengths of each of them given that each agent (or voter) is
allowed to express a preference ordering using, for simplicity, a
strong ordering. The orderings, $D$, is the set given in
\eqref{eq:axelrod:orderings}.

\begin{equation}
  \label{eq:axelrod:orderings}
  D = \{ a>b>c, a>c>b, b>c>a, b>a>c, c>b>a, c>a>b\} 
\end{equation}

If the features are $F = \langle a>b, a>c, b>c \rangle$ and the traits for
each feature are $T = \{1,-1\}$ then an agent preferring $a>b>c$ would have
a feature set of traits: $\langle 1,1,1 \rangle$, and an agent having
$a>c>b$ would have $\langle 1,1,-1 \rangle$. If these two agents interacted
they would quickly settle their differences on the relative merits of $c$
and $b$ and form a single variety of agent, similarly for agents preferring
$b$ and $c$.

If an agent preferring $a$ were to interact with an agent preferring $b$
then $\langle 1,1,1 \rangle$ meeting $\langle -1,-1,1 \rangle$ would allow
them to exchange their primary preference and the internal state of the
receiver would become inconsistent, so the criterion for interaction is
that the two agents must agree on two preferences before they can interact.

This interpretation is valid for a wider issue set and for weak orderings,
the choice of the number of traits on which to agree does become more
complicated. Referring to table \ref{tab:orders}, for three issues there
are 13 different weak orderings of those issues with respect to one
another. The features would be $F = \langle a>b, a>c, b>c \rangle$ and
the trait set would be $\{1, 0, -1\}$. An agent having a preference
ordering: $a>b>c$ would have a feature set: $\langle 1, 1, 1 \rangle$
and $a=b=c$ would be $\langle 0, 0, 0 \rangle$.

It is possible to be more liberal in this interpretation. If one has $n$
issues and the number of orderings is $O(n)$, as given by table
\ref{tab:orders}, if one then wants to simplify the orderings to some
sub--set, then one should set the number of features $\#(F)$ and the number
of traits $\#(T)$ so that $\#(F) \cdot \#(T) < O(n)$. The cultural model
then acts as a genetic algorithm, but of a peculiar kind: it crosses
varieties with one another, but does not mutate, and requires no global
fitness function.

One must then impose some kind of topology that allows each variety to
interact with every other. This need not be a mesh topology, because the
traits only have to find a migration path. A spanning tree for the
preference orderings will suffice and it is easy enough to set a size for a
useful spanning tree using \eqref{eq:axelrod:binom}, where $n$ is, once
again, the number of issues to be resolved.

\begin{equation}
  \label{eq:axelrod:binom}
  \Sigma_{i=1}^n \binom{n}{i}
\end{equation}

The cultural model can then be thought of as collective choice procedure
that forms a spanning tree of a well--ordered graph. Unlike the techniques
given in \myRef{sec:cycles-entropy}, it is a stochastically--driven
heuristic method. It attempts to reduce a prefereence ordering search space
of order $O(n)$ to a strongly--ordered subset of those orderings where some
of the issues have been merged by allowing a weak--ordering.

(The fact that some of the feature sets produced during the operation
of the cultural model may give rise to inconsistent states can be
justified in the same way as it is with genetic algorithms. It is
simply a transient state that allows new varieties to be developed.
This is argued more persuasively in \cite{ga:handbook}.)

\section{Formation of Cultures under Peer Pressure}

Axelrod's simple behavioural rule and the large--scale effects it appears
to introduce is discussed first.  A new rule is introduced and a system
using it is simulated and the results analysed.

\subsection{Egoistic Behaviour}

The small--scale behaviour that agents follow is described in the title of
figure \ref{fig:ego-1} as \textit{Egoistic}. Each agent can accrete a trait
from one of its neighbours regardless of the state of its other neighbours.

\paragraph{Increasing Heterogeneity}

\begin{figure}[htbp]
  \begin{center}
  \includegraphics{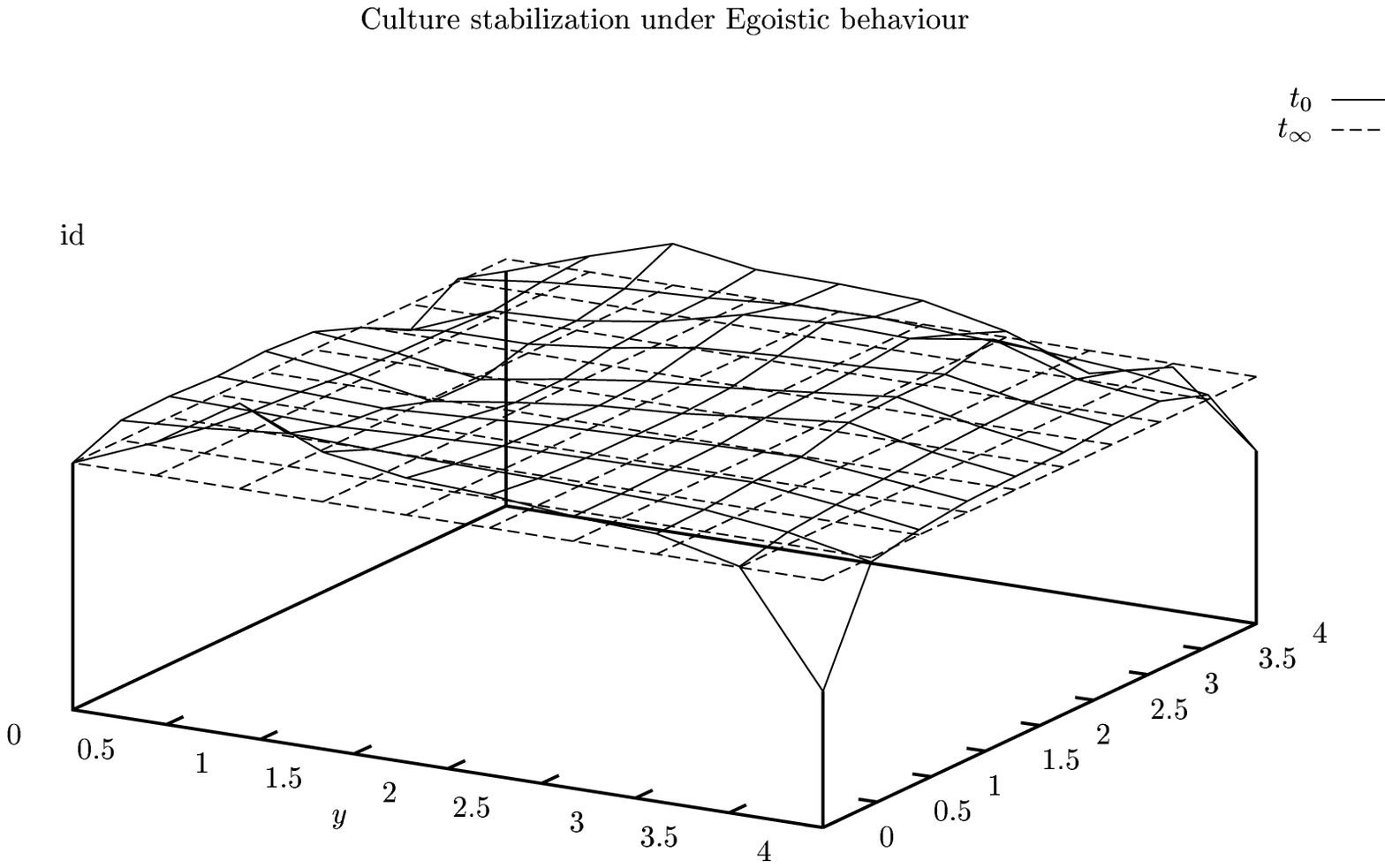}
  \caption{Culture that evolves to homogeneity}
  \label{fig:ego-2}
  \end{center}
\end{figure}

Some of the simulations conducted used a smaller playing field ($5$ by $5$)
with only $3$ features and $3$ traits per feature. These show that agents
can quickly agree to not differentiate amongst themselves, see figure
\ref{fig:ego-2}. (The axes are labelled in the same way as in
\ref{fig:ego-1}: identity, \textit{id}, on the vertical axis, location
within the playing surface on the \textit{x} and \textit{y} axes.)

It is unlikely that a wholly homogeneous culture should arise with a larger
more complex playing field, but it is usually the case that one variety of
agent wholly dominates the others. It might be desirable to control the
degree of variety.

\paragraph{Restricting the migration of traits}

\begin{figure}[htbp]
  \begin{center}
  \includegraphics{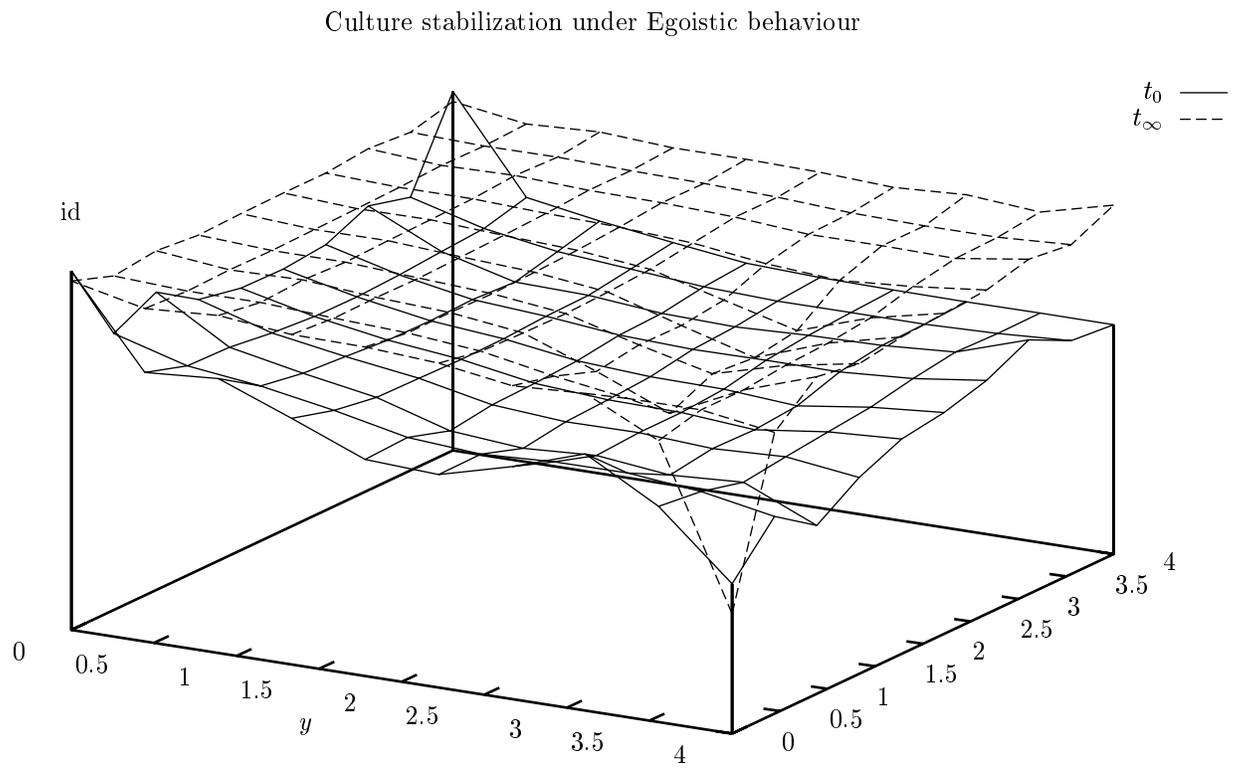}
  \caption{Culture unduly influenced by some agents}
  \label{fig:ego-3}
  \end{center}
\end{figure}

Of particular interest for information security is how a small
sophisticated group might be able to enforce a consensus that certain views 
should be globally accessible. Figure \ref{fig:ego-3} illustrates
this. Here, two agents placed at the origin and at $(4, 0)$ have seen their 
principal traits migrate across almost the whole of the surface.

It would be desirable that there be distinct regions having access to
particular views/traits, but it is not desirable that views/traits migrate
indiscriminately across regions.

\subsection{Peer Agreement}

Another simple behaviour that could be employed by an agent is to require
that one of its other neighbours also be compatible with the neighbour from
which it would accrete the trait. Unfortunately, to make this rule more
precise a modal logic is required. Modal logics are briefly discussed in
appendix \ref{cha:modal}.

The model under which an agent $x$ operates is denoted $\myModel$.
Containing a set of agents $\mathcal W$, which are its neighbours $\{y_1,
\dots, y_n\, z_1, \dots, z_n\}$ and a set of truths, the traits,
distributed amongst the worlds, $P = \{ P_0, \dots, P_N \}$.

\paragraph{Egoistic Behaviour}

Firstly, egoistic behaviour can be more formally specified. The
probabilistic fuzziness of $U(0,1)$ in \eqref{eq:adaptive:axelrod:1} is not
given in this formal specification. The form is of a schema, the conditions
above the line must appertain and the condition below the line can be
enforced.

\begin{equation}
  \frac{
    \begin{matrix}
      \myModelForAt{x} & P_i \wedge \neg P_j \wedge
      \possibility ( P_i \wedge P_j ) \\
    \end{matrix}
    } {
    \begin{matrix}
      \myModelForAt{x} & P_j \\
    \end{matrix}
    }
  \label{eq:adaptive:r1}
\end{equation}

$x$ has been selected and holds the trait $P_i$ but not $P_j$. There is
another agent in his neighbourhood where both $P_i$ and $P_j$ are held. $x$
accretes $P_j$\footnote{The class of modal logic employed has to be
  irreflexive. In particular any axiom which prevents $\possibility (P_i
  \wedge P_j) = \possibility P_i \wedge \possibility P_j$, \ie $P_i$ and
  $P_j$ must reside in the same agent.} (Incidentally, $i$ is not equal to
$j$ because $P_i \wedge \neg P_j$ would be false.)

\paragraph{Peer Agreement Behaviour}

A form of peer agreement behaviour can be expressed thus:

\begin{equation}
  \frac{
    \begin{matrix}
      \myModelForAt{x} & 
      P_i \wedge \neg P_j \wedge \possibility (P_i \wedge P_j) \\
      \myModelForAt{x} & \possibility ( P_j \wedge \neg P_i ) \vee
      \possibility (
      P_k \wedge \neg P_j \wedge \possibility (P_k \wedge P_j) 
      ) 
    \end{matrix}
    } {
    \begin{matrix}
      \myModelForAt{x} & P_j \\
    \end{matrix}
    }
  \label{eq:adaptive:r2}
\end{equation}

The first condition is the same as in \eqref{eq:adaptive:r1}. The second
has two parts, there is someone else in the neighbourhood who:

\begin{itemize}
\item Either: holds $P_j$ but not $P_i$.
\item Or: could also accrete $P_j$, \ie is compatible with $z$ in some
  other way.
\end{itemize}

This form of behaviour{\myDash}implemented as the class
\textsf{PeerPossible} in \cite{sft:aidan:cultural}{\myDash}is effectively
a membership rule. If $z$ is the holder of $P_j$ who also holds $P_i$, then
$x$ proposes $P_j$ and someone, $y$, seconds it.

In the former case, if the number of neighbours is limited to four (as they
are in the Axelrod square topology) then five may vote and a majority, $x$,
$y$ and $z$, have stated that they are compatible with $P_j${\myDash}$z$
votes ``for'' because it already holds $P_j$, the other two, $x$ and $y$,
because they hold something that $z$ also holds.

\subsection{Some Expectations}

\begin{enumerate}
\item Slower Trait Migration and More Probable Limit Cycles
  
  Clearly, one can expect the rate of trait migration to be slower. A trait
  not already extant in a neighbourhood will have to wait for a trait that
  is extant to join it, before it can propose itself. If trait migration is
  very slow and in pairs, it might be the case that traits repeatedly cross
  and re--cross the field without actually appearing in the same agent
  together. This could lead to very long limit cycles.

\item Edge Effects

  Under egoistic behaviour, if a trait becomes isolated at an edge, it had
  one less degree of freedom in the direction in which it could migrate.
  Trait migration on a square field tends to be from the centre to the
  edges. Under \textsf{PeerPossible} behaviour it should be the case that
  traits will be more difficult to dislodge from the edges, because the
  neighbourhood they belong to has one less voter, but, as proposer and a
  seconder are still needed, three out of four must concur; at the corners,
  the condition is even more stringent, requiring three out of three to
  concur.

\item Dormancy and Second Waves

  The edge effects might lead to agents being able to preserve traits at
  the edges and corners so that as the system stabilises, and the traits
  held at the centre migrate outward, the traits held at the edges would
  overcome the traits that originated from the centre. This would lead to a
  second wave of activity.

\end{enumerate}

\subsection{Some Results}

\begin{figure}[htbp]
  \begin{center}
  \includegraphics{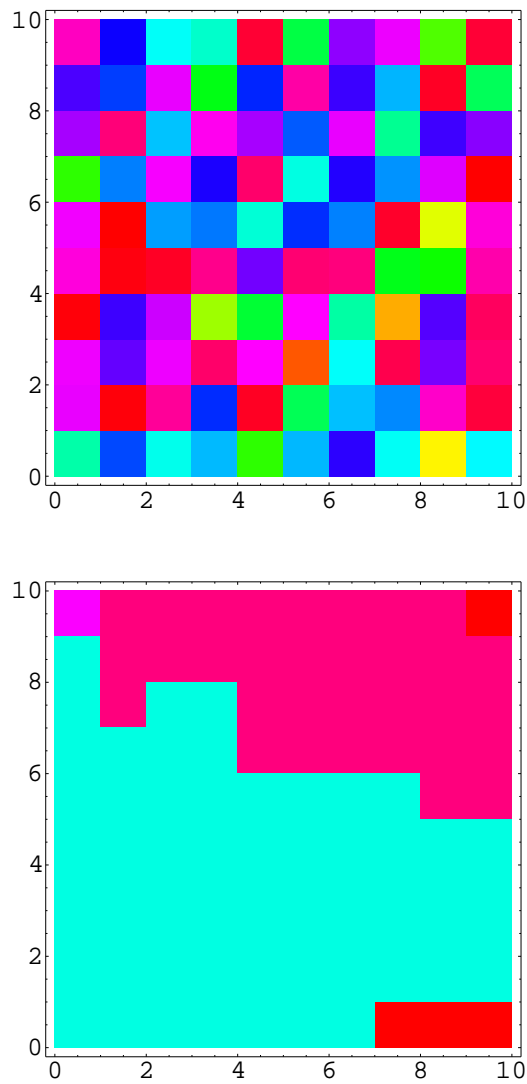}
  \caption{Peer Agreement Density Plot at $t_0$ and $t_\infty$}
  \label{fig:peer-1}
  \end{center}
\end{figure}

\begin{figure}[htbp]
  \begin{center}
    \includegraphics[angle=-90,keepaspectratio=1,totalheight=7in]{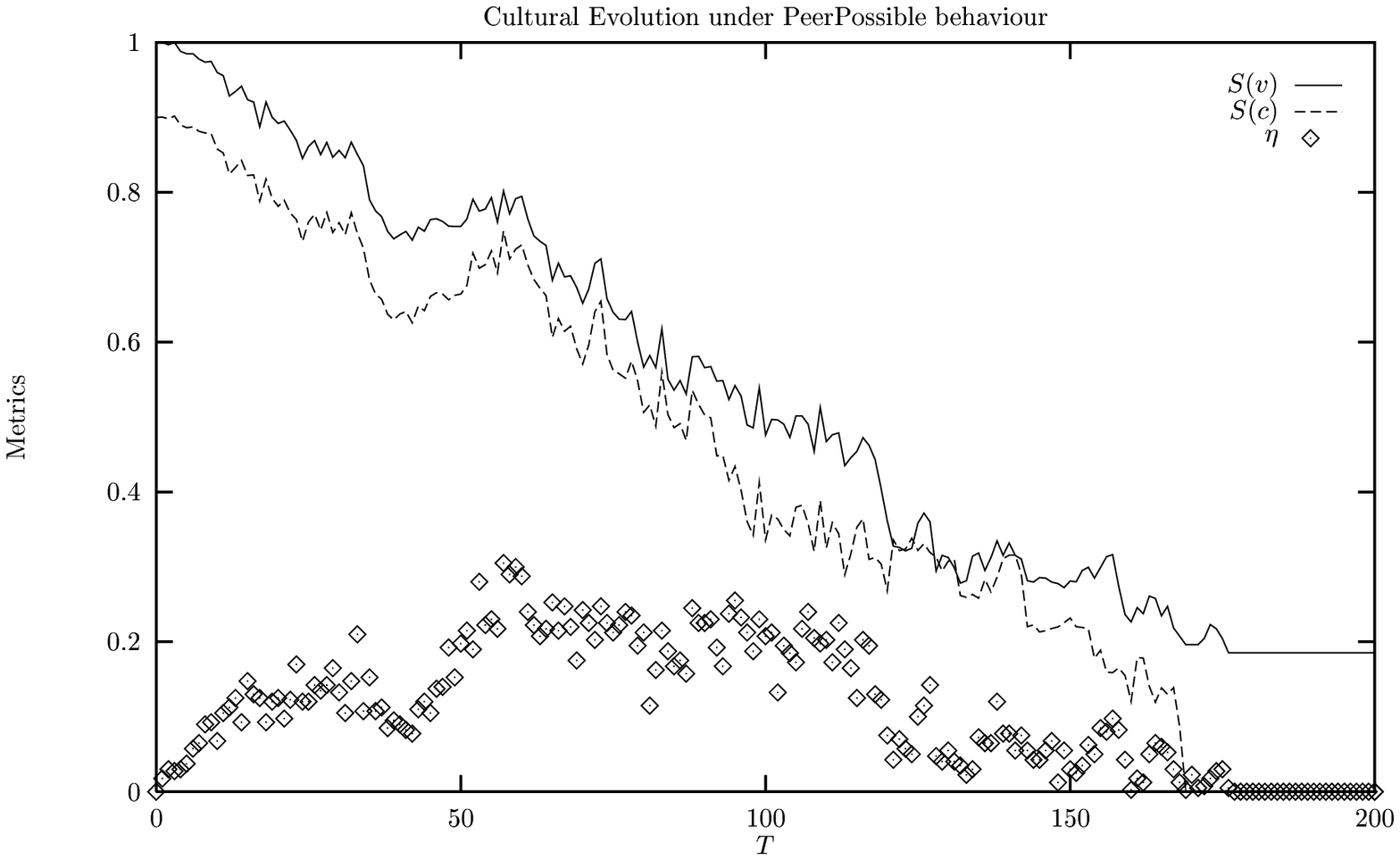}
  \caption{Activity and entropy for figure \ref{fig:peer-1}}
  \label{fig:peer-1a}
  \end{center}
\end{figure}

A set of simulations was undertaken using \textsf{PeerPossible} behaviour
instead of \textsf{Egoistic}. The model is susceptible to the effects of
different starting conditions{\myDash}the initial allocation of traits to
each individual, $\binom{10^5}{100}$ different configurations{\myDash}and
the number of different sequences of interactions{\myDash}$100^{200}$ for a
typical $200$ period run.  Nonetheless, some useful results were observed.
A typical pair of density plots and an activity plot appear as figures
\ref{fig:peer-1} and \ref{fig:peer-1a} respectively.

\paragraph{Large--scale behaviours}

\textsf{PeerPossible} is similar to \textsf{Egoistic} behaviour in that it
gives rise to diverse populations which can either be ultimately quiescent
or fall into a limit cycle. \textsf{PeerPossible} leads to limit--cycling
populations more often than \textsf{Egoistic}{\myDash}as was expected.
Unlike \textsf{Egoistic} behaviour these can be predicted and can remain
relatively stable. \textsf{PeerPossible} behaviour very often results in
limit cycles between comparably sized groups, but these are more or less
defined after $200$ periods (this is discussed in more detail later.)

This can be summarised:

\begin{itemize}
\item If the system does stabilise quickly, it invariably results in one
  dominant variety.
\item If the system takes longer to stabilise, then a limit cycle with a
  dominant variety varying in the number of members is the usual result.
\end{itemize}

An analogy to political systems might be useful here: systems that
stabilise rapidly to an authoritarian regime are similar in behaviour to
third world political systems{\myDash}an immutable consensus emerges;
systems that exhibit limit cycles are comparable to first world political
systems{\myDash}a variable consensus emerges.

Whether a population will stabilise quickly can be determined quite
reliably by the changes in its entropy characteristics as it evolves.

Figure \ref{fig:peer-1} shows different identities are usually attached to
an edge. This is also true of cultures produced by egoistic behaviour, but
it appears to be more marked for \textsf{PeerPossible} behaviour. This is a
result that was also expected.

An interesting and useful side--effect of slower trait migration is that
large--scale behaviour becomes more predictable because traits are more
likely to cluster in their original locations and individuals at the edges
tend to become the dominating variety. Referring to figure
\ref{fig:peer-1}, there are two large distinct regions:

\begin{itemize}
\item The turquoise lower left--hand side
  
  The colours in the lower left--hand side corner of the initial state are
  more often of the turquoise hue that will prove to be dominant. There are
  some agents on the edge, at $(0,0)$, $(0,3)$ and $(0,7)$, that are
  already of colour that will prove to be dominant in that corner. Note
  that $(0,7)$ and $(0, 10)$, already closely related to the dominant
  turquoise, have joined the red variety.

\item The garnet upper right--hand side

  This area appears to have been constructed in response to the turquoise
  area. There are no explicitly garnet individuals in the initial
  populations, the final colour appears to be a blend of red and the light
  puce coloured individuals. Notice that the individuals in the upper
  corners are unchanged throughout the evolution.

\end{itemize}

\paragraph{System Activity}

Referring to the four epochs that were characterised for egoistic
behaviour, there are some differences for \textsf{PeerPossible}:

\begin{enumerate}
\item Anarchy
  
  The anarchic period appears to last about twice as long as it does under
  egoistic behaviour, as one might expect, because the level of activity is
  about half. When both entropies fall to $0.6$ the collectivistic epoch
  commences. It also exhibits the dip associated with traits being
  annihilated at the edges.

\item Collectivism
  
  This is markedly different from egoistic behaviour. The entropy
  falls throughout the collectivistic epoch{\myDash}meaning that the system
  is organising itself faster. Other than that, it behaves in a similar
  manner: the difference between the variety and compatibility entropy
  reduces and the latter leads the former.

\item Oligarchy
  
  Under egoistic behaviour, this epoch is marked by a fall in entropy, and
  an increasing difference between variety and compatibility entropies
  which leads to a phase lead becoming a lag. Under \textsf{PeerPossible}
  behaviour only the phase change is noticeable, because the entropy has
  fallen to critical during the period of collectivism.

\item Authoritarianism

  The authoritarian epoch is the same under both egoistic and
  \textsf{PeerPossible} behaviour.

\end{enumerate}

Generally \textsf{PeerPossible} behaviour has a level of activity that is
$10\%$ lower than egoistic but takes about twice as long to stabilise. The
latter is commensurate with the requirement under \textsf{PeerPossible}
behaviour that an individual must gain a corroborating
neighbour{\myDash}suggesting that two agents, probabilistically, take twice
as long to agree as one{\myDash}but the level of activity is not half of
what it was under egoistic behaviour.  This would suggest that
\textsf{PeerPossible} behaviour is more efficient{\myDash}in that, the
interactions between agents are not as often undone.

\paragraph{Limit Cycles}

\begin{figure}[htbp]
  \begin{center}
  \includegraphics{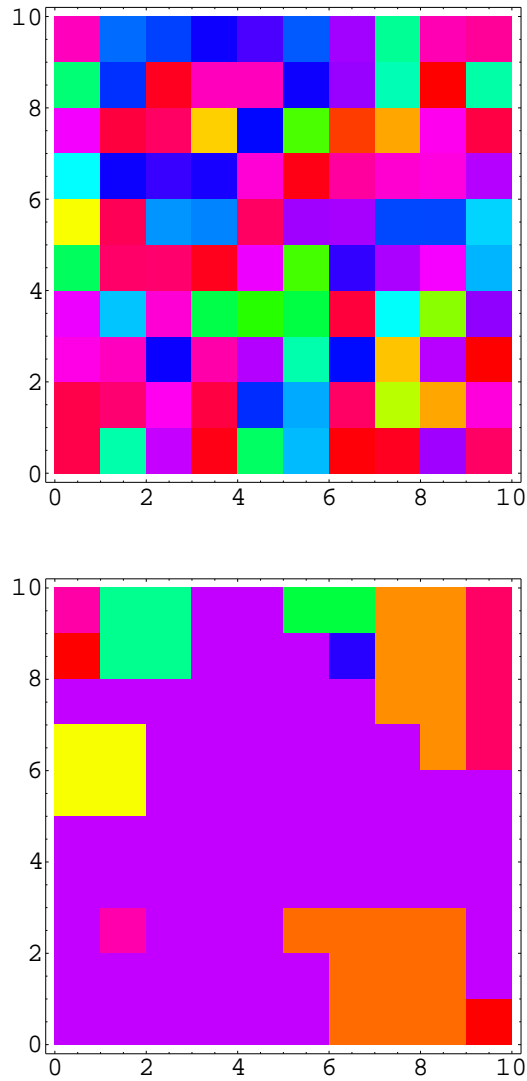}
  \caption{Peer Agreement Density Plot at $t_0$ and $t_{1100}$}
  \label{fig:peer-2}
  \end{center}
\end{figure}

\begin{figure}[htbp]
  \begin{center}
    \includegraphics[angle=-90,keepaspectratio=1,totalheight=7in]{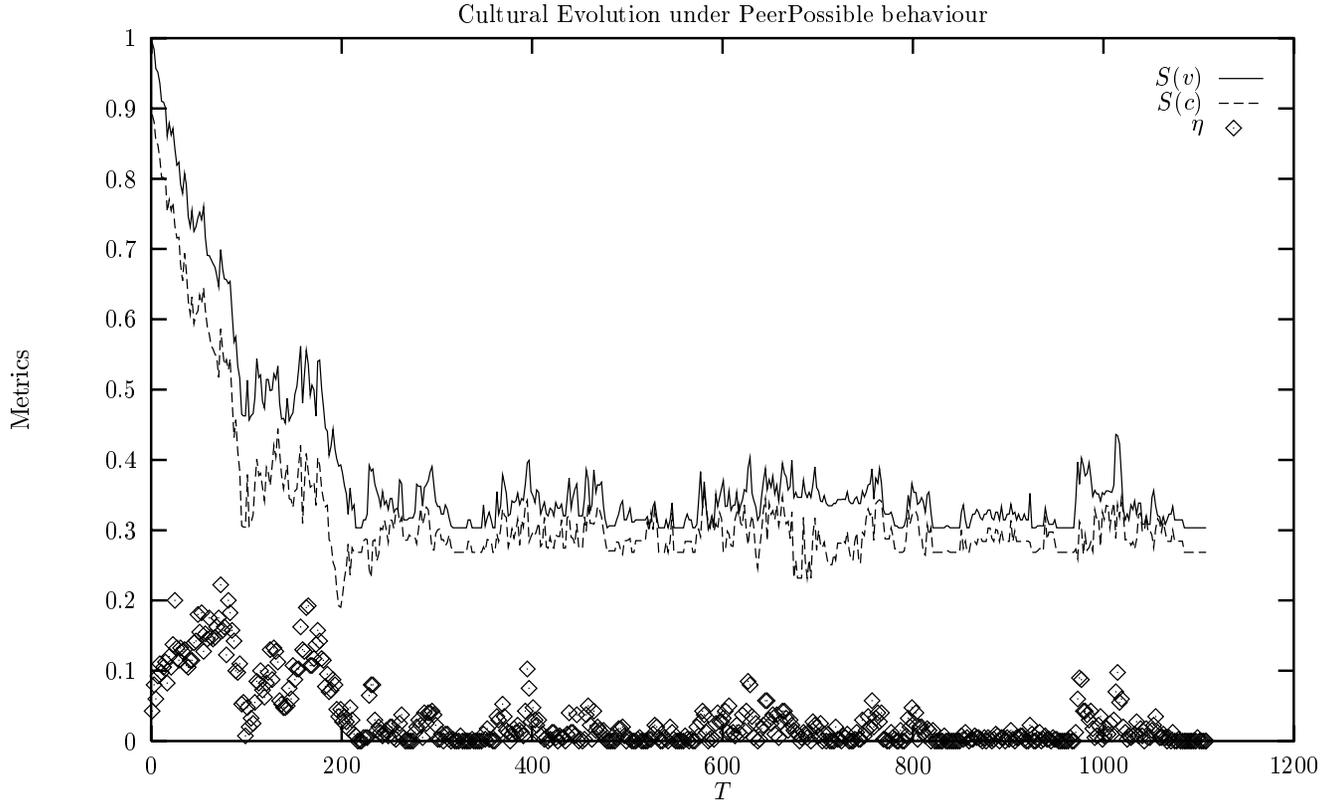}
  \caption{Activity and entropy for figure \ref{fig:peer-2}}
  \label{fig:peer-2a}
  \end{center}
\end{figure}

\textsf{PeerPossible} behaviour, as predicted, does suffer more from limit
cycles. In a set of 32 simulations only 11 reached stasis. It would seem
that the limit cycle is the preferred global behaviour for local
\textsf{PeerPossible} behaviour. A good example of a limit cycle's activity
appears in figure \ref{fig:peer-2a}. The state of the agents appears in
figure \ref{fig:peer-2}.

Although the system cannot reach stasis, which is arrived at when the
number of varieties is constant for 25 periods, the system has been more or
less stable since period 200, which is quite typical of
\textsf{PeerPossible} systems. Looking at the state of the agents, one
variety has dominated the others and, because of the lack of variation of
the variety entropy, has done so for some time.

One could be fairly confident in saying that when the variety entropy has
fallen below $50\%$ and the compatibility entropy is less than the variety
entropy, a system is probably stable, in that the dominant variety will
remain so.

It appears to be very rare for a system in a limit cycle with a dominant
variety to further evolve so that variety is no longer dominant.

\subsection{Some Conclusions}
\label{sec:peer:conc}

\textsf{PeerPossible} behaviour does seem to lead to more predictable
systems which, more often than not, avoids an authoritarian terminal state
and that trait accretions are less frequently reversed later.

\section{Predictability of Large--Scale Behaviour}

The simulation model has been used to determine emergent properties of
large--scale behaviour given different small--scale behaviours. In
\myRef{sec:adaptive:axelrod:large} it was seen that larger playing fields
led to more homogeneous cultures. The simulations used to demonstrate this
property were highly stochastic: individuals were given random traits, they
were then randomly located and then in each period randomly chosen to
interact with one another.

It is also hoped that this analysis of the evolution of cultures can give
us some assurance that, were a system allowed to organise itself, it would
consistently arrive at more or less the same set of cultures if the
individuals within it start with the same traits and behave in the same
way. This would mean that a safe access control management system would
arrive at very much the same allocations of access rights if the
individuals start with the same interests.

In the context of the simulation model developed above, some assurance must
be gained that the final state of the system is statistically independent
of the location of the individuals and when they are chosen to interact
with one another.

Before describing the experiments that were conducted to provide this
assurance, some insight will be gained from the analytic research that has
been conducted in this field.

\subsection{Analytic Research}

Essentially, some guidance is needed on how to construct a simulation model
whose final state will be almost wholly dependent on the initial states of
the individuals: their location with respect to one another and the
sequence they interact with one another is not important.

\subsubsection{Voter Models and Initial Distributions}

Axelrod pointed out in his paper \cite{Axelrod:culture}, that the
simulation model is a variant of the \emph{voter model} in which a particle
aligns itself with its neighbours based on whether they hold the same value
or not.

\paragraph{Consonant voting}

Bramson and Griffeath, \cite{voting:cluster:bramson}, have made a
comprehensive analysis of voter models having only one trait.  They quote
results showing that voters in one-- or two--dimensional space tend to
converge weakly to a majority, either for or against, which is dependent
only on the initial ratio of voters for and against. The voter model they
analysed was a \textit{consonant} voter model meaning that a voter aligned
himself to be the same as his neighbours.

Bramson and Griffeath's main interest was to establish conditions under
which the process would be ergodic, \ie under what conditions limit cycles
would not occur. They showed, analytically, that in one--dimensional
systems the consonant voting model for one trait individuals was ergodic,
but for more than two or more dimensions, \ie four neighbours or more, it
was not ergodic. 

They also showed that even though two-- or more dimensional systems might
not be ergodic, the ratio law still applied. In that, the probability of
the system attaining a state where the ratio of for and against voters was
reversed from the initial state of the system was ergodic: under a
consonant voting model, the same simple majority will be maintained. There
were a number of provisos to this. If the ratio was close to $\frac{1}{2}$
there was as a possibility of short excursions when the majority would be
reversed, but not indefinitely.

\paragraph{Dissonant voting}

Bramson and Griffeath's paper also analysed dissonant voting models and
discovered that they were unable to ascertain whether the ergodicity
theorems they had developed could be shown or not. Their analytical
technique was lacking because dissonance introduced cumulatively larger
probabilities of dissimilarity. It appears that, under dissonant behaviour,
chaotic behaviour can develop which can lead to very long limit cycles
which may not hold an initial majority in place.  This can be demonstrated
with reference to another of Axelrod's behavioural investigations.

The basis of Axelrod's simulation is a simple behavioural interaction,
which is best expressed in modal logic. The modal logic expression is a
useful formalisation, but it has proved difficult to extend it to describe
the dynamics of interacting systems. Axelrod's cultural model was preceded
by, and is, in some ways, an extension of, the Iterated Prisoners'
Dilemma{\myDash}IPD, see \cite{econ:ipd:1} for example. Each prisoner has
one neighbour, so it a very simple model under Bramson and Griffeath's
analysis, but the dynamics can be very complex. An analysis of the simple
interaction underlying the IPD was carried out by Mar \cite{ipd:chaos}. He
showed that this could lead to a system which possessed chaotic
self--similar behaviour if one of the prisoners acted consistently
\textit{dissonantly}.

Unfortunately, consonance and dissonance become non--bivalent concepts when
more than one trait is involved and the properties of the distance metric
and the behavioural rule that uses it become important.

Because of this property, it is not possible to make any useful predictions
about large--scale behaviour in dissonant systems. Axelrod's cultural model
can move from a disordered to an ordered state with predictable
large--scale behaviour, but it cannot move from an ordered state to a more
disordered one and remain predictable.

\subparagraph{The relative consonance of \textsf{PeerPossible}}

Referring to the behavioural models that have been investigated:
\textsf{PeerPossible} behaviour, it was noted in \myRef{sec:peer:conc},
gives rise to fewer trait reversals than \textsf{Egoistic}. This would
suggest that \textsf{PeerPossible} is a more consonant rule.

\subsubsection{Topologies and Initial Distributions}

Bramson and Griffeath's one trait voter models were immune to changes in
topology. There was no difference in large--scale behaviour if the voters
were laid out on squares, circles, cylinders, toroids or spheres. When the
voters have more than one trait, superposition effects occur which make
topology important. An analysis of the effects of topology leads to a
concept called meta--behaviour and suggests topologies that will be more
predictable.

\paragraph{Meta--behaviour and Topology}

As pointed out in the discussion of the \textsf{PeerPossible} behaviour,
agents located at the corners are more intransigent than those on the edges
who are more intransigent than those in the centre because corner agents
have only two neighbours, edge agents three and inner agents four.

It may be that the limiting distribution of identities is towards their
meta--behaviour determined by their intransigence which is, in turn,
determined by the number of neighbours they have.

\subparagraph{Squares}

This would mean that for a square topology, there would be three types of
meta--behaviour. The corner agents separate the groups of edge agents from
one another and \textit{vice versa}; this would suggest a mean of nine
varieties would evolve: four different corner varieties, four different
edge varieties and one variety for the inner agents. The inner agents would
outnumber the corner and edge agents when, for a square having sides of
length $L$: $(L-2)^2 > 4(L-1)$, \ie $L \geq 7$ and the number of agents is
$49$.

The inner agents would align themselves to have one variety and would then
separate the other types of agent from one another preventing them from
coalescing.

A useful analogy to a political system might help here: the United States
of America has more than 49 states and has a relatively stable political
spectrum. Changes to the constitution of the United States must be ratified
by 66\% of the legislative assemblies of its constituent states. This
closely approximates to the allocation of behaviours in the behavioural
model.

\subparagraph{Circles}

The number of meta--behaviours can only be changed by using a different
topology.

A circular topology could be constructed as a coiled helix{\myDash}like a
string of beads. The two end--agents would have two neighbours. The edge
agents would form one outer circle and the inner agents would be all the
agents within that circle: giving rise to three meta--behaviours. There
would then be four varieties: two types of end--agent, one type of edge
agent (they are now connected) and one type of inner agent. 

The two end--agents could then be connected to one another to give one
agent with three neighbours, \ie another edge agent. There would then be
only two meta--behaviours. The number of inner agents would exceed the
number of edge agents when $\pi (r-1)^2 > 4 \pi r$, \ie $r \geq 4$ or the
total number of agents is greater than $49$. This topology has been called
the \textit{M{\"o}bian circle} by Axelrod in \cite{Axelrod:1997}.

The inner agents would align themselves to have one variety and would then
be able to dominate the other group of edge agents. This circle is a more
responsive topology than the square, because the edge agents would be in
the majority given one defection by an inner agent. That is, it reduces to
a simple majority voting model.

The Axelrod cultural model can thus reduce a random selection of behaviours
to a choice between two aggregated behaviours:
\begin{itemize}
\item A conservative policy held by agents at the edge
\item A broad consensus policy held in the centre
\end{itemize}

Axelrod conducted a number of simulations to determine if this was the
emergent property for circular topologies and the result was more or less
in the affirmative, see \cite{Axelrod:1997}.

\subsection{Experimental Investigation}

From the analysis above, it would appear that this model would lead to more
predictable large--scale behaviour:

\begin{itemize}
\item \textsf{PeerPossible} small--scale behaviour
\item M{\"o}bian circle
\item More than 49 agents
\end{itemize}

The following experimental procedure was carried out in addition to
Axelrod's experiments: generate one set of agents and place them
randomly on a M{\"o}bian circle and allow the system to interact. When
stasis was reached, the final set of varieties of agent was recorded and
the experiment repeated with another random allocation of the same agents.

\paragraph{Regions}

These results agree closely with those of Axelrod's for the M{\"o}bian
circle where the simulations produced just two varieties for 70\% of the
simulations and these simulations reached an authoritarian state.  20\% of
simulations resulted in either, three varieties which were all mutually
incompatible and reached stasis, or, three varieties which remained
compatible but the system did not reach stasis. When three varieties
emerged it was invariably the case that two large varieties were
incompatible and separated by a small buffer region occupied by the third
small variety.  This buffer region invariably contained the two agents that
linked the outer edge with the inner core.

The remaining 10\% of the simulations seemed to be a variant of the buffer
region where there were two varieties in the buffer zone, which were
incompatible with one of its neighbouring zones.

No simulation resulted in more than four varieties. Clearly, this is very
consistent large--scale behaviour.

\paragraph{Varieties}

The simulations proved to be less decisive with regard to final
varieties.  The initial population of agents was seeded using a
binomial distribution of twelve traits for half of their features and
a uniform distribution of twelve for the remainder. The binomial
distribution was a throw of two six--sided dice\footnote{It was
  decided to operate in base twelve, because twelve has more divisors
  than ten which helped to simplify the calculations.}. The conditions
for the simulation were these:

\begin{itemize}
\item 144 agents were laid out in a M{\"o}bian circle
\item \textsf{PeerPossible} behaviour for the agents
\item There were twelve features, each having twelve traits
\item $n\%,\ n\geq 50\%$ of the agents had their lower six features
  assigned traits using the binomial and the upper six features using the
  uniform distribution.
\item The remaining $100-n\%$ had their upper six features generated using
  the binomial and the lower six generated using the uniform distribution.
\end{itemize}

It was seen that only when $n>66\%$, were the final varieties noticeably
similar across simulations with different initial distributions of the same
agents. When $50\% < n < 66\%$, one or more arbitrary traits from the
smaller group could establish themselves in the larger group. A similarity
between the varieties remained which did indicate a consensus had emerged.

It should be noted that $66\%$ is a statistically significant number. It is
one standard deviation of the normal distribution and the sum of a large
number of binomial distributions approximates to the normal.

This agrees with the Bramson and Griffeath's analysis: that the varieties
that emerged from the larger group were consistently in the majority.

\subsection{Sequences of Interaction}

It appears then that with a suitable choice of topology and an intrinsic
bias in the population a consistent consensus can be achieved. It was
decided that an investigation into the effects of the sequence of
interaction between agents was unnecessary. This may not be so easily
dismissed in a real culture where the choice of agents who may interact
with others may be biased towards particular individuals. This is worth
further investigation, but, for the time being, it is assumed that the
agents chosen to interact with one another can be safely assumed to be
uniformly random.

\subsection{Collective Choice}

Referring to the discussion of the theory of collective choice
\myRef{sec:pref:history}, the M{\"o}bian circle topology has the attractive
property of being able to reduce collections of issues to just two and thus
appears to reduce choice systems to one of simple majority and thereby
circumvents the limitations imposed by Arrow's \textit{Impossibility
  Theorem}, as summarised in \cite{scf:eaves}.

Referring again to the political structure of the United States, it would
appear to display the characteristics of this circular topology. There are
just two dominant political ideologies and the states neighbour each other
in different circular topologies on the different political issues
presented to them. The net effect is a superposition of pairs of different
behaviours all of which can be encompassed by the two political parties'
platforms. This construction bears a great deal of similarity to the
dynamic analysis of the Tiebout model by Kollman \etal
\cite{Miller:tiebout}.

\section{Summary}

\paragraph{Protocol adoption}

This chapter has shown that large systems where individual agents make
choices constrained by a simple, rational behaviour can lead to stable
behaviour for the system as a whole. This result, it is claimed
\cite{Axelrod:coal}, helps to explain the emergence of \textit{de facto}
standards. Axelrod's analysis was prompted by the evolution of different
\textit{Unix} standards, but it might be applied to different Internet
protocols.  For example, 90\% of Internet traffic is carried over TCP
connections rather than in UDP datagrams. TCP has very useful technical
advantages over UDP: it, unlike UDP, is rate--adaptive, does not require an
application programmer to fragment his own data, transparently recovers
from IP packet loss and the arrival of IP packets out of
sequence{\myDash}in summary, it provides a relatively simply session layer.
UDP is however to easier to manage: it requires no connection
management{\myDash}simply one listening endpoint from each party for each
connection{\myDash}it is therefore easily adaptable to multi--cast
protocols and it is easier to define rules for screening firewall routers.
Had it been the case that a sufficiently capable session layer were
available to application programmers early in the development of the
Internet, UDP might have become the \textit{de facto} standard for IP
communications rather than TCP.  Similar arguments can be made for the
domination of other protocols: the \textit{Sun Micro-systems} RPC protocols
based on the portmapper, could have been supplanted by the Domain Naming
Service based \textit{Hesiod} protocol from MIT.

\paragraph{Protocol adoption and behaviour}

The number of incompatible standards (meta--behaviours) that can emerge is
a function of topology; how long the system takes to arrive at a stable set
of standards is a function of the choices{\myDash}features and
traits{\myDash}available. The behaviour that each agent employs when making
choices controls the rate of migration of the traits and the number of
conflicts over their selection: \textsf{Egoistic} behaviour allows traits
to migrate quickly but introduces proportionally more conflicts to resolve,
\textsf{PeerPossible} behaviour the converse.

It has also been argued that \textsf{PeerPossible} should be a more
consonant voting behaviour than \textsf{Egoistic} and, when coupled with a
M{\"o}bian topology allows very homogeneous cultures to evolve. This
cultural system also has the attractive property that it does not remove
any intrinsic bias in population.

\paragraph{Engineering behaviours}

A point that has not been addressed is how behaviours like
\textsf{Egoistic} and \textsf{PeerPossible} would be engineered so that
they may be used to simplify policy choices in working systems. A simpler
example than access control policy evolution might be IP address
allocation. Looking at figure \ref{fig:peer-2}, it could be the case that a
hundred small separate networks at $t_0$ have interacted with one another
and joined each other's networks to yield a few larger networks by
$t_{1100}$. The criteria for the features they might employ would include
some of the following:

\begin{itemize}
\item Connections to different types of carrier network: some networks may
  consist almost entirely of single--homed 100BaseTX on the same
  subnetwork; some may have a number of dual--homed routers with access to
  ATM or SDH leased lines linking to other subnetworks.
  
\item Different protocols used for communication: some networks may make
  extensive use of multi--cast, point--to--point IP routes, or Generic
  Router Encapsulation (GRE) tunnels.
  
\item Traffic types: some networks may be simply web--browsers; some may use
  remote file-store.
  
\item Screening subnets: it may be the case that some networks must not be
  visible to one another.
\end{itemize}

There could be a very large set of traits for each of these and others.

For an operational protocol, one must consider resource--locking. Each
individual would operate autonomously from every other, but would need to
acquire locks on their neighbours when they are about to make the decision
to change their configuration. This is quite a difficult lock--acquiring
exercise since dead-- and livelock are distinct possibilities.

\paragraph{Population statistics}

The key point about the behaviours \textsf{Egoistic} and
\textsf{PeerPossible} is that they operate locally and can, consequently,
adapt very quickly to their neighbours. The entropy measures that have been
introduced are population measures and, in a working system, would be
expensive to compute. They are, as has been seen above, a very useful guide
to the operational state of the system{\myDash}whether anarchic,
collectivistic, oligarchic or authoritarian. A system administrator could
use the entropy measures to determine if a system has simplified itself
enough to be allowed to continue to fulfil its chosen function: more
efficiently and with less conflicts than before it organised itself. To
obtain an accurate statistic, it may be necessary for the administrator
to quell all interactions and request the status of all the individuals.

\paragraph{Migration of access rights}

The results obtained in this chapter give us confidence that a
self--organising system that allows access rights to be migrated from one
individual to another should be predictable. Some of the metrics developed
could be used to monitor the migration.

The cultural model gives us a reference model of behaviour. When one designs
a system, one can attempt to reduce its operation to that of the simulation
described in this chapter. This then gives us some expectations for its
behaviour.


\chapter{Self--Organising Permissions Policy System}
\label{cha:finally}

The findings of chapters \ref{cha:top} together with the small--scale
behaviours investigated in chapter \ref{cha:norms} can be used as the basis
for a self--organising permissions policy system. In chapter \ref{cha:top},
integrity checks and distance measures for preference hierarchies were
introduced.  In chapter \ref{cha:norms}, a cultural model illustrated how a
system of interacting agents with a fixed set of choices using a simple
behavioural rule based on a distance measure would have reasonably
predictable large--scale behaviour so that agents within cultures should
segregate themselves into large groups.

To assure ourselves that a system will behave like the cultures described in
\ref{cha:norms}, it must have the same construction:

\begin{itemize}
\item A small--scale behavioural rule
\item A fixed set of discernible features with traits
\item A distance measure for two feature sets
\item A fixed topology, preferably in two dimensions
\end{itemize}

Before discussing how a self--organising permissions system might work, a
simpler example of a self--organising set of newsgroups will be developed.

\section{Self--Organising Newsgroups}

One application of this system as proposed would be a set of
self--organising newsgroups or mailing lists. The problem with USENET
newsgroups is they have a very low ``signal to noise'' ratio. There are
lots of postings of dubious worth, some nothing more than advertisements.
Very often the quality of debate degenerates to a squabble.  Often cliques
of users develop threads of discussion which are of no interest to anyone
else. Cross--posting is another problem: subscribers send the same message
to a number of newsgroups simultaneously.

Newsgroups could be self--organising: so that squabblers will be moved to
their own groups as would persistent advertisers.  Cross--posting will be
limited by using managers who may choose to refer a posting to another
group rather than have it posted in their own.

There are three types of newsgroup management in place for the USENET.
\begin{description}
\item[Moderated] every posting to the newsgroup is vetted by a moderator.
\item[Managed] subscribers are only allowed to post if they have applied
  for permission to do so from the newsgroup manager. The newsgroup
  software then checks if each posting comes from an allowed subscriber.
\item[Unmanaged] Access is completely open.
\end{description}

Moderation is usually too burdensome. Managed newsgroups are quite rare
(but common for mailing lists), so the usual form is an unmanaged
newsgroup. The system proposed will be a sophisticated managed newsgroup.

To put newsgroup postings under some sort of control, a preference
hierarchy needs to be developed from some norms. This would then be used to
partition the subscribers to the newsgroup into sub--groups and to rank the
newsgroups with respect to one another. This section will continue with a
description of how such a self--managing newsgroup protocol would work and
an analysis of it as a cultural model like that in chapter \ref{cha:norms}.

\subsection{Newsgroup Operation}

\subsubsection{How it might work}

Newsgroups are organised by a set of news administrators who require a
group of people to issue a charter and have a number of newsgroup
subscribers sign it before a group is propagated within the USENET
hierarchy. Discussions around an operating system, such as \textit{Linux},
in the \textit{comp.os} hierarchy has been split into these groups:

\begin{verse}
  comp.os.linux.advocacy \\
  comp.os.linux.announce \\
  comp.os.linux.hardware \\
  comp.os.linux.software \\
  comp.os.linux.setup \\
  comp.os.linux.networking \\
  \etc
\end{verse}

Within the \textit{comp.os.linux.advocacy} newsgroup there will be a number
of postings comparing \textit{Linux} with the Microsoft \textit{Windows}
operating system. The \textit{Linux} \versus \textit{Windows} debate is an
interest which all of the \textit{comp.os.linux} newsgroups share, but only
\textit{comp.os.linux.advocacy} would usually discuss it, but a debate
comparing \textit{Windows} network interface card driver support \visavis
that of \textit{Linux} would be of interest to the
\textit{comp.os.linux.networking} newsgroup.

A self--organising set of newsgroups for \textit{comp.os.linux} should
finally evolve a structure similar to that above, but it would use who
takes part in which debates to evolve the structure, rather than having one
imposed on it.

The operation of a self--organising newsgroup has only a small amount of
information to use: postings from subscribers to particular threads. There
are four procedures involved in self--management:

\begin{enumerate}
\item Initial group allocation
\item Posting using subscriber referral and access management
\item \label{sec:news:1} Generating group allocations
\item Posting using group referral and access management
\item Repeat from \ref{sec:news:1}
\end{enumerate}

\paragraph{Protocol}

A protocol of some kind has to be imposed on the newsgroup to yield more
information from the postings.

\begin{itemize}
\item When a subscriber initiates a thread, it must be followed up for it
  to considered.
\item A subscriber cannot follow up himself.
\item A follow--up must be followed-up to be considered.
\item A follow--up is closed if it is acknowledged by the subscriber who
  initiated the thread.
\end{itemize}

This set of rules makes for civilised debate, a simple example of which
might be:

\begin{enumerate}
\item $p$ initiates thread $a$
\item $q$ follows up $a$
\item $p$ acknowledges $q$
\end{enumerate}

$p$'s initiation of the thread $a$ counts as a statement of his
preferences because $q$ followed it up. $q$'s follow up is counted as a
preference, because $p$ acknowledged it. A more complicated scenario might
be:

\begin{enumerate}
\item $p$ initiates thread $a$
\item $q$ follows up $a$, creating $a'$
\item $r$ follows up $a'$
\item $q$ or $p$ acknowledges $r$ on $a'$
\item $p$ acknowledges $q$ on $a$
\end{enumerate}

Probably, the simplest discipline is to allow the initiator to acknowledge
all contributors, except those he considers irrelevant. If any of the
contributors follows a thread that the initiator feels is irrelevant then
that subscriber who followed up the irrelevant thread can acknowledge it
and form a new thread. So, in the interaction above, if $p$ does not
acknowledge $r$, then $q$ must acknowledge his contribution and their
contributions count to the sub--thread $a'$.

\subsubsection{Initial Grouping}

This is the first phase of generating a preference hierarchy. It is used to
derive a partitioning of the subscribers to the newsgroup. As each
subscriber posts to a thread, it generates a preference. If, for example,
there are three threads: $a, b, c$ and an apathy thread is added to this so
that all the threads can be related, call it $\emptyset$, then if a
subscriber posts one or more times to threads $a$ and $b$ and not to thread
$c$ then the preference ordering is $ a = b > c = \emptyset $.

\paragraph{Newsgroup partitioning by thread interest}

A preference hierarchy for each of the subscribers would be generated. It
would then be aggregated using the techniques described in
\myRef{sec:unan:for}. The preference hierarchies would be just
two--ply{\myDash}those subscribed to and those not{\myDash}but it might be
the case that global indifferences emerge from the $I$--graphs. For
example, everyone who posts to $a$ will always post to $b$. Thread $a$
would then be merged with $b$ and would form a new sub--newsgroup.

In the aggregation there would almost certainly be cycles{\myDash}a cycle
such as $ a > b = c = \emptyset$, $b > a = c = \emptyset$ and $c > a = b =
\emptyset$ could arise. These would have to be removed by merging threads
or by removing subscribers who introduce cycles. Using some set
nomenclature, $\{s | <\text{ordering}>\}$ is a statement of set
membership for subscribers $s$ who have ordered the preferences in the
specified way.  For the example of three initial threads, all of the
possible new sub--newsgroups that could emerge are given below:

\begin{equation}
  \begin{aligned}[t]
    g_a & = \{ s | a>b=c \} \\
    g_b & = \{ s | b>a=c=\emptyset \} \\
    g_c & = \{ s | c>a=b=\emptyset \} \\
  \end{aligned}
  \droptext{and}
  \begin{aligned}[t]
    g_{ab} & = \{ s | a=b=\emptyset>c \} \\
    g_{ac} & = \{ s | a=c=\emptyset>b \} \\
    g_{bc} & = \{ s | b=c=\emptyset>a \} \\
  \end{aligned}
  \droptext{and}
  \begin{aligned}[t]
    g_{abc} & = \{ s | a=b=c>\emptyset \}
  \end{aligned}
  \label{eq:news:groups}
\end{equation}

There are only 7 rather than 13 groups that table \ref{tab:orders} would
suggest. This is because the simplification discussed in
\myRef{sec:axelrod:interp} has been used. These 7 groups represent all the
valid shades of opinion there might be. This has been achieved in the
newsgroup system by only allowing two--ply preferences for the subscribers.

In more colloquial terms the subscribers have been partitioned into ``one
interest only'', ``two--interests'' and ``interested in all'' groups.

\subparagraph{The Entry Group}

This group is used by all groups to post new threads and by people who wish
to join the newsgroup to see what threads are being discussed. A thread
that is never followed up would stay in the entry group, when there is a
follow--up, it would move to one of the sub--newsgroups. The group is
called the \textit{entry} group.

Every subscriber is a member of the entry group. Every subscriber may
initiate a thread in the entry group.

\paragraph{Ordering Subscribers and Appointing Managers}

For each sub--newsgroup generate an ordering of the subscribers based on
the number of times they have posted to the threads of their
sub--newsgroup. Form the most active members into a collective that acts as
the sub--newsgroup's access managers{\myDash}it might be the top 5\% of
subscribers for each group, for example. (This, incidentally, is one of the 
problems of this system's design: a sophisticated group of subscribers
can make themselves managers of groups by answering each other's
threads. This is similar to tactical voting. It will be seen that an
entropy measure can be used to determine how much support managers have.)

\subsubsection{New postings}

When an existing subscriber initiates a new thread, it would be posted to
his sub--newsgroup and also to the entry group. Followups to the new thread
would appear only in the sub--newsgroup. If a subscriber in another
sub--newsgroup wants to follow a new thread in a sub--newsgroup, he would
be able to review the postings for that new thread but not make any.

If he did want to make a posting to that thread, his mail would be sent to
the access managers. They would then decide whether to permit the posting
and would thus grant to the new subscriber membership of their
sub--newsgroup.  He would then be allowed to initiate a thread in that
sub--newsgroup as well as submit postings.

\subsubsection{Group Preference Hierarchy}

After a number of new threads have been started, the subscribers can be
re--partitioned and, additionally, a hierarchy of the groups can be
generated. This hierarchy expresses a norm of behaviour between the groups.

\paragraph{Norm of behaviour}

It is best to explain this by example: if a member of sub--newsgroup $g_a$,
the one containing all those who chose $a>b=c$, chooses, via the entry
group, to join group $g_b$ for two threads, ${b'}_1, {b'}_2$ and only takes
part in one thread in his own group ${a'}_1$ then he has the following
preference orderings:

\begin{itemize}
\item An ordering between threads as before: ${a'}_1 = {b'}_1 = {b'}_2 >
  \droptext{any others}$
\item An ordering between sub--newsgroups: $\operatorname{group}(g_{b}) >
  \operatorname{group}(g_a) > \operatorname{group}(\text{others})$.
\end{itemize}

The latter preference is formed because the subscriber has posted twice to
threads originating in group $g_b = \{ s | b>a=c=\emptyset\} $ and only once
in $g_a = \{ s | a>b=c=\emptyset \}$. The ordering between threads will be only
two--ply, as before, but that between newsgroups can be as long a chain as
there are newsgroups.

The ordering between threads represents a subscriber's current interests
and the ordering between newsgroups relates his current interests to his
past interests. (The relative strengths of his interests are not used in
forming his preference hierarchy, these are used in the aggregation across
the newsgroups.)

This forms a norm of behaviour. One would expect a subscriber in group
$g_a$ to post subsequently solely to that group, but if he posts to $g_b$
as well, this would suggest he should join $g_{ab}$.

\begin{figure}[htbp]
  \begin{center}
  \includegraphics[keepaspectratio=1, totalheight=2.5in]{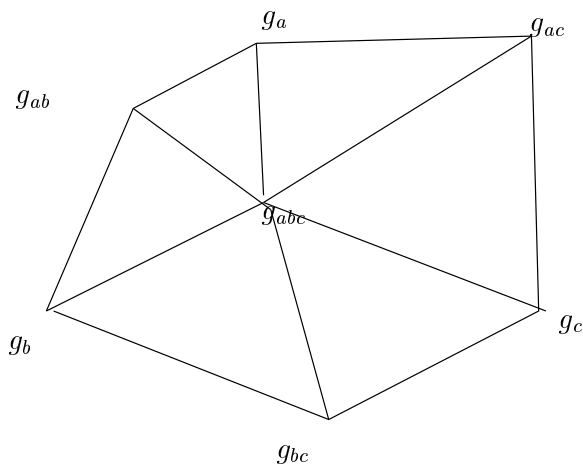} 
  \caption{Newsgroups: implied topology}
  \label{fig:news:1}
  \end{center}
\end{figure}

For all the members of a particular newsgroup, the ordering between
sub--newsgroups is aggregated using the maximum likelihood preference
ordering procedure described in \myRef{sec:top:max-likelihood}. This yields
an ordering across the newsgroups that is specific to each newsgroup. The
newsgroup effectively decides who its neighbours are. If one considers a
three thread system, the arrangement of the newsgroups will develop from
that given in figure \ref{fig:news:1}.

\subsubsection{New postings with a group hierarchy}

When a subscriber posts a new thread in his own group, it will be sent to
the access managers of the groups adjacent to his own group in the
preference hierarchy of groups. They can then choose to accept it or not.
If they do, then the interested members in the neighbouring groups will
post to the new thread and make their group more similar to the thread
initiator's group.

\subsection{Summary}

The preference hierarchy generated decides the order in which the groups
refer to one another in figure \ref{fig:news:1}. There is an order imposed:
when a subscriber posts to both $g_a$ and $g_b$ he indicates that he would
prefer to join a group $g_{ab}$.

The access managers are the most frequent message posters to their own
groups and they are responsible for vetting the subscribers allowed to join
a debate and thus a newsgroup. In this way, very active subscribers to a
particular set of threads will find themselves acting as access managers
for groups that discuss mostly their sort of interests.

The groups derived from the first set of postings are not tied to
discussing debates concerning the subject or subjects they first expressed
an interest in. Each group will be constantly redefining itself: both in
the subscribers it has and the issues it discusses.

The use of automatic referral between newsgroups adjacent in the hierarchy
makes access management easier and allows for the migration of traits.
Using the entry group to initiate a thread will be relatively rare.

This system is self--organising in a rather subtle way. The access managers
are representative of the subscribers to the group. In the example of a
permissions' policy management system, it will be seen that this authority
by which subscribers are allowed to join a group can be determined from
another preference hierarchy.

\subsection{The Cultural Model and Stability}

Now the self--organising newsgroup system has been defined, a structural
iso--morphism has to be made to ascertain which entity in the newsgroup
system fulfils which function in the cultural model.

When that is done it will be possible to make some predictions about the
behaviour of the system using the analysis of the dynamics of the cultural
model.

\paragraph{The Cultural Model}

Referring to Axelrod's model in \myRef{sec:adaptive:convergence} and figure
\ref{fig:news:1}, the simulation model has only three
entities{\myDash}agents, features and traits. An agent has a set of
features, each feature can take any one of a number of different trait
values. (Each trait is therefore an object of a particular feature class.)
The set of traits for the different features is the identity of the agent.

\begin{displaymath}
  \begin{CD}
    Agent @>{has}>> SetOfFeatures \\
    Trait @>{is-example-of}>> Feature \\
    SetOfTraits @>{is-kind-of}>> Identity
  \end{CD}
\end{displaymath}

These are equated with entities in the newsgroup system in the following
way.

\begin{displaymath}
  \begin{CD}
    Newsgroup @>{has}>> SetOfInterests \\
    Posting @>{is-example-of}>> Interest \\
    SetOfPostings @>{is-kind-of}>> Identity
  \end{CD}
\end{displaymath}

The agents would be the newsgroups referred to in \eqref{eq:news:groups}
not the subscribers. They are located with respect to one another given by
figure \ref{fig:news:1}.

\paragraph{Interests as meta--threads as features}

One might ask what is an \textit{interest} in the context of the
newsgroups. An interest within a newsgroup manifests itself as a thread of
discussion and a thread of discussion is nothing more than a set of related
postings{\myDash}they are related by their common interest. All the
newsgroups do have the same set of interests, whether they foster any
interest in any particular subset is what the postings determine.

\paragraph{Distance Function}

The access managers of the group and the existing subscribers determine
which postings to accept within a newsgroup. The access managers accept new
threads, the existing subscribers choose whether to follow them up.
Together they act as the distance function does in Axelrod's model
\eqref{eq:adaptive:axelrod:4}. The mechanics of the distance function are
very different: the access managers and the existing members vote and the
maximum--likelihood preference ordering is used to determine which of its
neighbours each newsgroup is compatible with.

\paragraph{Topology}

\begin{figure}[htbp]
  \begin{center}
  \includegraphics[keepaspectratio=1, totalheight=2.5in]{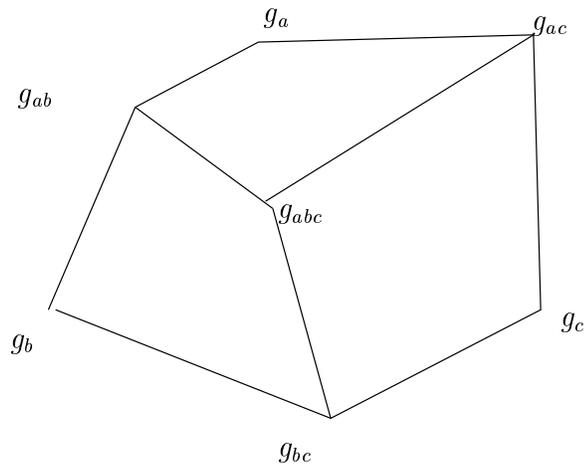} 
  \caption{Newsgroups: implied topology with fewer neighbours}
  \label{fig:news:2}
  \end{center}
\end{figure}

There are a number of possible topologies for the system. It can be
either as in figure \ref{fig:news:1}, which for three interests gives each
one three neighbours and the central point six. This could be simplified to
that given in figure \ref{fig:news:2}.

This latter topology implies that subscribers to the single interest group
only become interested in all three issues after they have become
interested in two of them. The number of neighbours is two, three and three
for the single--interest, two--interest and every--interest groups. It is a
more appealing topology because it gives each agent fewer neighbours and
can easily be formed by using a binary tree.

\paragraph{Dynamics}

The system described in figures \ref{fig:news:1} and \ref{fig:news:2} is
not particularly complex. Following the argument given in
\myRef{sec:axelrod:interp}, if one only allows seven different opinions to
be held regarding the preferences across three meta-interests then one need
only set $\#(F)=3$ for the meta--groups $g_a,g_b,g_c$ and have a trait set
of $\#(T)=2$. The playing field can have only 7 different newsgroups, using
\eqref{eq:axelrod:binom}. This would mean that every subscriber would be
allocated to one primary group which could be either a single--, dual-- or
every interest one.

This is a fairly simply model and it does usually reduce to just two or
three varieties regardless of which of the two topologies is used. The more
connected graph, figure \ref{fig:news:1}, usually reaches stasis faster. It
hardly ever develops a limit cycle. Typical patterns are fairly
predictable, whichever variety starting at $g_a,g_b,g_c$ asserts itself in
the middle point $g_{abc}$ usually asserts itself over at least one other
interest, so, for example, if the variety starting at $g_a$ reaches
$g_{abc}$ first, it might arise that $g_b$ takes on the same variety so
that, effectively, $a=b$.

\paragraph{More typical systems}

Usually newsgroups will have more than three interests, looking at the
\textit{comp.os.linux} hierarchy, the newsgroup administrators were
expecting to keep debate focused on about six broad subjects. This is a
more complex system, but easily derived using the spanning tree
construction of \myRef{sec:axelrod:interp} (which was used for figure
\ref{fig:news:2}). The size of the playing field would be 62:
$\Sigma_{i=1}^6 \binom{6}{i}$. These 62 newsgroups represent all the
different shades of opinion there are allowed to be.

With a feature set of $\#(F)=6$, we might say that subscribers have $3$
interests (they are hoping to pair the six meta--threads to form the
simpler spanning tree), this would give them a trait set of $\#(T)=13$,
from table \ref{tab:orders}, if we ask them to state their preferences on
the three pairs they have chosen using a weak ordering. This would mean
that when re--partitioning the subscribers, the preference ordering would
be a chain that is three--ply in length: \ie $a>b>c$.

This then gives quite similar parameters to the Axelrod cultural model
analysed in chapter \ref{cha:norms} and the dynamic behaviour can be
expected to be the same.

\paragraph{A \textsf{PeerPossible} System}

\textsf{Peer-Possible} behaviour can be implemented in effect in the
newsgroup system. As each subscriber makes a posting, if he is not already
a member of the group with the thread, it must be shown that he is a member
of a neighbouring group to the one he wishes to make the posting in.

This makes the access managers' job easier and should lend to the system
some of the properties that \textsf{PeerPossible} behaviour was deduced to
possess in the cultural model.

\section{Self--Organising Permissions Policy System}
\label{sec:sopps}

It now remains to apply what has been learnt from the newsgroups example to
a system to simplify preference hierarchies for access control systems to
produce a self--organising permissions policy system for healthcare.

\subsection{Newsgroups: Summarised}

The newsgroups example had the following similarities to the cultural
model.

\begin{itemize}
\item The subscribers were not the agents that interacted in the cultural
  model, but rather the groups that they belonged to.
\item The individual subscribers acted as traits for a feature. A group
  collected members which formed its identity.
\end{itemize}

\subsection{R\^ole--based Access Control System}

Little mention was made in the newgroups example of the target technology
for the access control system to the newgroups. In this discussion of the
self--organising permissions policy system a r\^ole--based access control
system will be the target technology. This is because it can be adapted to
suit all types of access--control system in current use and its information
model, \myRef{sec:informational}, has the key entities needed to manage the
system: the r\^oles and constraints objects.

\subsection{Healthcare}

Within the healthcare sector, the collegiate organisational model described
by Anderson \cite{Anderson:security} applies. Within each
college{\myDash}that is, every clinic or surgery{\myDash}the security
rights are well--defined, but they will almost certainly not be consistent
across the colleges.

\paragraph{An example: receptionists in different practices} The rights
given to receptionist in one general practice may only entitle them to view
the records of patients who are visiting their general practitioner on that
day{\myDash}such a system might be wholly electronic with access policies
defined within a relational database that holds all patient records;
another surgery may allow their receptionists to see all records of all
patients at all times{\myDash}a paper--based system where the receptionist
is given a key to the records room.

The difficulty here is that within their own colleges, the receptionists
have been assigned the same nominal r\^ole, but the r\^ole has different rights
and privileges within each college.

It would probably be the case that the surgery that has the paper--based
system needs more trustworthy receptionists because they would have access
to so much more information.  Consequently, the qualifications and
experience of the receptionists would need to be qualitatively better than
those of receptionists working in the surgery that has a better protected
electronic system. Because these better qualified receptionists are
considered more trustworthy, they might be given more rights and privileges
in other colleges.

\subsubsection{Simplification of the R\^ole Space} 

A system administrator, if he were to define this r\^ole across the
practices\footnote{Anderson makes it clear that this scenario is not one
  that would arise: the receptionist r\^ole would be constrained to allow
  only access to records in the receptionist's own practice. This scenario
  is though, in miniature, the problem that would be faced in allocating
  access rights across all parts of the healthcare sector.  }, would need
to know what the rights and privileges associated with the r\^oles are
\textit{and} the quality of the people assigned the r\^ole before he can
determine who should be given the federal r\^ole. The system administrator
would find it expedient to divide the r\^ole of receptionist into a number of
other sub--r\^oles. Some r\^oles would require that an individual assigned to
the r\^ole must meet certain requirements that would give one more confidence
that the individuals will be trustworthy.

A r\^ole, as a data structure, is probably best represented as an ordered
pair $\langle r_{\initiator}, r_{\acceptor} \rangle ${\myDash}the r\^ole of
the initiator and the r\^ole of the acceptor. The total number of r\^oles
in the federation is $N_{r} = \sum_{i=1}^N c_i$, where $c_i$ is the number
of r\^oles in one college of the $N$ there are. The total number of federal
r\^ole pairs will be less than $N_{r}^2$. The hope is that by simplifying
them and eliminating the irrelevant alternatives, it may be possible to
reduce them to a more manageable number.

\subsubsection{R\^oles and Interactions}

When one analyses the operation of the newsgroup system, one sees that each
posting to a newsgroup by a subscriber is an interaction between newgroups:
the newsgroup that the subscriber belongs to and the newsgroup the
subscriber posts to.

Within healthcare, interactions are between individuals in one of their
defined r\^oles: when a general practitioner refers a patient to a
consultant, he is known as the \textit{referring physician} and the
consultant is the \textit{consulting physician}; within a hospital a doctor
present at the treatment of a patient by another doctor is fulfilling the
r\^ole of \textit{attending physician}. There are a number of other such
r\^oles played by physicians within hospitals; these are currently being
codified for the proposed HL7 standard \cite{iso:hl7}. HL7 also defines
r\^oles that are not medical: administrative, auditing and clerical r\^oles are
also defined.

In addition to the r\^oles defined within HL7, there are other r\^oles that do
not fall into its remit, but one would expect them to be defined within the
federal system: in particular, medical researcher and comptroller.

With regard to the referral process: an interaction takes place when a
general practitioner refers a patient to a consultant, whether his referral
is accepted or rejected, an interaction has taken place. Similarly, when a
medical researcher requests access to a set of records in a database, this
constitutes an interaction.

\paragraph{Distance Measure}

The HL7 r\^oles are already well--defined, they would in fact form part of the
distance measure between r\^oles and would be axiomatic pre--conditions
for an interaction.

\subsubsection{Features, Traits and a Data Structure}

\begin{figure}[htbp]
  \begin{center}
  \includegraphics[angle=-90,keepaspectratio=1,totalheight=8in]{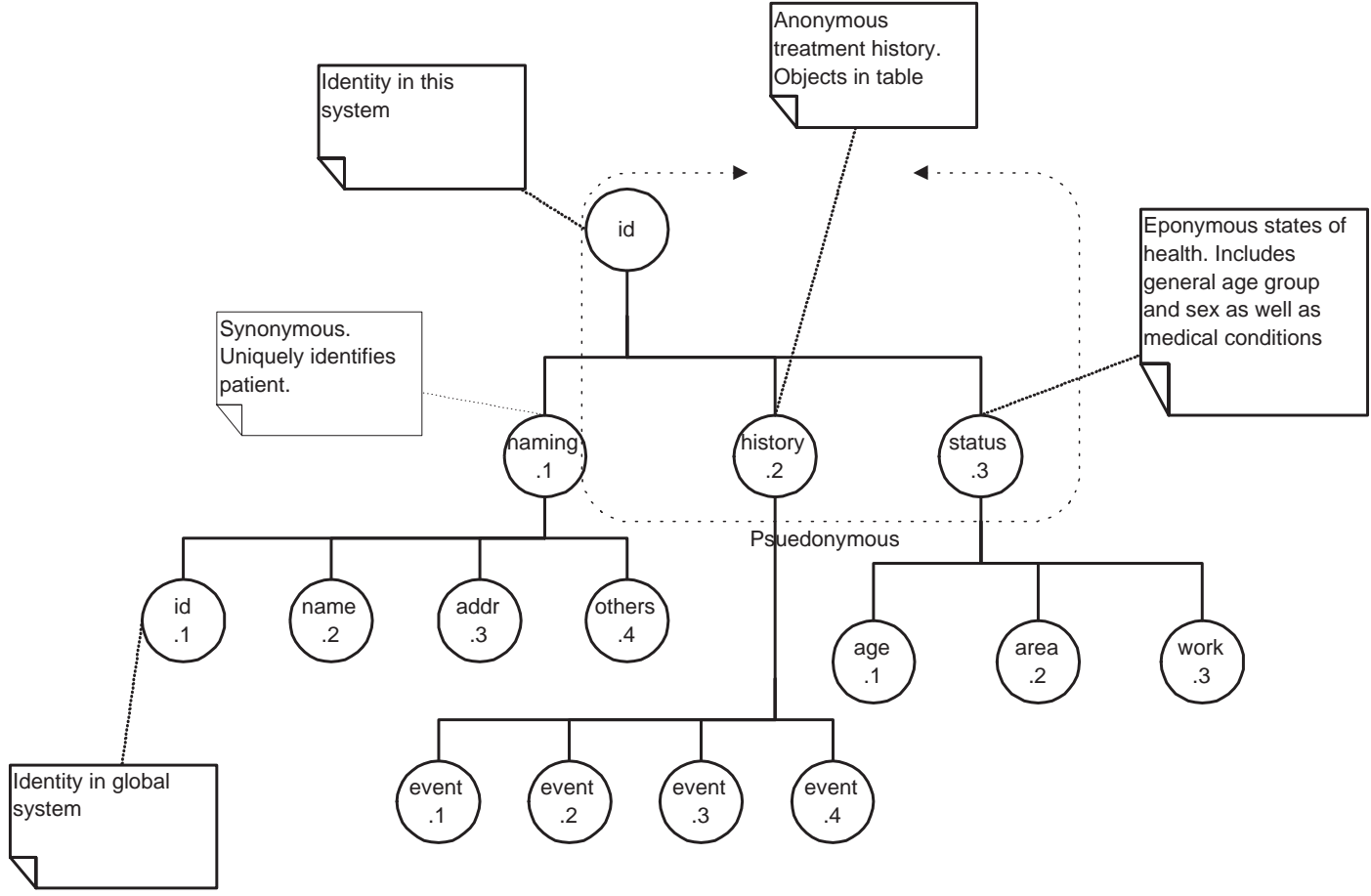}
  \caption{Tree structured patient record with features for classes of access}
  \label{fig:prec-tree}
  \end{center}
\end{figure}

Figure \ref{fig:prec-tree} illustrates a proposed data structure for a
patient record. Its design is intended to illustrate an instance of one of
the four broad classifications of access that were introduced in 
in \myRef{tab:identity}. 

They are constituted in the following way using three branches of the
patient record tree: \textit{naming}, \textit{history} and \textit{status}.
These correspond to synonymous, anonymous and eponymous in the following
way. Synonymous information is any combination of facts from the
\textit{naming} branch of the tree that would allow a patient to uniquely
identified. The anonymous branch, \textit{history}, contains the history of
the patient: these records are all of the same type which is generically
called \textit{event}. The eponymous branch, \textit{status}, contains
summary statements{\myDash}age group, sex, diabetic, allergies, body--mass
index and so forth. It should not be possible to identify a patient
uniquely using sets of anonymous and eponymous information. This means that
the inference threats that Denning describes in \cite{infr:schlorer} are
protected against, probably by a query system enabled in the way Denning
describes.

A fourth class of access is synonmous: this is viewed as being knowledge of
a set of anonymous facts and a set of eponymous facts, \textit{but} it is
also possible to uniquely identify the patient in question. This would be
used by another physician to ask the custodian of the patient's record for
additional information. This is typical of the referral process that
physicians employ; they initially discuss a patient under an assumed name,
passing on what they consider to be salient facts before the consulting
doctor assumes responsibility and is told the patient's identity.

Finally, the reason for choosing a numbered tree structure is that every
entity{\myDash}be it a field or linked historical record{\myDash}can be
given a unique vector relative to the \textit{id}. The type of a linked
record is given a unique vector identifier; this is concatenated with the
\textit{id} of the record of having the contents. The \textit{id} fields in
linked records would be informative. They might specify a date, in which
case it would be possible to limit historical access to dates within a
range.  Essentially, the tree structure is a linked structure of all the
normal forms\footnote{A record is decomposed into a composite record with
  links to records in other files. Each of these should be in a normal form
  which depends wholly upon each primary key for that file; all of fields
  within the record are either atomic or a link to another record, see
  \cite{Wiederhold:DD83} for a discussion of normal forms and the
  construction of database keys. The key would be the \textit{id} field
  described in the text.} of the records in a database that pertain to a
patient.

Between different r\^oles under different contexts different views of patient
records will be granted. As patient record views are defined they can be
compared with one another as trees: when one r\^ole grants access to another
they agree on a standard view; one of those r\^oles may in a later
interaction with another r\^ole decide that the current r\^ole is entitled to
the same view as the previous one. They may choose to standardise or
diverge, but a number of standard representations in different contexts
will emerge.

The definitions of anonymous, eponymous and synonymous access will differ
for each r\^ole. Their meanings in each context will be refined and generic
types will emerge between r\^oles.

The set of facts permitted for each class of access constitutes a trait for
that feature. This is the first set of features and traits defined for the
permissions policy system.

\subsubsection{Ontological Features and Traits}

In addition to the views that are granted between r\^oles, the context in
which the interaction takes place must be defined: these are the
ontological features:

\begin{description}
\item[Administrative] Some of these may be administrative: working in the
  same clinic, the same hospital, health district and so forth.
\item[Discipline] In healthcare, this would usually be derived from the
  medical clinic: p{\ae}diatric, oncological, genito--urinary and so
  forth. It may also be derived from affiliations to professional
  organisations.
\item[Procedural] The procedure that is being followed. Whether a referral,
  consultation, prescription, treatment and so forth. Many of these
  procedures are defined within HL7.
\end{description}

It may well prove expedient to introduce others.

\subsubsection{Inconsistencies and Distance}

In more conventional database access control systems, the problem that a
security rights administrator faces when a new object is put under his
jurisdiction is whether to grant to existing groups rights to access that
object or whether to sub--divide those groups to form a new group and grant
access rights only to it. The choice is usually determined by whether an
inconsistency results: if the right granted to an existing group gives them
information that would allow them to compromise the existing group
structure then they should not be granted that right. For example, if the
right to grant read access is given away too freely, then a number of
groups might disappear because they effectively become the same group. Or,
it might be the case, that one group is denied access to a object it cannot
function without.

For the tree data structure described above, it may prove the case that one
individual in one r\^ole grants access to a view which would make the system
inconsistent in a similar way. This can be determined by using the
take--grant method of analysis introduced in \myRef{sec:safety}.

If such an inconsistency would result, then this should contribute to the
distance measure and would introduce an axiomatic inferred rule to its
evaluation. (One might assign an unbroachable distance to it.)

Another contributing factor to the distance measure might be statistical
belief in the authority of an identity. This can be quantitatively stated
using the methods proposed by Maurer \cite{KohMau00,Maurer96c}.

\subsection{Collective Choice Expert System}

The operation of the permissions policy system is that of a
\textit{collective choice expert system}\footnote{This is a relatively new
  software system, Hubermann's \textit{Beehive} system,
  \cite{Hubermann:beehive}, is one such example.}. Whenever an interaction
takes place, one r\^ole grants a data view to another, the cultural model is
constructed and is allowed to evolve as a statistical experiment to
determine how stable it is; a desirable outcome may be specified and a
sample of outcomes might lead one to conclude that the proposed data view
should or should not be be granted, depending on whether it is more or less
likely that a desirable consistent configuration would be statistically
likely to arise.

\paragraph{Modelling, Statistical Experiment and Comparison}

The model is constructed using the current attributes of the individuals in
the healthcare business sector: their r\^oles and business relationships
are analysed and a set of r\^ole assignments is made.

The data view is imposed on the model and it is allowed to evolve under
stochastic inputs. A sample of evolutions would be used to analyse the
behaviour of the system.

The trajectory of each cultural evolution experiment would be recorded. The
parameters measured for the trajectory would be the variety and
compatibility entropies: $S(c)$ and $S(v)$,
\myRef{sec:adaptive:axelrod:large}, their rates of change, their phase
difference and the activity of the system. These trajectories of the
different samples could be compared to one another. They would also be
recorded for comparison with the real system as it evolves.

With this knowledge, it should be possible to show that, for a large enough
group of r\^oles, a degree of stability could be reached and the data views
would not allow inconsistencies and that the system could be shown to be
near pareto--optimal in moving to its next operational state.

\paragraph{Final States}

The cultural model statistical experiments will either degenerate and
develop an inconsistency or become static or continue to evolve in a
limit cycle. All of these outcomes can be determined from an analysis of
the cultural model's global state and statistics of its evolution under the
stochastic input. The frequency distribution of stable to unstable outcomes
is indicative of the stability of the real--world system.

The state that would normally be considered to be most satisfactory is a
static condition, such as that shown in table \ref{tab:culture:1}, where
the majority group is incompatible with all the others, their r\^ole is
isolated, but the other groups are able to continue to migrate views across
one another. The majority group here might be the comptroller r\^ole which is
unable to obtain anything more than accounting details in aggregate from
any of the other r\^oles.

A limit cycle condition may also be a desirable, if the dominant r\^oles are
policy--making ones: high--level custodians such as ethics committees for
hospitals and the professional organisations of physicians, as well as
representative bodies for groups of accessors\footnote{It may well be the
  case that cultural systems having limit cycles are exhibiting another
  behavioural phenomenon examined by Axelrod, the \textit{Tribute} system,
  \cite{Axelrod:1997}.}.

\paragraph{Dynamics}

Hopefully there will be a similarity in the trajectories of the cultural
model evolutions. The rate of change of the entropies and their rate of
change with respect to each other is an indicator of when the final state
will appear. These metrics may also serve as an indicator of how complex
that final state might be.

The rates of change of entropy can be measured in the permissions policy
system as it evolves; it might then be possible to show that it is
following a similar trajectory to a cultural model experiment.

That cultural experiment could then be re-run with revised starting
parameters to see how it evolves. The process of analysis and comparison
could be repeated with more statistical certainty.

\paragraph{Generating Rules}

If it becomes apparent that a large r\^ole will be dominant and will be
isolated from the others then it could be imposed as an axiomatic data view
within the system and be enforced through the distance function.

\paragraph{Second and Subsequent Phases of Resolution}

The process could then be repeated with the large isolated majority group's
r\^ole eliminated from the cultural model, but its presence within the
system contained as a constraint on the distance measure. It would then be
possible to repeat the process with another series of cultural model
experiments until all of the r\^oles are clearly defined with respect to one
another.

\subsection{Comparison with Newsgroups Operation}

The operation of the newsgroup system was as follows:

\begin{itemize}
\item Membership of a group entitled its members to make postings to that
  group.
\item All subscribers belonged to an entry group.
\item Subscribers were assigned to the group that they contributed to most
  often.
\item Access Managers determined if a new subscriber could contribute to a
  group and therefore be a member of it.
\item An access manager was given his position within a group because he
  contributed most often to a group.
\item A finite number of interests were assumed to be expressed.
\item The length of the preference ordering across those interests
  determined how many different opinions would be allowed.
\end{itemize}

In the newsgroup example, only three interests were chosen to make
illustration easier and a preference ordering that was only two--ply
limited the number of varieties of newsgroup to 7 (of a total of 13). It
was seen that it could be easily extended to cover six interests, but the
preference ordering was limited in length to three ply, which meant, in
effect, that only 62 (of 4683) varieties of opinion were allowed.

The system was bootstrapped by allowing the subscribers to make some
initial statements of interest and they were then allocated to a
group. After that, they were moved from group to group and groups could
refer members to one another.

\paragraph{Access Rights for R\^oles}

Access rule generation is very similar using the collective choice expert
system to assign r\^oles and define the inter--r\^ole data views.

\begin{itemize}
\item The accessors will be placed in \textit{access r\^oles}. These r\^oles
  will be the interacting agents of the system. A r\^ole is entitled to claim
  certain views from other r\^oles.
\item All accessors will have an entry r\^ole{\myDash}unprivileged enquirer.
\item Accessors will be assigned to the r\^oles which would give them the
  most utility, \ie access to as much information as they can.
\item Interactions take place as custodians determine if an accessor can be
  assigned a r\^ole by determining what data view they may have.
\item Custodians will be appointed by the subjects of the views available
  in a given r\^ole{\myDash}they are not chosen from the accessors.
\item A finite number of ontologies will be allowed to be expressed. The
  number of plies for the preference ordering of the r\^oles that accessors
  have been assigned by the custodians must also be set.
\end{itemize}

The treatment of the accessors is identical to that of subscribers in the
newsgroups example. It is the appointment of custodians which is different.
Custodians do not have to be chosen from the accessors, they are already in
place.

A very important difference in the operation of the two systems must be
made clear: in the permissions policy system being a member of an access
group does not entitle the member to use all of the views the other members
of the access group have acquired, it makes it more probable.

\paragraph{An Example of the Bootstrapping Procedure}

If accessor $a$ makes an access request for a particular view of a database
then if the custodians $C$ of that database view grant the access request,
they collectively become the custodians of a r\^ole created uniquely for that
accessor. Call the access group $C(a)$ meaning custodians $C$ for $a$.

If accessor $a$ makes another access request for a different view, then the
custodians of that data, $D$, are also able to take into consideration
$C(a)$. With this, the custodians could make the following access rule: %

\begin{equation}
  \label{eq:dec-rule}
  x\ :\ C(x) \Rightarrow D(x)  
\end{equation}

\paragraph{Re-organisation}

There is however no explicit need for the custodians to make such a rule
explicit because it is already implicit in the system. When the r\^oles are
reorganized, r\^oles $C(a)$ and $D(a)$ would be merged, because they have no
other members, so it just an inconvenience to have them appear distinct,
but if another accessor were to be granted access to $C(a)$ but not to
$D(a)$ then the groups must be made distinct. This would also imply that
there is an ordering between the groups, $C(a) > D(a)$, meaning that
membership of $C(a)$ grants more access than $D(a)$.

To fully exploit the implicit rules, it must be possible for the system to
accurately compare the trustworthiness of the two accessors without having
to call upon the custodians repeatedly. In the newsgroups example, this was
achieved in a democratic way by having the access managers of the newsgroup
be chosen from the subscribers. In the permissions policy system, the
custodians are outside of the system, but are able to compare accessors. As
was explained in chapter \ref{cha:reqs}, each accessor would have acquired
certificates from other permission policy systems and these could be
compared. These other policy management systems would also be able to order
the r\^oles the accessors were members of.

It would thus be quantitatively possible to compare the amount of trust
that has been placed in two different accessors based on the r\^oles that
have assigned to them.

\paragraph{Large--scale behaviour}

The net effect of this procedure will be that accessors will be classified
accurately and will be expected to behave in a particular way and they can
be reasonably safely compared. This will make the generation of explicit
access rules much easier. Explicit access rules will usually be more
abstract but essentially of the same form as \eqref{eq:dec-rule}. The
explicit access rules would in fact be specified in the system's operation
and would take on the character of axioms of the system.

\paragraph{Safety}

It is the responsibility of the custodians to ensure that accessors do not
belong to any access groups which they should not ethically be members of.
The self--organising permissions system only simplifies the management of
access rights for accessors; it does not itself apply the principle of
least privilege. It should not contravene it though. It is therefore
important that some meta--data rules be specified which state a preference
ordering across views.

There is an implicit ordering across database views if they were classified
as in table \ref{tab:identity}. In more concrete terms, one could say that
a view that contains an indication of a subject's age is more confidential
than one that does not and, therefore, anyone who has access to such a view
has had more trust bestowed upon them that someone who does not.

\paragraph{Summary}

It should be apparent now how different permissions policy hierarchies
inter--work to generate implicit access rules. The custodians are able to
compare accessors by using the permission policy orderings generated in
other systems. This is the procedure that underlies much of modern business
where it is often required that companies have to prove creditworthiness to
one another by using bank statements and asset holdings. The procedures
described above simply apply this principle on less quantitative
information than money.

\section{Conclusions}

It has been seen in this chapter that self--organising systems can be
easily specified and designed and should display the useful
self--stabilising dynamics of the Axelrod cultural model of interaction. In
the following chapter, the permissions policy management system will be
discussed in relation to the other requirements given in this dissertation.

\chapter{Discussion, Future Work and Conclusion}

In this final chapter, the systems described in the preceding chapter will
be more critically assessed against the requirements. Future work to
validate the concepts discussed in this dissertation will be outlined and a
final conclusion on its usefulness will be given.

\section{Discussion}

It has already been made clear that the permissions policy system makes use
of other permissions policy systems preference hierarchies to be able to
compare accessors against one another. The broad goals of the requirements
chapter{\myDash}the duties of custodians to their subjects{\myDash}cannot
be directly met by this procedure but it does make it possible to detect
any infringements by the custodians since they too should conform to a norm
of behaviour which could be placed in a preference hierarchy.

An outstanding question is how safe are the decisions that are made. This
hinges upon the granularity of the grouping. If there are too many groups,
then it will not be possible to establish a reliable norm of behaviour from
a preference hierarchy based upon it. If there are two few groups, the
preference hierarchy will not be able to generate enough implicit rules and
to produce a useful organisation. It is worthwhile restating the key
relationship upon which self--organisation is based.

\begin{description}
\item[Ontologies] The number of ontologies is fixed at the first
  organisation of the accessors. In the newsgroup example, the number of
  interests (or meta--threads) was fixed at 3 because that made
  illustrations easier. It could easily be set to any arbitrary number.
\item[Variety] Once the number of ontologies has been fixed, one must then
  decide how much variety is allowed. For the newsgroup example, three
  ontologies could be combined in seven different ways. But for larger
  ontologies the number of combinations becomes very large. For the example
  of six newsgroups, the variety was limited by restricting the number of
  newsgroups expressed across a smaller playing field of choices. The
  number of which is given by \eqref{eq:axelrod:binom}. The assumption in
  this enumeration of choices is that they are grouped by varying degrees
  of indifference and this leads to a binary tree structure.
\end{description}

The underlying relationship between the topology of agents and the
complexity of the preference orderings they hold is the key to
understanding organisational phenomena.

In chapter \ref{cha:norms}, an observation of Axelrod's in
\cite{Axelrod:culture} led to a series of investigations of behaviour,
topology of proximity and system size. It was concluded that some
topologies gave rise to more stable and predictable behaviours.

In the discussion above, it was made clear that the only configuration
parameters for a self--organising permission system seemed to be the size
of the issue space and its internal connectedness. A spanning tree leads to
systems which can simplify complex issues very quickly. This is the basis
of the work carried out by Miller \etal \cite{Miller:evolving}, but it is
well--known from political theory that binary--tree systems can be easily
subverted \cite{scf:black,scf:farquharson}.

To make a system safe, it should not be possible to subvert it, but it is
well-known that no collective choice procedure can be fair (and therefore
safe) from Arrow's \textit{Impossibility Theorem}. Arrow gives a
possibility theorem, but it may now be possible to have a confidence level
for a collective choice procedure based upon the entropy of the preference
hierarchies. What is a significant level of entropy is difficult to decide,
but the analysis of evolving cultures shows that it may be the point where
the phase reverses between the variety and compatibility entropies of an
evolving population.

The design of the self--organising permissions policy system only showed
how much interconnectedness is needed to make useful decisions on a
quantitative basis about abstract concepts, such as ``trustworthiness''.

\section{Future Work}

There are four outstanding problems with the design of self--organising
preference aggregation systems:

\begin{enumerate}
\item Critical entropy for fair collective choice
\item Topologies for self--organising systems
\item Entropy measure for spectral analysis of votes
\item Cross--certification metrics
\end{enumerate}

The last of these has not been discussed at any length within this
dissertation, but it would be fundamental to a high--security system. It
requires that different preference hierarchy systems be able to
cross-validate one another. There is already a means whereby \textit{X.509}
certification authorities can validate one another, but they all have the
same degree of mutual trust. Recently, research by Maurer \cite{KohMau00}
suggests that this may be quantified reliably.

A prototype database management system has been presented in this
dissertation which could make use of \textit{X.509} certification, chapter
\ref{cha:simple}, and a practicable system for self--organisation of access
was proposed in \myRef{sec:sopps}. There should be further practical
investigation into the following types of system.

\begin{itemize}
\item Secure self--organising mailing list
\item Licencing, surety and insurance systems
\item Permissions policy system based on insurance
\end{itemize}

The self--organising mailing list has already been presented and it would
be a good proving ground for analysing some of the methods described in
this thesis. The cross--certification metrics would ultimately have to be
based on risk and the evaluation of risk demands that liability can be
limited.  This would require that licences and sureties be obtainable in
the same way as information, \ie electronically. Finally, a
self--organising permissions policy system could be implemented using risk
as the basis for grading accessors.

Finally, an interpretation of the principles proposed by Anderson
\cite{anderson96:secclinic} as economic goals should be made. The system
that Anderson proposes for resolving the access decision issues of a
collegiate healthcare sector is a widely applicable model of an economic
market. They are essentially \textit{fair}, which might imply they are
pareto--optimal. A pareto--optimal system cannot be said to exist by
Arrow's impossibility theorem, but Arrow's analytic choice system is based
upon a memoryless system. It may be the case that a pareto--optimal system
exists if a best--of--$N$ vote is allowed. There are some interesting games
theory scenarios that suggest this may be so
\cite{banks.camerer.ea:experimental-analysis}.

\section{Conclusion}

This thesis has looked at the problem of providing a secure environment for
collaborative computing. This reduced itself to the problem of providing
safe and secure access to databases. Secure access is already mature
technology, but safe access relies upon access control. This can only be
decided with foreknowledge of the information flow. The information flow
can only be determined by classifying all subjects and objects within the
system. Such an information system would be very large and it could not be
reliably managed without using some self--organising technology. This would
require an adaptive discretionary control mechanism.

The method proposed in this dissertation would be based on aggregating the
information flows specified by different subjects for the objects they
own. Such a system has been shown to be self--organising because it is
essentially a voter model: these have been analytically proven to be
ergodic for one--ply information flows and simulations indicate they are
also self--stabilising for an arbitrary number of plys, but take
exponential time to stabilise.

It is well worth continuing the investigation of organisational structure
using quantitative techniques based on the degree of information flow. This
is fundamentally an entropy measure where the degree of uncertainty about
an individual's behaviour is the random variable in a maximum likelihood
estimate.


\newcommand{\etalchar}[1]{$^{#1}$}
\hyphenation{ Kath-ryn Ker-n-i-ghan Krom-mes Lar-ra-bee Pat-rick Port-able
  Post-Script Pren-tice Rich-ard Richt-er Ro-bert Sha-mos Spring-er The-o-dore
  Uz-ga-lis }

\appendix

\chapter{Glossary}

\label{cha:glossary}

\begin{description}

\item[Pareto--optimal] An allocation of goods amongst consumers which
  possesses the property that should any one consumer change his
  consumption level, then the effect upon all other consumers is
  undifferentiated, they may all gain or may all lose, but, with respect to
  one another, they are unchanged. Pareto--optimality can be developed
  quantitatively \cite{balasko:pareto} and used as a measure of fairness.
  
\item[Possibility and Impossibility Theorems] These have been well--known
  in quantitative disciplines and are, respectively, the proof of the
  existence of a mathematical function and its non--existence. In this
  dissertation, the most important possibility and impossibility theorems
  are due to Arrow \cite{inst:arrow}: in which he shows that under a
  weak--ordering there is a ``fair'' means of choosing between two
  different policies, but there is no ``fair'' means of choosing between
  three or more policies. The former is a possibility theorem, the latter
  an impossibility one.
  
\item[Preference Relations] A partially ordered set, \textit{poset}, where
  the relations between entities are $>, <, =$ . Whether $=$, used as an
  indication of indifference, is permitted is a specification of the
  ordering of the poset. A poset is the mathematical structure best used to
  represent authorisation chains or lattices, which are fundamental to
  access control systems.
  
\item[Strong and Weak Ordering] A strong ordering does not permit
  indifference, only $a>b$, $b>a$ are valid expressions in a two--policy
  system; under a weak ordering $a>b, a<b, a=b$ are all valid expressions.
  A weak ordering is more expressive, less decisive and, because it
  introduces so many more permutations of preference, harder to analyse.
  
\item[Normality] In a statistical context: normality implies that the sum
  of a set of independent identically distributed random variables will
  have a aggregate distribution that will tend toward the Normal Gaussian
  distribution $N(0,1)${\myDash}when properly scaled. This is the basis of
  statistical quality control and \textit{least mean square} estimation. A
  least mean square estimation assumes normality.  Normality can usually be
  assumed when the mean and the standard deviation are related %
  as in the Tchebysheff inequalities\footnote{See \cite{stoch:papoulis},
    \textit{pp. 540}.}.
  
\item[Maximum Likelihood and Entropy] Maximum likelihood is a method of
  statistical analysis which can be used to fit data to frequency
  distributions where the sample mean is not \textit{normally} related to
  the population mean\footnote{See \cite{stoch:papoulis}, \textit{pp.
      535}.}.  It is consequently ``non--linear'' because it cannot make
  the assumption that the relation between the mean and the standard
  deviation that the Tchebysheff inequalities require\footnote{See
    \cite{stoch:papoulis}, \textit{pp. 540} and \textit{pp. 113}.
    Non--linearity is an artifact of the form of the moment--generating
    function's.}. Essentially, the underlying probability distributions are
  either not independent or not identical or both. It is possible to
  decompose many probabilility distributions based on posets into the sum
  of a set of uniformly distributed indicator probability
  distributions\cite{Swanson:entropy}.  This then allows one to assume that
  all constituent probability distribution functions are sufficiently
  similar that only interdependence cannot be disregarded.
  
\item[Maximum Likelihood Preference Relations] An aggregation of similarly
  formed{\myDash}weakly or strongly ordered{\myDash}preference relations of
  voters is said to be maximally likely if the hamming distances between
  each and every set of individual preference relations with respect to the
  aggregate is symmetrically distributed.  Preference relations of the
  individuals in a population of voters on issues are usually not formed
  independently from one another.  Consequently, the techniques of
  statistical estimation based on normality are not valid. Thompson and
  Remage \cite{stats:pairwise:thompson} developed a method for determining
  maximum likelihood preference relations.
  
\item[Entropy or negentropy] Originally developed as a statistical measure
  of the order within a system; most famously embodied by Schr\"odinger in
  his function giving the quantum energy available within an atomic
  nucleus. This statistical measure has since been applied fruitfully in
  many fields and is the basis of information theory developed by Shannon.
  The most pertinent applications of entropy for this dissertation are
  topological entropy \cite{math:top:ent} and maximum likelihood estimation
  \cite{stats:pairwise:thompson}. Within information system design, entropy
  can be thought of as the degree of confusion within the system. This
  manifests itself in an information system when it becomes less and less
  decisive \cite{Swanson:entropy}. Irrelevant alternative policies
  proliferate introducing instability: behaviour is not conventionally
  predictable{\myDash}it cannot be assumed to be statistically
  deterministic or markovian.
  
\item[Collective Choice Expert System] An information system designed to
  aid group decision--making. A collective choice expert system would have
  many of the same functional requirements as a conventional expert
  system{\myDash}the ability to provide a justification and determinacy
  under the same fact set, for example \cite{unfound:yet-1}. It would be
  developed as a cross--disciplinary study of decision theory
  \cite{scf:fishburn:hist}, information theory, games theory and
  information technology.
  
\item[Access Control Systems] There are variety of these: mandatory access
  control, lattic--based access control and discretionary access control
  mechanisms. All of these systems are modelled to comprise objects, which
  are accessed, and subjects, who access them. Each access attempt is
  mediated and a decision as to whether access should be granted is based
  upon a lookup in an access control matrix.
  \begin{description}
  \item[Mandatory] Under mandatory access control. Each object has a
    pre-defined access control matrix, each operation, (read, write,
    execute \etc) is a row, each subject has its own column, but subjects
    are defined in a linear hierarchy, so that the form of the matrix can
    be made upper--triangular. Objects can be similarly organised into an
    hierarchy.
  \item[Lattice--based] Is an extension of mandatory access control.  The
    subjects and objects are arranged in a tree hierarchy and the access
    control matrix used to determine access is formed from the least--upper
    bound of the subject and the object in the tree hierarchy.
  \item[Discretionary] This is the most flexible and the most common access
    control mechanism. Subjects are allowed to grant access rights to other
    subjects for objects they own. The access control matrix is constructed
    as the system operates.
  \end{description}
  
\item[Role--based Access Control Systems] These were formalised by Sandhu
  \cite{Sandhu98}. They differ from other access control mechanisms because
  they have a more sophisticated information model: rather than just have
  subjects and objects, r\^ole--based systems allow subjects to take on a
  number of r\^oles simultaneously within a session and a r\^ole is a
  defined interface onto an object. These constructs allow some of the
  vagaries of discretionary access models to be more precisely specified
  and managed.  In particular, it is possible to define a different r\^ole
  for a subject when she administers the access rights of objects she owns,
  when she uses an object of her own and when she creates and destroys
  them. Sandhu presents a number of r\^ole--based systems which provide
  mandatory access control, lattice--based and discretionary systems.
  
\item[Collegiate Structure] Used by Anderson \cite{Anderson:security} to
  describe the organisation of the healthcare sector with respect to the
  allocation of access rights. Rights allocated within one college for a
  particular r\^ole are respected in other colleges. This is achieved by
  mapping a r\^ole defined within one college to a different r\^ole within
  another. This requires a federal superstructure that defines the r\^oles
  and fairly administers them.  Anderson argues that although individual
  colleges may apply any access control scheme they wish, the federal
  system will be a discretionary one: it is argued in this dissertation
  that the discretionary access model used within the federation will be a
  r\^ole--based model with constraints: r\^oles defined within each will
  map to a set of abstract r\^oles within the federation.
  
\item[R\^ole--based access control with constraints] R\^oles provide a
  convenient means of implementing constraints which are
  context--sensitive. These constraints can be used within collegiate
  structures of access systems. For example, general practitioners are
  entitled to prescribe drugs to any member of the general public; but a
  receptionist working in a general practice, although able to access
  medical records for patients within her own practice may not access
  medical records for patients in another practice. The general
  practitioner r\^ole has the right to prescribe in any general
  practice{\myDash}an unconstrained permission; the receptionist r\^ole
  only has the right to inspect records in her own practice{\myDash}a
  constrained permission.
  
\item[Custodian, Subject, Accessor and Owner] R\^oles that are defined
  within the federal superstructure of a collegiate structure of
  organisations. These are abstract r\^oles which are used to define the
  relationships between members of different colleges. A preliminary
  information model is given in \myRef{sec:computational}. In the example
  of a general practitioner prescribing drugs and a receptionist accessing
  records: the general practitioner is acting as an owner, creating drug
  presciption records which are then associated with subjects who have
  custodiants. The receptionist of the general practitioner acts as an
  accessor, but is constrained to only those subjects' records the general
  practioner she works for is the custodian of.
  
\item[Ontologies] Within a collegiate access control structure, ontologies
  are the naming systems used to define objects and subjects and define
  what r\^oles subjects can take and objects can offer. An ontology must be
  respected within two colleges before r\^oles in each college can be
  aligned to a r\^ole with federally defined constraints.  In the general
  practitioner and receptionist example, the doctor's right to prescribe is
  federal, the receptionist's right to interrogate a record is only
  enforceable in her own college. An example of an ontology in a field
  outside of medicine might be the Library of Congress cataloguing system.
  Within medicine the classification of diseases by the World Health
  Organisation is such a universally recognised classification system.
  Other classifications such as those developed for hospital
  administration{\myDash}the HL7 attribute model{\myDash}will prove
  invaluable in establishing a common understanding of r\^oles across
  medical specialities.

\item[Federation] A concept similar to the collegiate structure proposed by
  Anderson, but used in the discussion of the design of information
  processing systems \cite{ODP:desc}.
  
\item[Institutional Design] A branch of quantitative political science that
  aims to produce a set of government institutions which are optimal for
  the operation of a federation of entities submitting to a collective
  will. The methods of institutional design can be used to produce
  constitutions which specify the institutions of government
  \cite{inst:knight}, whether they are collectively--managed (a cabinet) or
  solely--managed (a president's office). Other codes of practice that are
  considered are: federal and state inter--relationships; bi--camerality;
  requirements for holding office; rights of appointment and right of veto
  to an appointment; duration of office and sequencing of elections;
  requirements for rescinding an appointment . Modal logics, in particular
  deontics, have been used to analyse ways in which institutions interact
  \cite{brown.carno:third,ryu.lee:defeasible}.  Econometric analysis and
  adaptive systems have been developed which show that different
  institutions emerge given different conditions
  \cite{econ:tiebout:tiebout,Miller:tiebout}, but a consensus forms around
  meta--norms of what are considered irrelevant alternatives. In this
  dissertation, it is hoped that a properly designed set of deontic
  interaction protocols can be used to \textit{evolve} a fair federal
  superstructure for a collective access control system.
  
\item[Irrelevant Alternatives] In the theory of collective choice, this is
  one of the susceptibilities a collective choice function may suffer from.
  The problem is that the presence of a particular policy generates
  confusion over the ordering of issues. This is, it is argued, the
  principle \textit{confusing} factor in the operation of an information
  system and can be thought of as an increase in the disorder, the entropy,
  of the system.
  
\item[Entropy and Consenus] One of the difficulties in collective choice is
  that an aggregate social choice of $a=b$ may indicate either most
  individuals have stated $a=b$ or that many individuals have chosen $a>b$
  and an equal number $b>a$. In the former, where $a=b$ is the choice of
  many, there is a consensus and, in a statistical experiment of selecting
  one voter and comparing his choice to the social choice, the outcome is
  predictable, so entropy is low, the system is highly--ordered. In the
  latter, where an equal number have chosen $a>b$ and $b>a$, the outcome is
  not predictable, the system not highly ordered. Low entropy indicates a
  high degree of order, and therefore consensus, within a population.
  
\item[Colliding Particle Systems] A useful model for the operation of a
  complex system originally developed to solve problems in statistical
  thermodynamics. The model of the system is of interacting particles which
  exchange some quantity between themselves. Heat conduction is a
  well--documented area in which this model is used. Colliding particle
  systems can be either consonant: the particles collide and become more
  like one another; or dissonant: colliding and becoming more unlike one
  another. In consonant systems, entropy decreases, the system becomes more
  ordered and may stabilise; in dissonant systems, entropy increases and
  the system is unlikely to find a stable state that is not some kind of
  limit cycle. In this dissertation, the colliding particle model is used
  as the basis for the simulation of the evolution of cultures: cultures
  that evolve around different norms of behavior{\myDash}norms of behaviour
  for the use of medical records. Analysis of colliding particle systems is
  very difficult \cite{prob:voting:matloff}; consequently, \textit{Monte
    Carlo} simulations using cellular automata are often more insightful in
  real systems.
  
\item[Cellular Automata] Originally proposed by von Neumann
  \cite{vonNeumann:1966:TSR}, these computational entities are now part of
  the field of evolutionary computation, see for example \cite{ipd:chaos}.
  An intuitive description of the different methods of evolutionary
  computation can be found in \cite{ga:handbook}, a pr\'ecis of which
  follows: the problem a statistician faces, in curve--fitting to empirical
  data, is comparable to finding the highest point in Australia using only
  kangaroos \textit{(sic)}.
  \begin{description}
  \item[steepest ascent] requires that the statistician trains the
    kangaroos to climb the steepest obstacles the kangaroos find themselves
    presented with.
  \item[simulated annealing] asks the kangaroos to do the same, but under
    the influence of alcohol, the effects of which wear off gradually.
  \item[genetic algorithms] asks the kangaroos to find high places; they
    are expected to interbreed and migrate. The algorithm periodically
    culls all those kangaroos below a certain altitude.
  \item[cellular automata] Following this analogy, one could argue that the
    cellular automata method operates in the same way as a genetic
    algorithm, but, as well as interbreeding, the kangaroos cull one
    another, with success more likely for those who live uphill. In this
    dissertation, a cellular automata playing field is developed which
    measures the degree of homogenisation within a consonant colliding
    particle system{\myDash}of very simple kangaroos whose interaction
    protocols are defined within complete a modal logic.
  \end{description}
  
\item[Modal Logic and Kripke's semantics] A modal logic introduces to first
  order existential logic a degree of relative existence{\myDash}a
  statement of prepositonal logic shows that something exists, a statement
  in modal logic says that something exists exists somewhere. Modal logics
  are widely used within computer science. Stirling \cite{modal:stirling}
  has developed a temporal modal logic which can be used to determined if a
  state exists at a certain point in the execution of a program. A modal
  logic is the basis of one of the most useful methods of formal proof of
  authentication protocols \cite{GoNeYa90}, which concerns itself with the
  relative position of secrets at different points in the execution of a
  protocol. Recently, authorisation mechanisms have been based on modal
  semantics \cite{sec:dist,ponder:dulay}. The Kripke semantics
  \cite{Kripke:63} provide a more intuitive way of visualising modal
  systems that can be more easily applied to distributed systems: the
  semantics are of a connected graph, where a modal truth is immediately
  recognised as an existential truth possessed by a neighbour node in the
  graph. The two modal systems described in this dissertation are
  \textit{egoistic} and \textit{peer pressure}. Both of these have very
  precise Kripke representations and have been implemented, operated and
  evaluated in a common programming language.
  
\item[Egoistic interaction protocol] This protocol is used within a
  cultural playing field to form a consonant colliding particle system. A
  cellular automaton accretes a characteristic that one of its neighbours
  possesses only if a Hamming distance measure reports they are already
  relatively consonant. Egoistic systems converge rapidly; rarely fall into
  limit cycles; but their final state is not readily predictable from their
  initial one. Within modal logic they can be characterised as
  permissive. An egoistic interaction protocol allows each cellular automata to
  evolve a \textit{norm} of behaviour. Similar automata emerge and their
  norm of behaviour becomes more common.
  
\item[Peer Pressure] This is an alternative to an egoistic interaction
  protocol. A cellular automaton accretes a trait from a neighbour, only if
  another neighbour possesses the same characteristic. Peer pressure systems
  are slower to converge; often fall into limit cycles; but seem to be
  easier to predict than egoistic ones. Within modal logic they can be
  characterised as permissive, but a peer veto is respected. A peer 
  pressure protocol forces cellular automata to evolve a norm of behaviour
  in the context of a \textit{meta--norm} of behaviour enforced by their
  peers.
  
\item[Norms and Meta--norms] These are best explained with an example: the
  tragedy of the commons. In Britain, after the acts of enclosure, most
  villages were given a piece of land which was a common grazing area.
  There were no explicit regulations that limited the amount of grazing a
  particular farmer could do on the common land. The norm of behaviour that
  emerged was to graze as much as one could, because someone else was
  almost certainly going to do the same. Consequently, common grazing land
  was constantly overgrazed and quickly became a liability: the tragedy of
  the commons. If a punishment system is introduced, but there is no
  obligation to punish, then systems emerge which either punished a great
  deal or not at all, but were still unfair in the use of the common land.
  If a further punishment were introduced which allowed farmers who did not
  punish to be punished for not doing so, then and only then did a system
  emerge which achieved a degree of fairness in the use of the common land.
  Systems display one of three emergent properties \cite{Axelrod:1997}:
  \begin{itemize}
  \item Overgrazing with very little informing{\myDash}farmers mutually
    agree to not make use of the punishment system.
  \item Unfair grazing with much informing{\myDash}farmers who overgraze a
    lot punish farmers who overgraze a little.
  \item Fair grazing with little informing{\myDash}farmers seem to respect
    one another and do not overuse the punishment system{\myDash}this
    meta--norm emerged only when the secondary punishment was introduced.
  \end{itemize}
  In the cultural playing field model developed in this disseration, the
  meta--norm is enforced by the interaction protocol, which is either an
  egoistic one or one that is subject to peer pressure.

\end{description}


\chapter{Aidan Project Software System} \label{cha:aidan}

\section{\Aidan Project Software System} 

This appendix describes the software that was prototyped as part of the
\Aidan project. The project's goals appear at\cite{sft:aidan}. The software
used and developed is archived on CD-ROM. 

\paragraph{Software Documentation}

\begin{enumerate}

\item A Process is an instance of an application program.
  
\item If a something is a prototype, it is under development and may only
  have a skeleton implementation.

\item Classes, and objects, are marked in the following manner:

\begin{itemize}
\item Imported Style: when the implementation has been imported, it is a
  standard \Java or \Jigsaw object, or class for an object, then it is
  marked thus: \myImport{Database-Manager}.
\item Exported Style: when the implementation can be exported, it has been
  developed for the \Aidan project, \myExport{Database-Manager}.
\end{itemize}

This should help in finding documentation for the classes. There is no
documentation for exported classes.

\item Attributes are marked in the following manner: \myAttr{Age}. There
  are some properties that attributes can possess.
  \begin{description}
  \item[class] This type of attribute belongs to the class and there is
    only one instance of it for all objects instantiated from that class.
  \item[object] This type of attribute belongs to each object and each
    object has an instance of it.
  \item[persistent] This type of attribute can be \emph{either} class
    \emph{or} object and indicates that the attribute can store its changed 
    value on destruction of the object or unloading of the class.
  \end{description}

\item Environment variables are presented thus: \myEnv{CLASSPATH}, and can
  have mixed case.

\end{enumerate}

\section{Application Programs and Their Configurations}

\subsection{Web-server}

The web-server used was \Jigsaw \cite{sft:prod:jigsaw}.
On-line documentation is available. The following are outputs of the
configuration information obtained by printing the \Jigsaw frames.

\begin{enumerate}

\item Version Information, see figure \ref{fig:aidan_8}.

\begin{table}[htbp]
  \begin{center}
    \begin{tabular}[left]{|r|p{3 in}|}
      \hline
      Attributes of General & w3c.jigsaw.http.GeneralProp \\
      \hline
      identifier: & general \\
      w3c.jigsaw.server: & Jigsaw/1.0a5 \\
      w3c.jigsaw.checkSensitivity: & true \\
      w3c.jigsaw.root: & /user/eepg/eepgwde/ src/java/Jigsaw/Jigsaw \\
      w3c.jigsaw.host: & \\
      w3c.jigsaw.port: & \\
      w3c.jigsaw.root.store: & true \\
      w3c.jigsaw.root.name: & root \\
      w3c.jigsaw.publicMethods: &  \\
      w3c.jigsaw.trace: &  true \\
      w3c.jigsaw.docurl: &  /User/Reference \\
      \hline
    \end{tabular}
  \caption{Version Information}
  \label{fig:aidan_8}
  \end{center}
\end{table}

\item \Aidan Directory Resource, see figure \ref{fig:aidan_1}.

\begin{table}[htbp]
  \begin{center}
    \begin{tabular}[left]{|r|p{3 in}|}
      \hline
      Existing resources of Aidan & \\
      Parent: & root \\
      \hline
      w3c.jigsaw.resources.DirectoryResource: & \\
      Putable: & \\
      QueryByNames: & \\
      Reference: & \\
      ResourceAdder: & \\
      UserRepository: & \\
      eg-1.html: & \\
      ladies.html: & \\
      new.html: & \\
      tester.html: & \\
      \hline
    \end{tabular}
  \caption{\Aidan Directory Resources}
  \label{fig:aidan_1}
  \end{center}
\end{table}

\begin{itemize}
\item \myImport{Putable} is the directory containing the results of user
  enquiries. One directory for each user.
\item \myExport{QueryByNames} is the only database enquiry resource. It
  allows queries to be submitted to the database using surname and,
  optionally, a fore-name.
\item \myImport{Reference} is an unused directory resource.
\item \myExport{UserRepository} This is the database that contains the
  names of all the legitimate \Aidan users and control information about
  them. 
\item \myExport{ResourceAdder} is a prototype resource that is
  intended to allow different database resources to be added and generate a
  new \myExport{QueryByNames}.
\item The files suffixed HTML are test files and can be ignored.
\end{itemize}

\item \myExport{QueryByNames} Resource, see figure \ref{fig:aidan_2}.

\begin{table}[htbp]
  \begin{center}
    \begin{tabular}[left]{|r|p{3 in}|}
      \hline
      Attributes of QueryByNames & \\
      Parent: & Aidan \\
      \hline
      Resource url: & /Aidan/QueryByNames \\
      identifier: & QueryByNames \\
      quality: & 1.0 \\
      title: & \\
      content-language: & \\
      content-encoding: & \\
      content-type: & text/plain \\
      last-modified: & Sat Dec 05 17:38:00 GMT+0 1997 \\
      icon: & \\
      maxage: & \\
      filename: &  \\
      putable: & false \\
      override: & false \\
      convert-get: & true \\
      DatabaseURL: & jdbc:postgres95://h2ws-03.brunel.ac.uk/aidan;user=eepgwde \\
      View: & select * from patients\_view \\
      Owner: & eepgwde \\
      Table Caption Fields: & SURNAME FORENAMES \\
      Table Title Fields: & SURNAME FORENAMES DATE\_OF\_BIRTH \\
      \hline
    \end{tabular}
  \caption{Database Enquiry Resource}
  \label{fig:aidan_2}
  \end{center}
\end{table}

\begin{itemize}
\item \myAttr{aidan.jigsaw.AidanQuery} is the name of the class on the
  \myEnv{CLASSPATH} of the \Java compiler.
\item \myAttr{Resource url} is the location of the resource relative to the
  root of the HTTP server.
\item \myAttr{DatabaseURL} this is the resource locator for the database.
  This is a relatively rare form of URL and can be read as follows: use
  protocol ``JDBC'' with sub-protocol ``Postgres95'' on host
  ``h2ws-03.brunel.ac.uk'', use database ``aidan'' and operate under user
  ``eepgwde''. A password could also be passed.
\item \myAttr{View} This is the view that will be used by object, when
  instantiated.
\item \myAttr{Owner} Owner of this class.
\item \myAttr{Caption Fields, File Title Fields} These are used for
  formatting the output.
\item Other attributes are inherited from the superclass and are either
  generated automatically or can be left unspecified.
\end{itemize}

\item \myExport{UserRepository} Resource, see figure \ref{fig:aidan_3}.

\begin{table}[htbp]
  \begin{center}
    \begin{tabular}[left]{|r|p{3 in}|}
      \hline
      Attributes of UserRepository & \\
      Parent: & Aidan \\
      \hline
      Resource url: & /Aidan/UserRepository \\
      identifier: & UserRepository \\
      quality: & 1.0 \\
      title: & Database Resource Listing Aidan Users \\
      content-language: & en \\
      content-encoding: & \\
      content-type: & text/html \\
      last-modified: & Sat Dec 06 17:49:59 GMT+0 1997 \\
      icon: & \\
      maxage: & \\
      DatabaseURL: & jdbc:postgres95://h2ws-03.brunel.ac.uk/aidan;user=eepgwde \\
      Table: &  users \\
      Owner: & Walter.Eaves@brunel.ac.uk \\
      RealmDirectoryName: & Aidan \\
      UsersDirectoryName: & Putable \\
      Realm: & Aidan \\
      headers: & user-agent accept referer Authorization ChargeTo \\
      \hline
    \end{tabular}
  \caption{Database of Users}
  \label{fig:aidan_3}
  \end{center}
\end{table}

\begin{itemize}
\item \myAttr{aidan.jigsaw.UserRepository} is the name of the class on the
  \myEnv{CLASSPATH} of the \Java compiler.
\item \myAttr{Table} This the table containing the user information.
\item \myAttr{RealmDirectoryName} This is the name of directory that will
  contains the directory having the name \myAttr{UsersDirectoryName}.
\item \myAttr{UsersDirectoryName} This the name of the directory that the
  will contain users' directories.
\item \myAttr{headers} These are accessible but currently are not used.
\item Other attributes are as for \myExport{QueryByNames}.
\end{itemize}

\item \myImport{filter-0} Resource, see figure \ref{fig:aidan_5}. This is
  a class that, when instantiated, can be used to provide filter requests
  using the basic authentication method of HTTP.

\begin{table}[htbp]
  \begin{center}
    \begin{tabular}[left]{|r|p{3 in}|}
      \hline
      filter-0 & \\
      Parent: & root \\
      \hline
      w3c.jigsaw.resources.DirectoryResource: & \\
      Identifier: & \\
      methods: & \\
      realm: Aidan & \\
      shared-cachability: & false \\
      private-cachability & false \\
      public-cachability & false \\
      users: & \\
      groups: & main \\
      \hline
    \end{tabular}
  \caption{Authorisation Filter}
  \label{fig:aidan_5}
  \end{center}
\end{table}

\begin{itemize}
\item \myAttr{group} This is the name of a group within the
  \myAttr{realm}.
\item \myAttr{realm} This is the name of the super-group.
\item Other attributes are as not, as yet, relevant.
\end{itemize}

\item \myAttr{eepgwde} is an instance of the user class used by the
  \myImport{GenericAuthFilter}. Figure \ref{fig:aidan_7} just illustrates
  that user information can be edited remotely. Users are added to a
  realm, so the realm information for this user is known.

\begin{table}[htbp]
  \begin{center}
    \begin{tabular}[left]{|r|p{3 in}|}
      \hline
      Attributes of eepgwde & \\
      \hline
      identifier: & eepgwde \\
      email: & Walter.Eaves@brunel.ac.uk \\
      comments: & Researcher \\
      ipaddress: & \\
      password: & \*\*\*\*\*\*\*\* \\
      groups: & main \\
      \hline
    \end{tabular}
  \caption{A user account}
  \label{fig:aidan_7}
  \end{center}
\end{table}

\begin{itemize}
\item \myAttr{ipaddress} Users can be limited to logging on from a
  web-browser running on a particular machine \myDash or through a particular
  firewall proxy host.
\item \myAttr{groups} The groups to which the user belongs.
\item Other attributes are self-evident.
\end{itemize}

\end{enumerate}

\subsection{Database}

Databases are added in the same way as other resources.

\subsection{JDBC Driver}

JDBC drivers are added as resources as well.


\chapter{Relations} \label{cha:relations}

These definitions are mostly from \cite{maths:encyclo}, except the
some of the relation types which are from \cite{deontic:aqvist}.

\begin{definition}[Relation] A relation $\myR$ on a set $S$ is a set
  of ordered pairs of elements of $S$. If $(a,b) \in \myR$, one also
  says that $R$ holds for the ordered pair $(a,b)$ and sometimes one
  writes $a \myR b$.
\end{definition}

\begin{definition}[Support, Range, Domain]
  \begin{align}
    \support S & \eqdef \{ x \in S | (x,y) \in \myR
    \droptext{for at least one $y$ in} S \} \tag{Support} \\
    \range S & \eqdef \{ x \in S | (y,x) \in \myR
    \droptext{for at least one y in} S \}  \tag{Range} \\
    \domain R & \eqdef \support R \cup \range R \tag{Domain} \\
    \domain R & \subseteq S \notag
  \end{align}
\end{definition}

\begin{definition}[Simple Types of Relations] These are standard
  definitions of a relation $\myR$. The variables $x$, $y$ and $z$
  range over a non-empty set $\mathcal{W}$. The statement $x \myR y$
  means that the the tuple $(x,y)$ exists in the relation $\myR$. The
  symbols $\vee$ and $\wedge$ are logical symbols meaning disjunction
  and conjunction of the logical existence conditions (not of
  underlying sets). $\neg$ means non-existence. Order of operators is
  (highest) $\neg$, then $\myR$, then $\vee$ and $\wedge$.
  \begin{align}
    \forall x \exists y
    & (
      x \myR y
    ) \tag{Serial} \\
    \forall x,y,z
    & ( x \myR y \wedge y \myR z
      \Rightarrow x \myR z
    ) \tag{Transitive} \\
    \forall x,y,z
    & ( x \myR y \wedge y \myR z \wedge x \neq z 
      \Rightarrow x \myR z
    ) \tag{Weak Transitive} \\
    \forall x,y,z
    & ( 
      x \myR y \wedge x \myR z \Rightarrow y \myR z 
    ) \tag{Euclidean} \\
    \forall x
    & (
      x \myR x ) \tag{Reflexive} \\
    \forall x
    & (
      \neg ( x \myR x ) ) \tag{Irreflexive} \\
    \forall x,y
    & (
      x \myR y \Rightarrow y \myR y
    ) \tag{Almost Reflexive} \\
    \forall x,y
    & ( 
      x \myR y \Rightarrow y \myR x
    ) \tag{Symmetric} \\
    \forall x,y
    & ( 
      \neg ( x \myR y \Rightarrow y \myR x )
    ) \tag{Asymmetric} \\
    \forall x,y
    & ( 
      x \myR y \wedge y \myR x \Rightarrow x = y 
    ) \tag{Anti-symmetric} \\
    \forall x,y,z
    & (
      x \myR y  \Rightarrow  ( y \myR z \Rightarrow z \myR y )
    ) \tag{Almost Symmetric} \\
    \forall x,y,z \exists w
    & (
      x \myR y \wedge x \myR z \Rightarrow
      y \myR w \wedge z \myR w 
    ) \tag{Incestuous} \\
    \forall x,y
    & (
      x \neq y  \Rightarrow  ( x \myR y \vee y \myR x )
    ) \tag{Connected} \\
    \forall x,y,z
    & (
      x \myR z \wedge y \myR z \Rightarrow x = y
    ) \tag{Left Unique} \\
    \forall x,y,z
    & (
      x \myR y \wedge x \myR z \Rightarrow y = z
    ) \tag{Right Unique}
  \end{align}
  It might appear that, in effect, a Euclidean relation is the same as a 
  transitive. This not so. For a group of three, $X$, $Y$ and
  $Z$, with an initial topology that $X$ can see $Y$ and $Y$ can see
  $Z$, then for a transitive relation $X$ must be given a view of $Z$
  as well; for a Euclidean relation there no change is needed.

  If $X$ can see $Y$ and $Z$, then to be Euclidean, $Y$ must be able
  to see $Z$ \emph{and} $Z$ must be able to see $Y$. For a transitive
  relation no change would be needed.
\end{definition}

\begin{definition}[Equivalence Relation, Partition, Class] An
  \emph{equivalence relation} on a set $S$ is a relation that is
  reflexive, symmetric, transitive and has a support $S$ \myDash also just
  reflexive and Euclidean. An equivalence relation $R$ on $S$ induces a
  partition of $S$ into classes, which consist of those elements
  between which the relation holds. A \emph{partition} if a set $S$ is
  a family $\textbf{P}$ of non-empty subsets of $S$, called the
  classes of the partition, with the following two properties:
  \begin{enumerate}
  \item Any two distinct classes are disjoint
  \item Every element of $S$ line one class
  \end{enumerate}
\end{definition}

\begin{definition}[Quasi-, Partial and Total Ordering Relations] A relation
  $R$ on a set $S$ is called a quasi ordering  on $S$ if $R$ is
  reflexive and transitive; partial if it also anti-symmetric and, if
  $R$ is also connected, it is called a total or linear ordering.
\end{definition}

%
%
%
%
%

\chapter{Current Technology} \label{cha:tech}

\section{Distributed Processing}

\subsection{Computational Design Guidelines}

These are common design guidelines, which one should bear in mind when
considering how set of entities in a system interact. Protocols are
designed at a computational level, but usually must consider the
engineering and technological dimensions.

Any emphasised phrases, such as \emph{binding}, are defined in
\cite{ODP:presc}.

\paragraph{Domains (or Paradigms)} A client or a server is a computational
object that represents an end-user in a processing \emph{domain}. A
processing domain is determined by the \emph{resources} it uses. A
``costed'' domain\myDash %
access to objects is costed\myDash %
uses money; a
confidential domain uses keys\myDash %
access to objects is granted on presentation of a key. The computational
object is an abstract object that behaves in a certain way in one domain.
Where it is located is an engineering issue, how it is implemented is a
technological issue. The computational object has to resolve the
engineering and technological \emph{conflicts} in the computational domain
before performing a computation in its own domain.

\paragraph{Properties} This is the set of properties that clients and
servers are defined by.

\begin{description}
\item[Association] A binding\cite{ODP:desc} between client and
  server. At a high-level it may be something like an account number, which 
  both share. At lower-levels, cryptographic keys and session
  numbers. Goodness of maintaining associations means that the client or
  server can be relied upon to instantiate a binding without have to create 
  it again. The most important entities shared between clients and servers
  are names.
\item[State] A binding with the environment. It may include all previous
  associations if the object having this state is operating under a name
  that has existed before. The environment binding is usually the end-user
  (for clients) and is effectively the identity of the end-user in
  processing system domain. Goodness of maintaining state means that the
  clients or servers can be relied to have the same state from
  instantiation to instantiation in the domain. There is an initial state
  which exists prior to an object making any bindings. There is often a
  default state which usually forms the initial state.
\item[Interface] A means of accessing an object that may change the state
  of an object. Interfaces are expected to be invariant if the
  implementation changes.
\item[Implementation] A piece of technology that implements the expected
  behaviour of an interface.
\item[Role] If an object provides an interface, it assumes a role in a 
  system, which is a prescribed behaviour.
\item[Autonomy] An object that is autonomous is accountable for its
  behaviour to itself alone. An object that is not autonomous has at least
  one controller. It is generally desirable for that object to have just
  one controller\myDash %
  to avoid conflicts in policy.
\item[Relocation] If an entity relocates, it moves from one environment to
  another. This is rarely done while computations are in progress, but it
  may be demanded in some domains.
\end{description}

\paragraph{Clients} Observations about the nature of clients. Clients\dots

\begin{enumerate}
\item outnumber servers.
\item are technologically heterogeneous in implementation.
\item do not align implementations with one another.
\item do not align interfaces with servers or with one another.
\item are autonomous.
\item freely relocate.
\item are specific to a user not one role.
\item have at least two roles: user as personal role, user in professional
  role.
\item are bad at maintaining state.
\item are bad at maintaining associations.
\end{enumerate}

\paragraph{Servers} This is a set of observations, which one would hope a
server should comply with. Servers \dots

\begin{enumerate}
\item are less numerous than clients.
\item are technologically less heterogeneous
\item do align implementations with one another
\item must align interfaces with one another
\item perform a specific role.
\item are not autonomous.
\item are good at maintaining state.
\item are good at maintaining associations.
\end{enumerate}

\paragraph{Agents} Are essentially servers to clients and clients to
servers. The point of an agent is to group clients and couple them to
servers. Agents\dots

\begin{enumerate}
\item can be as numerous as clients.
\item are technologically less heterogeneous than clients.
\item have two roles: server to the client and client to a server.
\item are not autonomous.
\item are bad at maintaining state when acting as a client, because they
  track their clients, but are good at maintaining state for their
  clients.
\item are good at maintaining associations.
\end{enumerate}



\chapter{Graph--Theoretical Treatment of Preference Lattices}
\label{cha:graph} 

This appendix describes how preference lattices can be represented with
graphs. It follows the treatment given by Miller in
\cite{graph:voting:miller}. Only the descriptive parts of Miller's text is
used. His treatment is based on that given by Harary \etal
\cite{graph:harary}. It also refers to Christofides' compendium of
algorithmic methods \cite{graph:christo} and to Skiena's very practical
treatment \cite{graph:skiena} for use with \emph{Mathematica}
\cite{graph:wolfram}, which was used to test some of the methods employed.
Skiena's nomenclature is used for Miller's descriptions.

\section{Definitions}

\begin{definition}[A Digraph and its Components] A \emph{directed graph} or
  \emph{digraph} is a collection of ``points''\myDash %
  also nodes or
  vertices\myDash %
  and of ``directed lines''\myDash %
  also arcs or edges\myDash %
  between these points. A graph that allows a number of edges between
  vertices is a multi--graph and the directed form is a multi--digraph.
  \begin{description}
  \item[Vertices] $V$ is the set of vertices, $\{ x, y, \dots, z, t, \dots,
      w \}$.
  \item[Edges] $E$ is the set of edges, which are expressed as tuples
    between vertices: $\{ (x,y), \dots, (t,u), \dots, (t, w) \}$.
  \end{description}
\end{definition}

\begin{definition}[Domination] If there is a directed edge from vertex $x$
  to vertex $y$, this can be written: $x>y$, $x \rightarrow y$ or
  $\vec{xy}$; and one says that $x$ \emph{dominates} $y$.

  The set of all vertices that $x$ dominates is written $D(x)$.

  A vertex $x$ is \emph{undominated} if no other vertex dominates it.
\end{definition}

\begin{definition}[Paths, Cycles and others] Some common definitions:

  \begin{description}
  \item[Path] is a sequence of vertices and directed edges from vertex to
    another: $x \rightarrow y \rightarrow z \rightarrow \dots \rightarrow
    t$. The length of the path is also called its cardinality. A path of
    cardinality zero is a path from the vertex to itself and is also known
    as a self--loop.
  \item[Complete (or Hamiltonian) Path] includes all the vertices in a
    digraph.
  \item[Reachable] A vertex $y$ is \emph{reachable} from another vertex
    $x$, if there exists a path from $x$ to $y$. Associated with each
    vertex is a reachable set, the set of all vertices that can be reached
    from that vertex.
  \item[Source] A vertex $x$ is said to be a \emph{source} if every other
    vertex is reachable from $x$.
  \item[Sink] A vertex $x$ is said to be a \emph{sink} if every other
    vertex can reach $x$.
  \item[Cycle] A pair of paths are said to form a \emph{cycle} if, for a
    pair of vertices, $x$, $y$: $y$ is reachable from $x$ and $x$ is
    reachable from $y$.
  \item[Complete Cycle] is a cycle that visits all the vertices of the
    graph.
  \item[Semi--path] is a path if the graph were undirected. That is, some
    of the directed edges point the wrong way, \viz $x \rightarrow z
    \leftarrow v \rightarrow y$ is a semi--path from $x$ to $y$.
  \item[Semi--cycle] is a pair of semi--paths from one vertex to another
    and back.
  \end{description}
\end{definition}

\nb If there is a path from $x$ to $y$, $x$ only dominates $y$ if the path
has one edge, \ie the two nodes are adjacent.

\begin{definition}[Types of Digraph and their properties] Definitions of 
  common properties of graphs and digraphs:
  \begin{description}
  \item[Connected] a digraph is said to be \emph{connected} if there exists
    a semi--path between every pair of vertices.
  \item[Tree] a digraph is said to be a \emph{tree}, if it has one source
    and no semi--cycles or, equivalently, one vertex is undominated and
    every other is dominated exactly one.
  \item[Complete] a digraph is \emph{complete}, if every vertex is
    adjacent to every other.
  \item[Asymmetric] a digraph is \emph{asymmetric}, if for every pair of
    vertices, $x$ and $y$, if $x \rightarrow y$ then not $y \rightarrow
    x$.
  \item[Tournament] is a digraph that is both complete and asymmetric.
  \item[Transitive] a digraph is \emph{transitive} if, $x \rightarrow y$
    and $y \rightarrow z$ it is also true that $x \rightarrow z$.
  \item[Strong Ordering] If graph is transitive and has no cycles, it is a
    \emph{strong ordering}.
  \end{description}
\end{definition}

\begin{definition}[Indifference] In addition to specifying that one vertex
  dominates another, it is possible to specify that a pair of vertices
  do not dominate one another, by stating that one is jointly indifferent
  to them: $x = y$.

  Indifference is symmetric and a graph that is transitive, has no cycles
  is a \emph{weak ordering}.
\end{definition}

\begin{definition}[Partially Ordered Set or Partial Order] If a set of
  elements is in some way ordered, then the set is
  \emph{partially--ordered}.
  
  The set can be represented as a graph and may prove to be either strongly
  or weakly ordered
\end{definition}

\begin{definition}[Transitive Closure] The \emph{transitive closure} $C(G)$ 
  of a graph $G$ contains an arc $\{u,v\}$ whenever there is a directed
  path from $u$ to $v$ in $G$.
\end{definition}

Graphs can be simplified by condensing them. Usually, the condensation
takes sets of connected vertices and treats them as one vertex.

\begin{definition}[Condensed Graph] A condensed graph reduces a set of
  vertices to one ``super--vertex''.
\end{definition}

\begin{definition}[Independent Set] 
  An independent set of a graph is a subset of the vertices such that no
  two vertices in the subset represent an edge in the
  graph\footnote{\cite{graph:skiena} p. 242.}. The maximum independent set
  is a maximal set of vertices that meet this criterion.
\end{definition}

\section{Conditions and Deductions}

\begin{theorem} A digraph with no undominated point has a cycle.
\end{theorem}

\begin{theorem} Every tournament has at least one complete path and every
  tournament has an odd number of such paths.
\end{theorem}

\begin{theorem}[Decomposition of Partial Orders] For any partial order, the
  maximum size of the anti--chain equals the minimum number of chains which
  partition the elements of the partial--order\cite{graph:dil50}.
\end{theorem}

\section{Representations}

\subsection{Illustration}

\begin{figure}[htbp]
  \begin{center}
  \begin{center}
    \includegraphics{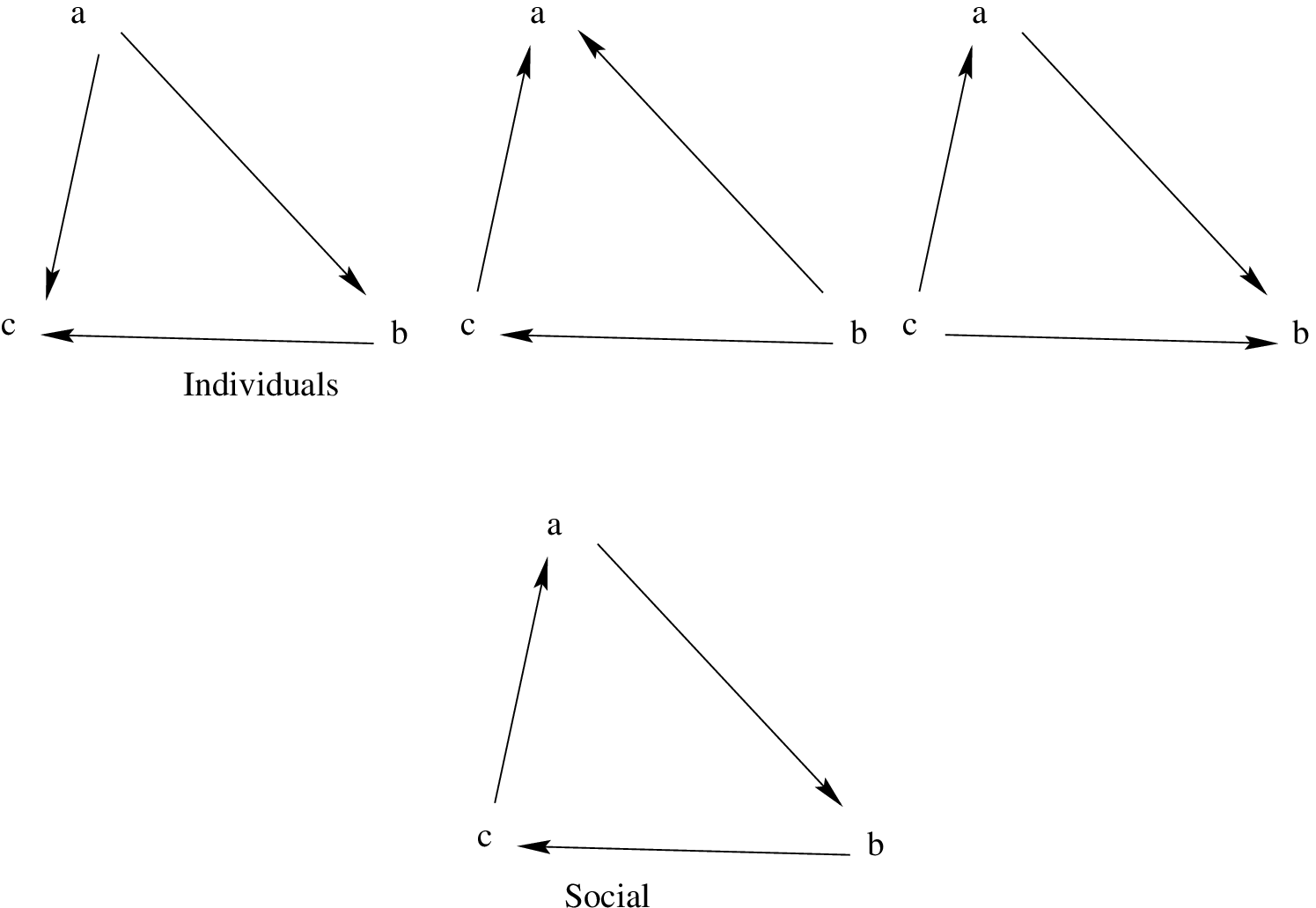}
    \caption{Preferences as Graphs}
  \end{center}
  \label{fig:choice:1}
  \end{center}
\end{figure}

A set of preferences, see table \ref{tab:choice:2} can be mapped to a
directed graph, see figure \ref{fig:choice:1}, if indifferent then no arc,
except for each policy with itself, \ie the central diagonal. The figure
contains three individual acyclic graphs which under simple majority rule
form a cyclic graph for the society. Each diagram is a transitive closure
of the preference relations.

\begin{table}[htbp]
  \begin{center}
    \begin{tabular}[left]{|l|l|}
      \hline
      Voter & Ranking \\
      A & $a>b>c$ \\
      B & $b>c>a$ \\
      C & $c>a>b$ \\
      \hline
    \end{tabular}
    \caption{Preferences}
    \label{tab:choice:2}
  \end{center}
\end{table}

\subsection{Adjacency Matrices}

Each one of the individual's graphs can be transformed to a matrix
representation using an adjacency matrix\footnote{\cite{graph:christo} 
 , p. 13}, which simply states if two vertices are connected by a directed
edge. For the graphs of figure \ref{fig:choice:1} these are:

\begin{equation*}
  \begin{aligned}[t]
    F_A = 
    \begin{pmatrix}
      1 & 1 & 1 \\
      0 & 1 & 1 \\
      0 & 0 & 1 
    \end{pmatrix}
  \end{aligned}
  \droptext{,}
  \begin{aligned}[t]
    F_B = 
    \begin{pmatrix}
      1 & 0 & 0 \\
      1 & 1 & 1 \\
      1 & 1 & 0 
    \end{pmatrix}
  \end{aligned}
  \droptext{,}
  \begin{aligned}[t]
    F_C = 
    \begin{pmatrix}
      1 & 1 & 0 \\
      0 & 1 & 0 \\
      1 & 1 & 1 
    \end{pmatrix}
  \end{aligned}
\end{equation*}

\subsection{Transition Matrices}

The system can be modelled using transition matrices. The preferences are
as before, table \ref{tab:choice:2}, but the direction of the edges of the
graph is to the state most preferred (\ie is reversed) and contains the
probability with which that transition takes place. For this simple choice
system, it is fairly clear what each voter's preferred position is, but
because of the transitivity, it is possible to go from the least--preferred
vertex to either of the two more preferred, unfortunately one has to make
a choice\myDash %
or one is describing a different system\myDash %
so the most preferred vertex is chosen as the one to transit to.

\begin{equation*}
  \begin{aligned}[t]
    F_A = 
    \begin{pmatrix}
      1 & 0 & 0 \\
      1 & 0 & 0 \\
      1 & 0 & 0 
    \end{pmatrix}
  \end{aligned}
  \droptext{,}
  \begin{aligned}[t]
    F_B = 
    \begin{pmatrix}
      0 & 1 & 0 \\
      0 & 1 & 0 \\
      0 & 1 & 0 
    \end{pmatrix}
  \end{aligned}
  \droptext{,}
  \begin{aligned}[t]
    F_C = 
    \begin{pmatrix}
      0 & 0 & 1 \\
      0 & 0 & 1 \\
      0 & 0 & 1 
    \end{pmatrix}
  \end{aligned}
\end{equation*}

\subsection{Reachability and Reaching Matrices}

\begin{definition}[Reachability Matrix]
  A \emph{reachability matrix}, $\textbf{R}$, is a square matrix having as
  many rows as there are vertices. The entry in the matrix at $r_{ij} = 1$
  if vertex $x_j$ is reachable from vertex
  $x_i$\footnote{\cite{graph:christo}, p.  18}.
\end{definition}

\begin{definition}[Reaching Matrix]
  A \emph{reaching matrix}, $\textbf{Q}$, is a square matrix having as many
  rows as there are vertices. The entry in the matrix at $q_{ij} = 1$ if
  vertex $x_j$ can reach vertex
  $x_i$\footnote{\cite{graph:christo}, p. 18}.
\end{definition}

\subsection{Chains and Anti--Chains}

(Note that Christofides uses the term chain for a semi--path.)

\begin{definition}[Chains and Anti--chains] These are complementary
  concepts\footnote{\cite[p. 243]{graph:skiena} for the definitions and the
    method of finding the largest anti--chain.} defined on a
  partially--ordered set:

  \begin{enumerate}
  \item Chain
    
    A chain in a partially ordered set is a set of elements $u_1, u_2,
    \dots, u_k$ such that $u_i$ is related to $u_{i+k},\ i<k$.
    
  \item Anti--Chain
    
    An anti--chain is a collection of elements no pair of which are
    related.
    
  \end{enumerate}
  
  Chains partition the elements of the partially--ordered set. There may be
  a number of anti--chains as well, but they do not partition the set
  unless the set is in some way degenerate. Each element of each
  anti--chain will be contained in one (and only one) chain.

\end{definition}


\chapter{Modal Logic} \label{cha:modal}

\section{Overview}
\label{sec:modal:ovw}

A modal logic can be used to express how truths are held between members of 
groups \cite{modal:chellas}. Modal logics have been studied extensively and
used for a number of purposes in computer science \cite{modal:stirling}.

\begin{remark}[Propositions] Modal logic usually concerns itself with
  propositional logic statements. These can be interpreted denoting the
  \emph{instantiations} of variables, \eg whether an entity 
  like $\cost(x, y)$ is held at all, and, if it is, whether it has the same
  value as that held elsewhere.
  
  When a set of variables has been instantiated, say $\cost(x, y) = 100$,
  then this can reduced to a statement in propositional logic, $A$, if need
  be using the following method. Firstly, a typing system needs to be
  established because the proposition involves a function; the function can 
  be written as having four parameters: the function signature, $\cost$,
  the result of the function call, $100$, then parameters, $f_1 = x, f_2 =
  y$. A logical proposition $A$ can then be defined as a statement in the
  type system of 2 argument functions: $A \eqdef \langle \cost, 100, x, y
  \rangle$; this in turn can be reduced to an untyped logical statement by
  stating: $\typeOf(A) = \text{two argument functions}, \text{signature} =
  \cost, \text{result} = 100, \text{arg1} = x, \text{arg2} =
  y$\footnote{This is a well--known technique in formal logic analysis
    which allows many seemingly more complex systems to be reduced to typed 
    argument comparisons, see Hodges contribution to
    \cite{philo:classic}}.
  
  Now $\neg A$ could mean any statement that is not $A$ in any way, for
  example: $\cost(y, x) = 100$ or even $\sin(a) = \frac{b}{4}$. For the
  purposes of this discussion it is best to think of $A$ as being a typed
  statement from a class of function results, something like $\cost(x, y)$,
  so that $\neg A$ would at least agree on the name of function, $\cost$,
  which arguments $x$ and $y$, but not the result. As will be seen, the
  point is to provide a language that can ensure that different entities
  have the same instantiations of variables.
\end{remark}

\begin{notation}[Modal Operators] These operators are:

  \begin{description}
  \item[$\possibility$] expressing the notion of \emph{possibly}.
  \item[$\necessity$] expressing the notion of \emph{necessarily}.
  \end{description}
  
  They can be explained, fairly informally, thus: an agent occupies a
  ``world of beliefs and values'' $x \in \mathcal{W}$ and can see other
  such worlds around him, $\mathcal{W'} \subseteq \mathcal{W}$. If the
  agent observes that the proposition $A$ is held in at least one world he
  observes, then $\possibility A$ is true and if it is held in all worlds
  he observes, then $\necessity A$.

  Only one of the operators $\possibility, \necessity$ need be primitive,
  it can then define the other, thus: $\possibility A \eqdef \neg
  \necessity \neg A$.
  
  These operators will be used to define inference rules and whether
  $\possibility$ or $\necessity$ is specified is a matter of choice for
  system designers, the notation $\isModal$ will be used to indicate this.
\end{notation}

\begin{notation}[Models] The concept of a model $\myModelName U$ and the
  $\models$ operator needs to be introduced. The details of its definition
  can be found in \cite{modal:chellas}. For now, these examples and their
  meanings should suffice:
  \begin{align}
    \begin{aligned}[t]
      \myModelFor & A
    \end{aligned} 
    \droptext{and} \quad
    \begin{aligned}[t]
      \myModelForByAt{\myModelName{U'}}{y} & A 
    \end{aligned}
    \label{eq:adaptive:model1}
    \\
    \begin{aligned}[t]
      \myModelValid & A
    \end{aligned} 
    \droptext{and}
    \begin{aligned}[t]
      \myClassValid & A
    \end{aligned} 
    \label{eq:adaptive:model2}
  \end{align}
  
  A model $\myModel$ contains a set of worlds $\mathcal{W}$ and a set of
  atomic propositions $\mathcal{P}$ distributed among the worlds. There can
  be different kinds of model and each model belongs to a class, such as
  $\myClassName{C}$, a class of models is differentiated from another class
  by the axioms that are in force.
  
  \begin{enumerate}
  \item \eqref{eq:adaptive:model1}

    \begin{enumerate}
    \item The first form means in model $\myModelName{U}$ in world $x$
      proposition $A$ is held.
    \item The second form means in model $\myModelName{U'}$ in world $y$
      proposition $A$ is held.
    \end{enumerate}
    
  \item \eqref{eq:adaptive:model2}

    \begin{enumerate}
    \item The first form means: is valid everywhere in this model. $A$ is
      valid in all worlds because of the distribution of the atomic truths.
    \item The second form means is valid in all models of this class. This
      means that $A$ is an axiom or theorem of the model and is not
      dependent on the distribution of truths.
    \end{enumerate}

  \end{enumerate}
\end{notation}

\nb The terminology of modal logic will be used, but bear in mind the
application of the modal logic in this context:

\begin{enumerate}
\item Agent or Organisation = World
\item Supra--organisation = Model
\item Behavioural type of supra--organisation = Class
\end{enumerate}

\section{Values and Beliefs}
\label{sec:modal:values}

\begin{notation}[Values and Beliefs] A proposition can be held in a number
  of ways in a modal logic:

  \begin{description}
  \item[Values] These are propositions that are local to a world, \eg
    $\myModelFor A$, $x$ believes $A$. These have no modal qualification in
    what is known as a normal form, \ie it is not possible to rewrite the
    truth as either $\possibility A$ or $\necessity A$.
  \item[Beliefs] These are based on how neighbouring worlds report their
    values. These can be written in their normal forms, which can, on
    reduction, be one of either $\myModelFor \possibility A$ or
    $\myModelFor \necessity A$.
  \item[Immutable Values] Propositions that are globally held. There are
    two forms of these:
    \begin{enumerate}
    \item $\myModelValid A$: propositions that are global to the model:
      these would be a set of atomic propositions which every world must
      possess and are immutable, and any theorems that are based solely
      on them.
    \item $\myClassValid A$: these are propositions and \emph{axioms} that
      are global to the class of models. There are just two propositions
      (the definitions of true and false), and a set of theorems, the
      axioms, that define the model, these may contain modal operators.
    \end{enumerate}
  \end{description}

  \begin{itemize}
  \item Values are private, in that only the world holding them knows
    whether they are true or false.
  \item Beliefs are jointly held by the worlds that observe the \emph{same}
    set of worlds.
  \end{itemize}
  
  \begin{figure}[htbp]
    \begin{center}
      \includegraphics[keepaspectratio=1,totalheight=4in]{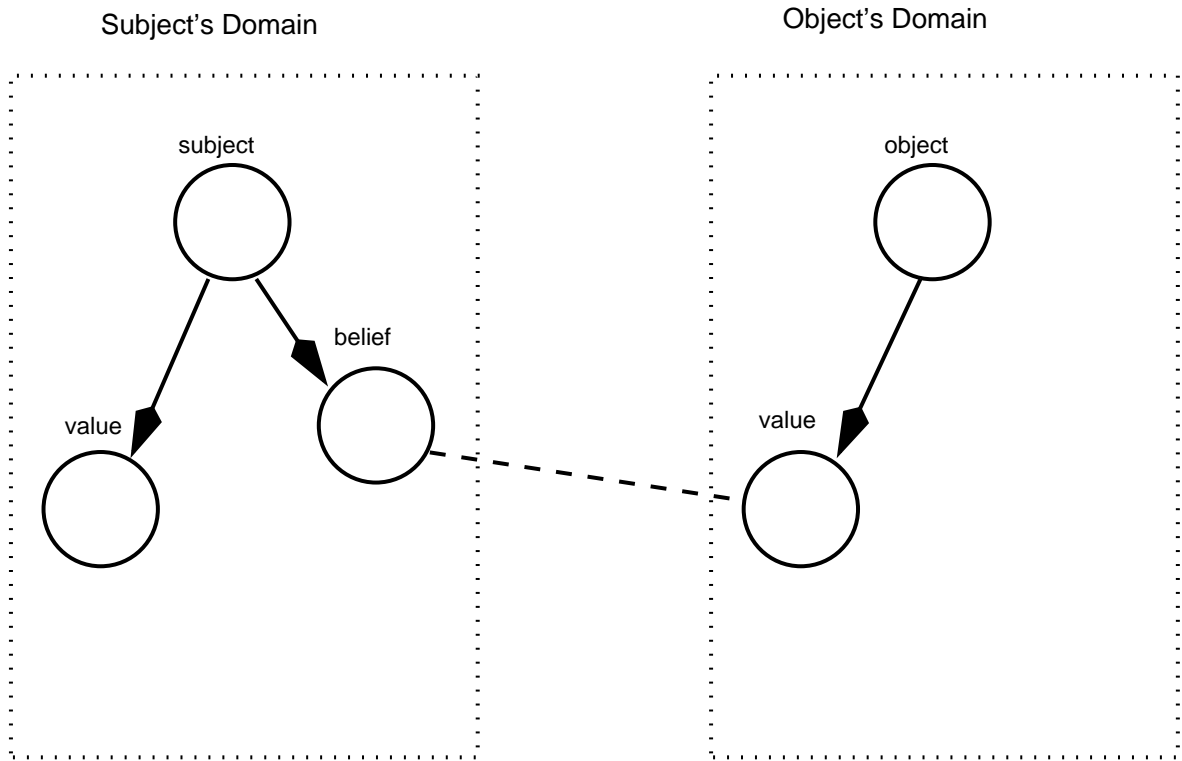}
      \caption{Relationship between a belief and a value}
      \label{fig:adaptive:beliefs-values}
    \end{center}
  \end{figure}

  One can also think of values and beliefs in terms of entity relationships
  between a subject and an object. ``A subject has a belief about the value 
  of an object'' is diagrammed in figure
  \ref{fig:adaptive:beliefs-values}: the subject owns a value; the object
  owns another value. The subject has a belief about the value that the
  object holds. The belief is in the subject's domain, but is associated
  with a value in the object's domain.

  A subject may never know the true nature of the value, but it can obtain
  corroborating beliefs from other sources, it can become more and more
  certain of the value the object holds. In figure
  \ref{fig:adaptive:beliefs-values-1}, the subject has obtained more
  beliefs\myDash %
  based on the beliefs of others\myDash %
  about the value the object holds.

  \begin{figure}[htbp]
    \begin{center}
      \includegraphics[keepaspectratio=1,totalheight=4in]{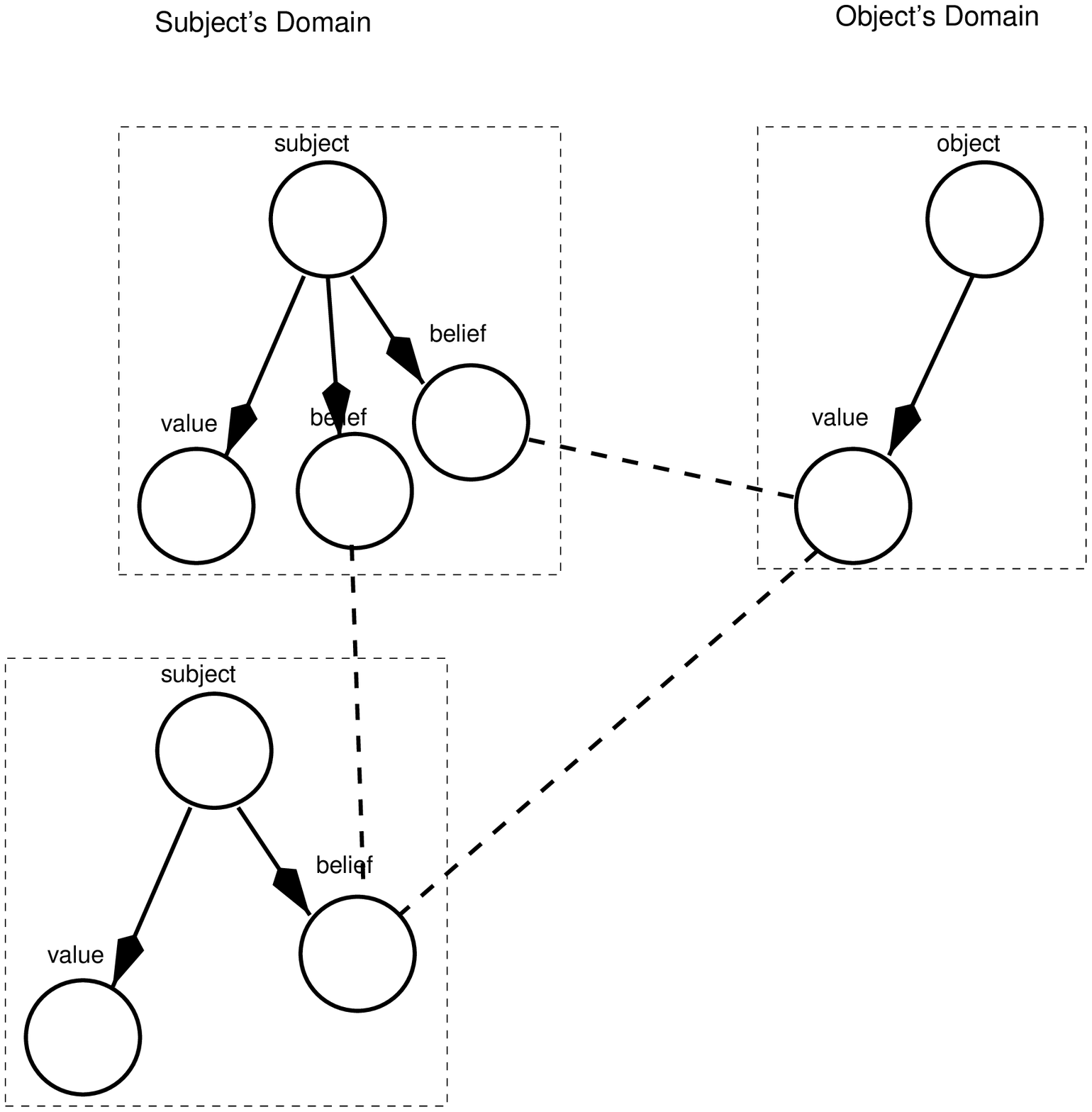}
      \caption{Corroborating beliefs about a value}
      \label{fig:adaptive:beliefs-values-1}
    \end{center}
  \end{figure}

\end{notation}

\begin{remark}[Observation and Lying] \label{rem:adaptive:observation}
  Worlds that observe others can only observe what the observed world
  admits to. In short, the observer poses a question and the observed
  answers it. There are two problems:

  \begin{enumerate}
  \item The observed world has no \textit{a--priori} value that it can
    return in answer to the question \label{sec:adaptive:observation}
  \item The observed world misrepresents itself and therefore:
    \begin{enumerate}
    \item Claims to have no value, in which case apply
      \ref{sec:adaptive:observation} above
    \item Negates the value it holds
    \end{enumerate}
  \end{enumerate}

  There is no way in which the observing world can ascertain the
  observation it makes truly reflects the values held by the observed
  world without obtaining some corroboration from elsewhere. This point
  will be addressed in more detail later, but for the time being one can
  assume that worlds report their truths truthfully.
\end{remark}

\section{Deontics}
\label{sec:modal:deontics}

Modal logic can be used to formalise laws of behaviour, in which case the
logics are part of the field of \emph{deontics}, see \cite{deontic:aqvist}
for a summary.

\begin{itemize}
\item Prohibition
\item Permission
\item Obligation
\end{itemize}

%
%
%
%
%


\chapter{KQML and KIF}
\label{cha:kqml} 

\section{Knowledge Query and Manipulation Language}
\label{sec:kqml:kqml} 

KQML is an unusual language because it provides services with a means of
interacting with one another that is similar to the way in which business
is conducted amongst business organisations: with KQML services may
advertise themselves, recommend one another and recruit one another. It 
also provides a high--level database access language, with primitives like
insert, delete and so forth. It can also be used for low--level data
streaming and redirection.

\paragraph{Performatives}

KQML is not a programming language or data access language, but a simple
set of primitives that are used more for their semantic connotation than
for their effect\myDash %
an invocation language. A list of the performatives
currently defined\myDash %
and their meaning\myDash %
is given in two tables: \ref{tbl:kqml:1} and \ref{tbl:kqml:2}. Some entries
refer to a VKB, which is a Virtual Knowledge Base, or a database of rules.

The performatives are invoked by a sender on a receiver. Performatives 
require parameters which are defined next.

\input{kqml-sum}

\paragraph{Parameters}

The parameters to KQML performatives are fixed, their meanings are given in
table \ref{tbl:kqml-1}.

\input{kqml-sum-1}

\section {Knowledge Interchange Format}
\label{sec:kqml:kif}

KIF provides a means whereby relationships and facts can be expressed as
meta-knowledge which can then be translated and passed to executive agents
- typically expert systems. Essentially KIF provides for: the
representation of knowledge \emph{about} the representation of knowledge;
representation of non-monotonic reasoning rules; the definition of objects,
functions and relations.

Its syntax is a cross between and Lisp and Prolog, but is used like Z, in
that it is only used to define entities. It has as its verbs all of the
logical and arithmetic operators.

\paragraph{Knowledge about the Representation of Knowledge}

KIF defines knowledge representation as lists.

\begin{verbatim}
    (defobject read-request := 
        (or (or (listof requestor target)) 
            (listof requestor requestor-location target)))
\end{verbatim}

Which defines an object that is a ``read-request'', which is either a list of
request-or identity and the target identity or the request-or identity, his
location and the target.

\paragraph{Non-monotonic Reasoning Rules}

Facts can be defined:

\begin{verbatim}
    (believes http-agent '(valid-user sybase))
    (believes http-agent '(non-interactive sybase))
    (believes http-agent '(read-access-only sybase))
\end{verbatim}

States that the HTTP-agent believes sybase is a valid user who will only
operate non-interactively and will only make use of read-only access. Such
facts can then be reasoned upon.

\begin{verbatim}
    (=> (believes http-agent ?p) (believes page-agent ?p))
\end{verbatim}

Which states that whatever the HTTP agent believes, the page agent
believes.

\paragraph{Definitions}

For example:

\begin{verbatim}
    (defrelation grant-right-to-read (?x ?o ?g) := 
        (and (member ?x ?g)
        (forall (?m) 
            (=> (member ?m ?g) (has-right-to-read (?m ?o))))))
\end{verbatim}

Defines a relation that grants to an entity ``x'' the right to read an object
``o'' if all of x's group are already able to read o. The relation
``has-right-to-read'' is defined elsewhere.


\end{document}

%% file: kqml-sum.tex
\begin{table}
\begin{center}
\raggedright
\begin{tabular}{|l|p{3.25in}|}
\hline 
Name & Meaning \\
\hline
achieve & S wants R to do/make something true in its environment \\
advertise & S is particularly suited to processing a performative \\
ask--about & S wants all relevant sentences in R's VKB \\
ask--all & S wants all of R's answers to a question \\
ask--if & S wants to know if the sentence is in R's VKB \\
ask--one & S wants one of R's answers to a question \\
break & S wants R to break an established pipe \\
broadcast & S wants R to send a preformative over all connections \\
broker--all & S wants R to collect all responses to a performative \\
broker--one & S wants R to get help in responding to a performative \\
deny & the embedded performative does not apply to S (anymore) \\
delete & S wants R to remove a specified sentence from its VKB \\
delete--all & S want R to remove all matching sentences from its VKB \\
delete--one & S wants R to remove one matching response from it VKB \\
discard & S will not want R's remaining responses to a previous performative \\
eos & end of stream of responses to an earlier query \\
error & S considers R's earlier message to be mal-formed \\
evaluate & S wants R to simplify the sentence \\
forward & S wants R to route a performative \\
generator & same as \emph{standby} of a \emph{stream--all} \\
\hline
\end{tabular}
\caption{KQML Performatives (A to G) for sender S and recipient R}
\label{tbl:kqml:1}
\end{center}
\end{table}

\begin{table}
  \begin{center}
    \raggedright
\begin{tabular}{|l|p{3.25in}|}
\hline 
Name & Meaning \\
\hline
insert & S asks R to add content to it VKB \\
monitor & S wants updates to R's response to a \emph{stream--all} \\
next & S wants R's next response to a previously mentioned performative \\
pipe & S wants R to route all further performative to another agent \\
ready & S is ready to respond to R's previously mentioned performative \\
recommend--all & S wants all names of agents who can respond to a
performative \\ 
recommend--one & S want the name of an agent who can respond to a
performative \\ 
recruit--all & S wants R to get all suitable agents to a performative \\
recruit--one & S wants R to get another agent to respond to a performative \\
register & S can deliver performatives to some named agent \\
reply & S communicates an expected reply to R \\
rest & S wants R's remaining responses to a previously-mentioned
performative \\ 
sorry & S cannot provide a more informative reply \\
standby & S wants R to be ready to respond to a performative \\
stream--about & multiple response version of \emph{ask--about} \\
stream--all & multiple response version of \emph{ask--all} \\
subscribe & S wants updates to R's response to a performative \\
tell & S admits to R that a particular sentence is in its VKB, usually 
issued in reply to an ask\\
transport--address & S associates a symbolic name with transport address \\
unadvertise & a \emph{deny} of an \emph{advertise} \\
unregister & a \emph{deny} of a \emph{register} \\
untell & S admits to R that a sentence is not in S's VKB, in response
to an ask \\
\hline
\end{tabular}
\caption{KQML Performatives (I to U) for sender S and recipient R}
\label{tbl:kqml:2}
\end{center}
\end{table}

%% file: kqml-sum-1.tex
\begin{table}
\begin{center}
\raggedright
\begin{tabular}{|l|p{3.25in}|}
\hline 
Name & Meaning \\
\hline
content & the information for which the performative expresses an attitude \\
force & whether the sender will ever deny the meaning of the performative \\
in-reply-to & the expected label in a reply \\
language & the name of the representation language of the \emph{content} parameter \\
ontology & the name of the ontology (e.g., set of term definitions) used in the \emph{content} parameter \\
receiver & the actual receiver of the performative \\
reply-with & whether the sender expects a reply and, if so, a label for the reply \\
sender & the actual sender of the performative \\
\hline
\end{tabular}
\caption{KQML Parameters for Performatives}
\label{tbl:kqml-1}
\end{center}
\end{table}